%% file: main.tex
\documentclass[9pt,twocolumn,twoside]{aap2018}
\usepackage{upgreek}
\usepackage{pdfpages}
\usepackage{authblk}

\usepackage{siunitx}
\usepackage{enumitem}

\usepackage{array}
\usepackage{makecell}
\usepackage{multirow}

\usepackage{wrapfig}
\usepackage{epsfig,epstopdf}
\usepackage{graphicx, color}
\usepackage{subcaption}
\usepackage[font={small,it}]{caption}

\templatetype{aap2018proceedings}

\begin{document}
\normalsize

% \begin{titlepage}\centering
% \vspace*{\fill}
% {\fontfamily{lmss}\bfseries\fontsize{22pt}{24pt}\selectfont Applied Antineutrino Physics 2018 Proceedings}
% \vspace{10mm}
% \includegraphics[width=0.75\textwidth]{2018-workshop-group.jpg}
% \vspace*{\fill}
% \end{titlepage}

\normalsize
\onecolumn
\title{Applied Antineutrino Physics 2018 Proceedings}

\author{}
\maketitle

\thispagestyle{firststyle}

 \vspace{15mm}
 
\begin{center}
 \includegraphics[width=0.75\textwidth]{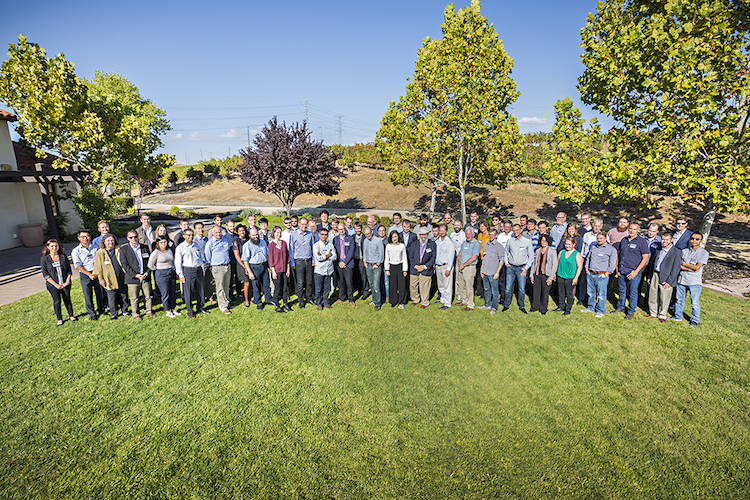}
\end{center}
\vspace*{15mm}

\dropcap{T}his was the 14$^{th}$ installment of Applied Antineutrino Physics (AAP) workshop series, and marked the second occasion the meeting has been held in Livermore, California. With more than 70 registered attendees, this iteration of the workshop generated great interest, speaking to the vitality and activity of this technical community. The program included many advances in the near and far field detection projects, new detection concepts, and discussion of flux predictions, the reactor anomalies, and antineutrino monitoring use cases. All presentations from the workshop are available at the following address \href{ https://neutrinos.llnl.gov/workshops/aap2018}{https://neutrinos.llnl.gov/workshops/aap2018}. 

We gratefully acknowledge support from \href{https://www.llnl.gov/}{Lawrence Livermore National Laboratory}\footnote{Lawrence Livermore National Laboratory is operated by Lawrence Livermore National Security, LLC, for the U.S. Department of Energy, National Nuclear Security Administration under Contract DE-AC52-07NA27344. LLNL-PROC-792757. Cover photo LLNL-PHOTO-760101.}, the  \href{https://nssc.berkeley.edu/}{Nuclear Science and Security Consortium}, and \href{https://www.hamamatsu.com}{Hamamatsu}.

The \href{http://spe.sysu.edu.cn/aap2019/}{AAP 2019 workshop} will be hosted by Sun Yat-sen University in Guangzhou, China during December of 2019. 

\clearpage
\begin{NoHyper}
\tableofcontents
\end{NoHyper}

\normalsize
\clearpage
\input{presentation-links.tex}
\twocolumn
%%%%%%%%%%%%%%%%%%%%%%%%%%%%%%%%%%%%%%%%%%%%%
% \clearpage
\makeatletter
\renewcommand{\AB@affillist}{}
\renewcommand{\AB@authlist}{}
\setcounter{authors}{0}
\makeatother
\clearpage
\normalsize
\addcontentsline{toc}{section}{The PROSPECT Experiment - H. P. Mumm}
\input{AAP-Mumm/aap2018-participant00x.tex}
% %%%%%%%%%%%%%%%%%%%%%%%%%%%%%%%%%%%%%%%%%%%%%
% \clearpage
\makeatletter
\renewcommand{\AB@affillist}{}
\renewcommand{\AB@authlist}{}
\setcounter{authors}{0}
\makeatother
% \clearpage
\normalsize
\graphicspath{{./AAP-Verstraeten/}}
\addcontentsline{toc}{section}{SoLid Neutrino Detector for Reactor Monitoring - 
M. Verstraeten}
\input{AAP-Verstraeten/aap2018-participant00x.tex}

% \clearpage
\makeatletter
\renewcommand{\AB@affillist}{}
\renewcommand{\AB@authlist}{}
\setcounter{authors}{0}
\makeatother
% \clearpage
\normalsize
\graphicspath{{./AAP-Park/}}
\addcontentsline{toc}{section}{Result of MiniCHANDLER - J. Park}
\input{AAP-Park/aap2018-participant00x.tex}

% \clearpage
\makeatletter
\renewcommand{\AB@affillist}{}
\renewcommand{\AB@authlist}{}
\setcounter{authors}{0}
\makeatother
% \clearpage
\normalsize
\addcontentsline{toc}{section}{Status of NEOS experiment - B. Han}
\input{AAP-Han/aap2018-participant00x.tex}

\clearpage
\makeatletter
\renewcommand{\AB@affillist}{}
\renewcommand{\AB@authlist}{}
\setcounter{authors}{0}
\makeatother
\clearpage
\normalsize
\graphicspath{{./AAP-Shitov/}}
\addcontentsline{toc}{section}{Status of the DANSS project - Y. Shitov}
\input{AAP-Shitov/aap2018-participant00x.tex}

\clearpage
\makeatletter
\renewcommand{\AB@affillist}{}
\renewcommand{\AB@authlist}{}
\setcounter{authors}{0}
\makeatother
\clearpage
\normalsize
\graphicspath{{./AAP-Serebrov/}}
\addcontentsline{toc}{section}{The first observation of effect of oscillation in
Neutrino-4 experiment on search for sterile
neutrino - A. P. Serebrov}
\input{AAP-Serebrov/aap2018-participant00x.tex}

\clearpage
\makeatletter
\renewcommand{\AB@affillist}{}
\renewcommand{\AB@authlist}{}
\setcounter{authors}{0}
\makeatother
\clearpage
\normalsize
\graphicspath{{./AAP-Onillon/}}
\addcontentsline{toc}{section}{Investigation of the ILL spectra normalization - A. Onillon}
\input{AAP-Onillon/aap2018-participant00x.tex}

\clearpage
\makeatletter
\renewcommand{\AB@affillist}{}
\renewcommand{\AB@authlist}{}
\setcounter{authors}{0}
\makeatother
\clearpage
\normalsize
\graphicspath{{./AAP-Karagiorgi/}}
\addcontentsline{toc}{section}{Overview and Status of Short Baseline Neutrino Anomalies - G. Karagiorgi}
\input{AAP-Karagiorgi/aap2018-participant00x.tex}

\clearpage
\makeatletter
\renewcommand{\AB@affillist}{}
\renewcommand{\AB@authlist}{}
\setcounter{authors}{0}
\makeatother
\clearpage
\normalsize
\graphicspath{{./AAP-Nakajima/}}
\addcontentsline{toc}{section}{Reactor neutrino monitor experiments in Japan - K. Nakajima}

\input{AAP-Nakajima/aap2018-participant00x.tex}

\clearpage
\makeatletter
\renewcommand{\AB@affillist}{}
\renewcommand{\AB@authlist}{}
\setcounter{authors}{0}
\makeatother
\clearpage
\normalsize
\graphicspath{{./AAP-Chimenti/}}
\addcontentsline{toc}{section}{Status of the Neutrinos Angra Experiment - P. Chimenti}
\input{AAP-Chimenti/aap2018-participant00x.tex}

\clearpage
\makeatletter
\renewcommand{\AB@affillist}{}
\renewcommand{\AB@authlist}{}
\setcounter{authors}{0}
\makeatother
\clearpage
\normalsize
\graphicspath{{./AAP-Coleman/}}
\addcontentsline{toc}{section}{The VIDARR $\bar{\nu}$-Detector - J. Coleman}
\input{AAP-Coleman/aap2018-participant00x.tex}

\clearpage
\makeatletter
\renewcommand{\AB@affillist}{}
\renewcommand{\AB@authlist}{}
\setcounter{authors}{0}
\makeatother
\clearpage
\normalsize
\graphicspath{{./AAP-Askins/}}
\addcontentsline{toc}{section}{Water Cherenkov Monitor for Antineutrinos
(WATCHMAN) - M. Askins}

\input{AAP-Askins/aap2018-participant00x.tex}

\clearpage
\makeatletter
\renewcommand{\AB@affillist}{}
\renewcommand{\AB@authlist}{}
\setcounter{authors}{0}
\makeatother
\clearpage
\normalsize
\graphicspath{{./AAP-MartiMagro/}}
\addcontentsline{toc}{section}{SuperK-Gd - L. Mart\'i-Magro}
\input{AAP-MartiMagro/aap2018-participant00x.tex}

\clearpage
\makeatletter
\renewcommand{\AB@affillist}{}
\renewcommand{\AB@authlist}{}
\setcounter{authors}{0}
\makeatother
\clearpage
\normalsize
\graphicspath{{./AAP-Hill/}}
\addcontentsline{toc}{section}{The Versatile Test Reactor Overview - T. Hill}
\input{AAP-Hill/aap2018-participant00x.tex}

\clearpage
\makeatletter
\renewcommand{\AB@affillist}{}
\renewcommand{\AB@authlist}{}
\setcounter{authors}{0}
\makeatother
\clearpage
\normalsize
\graphicspath{{./AAP-Carr/}}
\addcontentsline{toc}{section}{Nuclear explosion monitoring: Can neutrinos add
value to the global system? - R. Carr}
\input{AAP-Carr/aap2018-participant00x.tex}

\clearpage
\makeatletter
\renewcommand{\AB@affillist}{}
\renewcommand{\AB@authlist}{}
\setcounter{authors}{0}
\makeatother
\clearpage
\normalsize
\graphicspath{{./AAP-Johnston/}}
\addcontentsline{toc}{section}{Ricochet and Prospects for Probing New Physics
with Coherent Elastic Neutrino Nucleus
Scattering - J. Johnston}
\input{AAP-Johnston/aap2018-participant00x.tex}

\clearpage
\makeatletter
\renewcommand{\AB@affillist}{}
\renewcommand{\AB@authlist}{}
\setcounter{authors}{0}
\makeatother
\clearpage
\normalsize
\graphicspath{{./AAP-Mabe/}}
\addcontentsline{toc}{section}{Plastic Scintillator Development at LLNL - A. N. Mabe}
\input{AAP-Mabe/aap2018-participant00x.tex}

\clearpage
\makeatletter
\renewcommand{\AB@affillist}{}
\renewcommand{\AB@authlist}{}
\setcounter{authors}{0}
\makeatother
\clearpage
\normalsize
\graphicspath{{./AAP-Yeh/}}
\addcontentsline{toc}{section}{BNL Material Development - M. Yeh}
\input{AAP-Yeh/aap2018-participant00x.tex}

\clearpage
\makeatletter
\renewcommand{\AB@affillist}{}
\renewcommand{\AB@authlist}{}
\setcounter{authors}{0}
\makeatother
\clearpage
\normalsize
\graphicspath{{./AAP-OrebiGann/}}
\addcontentsline{toc}{section}{Large-Scale Water-Based Liquid Scintillator
Detector R\&D - G. D. Orebi Gann}
\input{AAP-OrebiGann/aap2018-participant00x.tex}

\clearpage
\makeatletter
\renewcommand{\AB@affillist}{}
\renewcommand{\AB@authlist}{}
\setcounter{authors}{0}
\makeatother
\clearpage
\normalsize
\graphicspath{{./AAP-Mendenhall/}}
\addcontentsline{toc}{section}{Near-surface backgrounds for ton-scale IBD
detectors - M. P. Mendenhall}
\input{AAP-Mendenhall/aap2018-participant00x.tex}

\clearpage
\makeatletter
\renewcommand{\AB@affillist}{}
\renewcommand{\AB@authlist}{}
\setcounter{authors}{0}
\makeatother
\clearpage
\normalsize
\graphicspath{{./AAP-Mulmule/}}
\addcontentsline{toc}{section}{Exploring anti-neutrino event selection and
background reduction techniques for ISMRAN - D. Mulmule}
\input{AAP-Mulmule/aap2018-participant00x.tex}
% Need to change the references because writer did not follow instructions

\clearpage
%\tableofcontents
\makeatletter
\renewcommand{\AB@affillist}{}
\renewcommand{\AB@authlist}{}
\setcounter{authors}{0}
% \makeatother
\normalsize
\graphicspath{{./AAP-Danielson/}}
\addcontentsline{toc}{section}{Directional Detection of Antineutrinos - D. L. Danielson}

\input{AAP-Danielson/aap2018-participant00x.tex}

\clearpage
\makeatletter
\renewcommand{\AB@affillist}{}
\renewcommand{\AB@authlist}{}
\setcounter{authors}{0}
\makeatother
\clearpage
\normalsize
\addcontentsline{toc}{section}{Truthiness and Neutrinos; A Discussion of
scientific truth in relation to neutrinos and their
applications - J. Learned}
\onecolumn
\input{AAP-JohnLearnedTalk/talk.tex} % Need to change the references because writer did not follow instructions

\clearpage

\leadauthor{References} 
\section*{References}
 
% \bibliography{bib}

\end{document}

%% file: presentation-links.tex
% \documentclass[9pt,lineno]{aap2018}

% \templatetype{aap2018proceedings} % Choose template 

\section*{Presentations at AAP 2018}

% \author{}
% \maketitle

\thispagestyle{firststyle}

\vspace{2mm}
\noindent{\bf Day 1}

\href{https://neutrinos.llnl.gov/content/assets/docs/workshops/2018/AAP2018-Intro-Lund.pdf}{Introduction: Antineutrino Engineering Approaches. J. Lund (SNL, Ret.)}. 

\href{https://neutrinos.llnl.gov/content/assets/docs/workshops/2018/AAP2018-PROSPECT-Mumm.pdf}{PROSPECT: a Precision Oscillation and Spectrum Experiment. H.P. Mumm (NIST)}. 

\href{https://neutrinos.llnl.gov/content/assets/docs/workshops/2018/AAP2018-Stereo-Halmazan.pdf}{The STEREO Experiment. H. Almazan (MPIK)}. 

\href{https://neutrinos.llnl.gov/content/assets/docs/workshops/2018/AAP2018-solid-Maja.pdf}{SoLid Reactor Neutrino Detector. M. Verstraeten (University of Antwerp)}. 

\href{https://neutrinos.llnl.gov/content/assets/docs/workshops/2018/AAP2018-CHANDLER-JaewonPark.pdf}{MiniCHANDLER Result. J. Park (Virginia Tech)}. 

\href{https://neutrinos.llnl.gov/content/assets/docs/workshops/2018/AAP2018-NEOS_Han.pdf}{Status of NEOS II. B. Han (KAERI)}. 

\href{https://neutrinos.llnl.gov/content/assets/docs/workshops/2018/AAP2018-DANSS-Shitov.pdf}{Status of the DANSS Experiment. Y. Shitov (JINR)}. 

% \href{https://neutrinos.llnl.gov/content/assets/docs/workshops/2018/AAP2018-Neutrino4_Serebrov.pdf}{The first observation of effect of oscillation in Neutrino-4 experiment. A. Serebrov (NRC KI PNPI)}. 

\href{https://neutrinos.llnl.gov/content/assets/docs/workshops/2018/AAP2018-NuclearTheory-Hayes.pdf}{Reactor Antineutrino Spectra. A. Hayes (LANL)}. 

\href{https://neutrinos.llnl.gov/content/assets/docs/workshops/2018/AAP2018-NuclearData-Sonzogni.pdf}{Reactor Antineutrino Flux Predictions - Nuclear Data. A. Sonzogni (BNL)}. 

\href{https://neutrinos.llnl.gov/content/assets/docs/workshops/2018/AAP2018-ILL-spectra-normalization-Onillon.pdf}{Investigation of the ILL Spectra Normalization. A. Onillon (CEA)}. 

\href{https://neutrinos.llnl.gov/content/assets/docs/workshops/2018/AAP2018-SBL-anomalies-Karagiorgi.pdf}{Overview and Status of Short Baseline Neutrino Anomalies. G. Karagiorgi (Columbia)}. 

\href{https://neutrinos.llnl.gov/content/assets/docs/workshops/2018/AAP2018-Japan_Nakajima.pdf}{Reactor Neutrino Monitor
Experiments in Japan. K. Nakajima (University of Fukui)}.

\href{https://neutrinos.llnl.gov/content/assets/docs/workshops/2018/AAP2018-Angra-Chimenti.pdf}{Status of the Neutrinos Angra Experiment. P. Chimenti (Universidade Estadual de Londrina)}.

\href{https://neutrinos.llnl.gov/content/assets/docs/workshops/2018/AAP2018-Vidarr-Coleman.pdf}{VIDARR. J. Coleman (University of Liverpool).}

\href{https://neutrinos.llnl.gov/content/assets/docs/workshops/2018/AAP2018-WATCHMAN-Askins.pdf}{AIT-WATCHMAN. M. Askins (LBNL and UCB)}

\href{https://neutrinos.llnl.gov/content/assets/docs/workshops/2018/AAP2018-JUNO-Wang.pdf}{The Design of JUNO and it's Current Status. W. Wang (Sun Yat-sen University).}

\href{https://neutrinos.llnl.gov/content/assets/docs/workshops/2018/AAP2018-SuperK-Gd-Marti.pdf}{SuperK-Gd. L. Marti-Magro (ICRR).}

\vspace{2mm}
\noindent{\bf Day 2}

\href{https://neutrinos.llnl.gov/content/assets/docs/workshops/2018/AAP2018-IAEA-Anzelon.pdf}{Safeguards Policy Overview. G. Anzelon (LLNL).}

\href{https://neutrinos.llnl.gov/content/assets/docs/workshops/2018/AAP2018-UseCase-Huber.pdf}{Antineutrino Detection Use Case Overview. P. Huber (Virginia Tech).}

\href{https://neutrinos.llnl.gov/content/assets/docs/workshops/2018/AAP2018-VTR-Overview-Hill.pdf}{Versatile Test Reactor (VTR) Overview. T. Hill (ISU/INL).}

\href{https://neutrinos.llnl.gov/content/assets/docs/workshops/2018/AAP2018-ExplosionConventional-Foxe.pdf}{ 	Nuclear Explosion Monitoring: An overview of the global monitoring system. 	M. Foxe (PNNL).}

\href{https://neutrinos.llnl.gov/content/assets/docs/workshops/2018/AAP2018-ExplosionAntineutrino-Carr.pdf}{ 	Explosion monitoring: What can neutrinos add to the global system? R. Carr (MIT).}

\href{https://neutrinos.llnl.gov/content/assets/docs/workshops/2018/AAP2018-COHERENT-CabreraPalmer.pdf}{COHERENT. B. Cabrera-Palmer (SNL).}

\href{https://neutrinos.llnl.gov/content/assets/docs/workshops/2018/AAP2018-CONUS-JHakenmueller.pdf}{CONUS. 	J. Hakenm\"uller (Max-Planck-Institut).}

\href{https://neutrinos.llnl.gov/content/assets/docs/workshops/2018/AAP2018-NobleLiquid-Xu.pdf}{Prospects for reactor monitoring using noble liquid detectors. J. Xu (LLNL).}

\href{https://neutrinos.llnl.gov/content/assets/docs/workshops/2018/AAP2018-CONNIE-Estrada.pdf}{CONNIE. J. Estrada (Fermilab).}

\href{https://neutrinos.llnl.gov/content/assets/docs/workshops/2018/AAP2018-Ricochet-Johnston.pdf}{Ricochet. J. Johnston (MIT).}

%A representative from the China Dual Phase Argon Working Group was not able to attend but provided the following slides on the status of that effort:\href{https://neutrinos.llnl.gov/content/assets/docs/workshops/2018/AAP2018-NobleLiquidChina-Han.pdf}{ The Development of Low Threshold Dual Phase Argon Detector for CE$\nu$NS.}

\href{https://neutrinos.llnl.gov/content/assets/docs/workshops/2018/AAP2018-NuLat-Learned.pdf}{NuLat. J. Learned (University of Hawaii).}

\href{https://neutrinos.llnl.gov/content/assets/docs/workshops/2018/AAP2018-LLNLMaterials-Mabe.pdf}{LLNL Materials Development. A. Mabe (LLNL).}

\href{https://neutrinos.llnl.gov/content/assets/docs/workshops/2018/AAP2018-BNLMaterials-Yeh.pdf}{BNL Materials Development. M. Yeh (BNL).}

\href{https://neutrinos.llnl.gov/content/assets/docs/workshops/2018/AAP2018-WbLSRD-OrebiGann.pdf}{Large-scale WbLS Detector R\&D. G. Orebi Gann (UC Berkeley and LBNL).}

\href{https://neutrinos.llnl.gov/content/assets/docs/workshops/2018/AAP2018-Background-Mendenhall.pdf}{On-Surface Background Studies. M. Mendenhall (LLNL).}

\href{https://neutrinos.llnl.gov/content/assets/docs/workshops/2018/AAP2018-Directionality-Danielson.pdf}{Antineutrino Directionality R\&D. Daine Danielson (LANL).}

\href{https://neutrinos.llnl.gov/content/assets/docs/workshops/2018/AAP2018-DistributedImaging-Gratta.pdf}{Distributed Imaging for Liquid Scintillation Detectors. G. Gratta. (Standford University)}

\href{https://neutrinos.llnl.gov/content/assets/docs/workshops/2018/AAP2018-SVSC-Brown.pdf}{SVSC: Development Towards a Compact Neutron Imager. J. Brown (SNL). }

\href{https://neutrinos.llnl.gov/content/assets/docs/workshops/2018/AAP2018-solar-dm2-Seo.pdf}{Constraining on the Solar $\Delta m^{2}$ using Daya Bay and RENO. S. Seo (IBS).}

\href{https://neutrinos.llnl.gov/content/assets/docs/workshops/2018/AAP2018-Summary-Maricic.pdf}{Meeting Summary. J. Maricic (University of Hawaii).}

\href{https://neutrinos.llnl.gov/content/assets/docs/workshops/2018/AAP2018-nextAAP2019.pdf}{Next AAP Workshop. W. Wang (Sun Yat-sen University).}

\vspace{2mm}
\noindent{\bf Material provided to AAP from participants unable to attend.}

\href{https://neutrinos.llnl.gov/content/assets/docs/workshops/2018/AAP2018-Neutrino4_Serebrov.pdf}{The first observation of effect of oscillation in Neutrino-4 experiment. A. Serebrov (NRC KI PNPI)}. 

%A representative from RED100 was unable to attend, but provided the following recent references: \href{}{Status of the RED-100 experiment}, \href{https://doi.org/10.1088/1748-0221/12/06/C06018}{The RED-100 two-phase emission detector}, \href{https://doi.org/10.1134/S1063785018070179}{Synthesis of Titanium Nanoparticles in Liquid Xenon by a High-Voltage Discharge}, \href{https://neutrinos.llnl.gov/content/assets/docs/workshops/2018/AAP2018-RED100-Poster.pdf}{The RED-100 experiment on CE$\nu$NS study}.

 %A representative from the China Dual Phase Argon Working Group was not able to attend but provided the following slides on the status of that effort: 
 \href{https://neutrinos.llnl.gov/content/assets/docs/workshops/2018/AAP2018-NobleLiquidChina-Han.pdf}{The Development of Low Threshold Dual Phase Argon Detector for CE$\nu$NS. China Dual Phase Argon Working Group}.
 
% \cleardoublepage

%% file: AAP-Mumm/aap2018-participant00x.tex
\graphicspath{{./AAP-Mumm/}}

% \documentclass[9pt,twocolumn,twoside,lineno]{aap2018}

% \templatetype{aap2018proceedings} % Choose template 
\newcommand{\ORNL}{Oak Ridge National Laboratory (ORNL) \renewcommand{\ORNL}{ORNL}}
\newcommand{\HFIR}{High Flux Isotope Reactor (HFIR) \renewcommand{\HFIR}{HFIR}}
\newcommand{\nue}{\ensuremath{\nu_{e}}}
\newcommand{\numu}{\ensuremath{\nu_{\mu}}}
\newcommand{\nuebarMUMM}{\ensuremath{\overline{\nu}_{e}}}
\newcommand{\gMUMM}{\gamma}

\title{The PROSPECT Experiment}

% Use letters for affiliations, numbers to show equal authorship (if applicable) and to indicate the corresponding author
\author[a,2]{H. Pieter Mumm on behalf of the PROSPECT Collaboration}

\affil[a]{National Institute of Standards and Technology}

% Please give the surname of the lead author for the running footer
\leadauthor{Mumm}

% Please include corresponding author, author contribution and author declaration information
% \authorcontributions{Please provide details of author contributions here.}
% \authordeclaration{Please declare any conflict of interest here.}
% \equalauthors{\textsuperscript{1}A.O.(Author One) and A.T. (Author Two) contributed equally to this work (remove if not applicable).}
\correspondingauthor{\textsuperscript{2}E-mail: prospect.collaboration@gmail.com}

% Keywords are not mandatory, but authors are strongly encouraged to provide them. If provided, please include two to five keywords, separated by the pipe symbol, e.g:
% \keywords{Keyword 1 $|$ Keyword 2 $|$ Keyword 3 $|$ ...} 

\begin{abstract}
The Precision Reactor Oscillation and Spectrum Experiment, PROSPECT, is designed to both perform a reactor-model independent search for eV-scale sterile neutrino oscillations at meter-long baselines and to make a precise measurement of the antineutrino spectrum from a highly-enriched uranium reactor.  To meet these goals, PROSPECT must be realized as a compact detector able to operate with little overburden in a high background environment. As such, PROSPECT also serves as an excellent test case for near-field reactor monitoring.   This proceeding briefly describes the design, first data, and characterization of the PROSPECT antineutrino detector from the monitoring perspective. 
\end{abstract}

% \dates{This manuscript was compiled on \today}
\doi{\url{https://neutrinos.llnl.gov/workshops/aap2018}}

% \begin{document}

\maketitle
\thispagestyle{firststyle}
\ifthenelse{\boolean{shortarticle}}{\ifthenelse{\boolean{singlecolumn}}{\abscontentformatted}{\abscontent}}{}

% If your first paragraph (i.e. with the \dropcap) contains a list environment (quote, quotation, theorem, definition, enumerate, itemize...), the line after the list may have some extra indentation. If this is the case, add \parshape=0 to the end of the list environment.
\dropcap{W}e now have a coherent picture of neutrino flavor mixing with recent measurements providing a precise determination of oscillation parameters in the 3-neutrino model. However, anomalous results in both the reactor \nuebarMUMM{} flux and spectrum have provided hints that this picture is incomplete.  Reactor \nuebarMUMM{} experiments observe a $\sim$6\,\% deficit in the absolute flux when compared to predictions~\cite{Mueller:2011nm,Huber:2011wv}. This deficit, the ``reactor antineutrino anomaly'', has motivated the hypothesis of oscillations involving a sterile neutrino state~\cite{Mention:2011rk,Kopp:2013vaa}. In addition, measurements of the reactor \nuebarMUMM{} spectrum by $\theta_{13}$ experiments observe notable spectral discrepancies compared to prediction ~\cite{An:2015nua,Abe:2014bwa,Seo:2016uom}, indicating deficiencies in current prediction methods and/or the nuclear data underlying them.  The Precision Reactor Oscillation and Spectrum Experiment, PROSPECT~\cite{Ashenfelter:2015uxt}, was designed to comprehensively address this situation by simultaneously searching for \nuebarMUMM{} oscillations at short baselines and making a precise \nuebarMUMM{} energy spectrum measurement from a highly-enriched uranium (HEU) compact reactor core.  A precision measurement of the $^{235}$U spectrum constrains predictions for a static single fissile isotope system, providing a measurement complementary to those at commercial power reactors with evolving fuel mixtures.  Determining the relative \nuebarMUMM{} flux and spectrum at multiple baselines within the same detector provides a reactor-model independent search for sterile neutrino driven oscillations in the parameter space favored by reactor and radioactive source experiments. These measurement goals necessitate that PROSPECT be realized as a compact detector able to operate with little overburden (< 1 mwe) in a high background environment.  As such, PROSPECT also serves as an excellent test case for near field reactor monitoring.

\section*{The PROSPECT Detector}
\begin{figure}[tbhp]
\centering
\includegraphics[width=.80\linewidth]{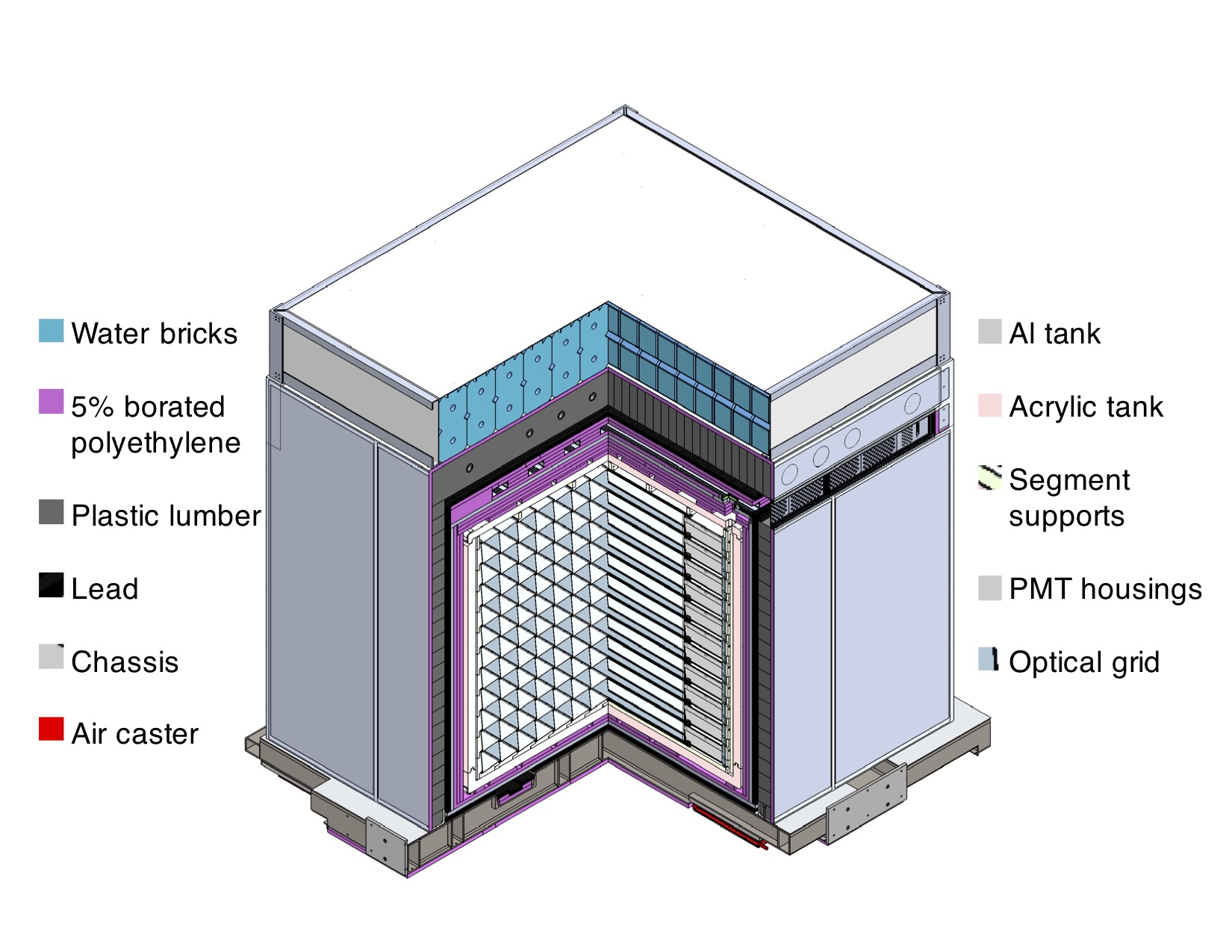}
\caption{A cutaway view of the PROSPECT detector. The inner detector, inside the acrylic tank (rose), is segmented into grid by reflective optical separators. Each segment viewed by PMT housings (beige) on either end. The housings and grid are supported by acrylic segment supports (light green). The acrylic tank is surrounded by borated polyethylene (purple) and a secondary aluminum  tank (light gray).}
\label{Mumm:Figure1}
\end{figure}

PROSPECT is located at the \HFIR\ at \ORNL\.  \HFIR\ is a HEU reactor with a nominal power of 85 MW and a very compact geometry (diameter of 0.435~m and height of 0.508~m) making it an excellent match to the PROSPECT physics goals~\cite{HFIR_model}.  The PROSPECT detector consists of a single 2.0~m\,$\times$\,1.6~m\,$\times$\,1.2~m rectangular volume containing 4000 liters of $^6$Li-doped liquid scintillator (LS) accessing baselines in the range 7~m to 13~m from the reactor core.  The active LS volume is divided into 154 (14 by 11) equal volume segments; each segment is 117.6~cm in length and has a 14.5~cm\,$\times$\,14.5~cm square cross-sectional area.  This optical grid is formed from low-mass, highly specularly reflective optical panels held in position by white 3D-printed support rods.  The hollow center of these rods provides calibration-source access throughout the detector volume.  Each segment is viewed on each end by a single 5 inch Photomultiplier Tube (PMT)-light concentrator assembly enclosed in a mineral oil filled acrylic housing.  The design of the separators and PMT housing limits optical cross talk between segments to less than 1\,\%.  PMT signals are recorded using a 250 MHz 14-bit waveform digitizer.  In the nominal operating mode an above-threshold ($\approx$5 photoelectron) signal from both PMTs in a single segment is required to trigger zero-suppressed readout of the full detector. Trigger rates of roughly 30~kHz and 5~kHz are observed during reactor-on and reactor-off running respectively. Surrounding the active region of the detector are multiple layers of hydrogenous material, high density polyethylene (HDPE) and water bricks, boron doped HDPE, and lead shielding to reduce both local reactor related backgrounds and, importantly, those originating from cosmogenic sources.  A representation of the PROSPECT detector is shown in Fig.~\ref{Mumm:Figure1}.  A detailed description of the PROSPECT experiment can be found here~\cite{Ashenfelter:2018zdm}. 

PROSPECT detects antineutrinos via the inverse beta-decay (IBD) reaction on protons in the liquid scintillating target. The positron carries most of the antineutrino energy and rapidly annihilates with an electron producing a prompt signal with energy ranging from 1~MeV to 8~MeV. The neutron, after thermalizing, captures on a $^6$Li or H nucleus,  with a typical capture time of 40\,$\mu s$.  The correlation in time and space between the prompt and delayed signals provides a distinctive $\bar{\nu}_e$ signature, greatly suppressing backgrounds.  $^6$Li  was chosen in favor of the more established gadolinium, because PROSPECT is a compact highly-segmented detector, $\gMUMM$-rays will generally interact in multiple segments or escape the detector entirely leading to detection efficiency variations.  Similarly, because of the need to operate in a high background environment, a capture signal composed entirely of $\gMUMM$-rays would lead to a lack of discrimination from high $\gMUMM$-ray backgrounds.  In contrast, neutron captures on $^6$Li produce well localized energy depositions via an alpha and triton.  As this capture only yields heavy charged particles, a pulse-shape discriminating $^6$Li liquid scintillator is able to separate neutron captures from background $\gMUMM$-ray events reducing the likelihood of random coincidences. To fully take advantage of these features, the PROSPECT collaboration developed a novel 0.08\,\% $^6$Li by mass-doped liquid scintillator (LiLS). In addition to the above event selection criteria, topology cuts vetoed events with extra energy deposits not associated with the segments containing the positron and neutron signals. However, such cuts lose effectiveness near the edge of the detector due to events physically spanning the active boundary.  As a result, the rate of IBD-like backgrounds that pass all cuts in the outermost segments is 10-100 times that of the innermost segments.  This motivates the use of a "fiducial" region such that accepted IBD events must originate in an inner volume (removing the outermost segments and ends of each segment close to the photomultipliers (PMTs). The combination of time and space correlations, localized capture signal, particle identification of both the prompt and delayed signal, and topology provides PROSPECT with excellent background rejection capabilities and makes possible the demonstration of an excellent signal to background in a minimally-shielded surface-deployed neutrino detector.  
 
\section*{Characterization and Performance}
PROSPECT has been extensively characterized through a combination of calibration sources deployed within the optical grid support rods, i.e.  $^{137}$Cs, $^{22}$Na, and $^{60}$Co, and $^{252}$Cf, and natural backgrounds such as detector-intrinsic ($^{219}$Rn, $^{215}$Po) correlated decays from $^{227}$Ac deliberately dissolved in the LS, ($^{214}$Bi, $^{214}$Po) correlated decays from $^{238}$U, and background neutron captures on hydrogen an chlorine.  Detector response stability and uniformity are demonstrated via examination of reconstructed physics quantities as a function of time and segment number.  Reconstructed energy and energy resolutions are seen to be stable to within $\approx$1\% and $\approx$10\%, respectively, over all times and segments.  Similarly,  reconstructed longitudinal positions and uncertainty are stable to within $\approx$5 cm and $\approx$10\% respectively.

\begin{figure}[tbhp]
\centering
\includegraphics[width=.75\linewidth]{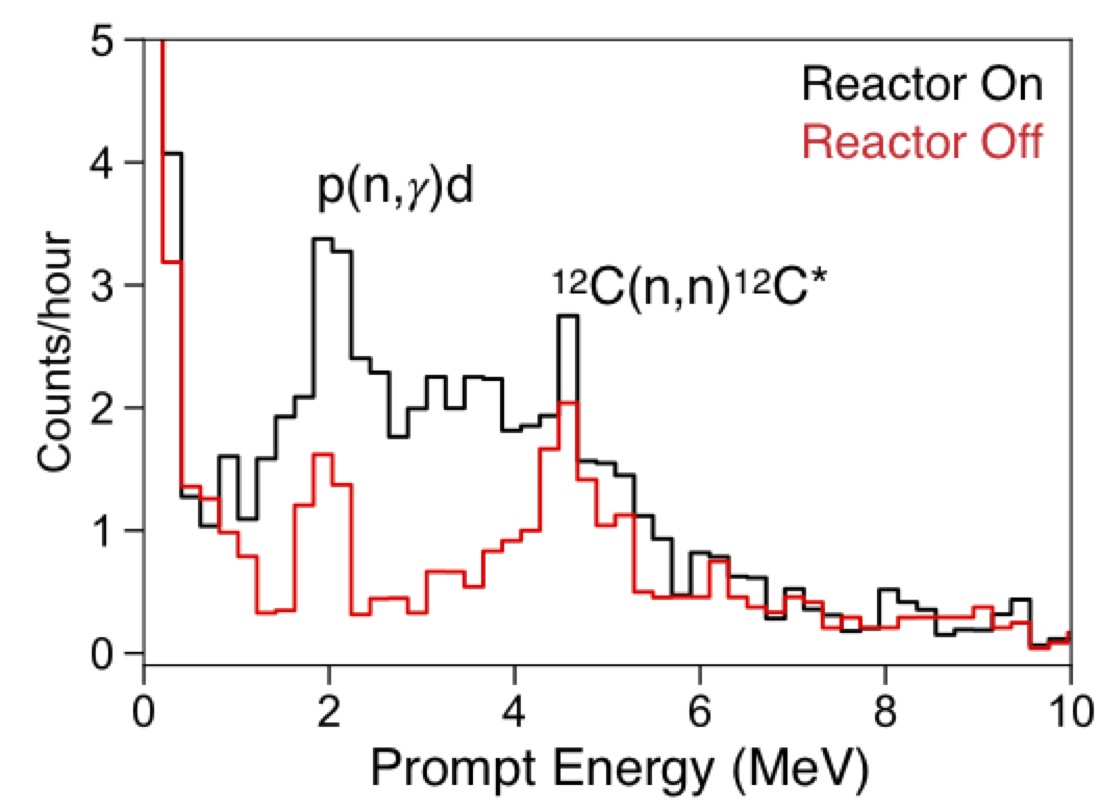}
\caption{The prompt energy spectra for the first 24 hours each of data with reactor on and off.  Both spectra show prominent structure related to cosmogenic backgrounds, but the  difference between the two data sets has the expected general shape of a reactor \nuebarMUMM{} spectrum and illustrates the excellent signal to background achieved.}
\label{Mumm:Figure2}
\end{figure}

The energy scale, nonlinearity, and resolution are established via a simultaneous fit to the measured spectra of the $^{137}$Cs, $^{22}$Na, and $^{60}$Co sources deployed at segment centers throughout the detector in combination with a convenient high-energy $\beta$ spectrum from the $\gMUMM$-ray correlated decay of cosmogenically produced $^{12}$B that is uniformly distributed throughout the detector.  More recently the multiplicity of segment hits has been incorporated into the fit as well.   The fit uses an energy response model with two LS nonlinearity parameters, one photo-statistics resolution parameter, and one absolute energy scale parameter.  Nonlinearities for the best-fit model are $\approx$20\% over the relevant energy range and, in spite of being a highly-segmented detector, an excellent photostatistics energy resolution of 4.5\% at 1~MeV is achieved.

To extract IBD event rates and spectra, accidental coincidences are subtracted run-by-run during both reactor-on and reactor-off periods with little statistical uncertainty using a pre-prompt window from 12\,ms to 2\,ms.  Correlated cosmogenic backgrounds, dominated by fast neutron interactions, are determined using statistically balanced background data acquired during reactor-off periods.  Spectra are scaled by less than 1\% to account for variations in atmospheric pressure.   Between prompt reconstructed energies of 0.8 MeV and 7.2 MeV the reactor-on data yields 771 detected IBDs per day, with a signal-to-background ratio (S:B) of 2.20 and 1.32 for accidental and correlated backgrounds, respectively.  This excellent signal to background is well illustrated in Fig.~\ref{Mumm:Figure2}.

%The layout of the experiment is shown in Fig.~\ref{fig-layout}. 
%PROSPECT combines competitive exposure, baseline mobility for increased physics reach and systematic checks, good energy and position resolution, and efficient background discrimination.
%PROSPECT has already demonstrated  a signal over correlated background ratio of $\gtrsim 1:1$~\cite{Ashenfelter:2018osc} and set new limits on sterile neutrino oscillations 
%based on its first 33 days of reactor operation.
%that yields a signal over background of $\gtrsim 1:1$; sufficient to achieve the stated goals. 

\section*{Conclusion}

The surface-deployed PROSPECT experiment has observed reactor \nuebarMUMM\ produced by a nearly pure $^{235}$U fission reactor over several reactor-on cycles. The current signal selection criteria provide a ratio of 1.32 \nuebarMUMM\ to IBD-like cosmogenic backgrounds, as well as the capability to identify reactor state transitions to 5-$\sigma$ statistical confidence level within 2 h.  The dimensions and design of the PROSPECT detector are such that they will inform fully portable and scalable designs.  These characteristics and performance demonstrate the feasibility of on-surface reactor \nuebarMUMM\ detection and the potential utility of this technology for reactor power monitoring~\cite{Bernstein:2009ab,Bernstein:2001cz,Carr:2018tak}.  Furthermore, PROSPECT will serve as a platform for detailed background studies into the future.   Finally, a comparison of measured IBD prompt energy spectra between detector baselines has provided no indication of sterile neutrino oscillations and disfavors the reactor antineutrino anomaly best-fit point at 2.2$\sigma$ confidence level and constrains significant portions of the previously allowed parameter space at 95\% confidence level~\cite{Ashenfelter:2018iov}.

\acknow{For full acknowledgements please see https://prospect.yale.edu}

\showacknow{} % Display the acknowledgments section

% % Bibliography
% \bibliography{aap2018-participant00x}

% \end{document}

%% file: AAP-Verstraeten/aap2018-participant00x.tex
% \documentclass[9pt,twocolumn,twoside,lineno]{aap2018}
%\documentclass{article}
% \usepackage[utf8]{inputenc}

% \title{AAP-PersonalizedTemplate}
% \author{bergevin1 }
% \date{September 2018}

% \begin{document}

% \maketitle

% \section{Introduction}

% \end{document}

% \documentclass[9pt,twocolumn,twoside,lineno]{pnas-new}
% Use the lineno option to display guide line numbers if required.

% \templatetype{aap2018proceedings} % Choose template 
% {pnasresearcharticle} = Template for a two-column research article
% {pnasmathematics} %= Template for a one-column mathematics article
% {pnasinvited} %= Template for a PNAS invited submission

\title{SoLid Neutrino Detector for Reactor Monitoring}

% Use letters for affiliations, numbers to show equal authorship (if applicable) and to indicate the corresponding author
\author[a,1]{Maja Verstraeten}

\affil[a]{Universiteit Antwerpen, Antwerpen, Belgium}

% Please give the surname of the lead author for the running footer
\leadauthor{Verstraeten}

% Please include corresponding author, author contribution and author declaration information
%\authorcontributions{On behalf of the SoLid collaboration.}
% \authordeclaration{Please declare any conflict of interest here.}
% \equalauthors{\textsuperscript{1}A.O.(Author One) and A.T. (Author Two) contributed equally to this work (remove if not applicable).}
\correspondingauthor{\textsuperscript{1}On behalf of the SoLid collaboration.\\ Correspondence: maja.verstraeten@uantwerpen.be}

% Keywords are not mandatory, but authors are strongly encouraged to provide them. If provided, please include two to five keywords, separated by the pipe symbol, e.g:

\begin{abstract}
The SoLid experiment will measure the dependence of the $\bar{\nu}_e$ flux to distance and energy, at very short baseline from the reactor core. This to deepen our knowledge of the reactor neutrino spectrum and to asses the reactor neutrino anomaly. SoLid is operating a 1.6 ton, highly segmented detector at 6-9m stand off from the compact core of the 60 MW BR2 reactor of the Belgian Nuclear Research Centre. To accomplish the challenging measurement in the high radiation environment - close to the nuclear reactor core and at the earth’s surface - a novel detector design was developed. An innovative, hybrid scintillator technology combines PVT and $^6$LiF:ZnS scintillators into a unit cell of 5x5x5 cm$^3$, in order to reach excellent particle identification and energy reconstruction. The detector technology can serve non-proliferation purposes. A sensitive monitoring of the neutrino flux allows assessing the core composition and thermal power output of a nuclear reactor, without interfering with its operation.
\end{abstract}

% \dates{This manuscript was compiled on \today}
\doi{\url{https://neutrinos.llnl.gov/workshops/aap2018}}

% \begin{document}

\maketitle
\thispagestyle{firststyle}
\ifthenelse{\boolean{shortarticle}}{\ifthenelse{\boolean{singlecolumn}}{\abscontentformatted}{\abscontent}}{}

% If your first paragraph (i.e. with the \dropcap) contains a list environment (quote, quotation, theorem, definition, enumerate, itemize...), the line after the list may have some extra indentation. If this is the case, add \parshape=0 to the end of the list environment.
\dropcap{N}eutrinos possess unique capacities concerning nuclear safeguard monitoring; neutrinos cannot be contained to a nuclear reactor site, they are unaffected by particular test conditions and they are specific to the type of fission. Measuring the reactor neutrinos gives critical insight in the operation conditions of a nuclear reactor. The composition of the reactor fuel after refueling, change in the operational status over time and production of isotopes can be deduced. Even if access to the reactor facility is denied. Precise measurement of the fuel spectrum requires adequate and convenient neutrino detectors.

\section*{SoLid neutrino detector}

Reactor monitoring, and more specifically safeguard monitoring in hazardous areas, demands compact and highly efficient neutrino detectors. High resolution of the energy spectrum is required to assess the fuel composition. The construction of the detector itself has to be inert and robust. Smooth operation and remote monitoring should limit and simplify intervention with the detector to the minimum.

\begin{figure}[tbhp]
\centering
\includegraphics[width=0.90\linewidth]{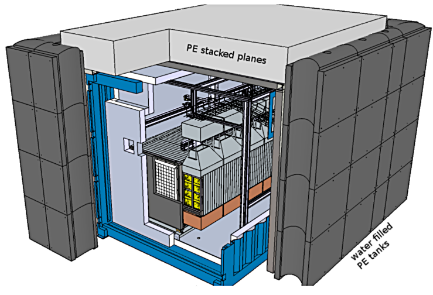}
\caption{Schematic view of the detector and passive shielding in Geant4 \cite{geant4}.}
\label{detector}
\end{figure}

The SoLid collaboration designed and built a highly voxelized, hybrid scintillator, antineutrino detector - a novel detector technology \cite{JINST_12_P04024}. The complete detector is enclosed in a shipping container which facilitates transport (see figure \ref{detector}). The detection is based on non flammable, solid scintillator technology. High segmentation of the scintillators renders a high position -and energy resolution of the neutrinos. An in-situ, fully-automated calibration robot allows to make an absolute measurement of the detection efficiency and enables to calibrate the energy scale at percent level. The SoLid detector proves itself useful for reactor monitoring. The experiment will provide a reference measurement of the neutrinos from highly enriched $^{235}$U.

\section*{BR2 reactor at SCK•CEN}

To determine a reactor's operational status, based on the characteristics of the neutrino output, a detailed reference measurement of the neutrino energy spectrum is required. 
Currently, unexpected spectral features around 5MeV were observed by long baseline reactor experiments using common fuels ($^{235}$U, $^{238}$U, $^{239}$Pu, $^{241}$Pu), which are correlated with reactor power and fuel composition \cite{Huber2016xis}, stressing our lacking understanding of the reactor neutrino spectrum,.

The SoLid detector is operated near the Belgian reactor 2 (BR2) at the SCK•CEN. BR2 has an uncommon fuel of highly enriched (>90\%), pure $^{235}$U. This single fuel isotope is of particular interest for the nuclear physics community to asses the 5 MeV distortion.
The research reactor is highly suited for a short baseline oscillation search \cite{Dentler2017}. The twisted core design results in a small core diameter ($\approx$ 0.5 m), ensuring very little position smearing. The detector is positioned on axis with as closest stand off only 6,4 m from the compact core. We cover a baseline up to 9 meter.

The  simulation  of  the reactor core is  developed, in close cooperation between BR2 reactor team and SoLid. The core's $\bar{\nu}_e$ spectrum for each cycle, i.e. for  a  given  fuel  loading  map  and  operation  history is required. The SCK•CEN team modeled the BR2 core with the Monte Carlo transport code MCNPX coupled to CINDER90 \cite{MCNPX}. The SoLid working group on the reactor $\bar{\nu}_e$ spectrum adds strong expertise on calculation of the antineutrino spectrum using both conversion and summation method \cite{pred}. Systematic errors will be associated with the emitted antineutrino spectrum.

The space in the reactor hall is sufficient for a relatively compact, above-ground detector and modest passive shielding. With 10 m.w.e. overburden, the atmospheric backgrounds are challenging. Radioactivity measurements show that the intrinsic background is low compared with other candidate sites. 
The reactor is powered about half the year around 60 MW, in 1 month cycles. The intermittent reactor off periods allow for an accurate background determination and calibration campaigns.
Phase 1 of the experiment, which has 1,6 ton active mass, is
scheduled to run for around 3 years. Efficient signal tagging is required in order to reach the experiment's physics aim in this period.

\section*{Detector specifications}

\begin{figure}[tbhp]
\centering
\includegraphics[width=0.90\linewidth]{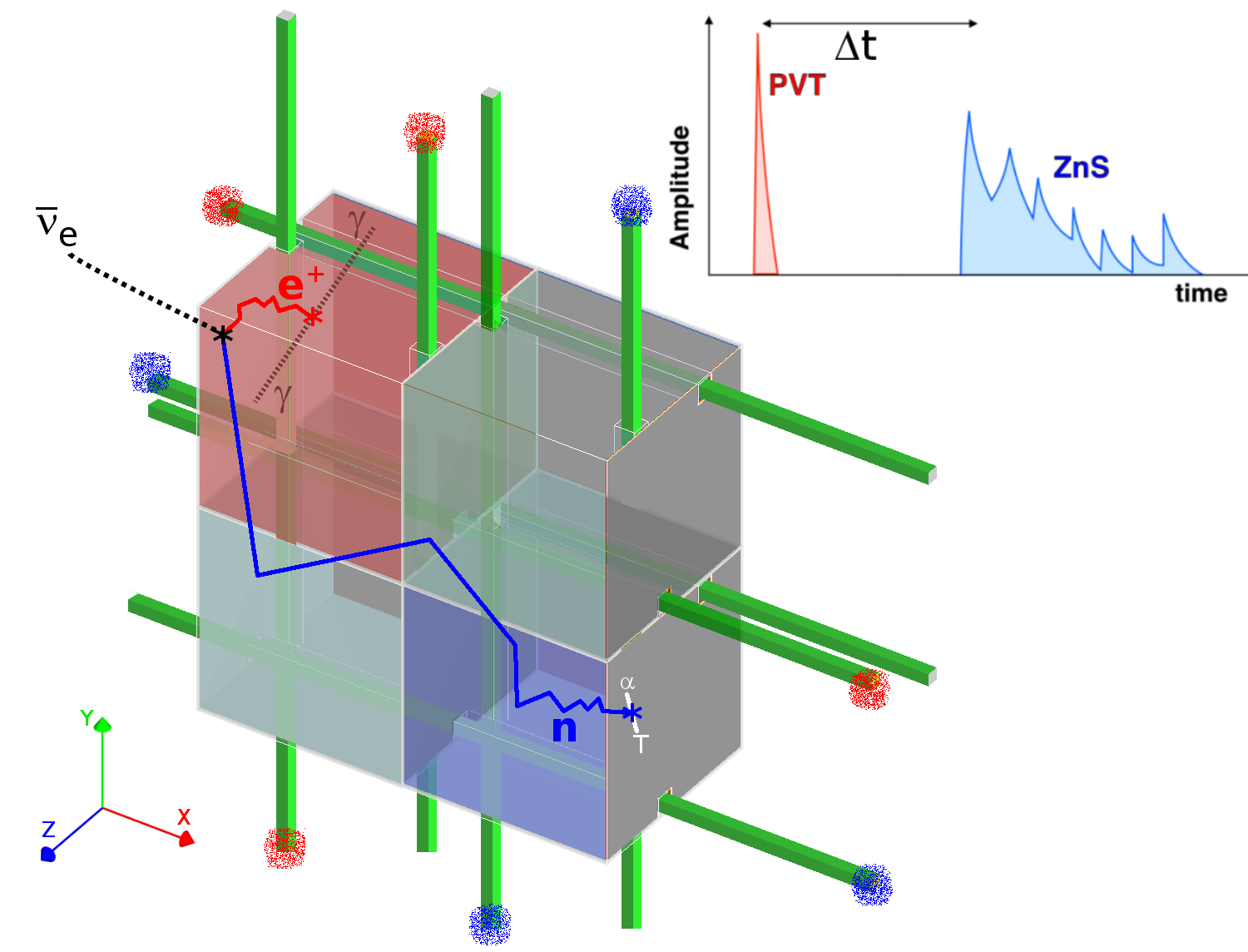}
\caption{Principle of $\bar{\nu}_e$ detection in cells of combined scintillator. wavelength shifting fibres placed in perpendicular orientations collect the scintillation light.}
\label{interaction}
\end{figure}

Neutrinos interact with the detector volume via inverse beta decay (IBD), resulting in a positron and a neutron that are correlated in time and space. To optimally detect and discriminate both particles, two solid scintillators are joined \cite{JINST_13_P05005}. Cubes of Polyvinyl-toluene (PVT) act as a scintillator for the positron prompt signal (see figure \ref{interaction}). PVT offers high light output and a linear energy response, from which both the location and the energy of the neutrino interaction can be determined.\\ 

Sheets of $^6$LiF:ZnS(Ag) are placed on two faces of each cube to detect the neutron. After thermalization, the neutron can be captured by a $^6$Li nucleus. This reaction produces an alpha and a tritium nucleus, sharing 4.78 MeV of kinetic energy. Both are highly ionizing and deposit all their energy within the sheet, scintillating in the ZnS(Ag) microcrystals. 
Crucially, this scintillation timescale is considerably slower, at O(1) $\mu$s, than all other scintillation signals in the detector, at O(1) ns. The ZnS(Ag) signals are characterised by a set of sporadic pulses emitted over several microseconds. With a waveform sampling of 40MHz, Pulse shape discrimination (PSD) can be used to identify and discriminate the signals with high efficiency and purity.\\

Each cell of a $5\times5\times5$ cm$^3$ PVT cube with two $^6$LiF:ZnS(Ag) sheets is wrapped in reflective Tyvek for optical insulation \cite{JINST_13_P09005}. The cells are arranged in 50 planes of 16 by 16 cells, split in five modules of ten planes each. The scintillation light is guided from the cells towards sensors by an orthogonal grid of wavelength shifing fibers. One end of the fiber is coupled to a mirror, whilst the other is connected to a second generation Hamamatsu silicon photomultiplier. The 3200 readout channels result in an enormous datarate of 3 tbps \cite{RO}. It is handled by neutron triggers, which count peaks in a rolling time window, and a sophisticated online data reduction.
The 3D topological information obtained by the cubes is the strength of the SoLid detector. The segmentation gives the high position and energy resolution, needed for precise oscillation pattern measurements. Due to the neutron thermalization, the IBD's positron and neutron signal have a characteristic position and time difference. The cubes allow discrimination of the IBD signature over the prominent backgrounds.

\section*{Automatic Callibration}
 
 Before installation, the performance and quality of each detector plane was checked with an automated calibration system, which placed gamma -and neutron sources in front of each cube separately \cite{QA}. The channels' amplitude response is calibrated to high quality, with a spread of $\sim$1,4\%. Different sources prove a high linearity over a wide energy range, which combined with the pure $^{235}$U fuel gives a strong handle on the neutrino energy spectrum measurement.
 
 In situ, a second calibration robot is installed in the container. The robot sits on a rail system above the detector. Between two modules a gap can be opened. The robot can freely manoeuvre a calibration source in the gap, such that the whole module can be calibrated.
 A homogeneous response is achieved for the highly segmented detector. The lightyield is higher than expected with more than 70 pixel avalanches per MeV deposited. The homogenous neutron reconstruction efficiency is above 75\% during commissioning. 

\section*{Data taking}

Since February, SoLid is in highly stable data taking mode, for both reactor on and off periods. Physics variables are available online which allow complete remote monitoring. The SoLid Data Quality Management (SDQM) is automated. Updates of monitoring variables are regularly sent by the automatic Solid Message System (SMS). In case stable data taking is obstructed, alerts are prompted to contact persons. The detector shifts are very convenient and user friendly.
With ongoing data taking, preliminary rate monitoring is performed. A significantly higher IBD-like rate during reactor on is observed. The rate of accidentals, comprising the background is very low. For IBD like events, the time difference between prompt and delayed signal is consistent with the capture of the thermalised neutron and the spatial separation is as expected.

\section*{Conclusion}

SoLid successfully deployed a new detector technology. A \mbox{1,6 ton} detector was commissioned end of 2017. The container design is well suited for rapid deployment. The operation is smooth and remote shifts simplified to the minimum. Automatic calibration with radioactive sources provides precision data for sterile search and spectrum measurement. SoLid is taking good quality physics data and observes IBD-like events. The analyis is further developed. The detector technology is applicable for non proliferation purposes like non invasive reactor monitoring.

% \section*{References}

% Bibliography
% \bibliography{aap2018-participant00x}

% \end{document}

%% file: AAP-Park/aap2018-participant00x.tex
% \documentclass[9pt,twocolumn,twoside,lineno]{aap2018}
%\documentclass{article}
% \usepackage[utf8]{inputenc}

% \title{AAP-PersonalizedTemplate}
% \author{bergevin1 }
% \date{September 2018}

% \begin{document}

% \maketitle

% \section{Introduction}

% \end{document}

% \documentclass[9pt,twocolumn,twoside,lineno]{pnas-new}
% Use the lineno option to display guide line numbers if required.

% \templatetype{aap2018proceedings} % Choose template 
% {pnasresearcharticle} = Template for a two-column research article
% {pnasmathematics} %= Template for a one-column mathematics article
% {pnasinvited} %= Template for a PNAS invited submission

\title{Result of MiniCHANDLER}

% Use letters for affiliations, numbers to show equal authorship (if applicable) and to indicate the corresponding author
\author[a,1]{Jaewon Park for CHANDLER collaboration}
%\author[a]{Alireza Haghighat}
%\author[a]{Patrick Huber}
%\author[a]{Shengchao Li}
%\author[a]{Jonathan M. Link}
%\author[a]{Camillo Mariani}
%\author[a,1]{Jaewon Park}
%\author[a]{Tulasi Subedi}

\affil[a]{
 Center for Neutrino Physics \\
 Department of Physics \\
 Virginia Tech, Blacksburg, VA
}%

% Please give the surname of the lead author for the running footer
\leadauthor{Park}

% Please include corresponding author, author contribution and author declaration information
% \authorcontributions{Please provide details of author contributions here.}
% \authordeclaration{Please declare any conflict of interest here.}
% \equalauthors{\textsuperscript{1}A.O.(Author One) and A.T. (Author Two) contributed equally to this work (remove if not applicable).}
\correspondingauthor{\textsuperscript{2}To whom correspondence should be addressed. E-mail: \href{mailto:jaewon.park@vt.edu}{jaewon.park@vt.edu}}

% Keywords are not mandatory, but authors are strongly encouraged to provide them. If provided, please include two to five keywords, separated by the pipe symbol, e.g:
% \keywords{Keyword 1 $|$ Keyword 2 $|$ Keyword 3 $|$ ...} 

\begin{abstract}
There have been hints of sterile neutrino from reactor and source experiments. Reactor neutrino oscillation can be observed at 5-10 m baseline if such sterile neutrino exists. CHANDLER was designed to detect inverse beta decay (IBD) using $^{6}$Li-loaded ZnS sheet and scintillator cubes. Such design gives clear neutron identification and detector granularity, which are critical to observe IBD in the high background environment. MiniCHANDLER is a prototype detector, that collected data from North Anna Power Plant in 2017. The result of MiniCHANDLER was presented. It has also demonstrated the CHANDLER technology can be used for reactor monitor for nuclear nonproliferation.
\end{abstract}

% \dates{This manuscript was compiled on \today}
\doi{\url{https://neutrinos.llnl.gov/workshops/aap2018}}

% \begin{document}

\maketitle
\thispagestyle{firststyle}
\ifthenelse{\boolean{shortarticle}}{\ifthenelse{\boolean{singlecolumn}}{\abscontentformatted}{\abscontent}}{}

% If your first paragraph (i.e. with the \dropcap) contains a list environment (quote, quotation, theorem, definition, enumerate, itemize...), the line after the list may have some extra indentation. If this is the case, add \parshape=0 to the end of the list environment.
%\dropcap{T}his template is provided to help you summarize your work in the correct format.  Instructions for use are provided below. 

\section*{Introduction}

Antineutrinos from nuclear reactor undergo inverse beta decay (IBD), $\bar\nu_e+p \rightarrow n + e^+$, where neutron is thermalized and is captured later.   Reactor antineutrino experiments utilize the coincidence of prompt positron and delayed neutron signals. Sterile neutrino search or reactor monitoring detector is located at $\approx$ 5 -- 15 m from reactor core. The main challenge is overwhelming backgrounds from fast neutron since the detector at the surface is minimally shielded.
In order to defeat the high backgrounds, fine detector granularity and pure neutron tagging are demanded. CHANDLER uses $^{6}$Li for neutron capture target. Neutron capture on $^{6}$Li produces a triton and an alpha that range out in a very short distance. Thin sheets of $^{6}$Li-loaded ZnS scintillator are alternated with scintillator cube layers. Neutron signal is identified by pulse shape discrimination (PSD) because ZnS has $\approx$20 times longer scintillator decay time than scintillator cube. 

Scintillator cube has a dimension of $6.1 \times 6.1 \times 6.1$ cm$^{3}$. Such granularity allows us to use the tight spatial correlation between prompt and delayed signals to reject backgrounds. Most positron from IBD stops in a cube and prompt and delay signals are mostly within 1 or 2 cube distance.
The fine detector granularity furthermore helps to look for two 511 keV gammas from positron annhiliation.

\section*{MiniCHANDLER Detector}

MiniCHANDLER is a prototype detector of CHANDLER. It uses Eljen EJ-260\cite{EJ-260} wavelength shifting (WLS) scintillator for scintillator cube. A layer of tightly packed scintillator cubes forms a Raghavan Optical Lattice (ROL)\cite{PhysRevD.75.093006}. MiniCHANDLER has 8 $\times$ 8 cubes in a layer. Light produced in a cube is transported by total internal reflection in X and Y directions along the row and column of cubes. The MiniCHANDLER consists of 5 layers of 8 $\times$ 8 cube array. Total 80 two-inch PMTs are located in X and Y sides of detector to read light from ROL.

Neutron detection sheet is Eljen EJ-426\cite{EJ-426}, which is a homogeneous mixture of lithium6-flouride ($^{6}$LiF) and zinc sulfide phosphor (ZnS:Ag). The neutron sheets are placed between each scintillator layers and top and bottom of the detector. $^{6}$Li-loaded ZnS sheet is made as thin sheet so the scintillator light produced in the sheet can easily escape from the sheet. After this light entering into the scintillator cube, it is absorbed and reemitted by WLS process, then it is transported to PMTs by total internal reflection. Since the sheet is placed between scintillator layers, the light can propagate both sides of the sheet. A neutron capture on the sheet can be seen as either single or double layer light. PMT pulses are further amplified and get widened using 25 ns shaping time before they are fed into 16 ns waveform digitizer (CAEN V1740).

All the electronics and the MiniCHANDLER are installed in a small trailer, also known as Mobile Neutrino Lab. This detector technique with mobility and fast deployment can be used in reactor monitor for nuclear nonproliferation.
 Mobile Neutrino Lab was deployed at North Anna Nuclear Power Plant from June 15, 2017 to November 2, 2017, which yields 48 days of reactor-on and 24 days of reactor-off data for IBD analysis.
 
 \section*{Calibration}
PMT HVs are tuned based on muon peak positions before taking physics data. Cosmic muon makes about 12 MeV energy deposit when it transverses the cube. Since the background rate is low in this energy region and the shape of spectrum is a slowly varying exponential distribution, a distinctive peak from cosmic muon appears in the spectrum of each channel. Data is taken as 59 minute long run repeatedly for monitoring, calibration and processing convenience. Small variation of PMT gain is observed via muon peak position in each run, and it is later corrected on offline data.

Vertical muon sample from whole data is used to understand the light profile of ROL. Position of hit cube is well determined from vertical muon. Attenuation curves are obtained from all 64 cube locations and later used for energy reconstruction. These attenuation curves are also cross-checked with ones from sodium source Compton edge study. In the light profile of the vertical muon, small signal can be seen in neighboring channels due to unchanneled light and electronics crosstalk. The light profile from vertical muon is used for energy reconstruction as well as smearing simulation data.
 
\section*{Event Reconstruction}
High purity neutron tagging is one of critial requirements to observe IBD event in high background environment. Basic neutron PID is a ratio of area under pulse to pulse height. High value of neutron PID will select neutrons but PMT flashes occasionally mimic neutron signals. Waveforms are compared with electron-like and neutron-like waveform templates to reject PMT flashes. Two corresponding $\chi^{2}_n$ and $\chi^{2}_{\gamma}$ are calculated using coarse 6 regions of a waveform.

From IBD simulation, IBD events are mostly low cube multiplicity events. Prompt energy reconstruction is performed per scintillator layer looking for cube energies that are consistent with observed PMT pulse heights.
The energy reconstruction starts with one cube and keeps adding one cube at time until it finds a consistent solution based on log-likelihood maximization method.

\section*{IBD Analysis}
Once neutron is tagged and prompt energy is reconstructed, $\Delta t$ distribution of prompt-delayed signals is made for each 1 MeV prompt energy bin. Then, $\Delta t$ fit is performed to subtract the random accidental backgrounds. $\Delta t<40~\mu$s is excluded from correlated events due to electronics effect. Reactor-off data is normalized using high energy tail above 8 MeV since there is no IBD event above the limit.
%\cite{1674-1137-41-1-013002}.

\begin{figure}[tbhp]
\centering
\includegraphics[width=0.98\linewidth]{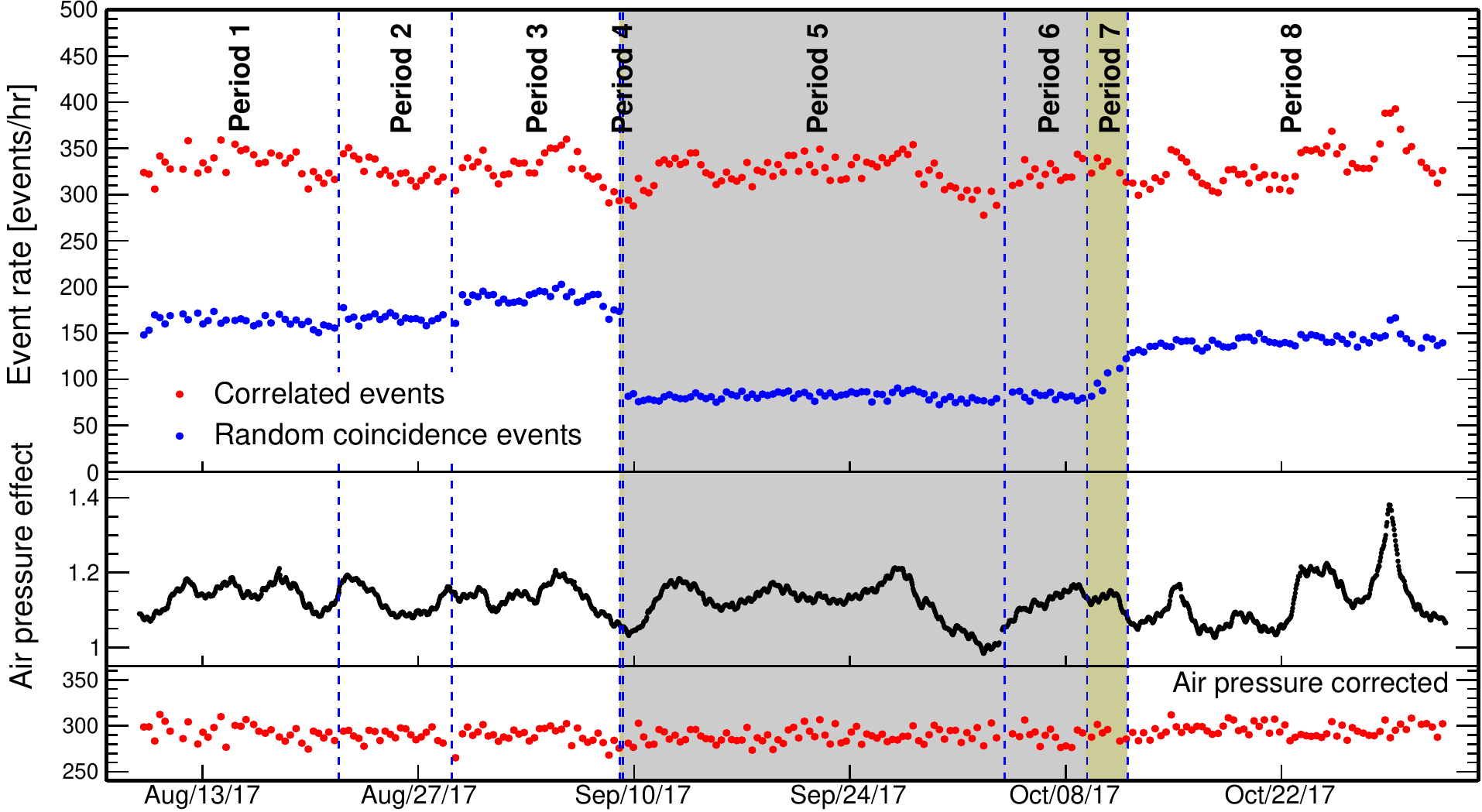}
\caption{Event rate stability. Gray shaded area is reactor-off period. Yellow shaded area is reactor-on/off transition period. Different period represents change in detector or reactor condition }
\label{park:evtstability}
\end{figure}

Event rate stability was checked and no unexpected behavior was found as shown in Figure \ref{park:evtstability}. The random coincidence event rate dropped during reactor-off period where there is no thermal neutrons from nuclear reactor. Major background of correlated events is the cosmogenic fast neutron that produces recoil protons. Observed rate of correlated events shows anti-correlation with atmospheric air pressure. If air pressure correction is applied, more stable event rate is observed.

\begin{figure}[tbhp]
\centering
\includegraphics[width=.8\linewidth]{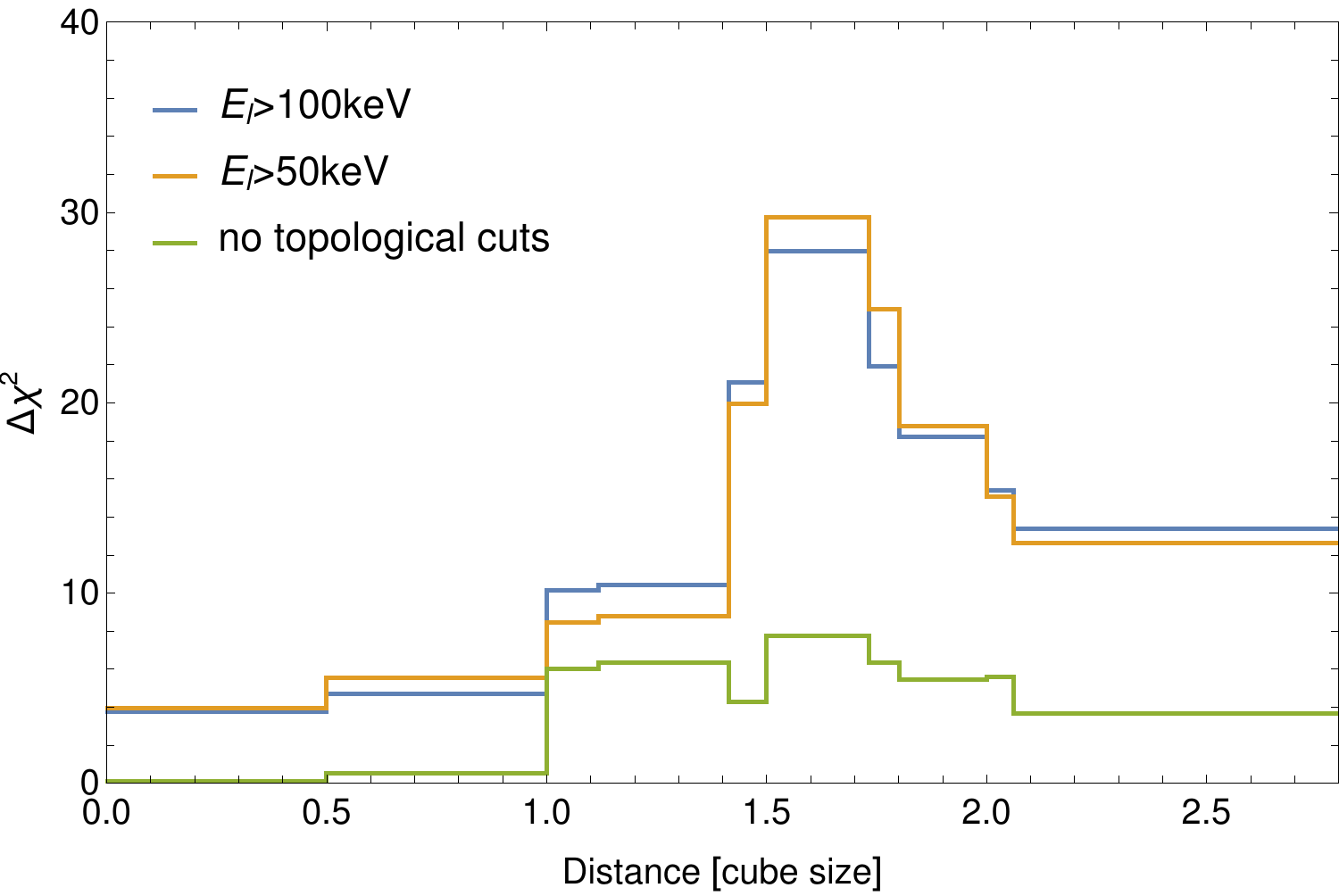}
\caption{Optimization of distance cut and topological selection energy threshold}
\label{park:evtselopt}
\end{figure}

Two major event selections are used to reject the backgrounds. First, prompt-delayed distance is required to be less than or equal to $\sqrt{3}$. It effectively rejects fast neutron backgrounds, since prompt and delayed signals from IBD are tightly localized in space. Figure \ref{park:evtselopt} shows the optimization of distance cut. 
Second, to further reduce the backgrounds, it uses a topological event selection that looks for Compton scatterings from two 511 keV gammas. For this selection, it requires at least one cube of energy 0.05 -- 0.511 MeV, and total energy outside primary cubes is less than 2 $\times$ 0.511 MeV. Figure \ref{park:evtselopt} shows the importance of the topological event selection. Large signal significance can not be achieved without  the topological event selection.

\section*{Result}
\begin{figure}[tbhp]
\centering
\includegraphics[width=.8\linewidth]{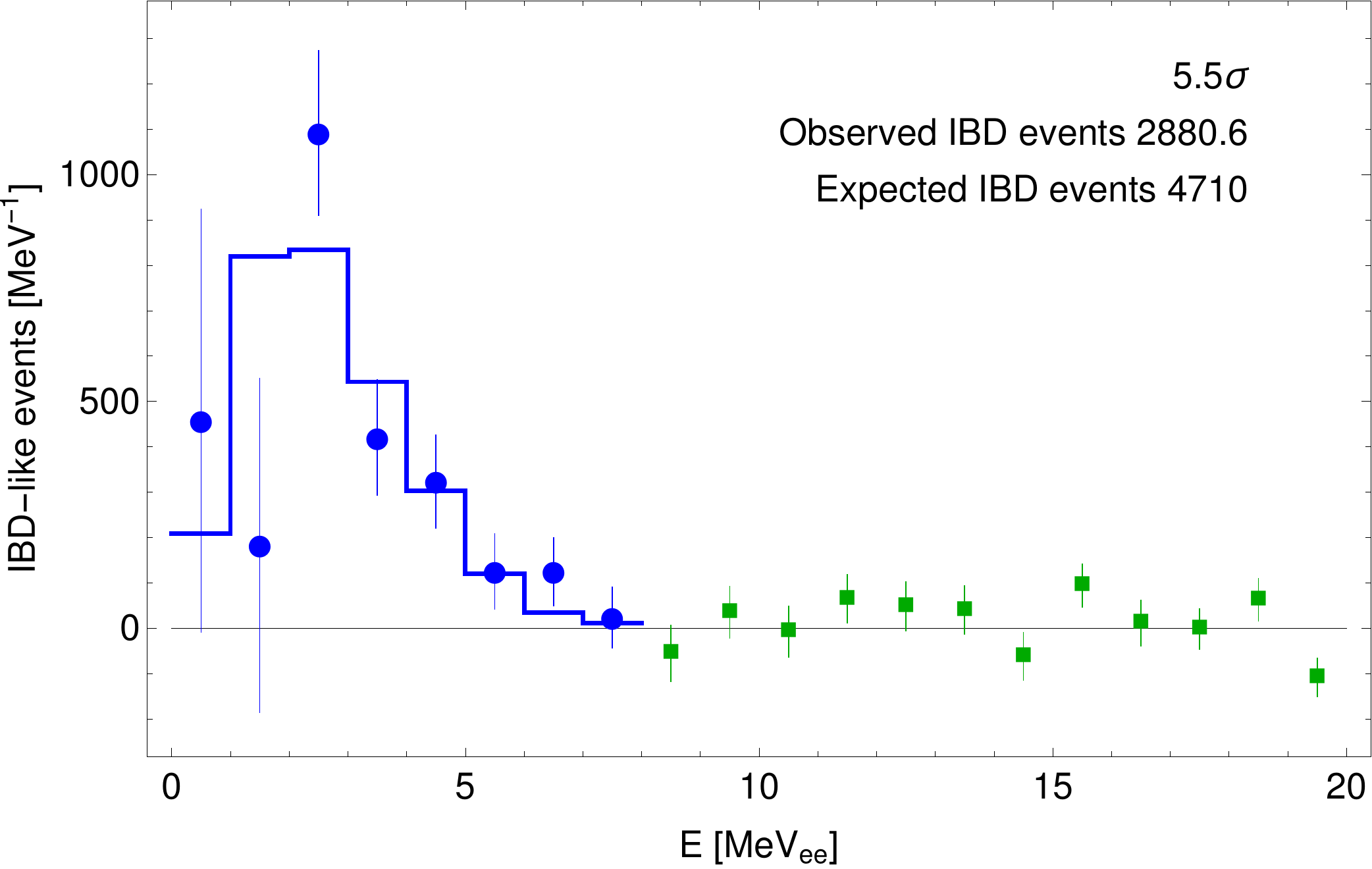}
\caption{Observed IBD spectrum}
\label{park:ibdspectrum}
\end{figure}

IBD spectrum is obtained from subtracting reactor-off spectrum from reactor-on as shown Figure \ref{park:ibdspectrum}. About 2900 IBD events have been observed and it corresponds to $5.5\sigma$ signal significance. Reconstructed spectrum of IBD simulation was normalized to data distribution. Signal-to-background ratio in the final sample was 1:60. Successful measurement of IBD signal owes prompt-delay distance cut and topological event selection that suppressed the background events.  

\section*{Conclusion}
MiniCHANDLER has successfully measured $>5\sigma$ IBD signal from North Anna Nuclear Power Plant. It demonstrated the fast-deployable antineutrino detector technology for possible nuclear nonproliferation application. It gives valuable feedbacks on improving potential full CHANDLER detector.

\acknow{This work was supported by the National Science
Foundation, under grant number PHY-1740247; Virginia Tech’s Institute
for Critical Technology and Applied Science; Virginia Tech’s College
of Science; the Office of the Vice President of Research and
Innovation at Virginia Tech; Virginia Tech’s College of Engineering;
and Virginia Tech’s Institute for Society, Culture and Environment.
We are grateful for the cooperation and support of Dominion Energy,
and the staff of the North Anna Power Station.}

\showacknow{} % Display the acknowledgments section

% Bibliography
% \bibliographystyle{apsrev}
% \bibliography{aap2018-participant00x}

% \end{document}

%% file: AAP-Han/aap2018-participant00x.tex
% \documentclass[9pt,twocolumn,twoside,lineno]{aap2018}

% \templatetype{aap2018proceedings} % Choose template 

\title{Status of NEOS experiment}

% Use letters for affiliations, numbers to show equal authorship (if applicable) and to indicate the corresponding author
\author[a,1]{Bo-Young Han}
\author[a]{Gwang-Min Sun} 
\author[b]{Eunju Jeon}
\author[c]{Kyungkwang Joo}
\author[c]{Ba Ro Kim}
\author[d]{Hongjoo Kim}
\author[e]{Hyunsoo Kim}
\author[e]{Jinyu Kim}
\author[f]{Siyeon Kim}
\author[e]{Jinyu Kim}
\author[b]{Yeongduk Kim}
\author[f]{Youngju Ko}
\author[b]{Moo Hyun Lee}
\author[b]{Jaison Lee}
\author[d]{Jooyoung Lee}
\author[b]{Yoomin Oh}
\author[b]{Hyangkyu Park}
\author[b]{Kang Soon Park}
\author[e]{Kyungmin Seo}

\affil[a]{Korea Atomic Energy Research Institute, Deajeon, Korea}
\affil[b]{Center for Underground Physics, Institute for Basic Science (IBS), Daejeon, Korea}
\affil[c]{Chonnam National University, Gwangju, Korea}
\affil[d]{Kyungpook National University, Daegu, Korea}
\affil[e]{Sejong University, Seoul, Korea}
\affil[f]{Chung-ang University, Seoul, Korea}

% Please give the surname of the lead author for the running footer
\leadauthor{Han}

% Please include corresponding author, author contribution and author declaration information
% \authorcontributions{Please provide details of author contributions here.}
% \authordeclaration{Please declare any conflict of interest here.}
% \equalauthors{\textsuperscript{1}A.O.(Author One) and A.T. (Author Two) contributed equally to this work (remove if not applicable).}
\correspondingauthor{\textsuperscript{2}Corresponding author E-mail: byhan\@kaeri.re.kr}

% Keywords are not mandatory, but authors are strongly encouraged to provide them. If provided, please include two to five keywords, separated by the pipe symbol, e.g:
% \keywords{Keyword 1 $|$ Keyword 2 $|$ Keyword 3 $|$ ...} 

\begin{abstract}
Neutrino Experiment for Oscillation at Short baseline (NEOS) experiment had been conducted to interpret reactor anti-neutrino anomaly \cite{Mention:2011rk} and has carried out the feasibility study for monitoring the burning process of a nuclear power reactor \cite{Christopher}. Here we describe results of NEOS phase I data taken from summer 2015 to spring 2016 at Hanbit power plant in ROK and status of NEOS phase II experiment. 
\end{abstract}

% \dates{This manuscript was compiled on \today}
\doi{\url{https://neutrinos.llnl.gov/workshops/aap2018}}

% \begin{document}

\maketitle
\thispagestyle{firststyle}
\ifthenelse{\boolean{shortarticle}}{\ifthenelse{\boolean{singlecolumn}}{\abscontentformatted}{\abscontent}}{}

% If your first paragraph (i.e. with the \dropcap) contains a list environment (quote, quotation, theorem, definition, enumerate, itemize...), the line after the list may have some extra indentation. If this is the case, add \parshape=0 to the end of the list environment.
%\dropcap{T}his template is provided to help you summarize your work in the correct format.  Instructions for use are provided below. 

\section*{Results of NEOS Phase I \cite{Ko:2016owz}}

\begin{figure}[tbhp]
\centering
\includegraphics[width=1.0\linewidth]{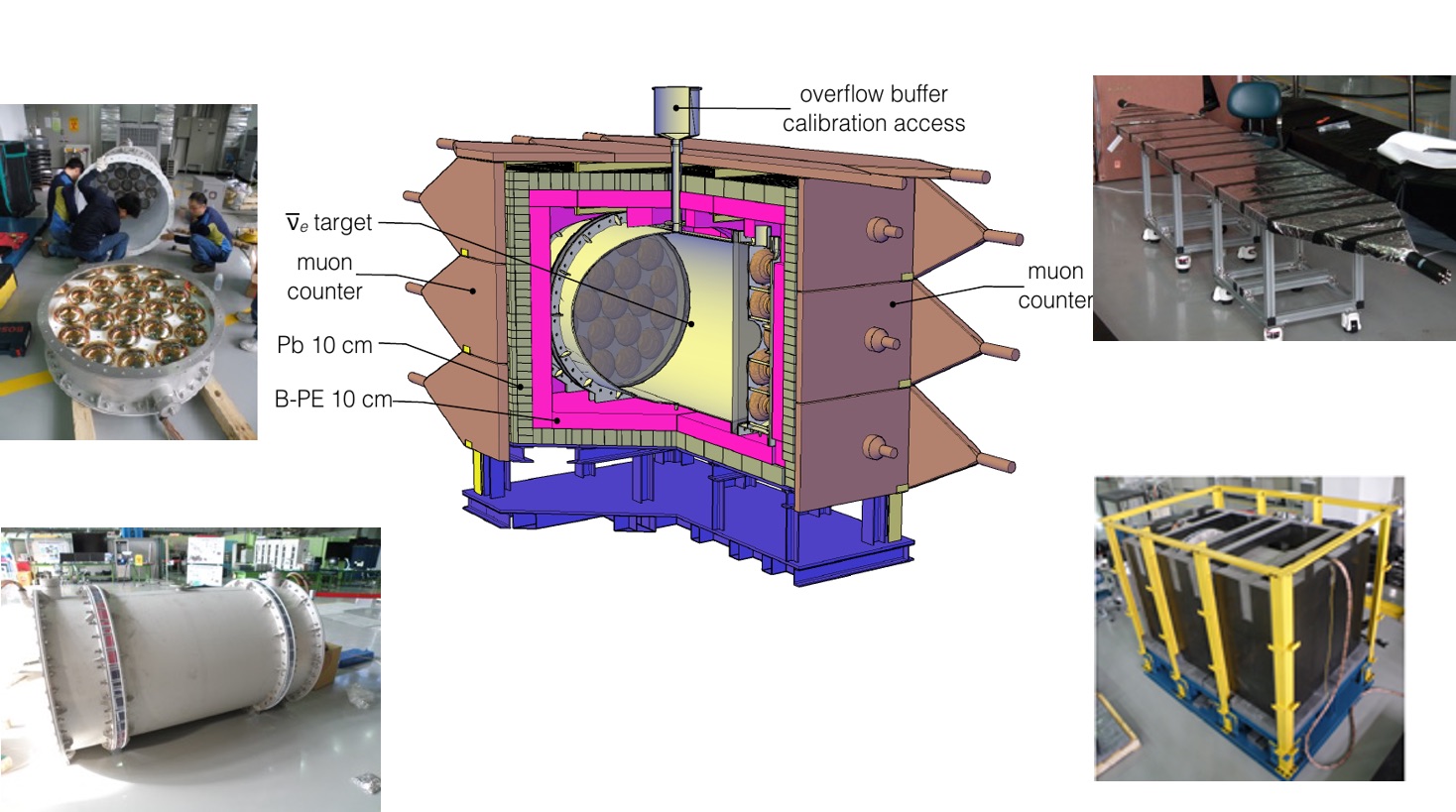}
\caption{NEOS detector.}
\label{fig:detector}
\end{figure}

In the NEOS experiment, the inverse beta decay (IBD) of anti-neutrinos in the Gd-loaded liquid scintillator refers to the process; $\bar{\nu} + p \rightarrow e^{+} + n$. The anti-neutrino reacting with a proton (Hydrogen atom) decays into a positron and a neutron. A prompt light signal is produced from the positron and the thermalized neutron travels for $\sim$30$\mu$s and being captured on the Gd in liquid scintillator and then, a gamma cascade of mean energy $\sim 8$ MeV is generated by the radioactive capture. The kinematical threshold of the IBD reaction due to the mass excess of the final state is 1.8 MeV for the antineutrino energy. The NEOS detector (Fig.\ref{fig:detector}) was designed with a steel cylindrical tank target filled with about 1008 L of 0.5\% Gd-loaded liquid scintillator (LAB + Ultima Gold F (DIN) 9:1) \cite{Kim:2015pba}. Nineteen of 8-inch PMTs are located at the left and right sides of the target. 100 mm of Pb and 100 mm of borated polyethylene layers are covering the target and shielding from backgrounds. A muon-veto detector with 50 mm thick plastic scintillators covers the steel structure and located in tendon gallery which is about 23 m baseline from the reactor core and $\sim$20 m.w.e overburden ($\sim$10 m below ground) for a good background shielding.   

\begin{figure}[tbhp]
\centering
\includegraphics[width=.8\linewidth]{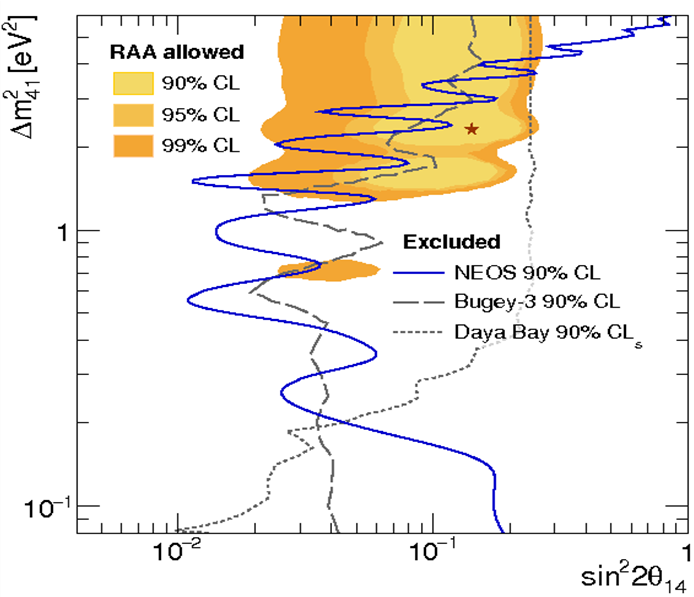}
\caption{Exclusion curves for 3+1 neutrino oscillations.}
\label{fig:sensitivity}
\end{figure}

NEOS detector simulation was developed with GEANT4 program \cite{Agostinelli:2002hh}. This simulation is based on exact physics models and an enormous particle database describes the signature of detector interacting with particles. The best parameters were tuned with data measured by the detector. In particular, the optical characteristics such as the light yield and attenuation length of liquid scintillator were optimized using the energy distributions of gammas. The optical characteristics strongly depend on the signal source positions and the parameters related with the source positions have to be empirically optimized. $^{137}$Cs, $^{60}$Co, $^{252}$Cf sources are used to calibrate the detector every week. Energy resolution ($\sigma/E_{\gamma} \sim 5\%$ at 1 MeV), gamma-ray escaping effect, and the nonlinear $Q$ to $E_{\gamma}$ response are well addressed to reconstruct energy spectra. The IBD measurement data were recorded and analyzed with an identical program frame of the simulation. Cosmogenic neutrons mimic the IBD candidate behavior with random coincidences of gammas (prompt signal). For a detector placed $\sim 10 m$ below ground, the overburden of reactor structure plays an important role in the background reduction. Additionally, a plastic muon veto detector surrounding the target is used to tag the induced background. Untagged fast neutrons generated by cosmic muons can be rejected using a pulse shape discrimination (PSD) in the liquid scintillator. 
The detector was tested during July-August 2015 and the unit 5 reactor was paused from Aug. 15 to Sep. 15 because of a nuclear reactor fuel replacement. The reactor restarted the ramping up from Sep. 16 and the anti-neutrino data were taken for 226 days. To reconstruct neutrino events, first of all, all events are vetoed for 150 s after accepting muon signals from the muon detector. The positron candidates are selected for 1 MeV $< E_{e^{+}} <$ 10 MeV, where $E_{e^{+}}$ is the positron energy and then the neutron signal is required to be within a time interval of 1 $< \Delta t <$ 30 s and its energy to be 4 MeV $< E_{n} <$ 10 MeV. The IBD count rate was 1946 ± 8 antineutrino interactions per day with the signal to noise ratio of $\sim$23. The p value of the $\chi^{2}$ difference between the 3$\nu$ hypothesis and the best fit for the 3 + 1 $\nu$ hypothesis is estimated to be 22\% using a large number of Monte Carlo data sets with statistical and systematic fluctuations. As a result, no significant evidence for the 3 + 1 $\nu$ hypothesis is found in Fig.\ref{fig:sensitivity}.

\section*{Status of NEOS Phase II}

\begin{figure}[tbhp]
\centering
\includegraphics[width=.9\linewidth]{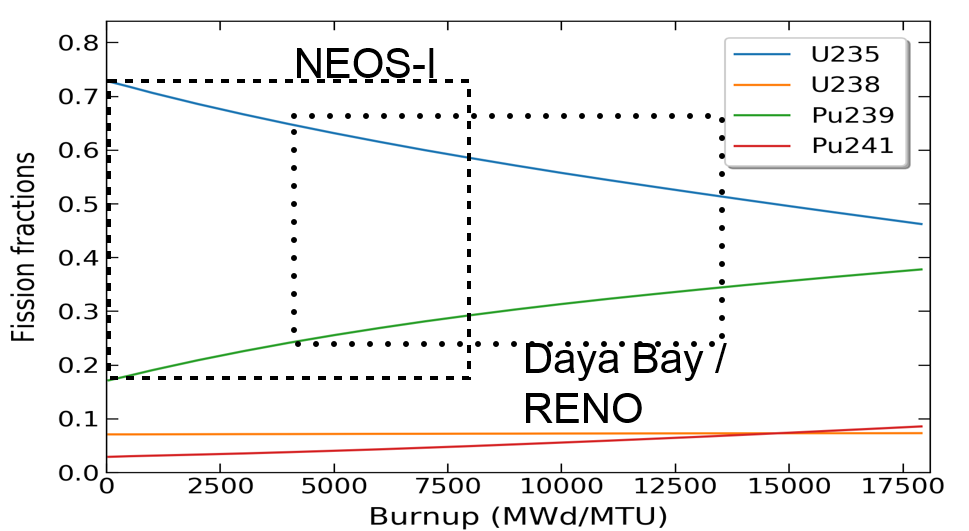}
\caption{Fission fractions of the fissile isotopes for one fuel cycle.}
\label{fig:phase2}
\end{figure}

In May 2016, NEOS detector was paused and removed from tendon gallery due to reactor regular inspection. New phase of NEOS experiment (NEOS II) was proposed with same detector and site for one full fuel cycle. Finally, NEOS II run resumed in Sep. 2018 with new liquid scintillator and minor changes in detector structure.
Antineutrinos from nuclear reactors are produced by the $\beta$-decay of fission fragments into more stable nuclei: The two main fissile isotopes contained in the fuel of nuclear reactor are $^{235}$U and $^{239}$Pu. The $^{239}$Pu is produced by neutron captures in the original $^{238}$U followed by two consecutive $\beta$-decays: $^{239}$U $\rightarrow$ $^{239}$Np $\rightarrow$ $^{239}$Pu. The relative contribution to the total number of fissions induced by these two isotopes changes over time: it increases for $^{239}$Pu while decreasing for $^{235}$U. This is called the “burn-up” effect. The remaining fissions of $^{241}$Pu and fast neutron induced fissions of $^{238}$U share about 10\% of the reactor power. Because the number of emitted neutrinos and their mean energy depend on the fissile isotopes, the differential energy cross-section of emitted neutrinos can provide a direct information of the burn-up for a nuclear reactor. NEOS Phase II has identical detector and experimental spot to the Phase I. The data of NEOS phase II experiment will be taken at least for one full fuel cycle ($\sim 500$ days) from Sep. 2018 (Fig. \ref{fig:phase2})and further study for spectrum evaluation with the fissile isotopes changes can be performed as a clue to address the observed 5 MeV excess and fine structure of energy spectrum.

\acknow{This work was supported by the National Research Foundation of Korea (NRF) Grant funded by the Korea goverment (MSIP) (NRF-2017M2A2A6A05018529).}

%\subsection*{References}

%References should be cited in numerical order as they appear in text; this will be done automatically via bibtex, e.g.  

% \matmethods{Please describe your materials and methods here. This can be more than one paragraph, and may contain subsections and equations as required. Authors should include a statement in the methods section describing how readers will be able to access the data in the paper. 

% \subsection*{Subsection for Method}
% Example text for subsection.
% }

% \showmatmethods{} % Display the Materials and Methods section

\showacknow{} % Display the acknowledgments section

% Bibliography
% \bibliography{aap2018-participant00x}

% \end{document}

%% file: AAP-Shitov/aap2018-participant00x.tex
% \documentclass[9pt,twocolumn,twoside,lineno]{aap2018}

% \templatetype{aap2018proceedings} % Choose template 
% {pnasresearcharticle} = Template for a two-column research article
% {pnasmathematics} %= Template for a one-column mathematics article
% {pnasinvited} %= Template for a PNAS invited submission

\title{Status of the DANSS project}

% Use letters for affiliations, numbers to show equal authorship (if applicable) and to indicate the corresponding author
\author[a]{Yury Shitov}
\author[ ]{the DANSS collaboration}

%\author[b,1,2]{Author Two} 
%\author[a]{Author Three}

\affil[a]{Joint Institute for Nuclear Research, Dubna, Russia}
%\affil[b]{Affiliation Two}
%\affil[c]{Affiliation Three}

% Please give the surname of the lead author for the running footer
\leadauthor{Shitov} 

% Please include corresponding author, author contribution and author declaration information
% \authorcontributions{Please provide details of author contributions here.}
% \authordeclaration{Please declare any conflict of interest here.}
% \equalauthors{\textsuperscript{1}A.O.(Author One) and A.T. (Author Two) contributed equally to this work (remove if not applicable).}
\correspondingauthor{\textsuperscript{2}E-mail: shitov@jinr.ru}

% Keywords are not mandatory, but authors are strongly encouraged to provide them. If provided, please include two to five keywords, separated by the pipe symbol, e.g:
% \keywords{Keyword 1 $|$ Keyword 2 $|$ Keyword 3 $|$ ...} 

\begin{abstract}
The DANSS collaboration has built a highly segmented compact m$^{3}$ neutrino spectrometer (using PS-Gd technology), which is able to work safely close to an industrial nuclear reactor (NR). It is used to monitor the NR (applied task),  as well as to probe short baseline oscillations to the sterile neutrino state (fundamental research). The first results of the oscillation analysis are presented.

\end{abstract}

% \dates{This manuscript was compiled on \today}
\doi{\url{https://neutrinos.llnl.gov/workshops/aap2018}}

% \begin{document}

\maketitle
\thispagestyle{firststyle}
\ifthenelse{\boolean{shortarticle}}{\ifthenelse{\boolean{singlecolumn}}{\abscontentformatted}{\abscontent}}{}

% If your first paragraph (i.e. with the \dropcap) contains a list environment (quote, quotation, theorem, definition, enumerate, itemize...), the line after the list may have some extra indentation. If this is the case, add \parshape=0 to the end of the list environment.

\dropcap{M}ost of the neutrino oscillation results are well covered by the three-component neutrino theory (PMNS-matrix). However several anomalies in short baseline oscillation data (Reactor Neutrino Anomalies ({\bf RAA})~\cite{Mueller:2011nm}, controversial results in the GALLEX and the SAGE (Gallium Anomaly, {\bf GA})~\cite{Abdurashitov:2005tb,Giunti:2010zu}, the LSND~\cite{Athanassopoulos:1996jb}, and the MiniBoone~\cite{Aguilar-Arevalo:2013pmq} experiments) could be interpreted by invoking a hypothetical fourth neutrino. This fourth neutrino is not involved in the standard interactions (hence the term "sterile"), but mix with the others. Expected phase space range of this mixing $\Delta m^2$ $\in$ [0.1-10] eV$^2$, sin$^2$(2$\theta$) $\in$ [0.001-0.01] was determined from the global fit of the available experimental data with the best value $\Delta m^2$ $\sim$ 2 eV$^2$, sin$^2$(2$\theta$) $\sim$ 0.1. Recent experimental hints of possible existence of this new fundamental physics have boosted a huge activity toward experiment tests of this hypothesis in different directions of physical researches.

The DANSS~\cite{Alekseev:2018ijk} is one of the leading short baseline (SBL) reactor projects, which is measuring antineutrinos by the inverse beta decay (IBD) method at distances of 10.7-12.7 m from the industrial reactor, taking data since October 2016. The search for oscillations in sterile neutrino is carried out through the analysis of the ratios of the IBD-positron spectra collected at different distances from the reactor. This relative method is free from systematic errors associated with the calculation of the reactor antineutrino spectra. This paper presents the preliminary results of the analysis of annual measurement statistics (almost 1 million of antineutrino events).

\section*{The DANSS detector}

The highly segmented (2500 1 x 4 x 100 cm$^3$ strips made of plastic scintillator viewed by 2500 SiPMs and 50 PMTs) compact DANSS detector covered by multilayer passive shield and active $\mu$-veto (a detailed description is in~\cite{Alekseev:2016llm}) is mounted under the Unit~No.4 (3.1 GW$_{th}$) of the Kalinin NPP (KNPP) on a mobile platform. Data are taken at three distances 10.7 m (Up), 11.7 m (Middle), and 12.7 m (Down) from the reactor (center to center) changed sequentially with a full cycle of passage through 3 positions in a week.

\subsection*{Background condition and monitoring}

Permanent background monitoring is carried out for $\gamma$-flux (by four 3 ’x 3’ NaI detectors: one inside and three outside the DANSS shield) and for thermal neutron flux (by three $^3$He counters: one inside and two outside the DANSS shield). In addition to this, periodically the $\gamma$-flux is measured by the HPGe-detector and the ($\theta$,$\phi$) 2D-map of intensity of $\mu$-flux is measuring using a specially designed $\mu$-meter. It is important that the reactor and water storage for spent fuel under the DANSS spectrometer provide $\sim$ 50 m.w.e. protection against cosmic rays.

\subsection*{Signal signature}

The IBD process  $\bar{\nu} + p \rightarrow e^{+} +n + 1.81 MeV$ is using to detect the antineutrinos. The positron gives the first (fast, prompt) hit, followed by the second (delayed) signal in [0-100]~$\mu$s window from the thermalized neutron captured by gadolinium, introduced into the strip coating ($\sim$~0.35~\% w.r.t. the whole mass). Double IBD-signature provides excellent background suppression.

\subsection*{Calibrations}

Various time and energy calibrations are performed regularly using a number of sources: cosmic muons, $^{22}$Na, $^{60}$Co, $^{137}$Cs, and $^{248}$Cm.

\section*{Data analysis and results}

\subsection*{The cuts}

The detailed description of the selection criteria is presented elsewhere~\cite{Alekseev:2018ijk}, the basic cuts are: the prompt signal is $\geq$~1 MeV, the delayed signal is in [2,50]~$\mu$s window and $\geq$~3.5 MeV, and no muon veto signal in 60 $\mu$s before the prompt signal.

\begin{figure}[tbhp]
\centering
\includegraphics[width=.95\linewidth]{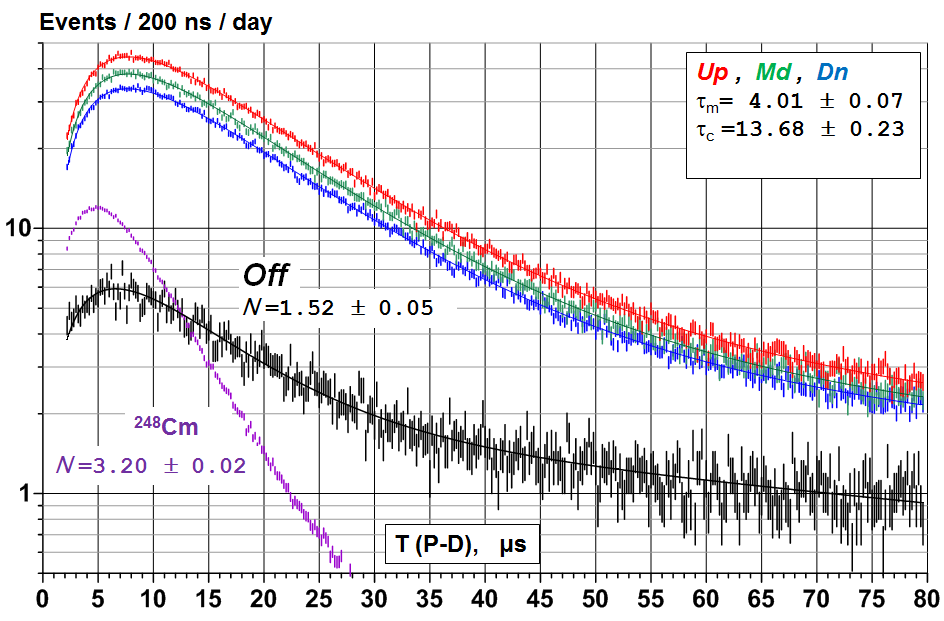}
\caption{Time between fast and delayed signal in the DANSS spectrometer: reactor ON data at up (red), middle (green), and down (blue) positions, as well as reactor OFF (black) and neutron calibration (magenta) runs with $^{248}$Cm source.}
\label{Shitov:Figure1}
\end{figure}

\subsection*{The time spectra}

The time curves between the fast and the delayed signals are shown in Fig.\ref{Shitov:Figure1}. The shapes are identical for the three positions of the reactor ON data and differ from the reactor OFF and neutron calibration data.

\subsection*{The random coincidences}

The accidental background is measuring directly by selecting positron-like and neutron-like signals outside the time coincidence window and is subtracted from the total IBD-spectra of reactor ON/OFF data.

\subsection*{The energy spectra}

\begin{figure}[tbhp]
\centering
\includegraphics[width=.8\linewidth]{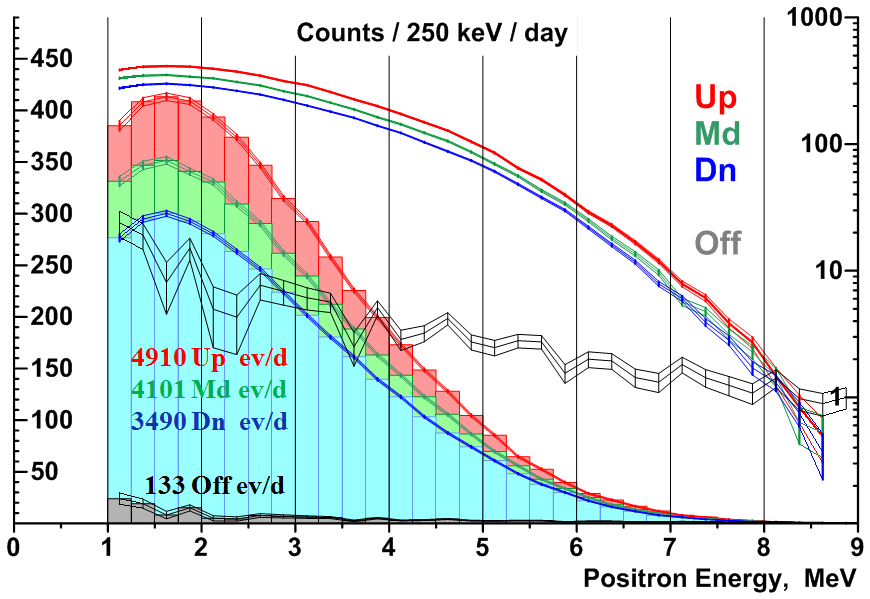}
\caption{Positron energy spectra taken by the DANSS spectrometer in linear/log scales (left/right vertical axis respectively): reactor ON data at up (red), middle (green), and down (blue-cyan) positions, as well as reactor OFF (black) data. Statistical errors are given only. The accidental background is subtracted.}
\label{Shitov:Figure2}
\end{figure}

\begin{figure}[tbhp]
\centering
\includegraphics[width=.8\linewidth]{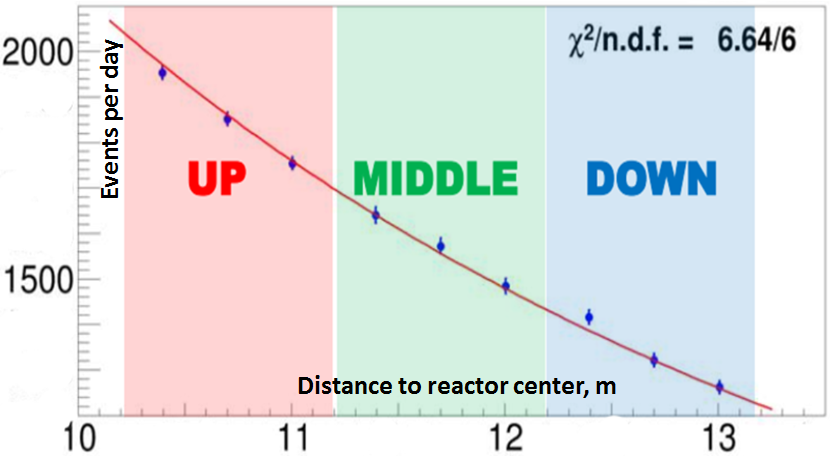}
\caption{The intensity of the IBD-like events vs. distance.}
\label{Shitov:Figure3}
\end{figure}

The energy spectra of positrons are shown in Fig.~\ref{Shitov:Figure2}
As can be seen from Fig.~\ref{Shitov:Figure3}, the IDB-signal intensity follows reasonably well to the expected dependence 1/L$^2_{EFF}$. Where the effective distance L$_{EFF}$ is calculated taking into account the burning profile of the reactor core, monitored with an accuracy of 10~cm every 30 minutes.

\begin{figure*}[b]
\centering
\includegraphics[width=.8\linewidth]{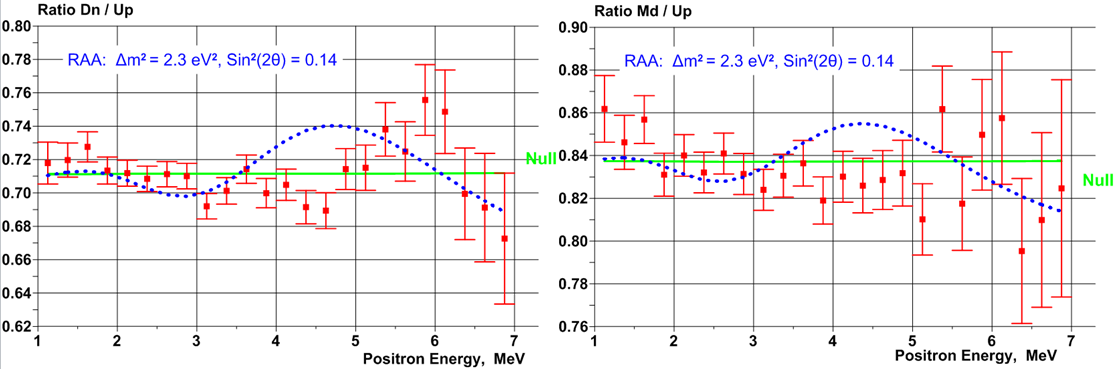}
\caption{Ratios of positron energy spectra measured at different positions: Down/Top (left) and Middle/Top. While the green line represents no oscillation case, the blue dotted curve corresponds to 3+1 sterile mixing mode for the RAA best fit point~\cite{Mueller:2011nm}.}
\label{Shitov:Figure4}
\end{figure*}

\subsection*{The oscillation test}
\begin{figure}[h]
\centering
\includegraphics[width=.95\linewidth]{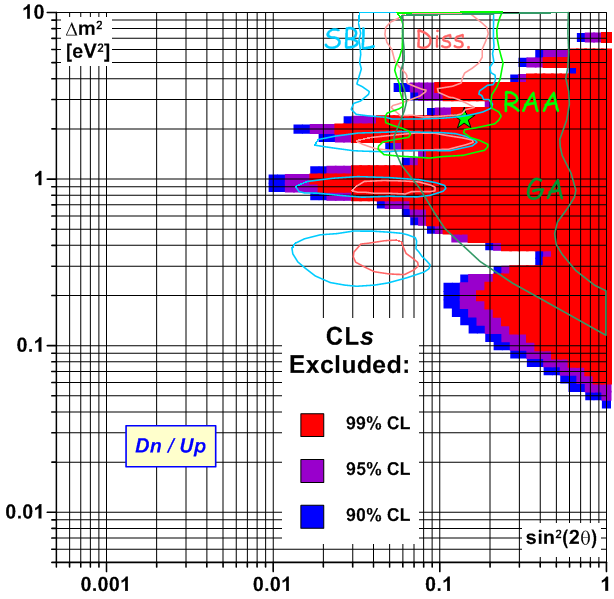}
\caption{CL exclusion plot in the sterile neutrino 3+1 model phase space. The shaded exclusion area is set by this analysis. Curves show allowed oscillation regions~\cite{Mention:2011rk} with the green star pointed the RAA+GA best fit point~\cite{Kopp:2013vaa}.}
\label{Shitov:Figure5}
\end{figure}

In order to test the hypothesis of oscillations into sterile neutrinos, we have analyzed the ratios of positron spectra measured at different distances (Fig.\ref{Shitov:Figure4}). This allowed us to exclude a significant part of the phase space of oscillations (Fig.\ref{Shitov:Figure5}), including the RAA+GA best fit point~\cite{Kopp:2013vaa}. Our best fit point $\Delta m^2 = 1.4 eV^2, sin^2(2\theta)=0.05$ is not yet significant enough and will be further validated on higher statistics.

\acknow{The DANSS collaboration is grateful to the directorates of ITEP, JINR, KNPP administration and Radiation and Nuclear Safety Departments for constant support of this work. The detector construction was supported by the Russian State Corporation ROSATOM (state contracts H.4x.44.90.13.1119 and H.4x.44.9B.16.1006). The operation and data analysis is partially supported by the Russian Science Foundation, grant 17-12-01145.}

\showacknow{} % Display the acknowledgments section

% % Bibliography
% \bibliography{aap2018-participant00x}
  
% \end{document}

%% file: AAP-Serebrov/aap2018-participant00x.tex
% \documentclass[9pt,twocolumn,twoside,lineno]{aap2018}

% \templatetype{aap2018proceedings} % Choose template 
% % {pnasresearcharticle} = Template for a two-column research article
% % {pnasmathematics} %= Template for a one-column mathematics article
% {pnasinvited} %= Template for a PNAS invited submission

\title{The first observation of effect of oscillation in Neutrino-4 experiment on search for sterile neutrino}

% Use letters for affiliations, numbers to show equal authorship (if applicable) and to indicate the corresponding author
\author[1,*]{A.P. Serebrov}
\author[1]{V.G. Ivochkin} 
\author[1]{R.M. Samoilov}
\author[1]{A.K. Fomin}
\author[1]{A.O. Polyushkin}
\author[1]{V.G. Zinoviev}
\author[1]{P.V. Neustroev}
\author[1]{V.L. Golovtsov}
\author[1]{A.V. Chernyj}
\author[1]{O.M. Zherebtsov}
\author[1]{M.E. Chaikovskii}
\author[2]{V.P. Martemyanov}
\author[2]{V.G. Tarasenkov}
\author[2]{V.I. Aleshin} 
\author[3]{A.L. Petelin}
\author[3]{A.L. Izhutov}
\author[3]{A.A. Tuzov}
\author[3]{S.A. Sazontov}
\author[3]{M.O. Gromov}
\author[3]{V.V. Afanasiev}
\author[1, 4]{M.E. Zaytsev}
\author[1]{A.A.Gerasimov}
\author[4]{D.K. Ryazanov}

\affil[1]{NRC “KI” Petersburg Nuclear Physics Institute, Gatchina}
\affil[2]{NRC “Kurchatov Institute”, Moscow}
\affil[3]{JSC “SSC Research Institute of Atomic Reactors”, Dimitrovgrad, Russia}
\affil[4]{Dimitrovgrad Engineering and Technological Institute MEPhI, Dimitrovgrad, Russia}
% Please give the surname of the lead author for the running footer
\leadauthor{Serebrov}

% Please include corresponding author, author contribution and author declaration information
% \authorcontributions{Please provide details of author contributions here.}
% \authordeclaration{Please declare any conflict of interest here.}
% \equalauthors{\textsuperscript{1}A.O.(Author One) and A.T. (Author Two) contributed equally to this work (remove if not applicable).}
\correspondingauthor{\textsuperscript{*}serebrov\_ap@pnpi.nrcki.ru}

% Keywords are not mandatory, but authors are strongly encouraged to provide them. If provided, please include two to five keywords, separated by the pipe symbol, e.g:
% \keywords{Keyword 1 $|$ Keyword 2 $|$ Keyword 3 $|$ ...} 

% \dates{This manuscript was compiled on \today}
\doi{\url{https://neutrinos.llnl.gov/workshops/aap2018}}

% \begin{document}
\begin{abstract}
Model independent analysis of Neutrino-4 experiment spectra exclude area of reactor and gallium anomaly at C.L more than $3\sigma$ for $\Delta m_{14}^2 < 4 \text{eV}^2$. However, we observed an oscillation effect at $3\sigma$ C.L. in vicinity of $\Delta m_{14}^2\approx7.34 \text{eV}^2$ and $\sin^22\theta_{14}\approx0.4$. 
\end{abstract}
\maketitle
\thispagestyle{firststyle}
\ifthenelse{\boolean{shortarticle}}{\ifthenelse{\boolean{singlecolumn}}{\abscontentformatted}{\abscontent}}{}

\begin{figure*}
\centering
\includegraphics[width=17.8cm]{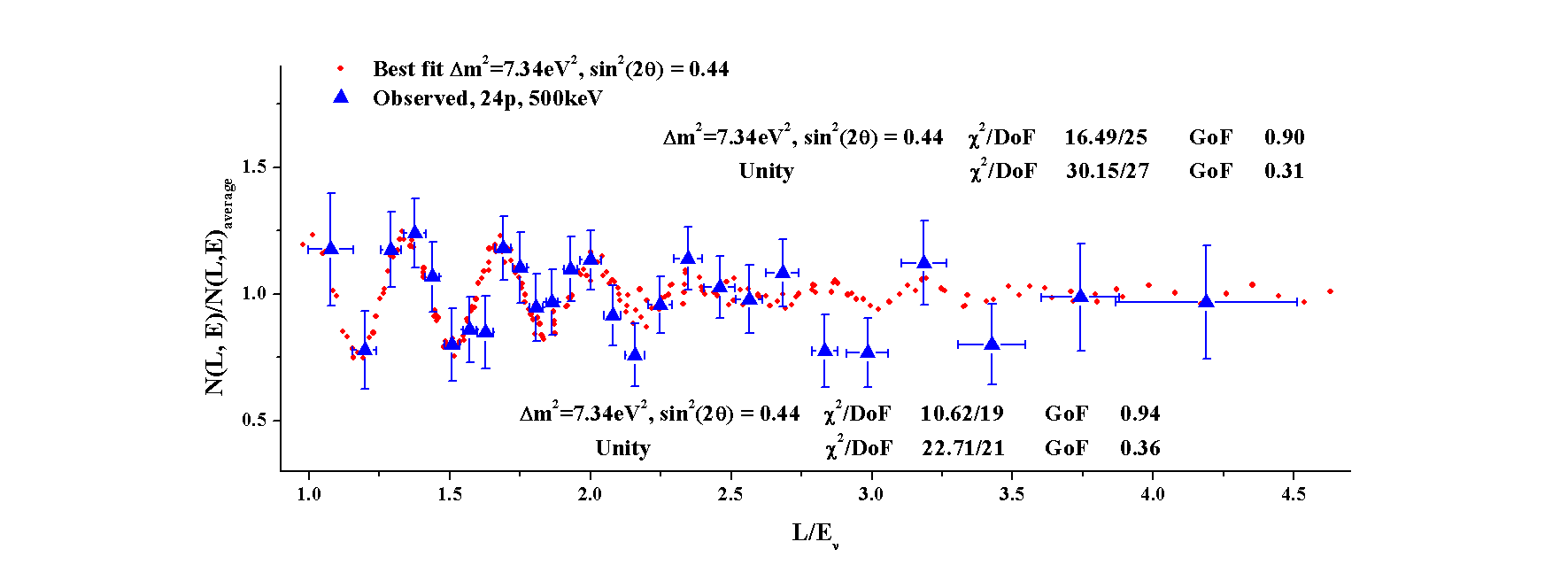}
\caption{Coherent addition of the experimental result with data selection by variable L/E for direct observation of antineutrino oscillation. Comparison of left (blue triangles) and right (red dots, with optimal oscillation parameters) parts of equation (\ref{eqn:one}).}
\label{fig:cohAdd}
\end{figure*}

% If your first paragraph (i.e. with the \dropcap) contains a list environment (quote, quotation, theorem, definition, enumerate, itemize...), the line after the list may have some extra indentation. If this is the case, add \parshape=0 to the end of the list environment.
\dropcap{O}ur experiment focuses on the task of exploring Reactor Antineutrino Anomaly \cite{Mueller:2011nm} and possible existence of a sterile neutrino \cite{Mention:2011rk,Gariazzo:2017fdh} at certain confidence level. The hypothesis of oscillation can be verified by direct measurement of the antineutrino flux and spectrum vs. distance at short 6 – 12m distances from the reactor core. \\
Due to small core size ($42\times42\times35\text{cm}^3$), high power (100MW) and other peculiar characteristics reactor SM-3 provides the most favorable conditions to search for neutrino oscillations at short distances \cite{Serebrov:2015txp,Serebrov:2015ros}. However, it is located on Earth surface, therefore cosmic background is major problem for the experiment.\\
The detector has sectional structure (5x10 sections). Gadolinium (0.1\% concentration) loaded liquid scintillator is using. For carrying out measurements, the detector has been moved to various positions at the distances divisible by section size. As a result, different sections can be placed at the same coordinates with respect to the reactor except for the edges at closest and farthest positions.\\
Measurements with the detector have started in June 2016. Measurements with the reactor ON were carried out for 480 days, and with the reactor OFF for 278 days. In total, the reactor was switched on and off 58 times. We carried out model independent analysis using equation (\ref{eqn:one}), where numerator is the rate of antineutrino events with correction to geometric factor $1/L^2$ and denominator is its value averaged over all distances:
\begin{figure}[ht!]
\centering
\includegraphics[width=.7\linewidth, height=0.53\textheight]{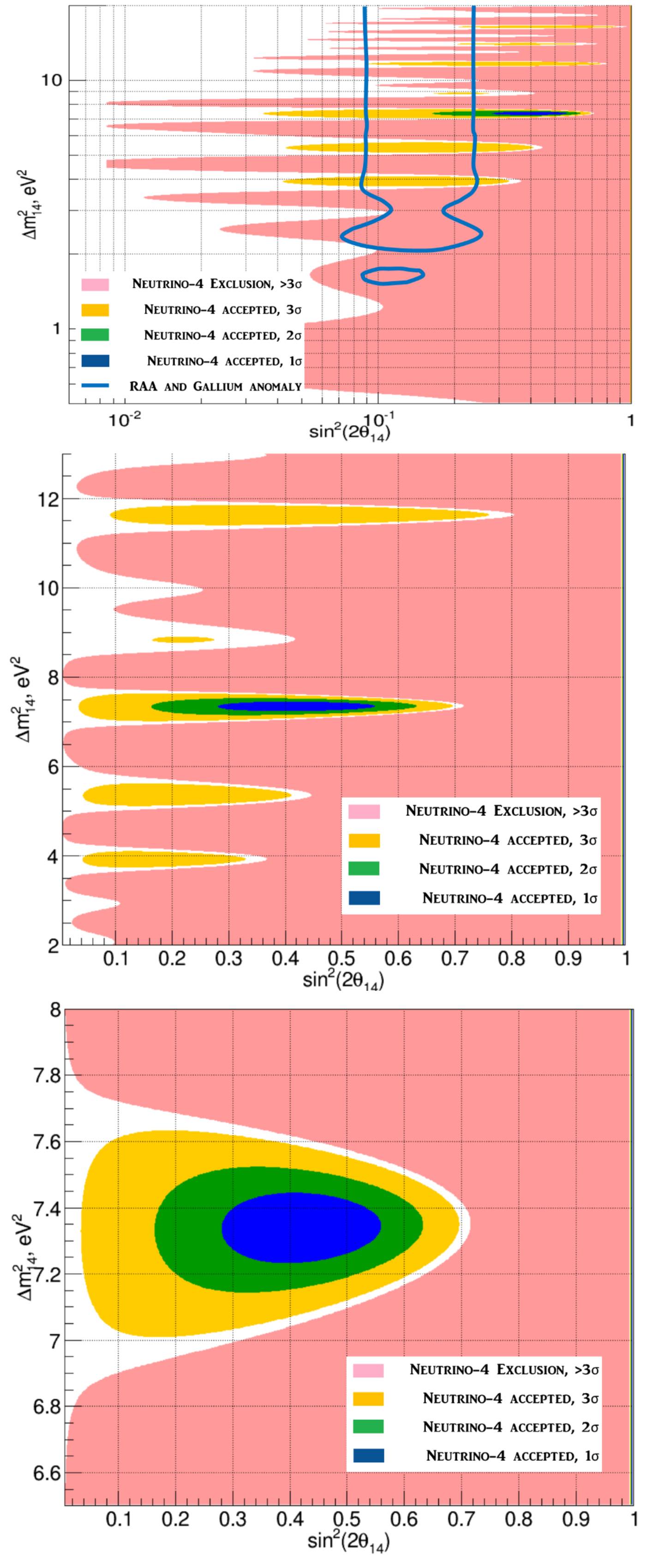}
\caption{top – Restrictions on parameters of oscillation into sterile state with 99.73\% CL (pink), area of acceptable with 99.73\% C.L. values of the parameters (yellow), area of acceptable with 95.45\% C.L. values of the parameters (green), area of acceptable with 68.30\% C.L. values of the parameters (blue).  middle – Area around central values in linear scale and significantly magnified, bottom – even further magnified central part}
\label{fig:dChiSq}
\end{figure}\
\begin{eqnarray}
\frac{N_{i,k} \pm \Delta N_{i,k}L_{k}^{2}}{K^{-1}\sum\limits_k^K(N_{i,k} \pm \Delta N_{i,k})L_{k}^{2}} = \nonumber\\
\frac{1-\sin^22\theta_{14}\sin^2(1.27\Delta m^2_{14}L_k/E_i)}{K^{-1}\sum\limits_k^K(1-\sin^22\theta_{14}\sin^2(1.27\Delta m^2_{14}L_k/E_i))}
 \label{eqn:one}
\end{eqnarray}
Equation (\ref{eqn:one}) is model independent because left part includes only experimental data $k=1\dots K, K = 24$ for all distances in range 6.5-11.7m; $i=1\dots9$ corresponding to 500keV energy intervals in range 1.5MeV to 6.0MeV. The right part is the same ratio obtained within oscillation hypothesis. In right part of the equation energy spectrum is completely canceled out. It should be emphasized, that spectrum shape does not influence the expression, because it appears in equation (\ref{eqn:one}) in numerator and denominator. The results of the analysis of optimal parameters $\Delta m^2_{14}$ and $\sin^22\theta_{14}$, using  method $\chi^2$ are shown in fig.\ref{fig:dChiSq}.\\
% \begin{figure}[h!]
% \centering
% \includegraphics[width=.7\linewidth, height=0.53\textheight]{deltaChiSq.png}
% \caption{top – Restrictions on parameters of oscillation into sterile state with 99.73\% CL (pink), area of acceptable with 99.73\% C.L. values of the parameters (yellow), area of acceptable with 95.45\% C.L. values of the parameters (green), area of acceptable with 68.30\% C.L. values of the parameters (blue).  middle – Area around central values in linear scale and significantly magnified, bottom – even further magnified central part}
% \label{fig:dChiSq}
% \end{figure}
The area of oscillation parameters colored in pink are excluded with C.L. more than 99.73\% ($>3\sigma$). However, in area $\Delta m_{14}^2\approx7 \text{eV}^2$ and $\sin^22\theta_{14}\approx0.4$ the oscillation effect is observed at C.L. 99\% ($3\sigma$), and it is followed by a few satellites. Minimal value $\chi^2$ occurs at $\Delta m_{14}^2\approx7.34 \text{eV}^2$ and $\sin^22\theta_{14}\approx0.44$. \\
It is beneficial to make experimental data selection using L/E parameter. This method we call the coherent summation of the experimental results with data selection using variable L/E and it provides direct observation of antineutrino oscillation.\\
For this purpose, we used 24 distance points (with 23.5 cm interval) and 9 energy points (with 0.5MeV interval). The selection for left part of equation (\ref{eqn:one}) (of total 216 points each 8 points are averaged) is shown in fig. \ref{fig:cohAdd} with blue triangles.\\
Same selection for right part of equation (\ref{eqn:one}) with most probable parameters $\Delta m_{14}^2\approx7.34 \text{eV}^2$ and $\sin^22\theta_{14}\approx0.44$ is also shown in fig. \ref{fig:cohAdd} with red dots. Fit with such parameters has goodness of fit 90\%, while fit with a constant equal to one (assumption of no oscillations) has goodness of fit only 31\%. It is important to notice that attenuation of sinusoidal process for red curve in area L/E > 2.5 can be explained by taken energy interval 0.5MeV. Considering the smaller interval 0.25MeV we did not obtain increasing of oscillation area of blue experimental, because of insufficient energy resolution of the detector in low energy region. Thus, the data obtained in region L/E > 2.5 do not influence registration of oscillation process. Using first 21 points in analysis, we obtained new goodness of fit which are shown under the curve in fig. \ref{fig:cohAdd}. As a result, goodness of fit increased to 94\%.\\ 
The result of presented analysis can be summarized in several conclusions. Area of reactor and gallium anomaly for $\Delta m_{14}^2 < 4 \text{eV}^2$  and $\sin^22\theta_{14} > 0.1$ is excluded at C.L. more than 99.7\% ($>3\sigma$). However, oscillation effect is observed in area $\Delta m_{14}^2\approx7.34 \text{eV}^2$ and $\sin^22\theta_{14}\approx0.44$  with 99.7\% C.L ($3\sigma$) and it is located in upper area of reactor and gallium anomaly. In general, it seems that the effect predicted in gallium and reactor experiments is confirmed but at sufficiently large value of $\Delta m_{14}^2$. However, confidence level is not sufficient. Therefore, increasing of experimental accuracy is essential as well as additional analysis of possible systematic errors of the experiment.\\
Experiment Neutrino-4 has some advantages in sensitivity to big values of  $\Delta m_{14}^2$ owing to a compact reactor core, close minimal detector distance from the reactor and wide range of detector movements. Next highest sensitivity to large values of  $\Delta m_{14}^2$ belongs to PROSPECT\cite{Ashenfelter:2018zdm} experiment. Currently its sensitivity is two times lower than Neutrino-4 sensitivity, but it recently has started data collection, so it possibly can confirm or refute our result.
Below we discuss the future prospects of Neutrino-4 experiment. Increasing of experimental accuracy is required. For that reason, the improvement of current setup and creation of new neutrino lab with new detector system at SM-3 reactor is planned.\\
Firstly, the improvement of current setup requires replacing of currently used scintillator with a new highly efficient liquid scintillator with capability of pulse-shape discrimination, and with an increased concentration of gadolinium up to 0.5\%. It is expected that the accidental coincidence background will be reduced by factor of 3 and measurement accuracy will be doubled. Moreover, anti-coincidence shielding will be increased. The project is planned to be implemented with participation of colleagues from Joint Institute for Nuclear Research (JINR) and NEOS collaboration. 

% Uncomment following for example of equation spanning two columns :
% \begin{figure*}[bt!]
% \begin{align*}
% (x+y)^3&=(x+y)(x+y)^2\\
%       &=(x+y)(x^2+2xy+y^2) \numberthis \label{eqn:example} \\
%       &=x^3+3x^2y+3xy^3+x^3. 
% \end{align*}
% \end{figure*}

% \matmethods{Please describe your materials and methods here. This can be more than one paragraph, and may contain subsections and equations as required. Authors should include a statement in the methods section describing how readers will be able to access the data in the paper. 

% \subsection*{Subsection for Method}
% Example text for subsection.
% }

% \showmatmethods{} % Display the Materials and Methods section

\acknow{The authors are grateful to the Russian Foundation of Basic Research for support under Contract No. 14-22-03055-ofi\_m. Authors are grateful to Y.G.Kudenko, V.B.Brudanin, V.G.Egorov, Y.Kamyshkov and V.A.Shegelsky for beneficial discussion of experimental results. The delivery of the scintillator from the laboratory headed by Prof. Jun Cao (Institute of High Energy Physics, Beijing, China) has made a considerable contribution to this research. }

\showacknow{} % Display the acknowledgments section

% Bibliography
% \bibliography{aap2018-participant00x}

% \end{document}

%% file: AAP-Onillon/aap2018-participant00x.tex
% \documentclass[9pt,twocolumn,twoside,lineno]{aap2018}
% \usepackage{array}
% \usepackage{makecell}
% \usepackage{multirow}

% \templatetype{aap2018proceedings} % Choose template 

\title{Investigation of the ILL spectra normalization}

\author[a,b,1]{A. Onillon}
\author[a,2]{A. Letourneau} 

\affil[a]{Irfu, CEA, Université Paris-Saclay, F-91191 Gif-sur-Yvette, France}

\affil[b]{Den-Service d'\'etudes des r\'eacteurs et de math\'ematiques
appliqu\'ees (SERMA), CEA, Universit\'e Paris-Saclay, F-91191,
Gif-sur-Yvette, France}

% Commissariat à l’Energie Atomique et aux Energies Alternatives

\leadauthor{Onillon} 
\correspondingauthor{\textsuperscript{1}anthony.onillon@cea.fr, \textsuperscript{2}alain.letourneau@cea.fr}

\begin{abstract}

In 2011, two independent reevaluations of the reference reactor $\bar\nu_e$ spectra were published \cite{Mueller:2011nm,Huber:2011wv}. 
The comparison of the reactor $\bar\nu_e$ flux measured by short baseline experiments to the predictions using the new spectra leads to the so-called Reactor Antineutrino Anomaly or RAA \cite{Mention:2011rk}: 
an overall (5.7\,$\pm$\,2.3)\% flux deficit between the measurements and the expectations. 
In this proceeding we present an on-going work investigating the normalization of the integral beta spectrum of fissionable isotopes performed at the ILL and from which the reference $\bar\nu_e$ spectra are derived. 
Preliminary results of this investigation using MCNPX-2.5 \cite{MCNPX} and TRIPOLI-4$^{\mbox{\scriptsize{\textregistered}}}$ \cite{Brun:2015aul} simulations of the HFR reactor are presented. 

\end{abstract}

% \dates{This manuscript was compiled on \today}
\doi{\url{https://neutrinos.llnl.gov/workshops/aap2018}}

% \begin{document}

\maketitle
\thispagestyle{firststyle}
\ifthenelse{\boolean{shortarticle}}{\ifthenelse{\boolean{singlecolumn}}{\abscontentformatted}{\abscontent}}{}

% If your first paragraph (i.e. with the \dropcap) contains a list environment (quote, quotation, theorem, definition, enumerate, itemize...), the line after the list may have some extra indentation. If this is the case, add \parshape=0 to the end of the list environment.

\dropcap{R}eactor $\bar\nu_e$ experiments use the huge and pure $\bar\nu_e$ flux emitted by the successive beta decay of unstable fission products following the fission of uranium and plutonium elements. 
In experiments using research reactors, the flux is dominated by the sole fission of $^{235}$U isotope while in experiments using commercial reactors the flux is dominated (>80\%) by the fission of $^{235}$U and $^{239}$Pu isotopes with a smaller contribution  coming from the $^{238}$U and $^{241}$Pu isotopes.

Reactor flux models consist in the convolution of a detector model and a reactor $\bar\nu_e$ flux prediction. The $\bar\nu_e$ fission spectra used as inputs for the prediction \cite{Mueller:2011nm,Huber:2011wv} derived from the conversion of the integral beta spectra measurements performed in the 80's at the HFR reactor for the isotopes of $^{235}$U, $^{239}$Pu and $^{241}$Pu \cite{Schreckenbach:1981wlm,VonFeilitzsch:1982jw,Schreckenbach:1985ep,Hahn:1989zr}. For $^{235}$U, two measurements were performed. In \cite{Mueller:2011nm,Huber:2011wv}, the second measurement is used as a reference as it exhibits smaller uncertainties. For $^{238}$U, the prediction can be done using summation calculations \cite{Mueller:2011nm} or conversion calculations of the integral beta spectra measurement performed in 2013 at the FRM-II reactor \cite{Haag:2013raa}.
Normalization uncertainties associated to the reactor flux model are dominated by the uncertainties associated to the reference $\bar\nu_e$ spectra.

Since it was pointed out, there has been much debate about the origin of the RAA \cite{Mention:2011rk}. 
This deficit between the measurements and the expected reactor $\bar\nu_e$ rates can be either explained in terms of sterile neutrinos or by an underestimation of the uncertainties associated to the reference $\bar\nu_e$ spectra. 
Recently, the Daya Bay experiment observed with a high precision the reactor fuel burnup \cite{An:2017osx} by measuring a decrease of the $\bar\nu_e$ rate with the accumulation of plutonium isotopes in the depleted fuel. 
This analysis pointed out another anomaly consisting in an incompatible decrease of the observed rate compared to the expectation. 
This new anomaly can be interpreted as a inconsistency between the relative normalization of the reference $\bar\nu_e$ spectra with the measurement.

\subsection*{Reference ILL spectra}

The \textit{High Flux Reactor} (HFR) is installed at the Institut Laue-Langevin (ILL). It is a 58\,MW research reactor using highly enriched $^{235}$U fuel (93\%) and using heavy water as a neutron moderator and reflector. 
This combination leads to the world's most intense thermal neutron flux of $\sim1.5\times10^{15}$ neutrons.cm$^{-2}$.s$^{-1}$.
The experiments from which are derived the reference $\bar\nu_e$ spectra consisted in irradiating thin layers of uranium and plutonium placed in a vacuum beam tube inserted at 80\,cm of the reactor z-axis. The layers were covered by Ni foils in order to contain the fission products. 
The cumulative $\beta^{-}$ spectra of the fission products of the irradiated isotopes were measured with a high precision by guiding the $e^{-}$ to the BILL magnetic spectrometer \cite{BILL}.

The neutron flux spectra knowledge, required to normalize the absolute rate of fission in the target and thus the number of beta particle collected was not precisely known. 
The absolute normalization was thus determined via a relative approach consisting in irradiating a calibration target with a well known partial (n, $e^{-}$) cross-section for thermal neutron in place of the fission target. 
The number of $\beta^{-}$ particle in the fission target were normalized by measuring the intensities of the internal conversion electron lines or of the $\beta^{-}$ decay rates following the neutron capture in the calibration target via the formula:
\begin{eqnarray}
N_{\beta} (fission^{-1}) 
= 
\frac{N_{f}}{N_{c}}  
\frac{\alpha  \left\langle \sigma_{c}(n,\gamma) \right\rangle}
     {        \left\langle \sigma_{f}(n,f)       \right\rangle}  
\frac{n_{c}}{n_{f}}
\end{eqnarray} 
where $N_{f}$ and $N_{c}$ are the measured counting rates for the
fission and the calibration targets respectively. For a $\beta^{-}$ decay, $\alpha$ represents the branching ratio of the relevant state and for an internal conversion it represents the internal conversion coefficient (ICC). 
$n_{c}$ and $n_{f}$ are respectively the number of atoms of the calibration and fission targets. 
Finally, $\left\langle \sigma_{c}(n,\gamma) \right\rangle$ and $ \left\langle \sigma(n,f) \right\rangle$ respectively represent the average cross-section of the considered calibration and fission reaction defined as:
$\left\langle \sigma \right\rangle = \int \sigma(E) \phi(E)dE \slash \int \phi(E)dE$, where $\sigma(E)$ and $\phi(E)$ are respectively the energy dependent differential cross-section of the considered reaction and the neutron flux spectrum crossing the target.

The first $^{235}$U measurement and the one of the $^{239}$Pu were calibrated  using the reactions:
\begin{itemize}[noitemsep,topsep=0pt]
\item[-] $^{197}$Au(n,e$^{-}$)$^{198}$Au: absolute calibration between 6 and 6.5\,MeV (E1 transitions: 6.251, 6.265, 6.276 and 6.319\,MeV).
\item[-] $^{207}$Pb(n, e$^{-}$)$^{208}$Pb: absolute calibration at 7.37\,MeV (E1 transition with $ \alpha = 9.25\times$10$^{-5}$).
\item[-] $^{116gs}$In: $\beta^{-}$ decay following $^{115}$In(n, $\gamma$)$^{116gs}$In: relative calibration between 0 and 3.3 MeV.

\end{itemize}
For the first $^{235}$U measurement, only the $^{197}$Au(n,e$^{-}$)$^{198}$Au reaction was used for the absolute normalization. 
The normalization uncertainty was quoted to be 5\% (90\% C.L.) 
For the $^{239}$Pu measurement, the precision on the $^{197}$Au(n,$\gamma$)$^{198}$Au reaction was reduced to 3.9\% and an uncertainty of 2.9$\%$ (90$\%$ C.L.) was taken for the $^{116gs}$In $\beta^{-}$ decay.

The second $^{235}$U measurement and the $^{241}$Pu were calibrated using the reactions:
\begin{itemize}[noitemsep,topsep=0pt]
\item[-] $^{207}$Pb(n, e$^{-}$)$^{208}$Pb: absolute calibration at 7.37\,MeV (E1 transition with $\alpha = (9.25 \pm 0.09)\times$10$^{-5}$ ).
\item[-] $^{115}$In(n, e$^{-}$)$^{116m}$In: absolute calibration at 1.29\,MeV (E2 transition with $\alpha = (6.47 \pm 0.07)\times$10$^{-4}$ ).
\item[-] $^{113}$Cd(n, e$^{-}$)$^{114}$Cd: relative calibration between 0.5 to 9\,MeV.
\end{itemize}

For the second $^{235}$U measurement and the $^{241}$Pu measurement, the uncertainties on the absolute normalization were evaluated to be of 3.1\% (90$\%$ C.L.) at 7.4\,MeV and of 2.8\% (90$\%$ C.L.) at 1.3\,MeV.
In the original publication, it is not clear how the average cross-sections were calculated. 
It is stated that the cross-sections for thermal neutrons were used with a correction to take into account the non 1/v  maxwellian neutron spectrum in heavy water.

\subsection*{Normalization investigation}

In this study we investigate the $\alpha \left\langle \sigma_{c}(n,\gamma) \right\rangle \slash \left\langle \sigma_{f}(n,f) \right\rangle$ ratio by using updated nuclear data for the ICC coefficients and new estimations of the average cross-sections of the calibration and fission targets. These estimations are performed using simulations of the HFR developed at CEA with the two well known and validated Monte-Carlo code MCNPX2.5 and TRIPOLI-4$^{\mbox{\scriptsize{\textregistered}}}$. Such simulations take advantage of being able to estimate the cross-section by taking into account the energy dependent shape of the neutron flux without relying on an external parametrization.

Table \ref{tab:tab1} presents a summary of the ICC values used in the original and the reviewed calculation. 
New ICC values are taken from the BrIcc v2.3S code \cite{BrIcc}. 
New values exhibit higher uncertainties. 
A good agreement is observed for the $^{116}$Sn and $^{198}$Au while a much higher difference is observed for $^{208}$Pb with a relative difference of (10.5\,$\pm$\,1.8)\% between the old and the new value.

\begin{figure}
\centering
\includegraphics[width=.95\linewidth]{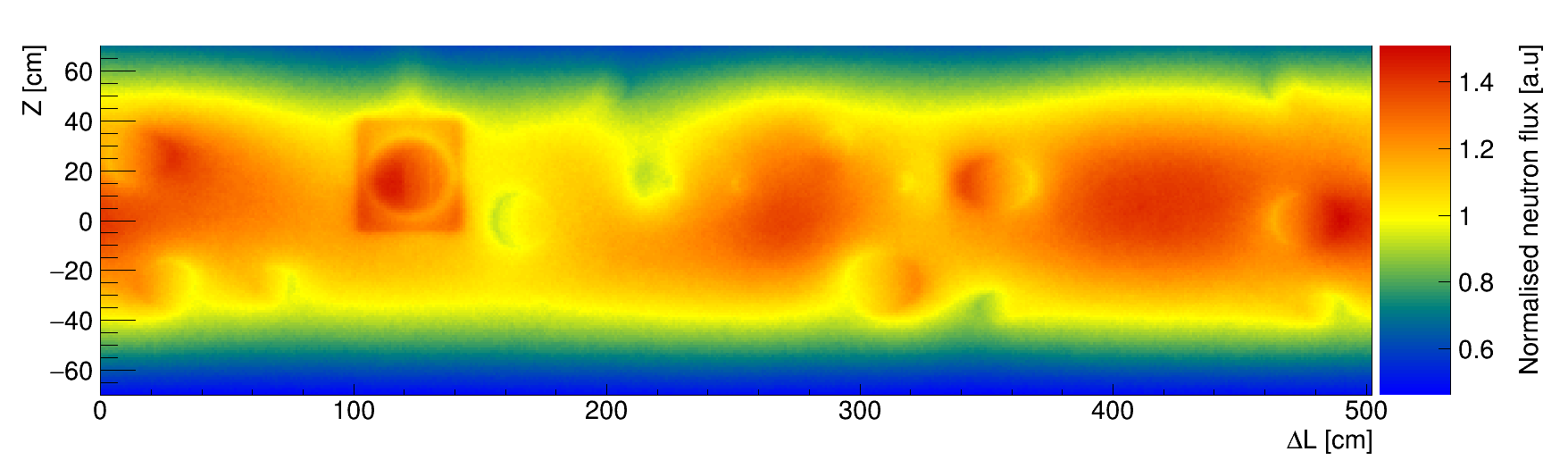}
\includegraphics[width=.95\linewidth]{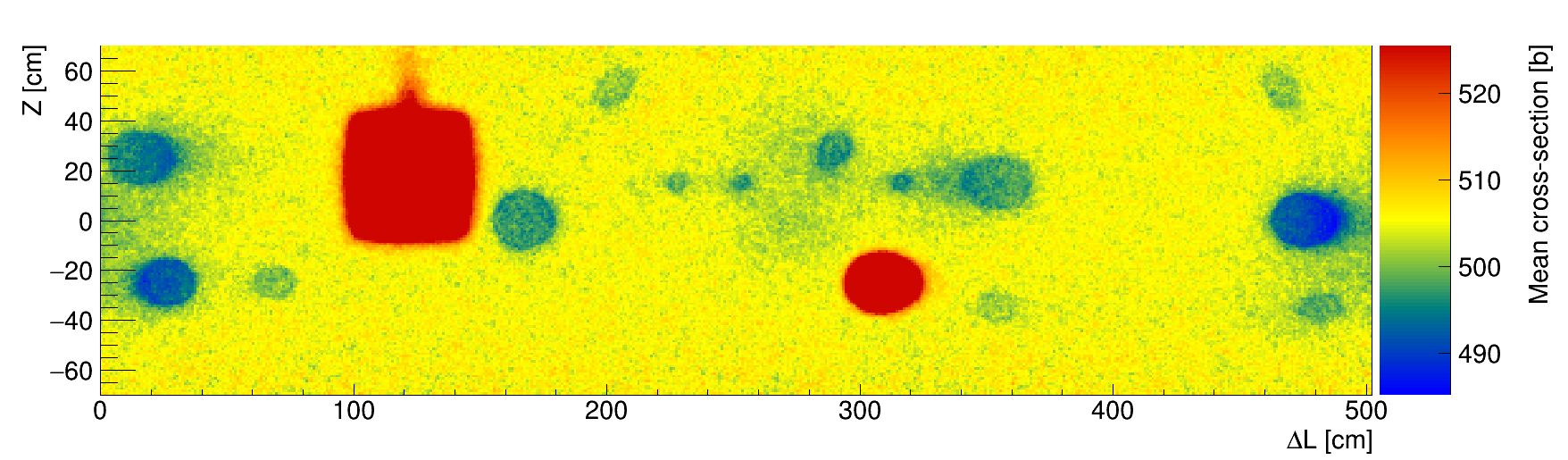}
\caption{Total neutron flux (top) and average cross-section distribution of the $^{235}$U(n,f) reaction (bottom)  at 80\,cm of the reactor z-axis from the TRIPOLI-4$^{\mbox{\scriptsize{\textregistered}}}$ simulation. When the cutting plane crosses cold sources (beam tubes), higher (lower) average cross-section values are observed.}
\label{fig:flux_rr}
\end{figure}

In both simulations a precise description of the fuel and main reactor geometry components is implemented. 
In the TRIPOLI-4$^{\mbox{\scriptsize{\textregistered}}}$ simulation, additional geometrical elements related to the beam tubes dedicated to other irradiation experiments are also modeled.
While we know that the beam tube used to irradiate the target was located at 80\,cm of the reactor z-axis, we do not know exactly at which position the BILL beam tube was inserted into the reactor neither precisely at which z position the targets where positioned into the beam tube.
An illustration of the total neutron flux, and of the average cross section of the $^{235}$U at 80\,cm of the reactor z-axis is presented in figure \ref{fig:flux_rr}. 
In a first approach, we estimated the average cross-section by using the average neutron spectrum by integrating all positions in the water located at 80\,cm of the reactor z-axis with z$\in$[-5,5]\,cm. 
Cross-section ratio used in the original ILL publications are reported in table \ref{tab:tab2} and compared with the preliminary results obtained with the MCNP simulation.

For the first $^{235}$U measurement, the normalization estimated with MCNP is  2.1\% lower than the ILL one, while a good agreement is obtained for the $^{239}$Pu.
For the second $^{235}$U and the $^{241}$Pu measurements, the normalization
is in good agreement at low energy (normalization on In reaction).
When using the JEFF-3.3 database, the normalization at high energy (normalization on Pb reaction) seems underestimated but when using the $^{207}$Pb cross section from \citep{Schill}, the disagreement disappears between the low and the high energy part of the calibration and a good agreement is found for the normalization of the second $^{235}$U energy spectrum whereas $^{241}$Pu seems to be slightly underestimated by about 1.7\%.

\subsection*{Conclusion}
In this on-going work, we have started to investigate the normalization of the integral beta spectra measurement performed at the ILL. 
We reported results using updated nuclear data and dedicated MCNPX-2.5 simulation of the HFR reactor. From these preliminary results, there is no clear evidence for a bias in the normalization of the ILL energy spectra. The obtained results are still preliminary and refined results including improved modeling of the experiments, propagation of nuclear data uncertainties to the results and inter-comparison of MCNPX-2.5 and TRIPOLI-4$^{\mbox{\scriptsize{\textregistered}}}$ results are under progress and necessary before concluding on the reviewed normalization.

\acknow{}
\showacknow{}  
This work was done in the framework of the NENuFAR project, which is supported by the direction of cross-disciplinary programs at CEA.

\begin{table}[]
\centering
\caption{ICC values used in the original calculations and recent ones taken from the BrIcc code \cite{BrIcc}. The digits in parenthesis represent the 1$\sigma$ uncertainties and applies on the last digits of the central value.}
\begin{tabular}{l|rrrr}
                      & \makecell{ICC - ILL \\ ($\times 10^{-4}$)}  & \makecell{ICC - BrIcc \\ ($\times 10^{-4}$)} & new/old &  \\ 
                      \cline{1-4}  
$^{116}$Sn (1.29 MeV) & 6.47(7)                       & 6.48(9)                                                                    & 1.002(14)      &  \\
$^{198}$Au (6 MeV)    & 1.092(10)                  & 1.071(15)                                                                  & 0.981(16)      &  \\
$^{208}$Pb (7.37 MeV) & 0.925(9)                      & 1.022(14)                                                                  & 1.105(18)      &  \\
%\bottomrule
\bottomrule
\bottomrule
\end{tabular}
\label{tab:tab1}
\end{table}

\begin{table}[]
\centering
        \caption{Preliminary average cross-section ratios used in the original ILL calculations (left) \cite{Schreckenbach:1981wlm,VonFeilitzsch:1982jw,Schreckenbach:1985ep,Hahn:1989zr} and ratios of theses values with those estimated with MCNPX2.5 using the JEFF-3.3 database (right). For the $^{207}$Pb isotope, additional results obtained using the cross-section estimated with recent measurement from \cite{Schill} (not yet integrated in the evaluation process) are presented and reported as a).
}                                               
\begin{tabular}{r|rl}
 & $\alpha \left\langle \sigma_{c}\right\rangle\slash\left\langle \sigma_{f} \right\rangle$ (ILL)    & \multicolumn{1}{c}{MCNP$\slash$ILL}       \\ \hline
$^{197}$Au(n,e$^{-}$)\slash$^{235}$U(n,f)                  & $(1.91\pm0.02)\times10^{-5}$                  & $0.979\pm0.010$       \\ \hline
$^{115}$In(n,e$^{-}$)\slash$^{235}$U(n,f)                  & $(1.60\pm0.03)\times10^{-4}$                  & $0.998\pm0.028$       \\ \hline
\multirow{2}{*}{$^{207}$Pb(n,e$^{-}$)\slash$^{235}$U(n,f)} & \multirow{2}{*}{$(1.16\pm0.02)\times10^{-7}$} & $0.955\pm0.026$       \\ 
                                                        &                                               & $0.999\pm0.026 ^{a)}$ \\ \hline
$^{197}$Au(n,e$^{-}$)\slash$^{239}$Pu(n,f)                  & $(1.35\pm0.02)\times10^{-5}$                  & $0.998\pm0.014$       \\ \hline
$^{115}$In(n,e$^{-}$)\slash$^{241}$Pu(n,f)                  & $(8.45\pm0.19)\times10^{-5}$                  & $1.014\pm0.004$       \\ \hline
\multirow{2}{*}{$^{207}$Pb(n,e$^{-}$)\slash$^{241}$Pu(n,f)} & \multirow{2}{*}{$(6.1\pm0.1)\times10^{-8}$}   & $0.972\pm0.033$       \\
                                                        &                                               & $1.018\pm0.033 ^{a)}$ \\ \hline \hline
\end{tabular}
\label{tab:tab2}
\end{table}

%% file: AAP-Karagiorgi/aap2018-participant00x.tex
% \documentclass[9pt,twocolumn,twoside,lineno]{aap2018}

% \templatetype{aap2018proceedings} % Choose template 

\title{Overview and Status of Short Baseline Neutrino Anomalies}

% Use letters for affiliations, numbers to show equal authorship (if applicable) and to indicate the corresponding author
\author[a,1]{Georgia Karagiorgi}
%\author[b,1,2]{Author Two} 
%\author[a]{Author Three}

\affil[a]{Department of Physics, Columbia University, New York, New York, 10027}
%\affil[b]{Affiliation Two}
%\affil[c]{Affiliation Three}

% Please give the surname of the lead author for the running footer
\leadauthor{Karagiorgi}

% Please include corresponding author, author contribution and author declaration information
% \authorcontributions{Please provide details of author contributions here.}
% \authordeclaration{Please declare any conflict of interest here.}
% \equalauthors{\textsuperscript{1}A.O.(Author One) and A.T. (Author Two) contributed equally to this work (remove if not applicable).}
\correspondingauthor{\textsuperscript{1}E-mail: georgia\@nevis.columbia.edu}

% Keywords are not mandatory, but authors are strongly encouraged to provide them. If provided, please include two to five keywords, separated by the pipe symbol, e.g:
% \keywords{Keyword 1 $|$ Keyword 2 $|$ Keyword 3 $|$ ...} 

\begin{abstract}
This article provides a brief overview of anomalous signatures observed at ``short-baseline'' neutrino experiments, along with suggested interpretations put forth by the wider community. Particular focus is paid to the interpretation involving light sterile neutrino oscillations, which seems, however, to be in conflict with the lack of $\nu_\mu$ or $\bar\nu_\mu$ disappearance signals at short baselines. A number of experiments are planning to further test this hypothesis, including reactor-based experiments presented in these proceedings. 
\end{abstract}

% \dates{This manuscript was compiled on \today}
\doi{\url{https://neutrinos.llnl.gov/workshops/aap2018}}

% \begin{document}

\maketitle
\thispagestyle{firststyle}
\ifthenelse{\boolean{shortarticle}}{\ifthenelse{\boolean{singlecolumn}}{\abscontentformatted}{\abscontent}}{}

% If your first paragraph (i.e. with the \dropcap) contains a list environment (quote, quotation, theorem, definition, enumerate, itemize...), the line after the list may have some extra indentation. If this is the case, add \parshape=0 to the end of the list environment.
\dropcap{O}ver the last two decades, a number of experimental neutrino signatures performed at relatively short baselines ($L/E\sim1$~m/MeV) have been found inconsistent with the three-neutrino oscillation framework \cite{Tanabashi:2018oca}, and have been referred to as ``short-baseline anomalies;'' for an extended review, see~\cite{Abazajian:2012ys}. The signatures are (anti)$\nu_\mu\rightarrow\ $(anti)$\nu_e$ appearance-like and (anti)$\nu_e$ disappearance-like in nature, and each one has been individually found consistent with two-neutrino oscillations at anomalously large $\Delta m^2$ of $\sim$0.1-10~eV$^2$, and therefore point to physics beyond the standard three-neutrino framework. Each signature is described further below. %The experiments historically contributing to these observations are summarized in Tab.~\ref{Karagiorgi:Table1}, and are described further below. 

%\begin{table}[thbp]
%\centering
%\caption{Short-baseline neutrino experiments which have observed signatures consistent with two-neutrino oscillation at anomalously large $\Delta m^2$. The channel in which the oscillation signature was observed and the significance of each anomalous excess or deficit are also provided.}
%\begin{tabular}{lcc}
%Experiment & Channel & Significance \\
%\midrule
%1. LSND \cite{Athanassopoulos:1996jb,Athanassopoulos:1997pv} & $\bar{\nu}_\mu\rightarrow\bar{\nu}_e$ & 3.8$\sigma$ \\
%2. MiniBooNE \cite{Aguilar-Arevalo:2018gpe} & $\nu_\mu\rightarrow\nu_e$ and $\bar{\nu}_\mu\rightarrow\bar{\nu}_e$ & 4.8$\sigma$ \\
%3. Reactor Anomaly \cite{Mention:2011rk} & $\bar{\nu}_e\rightarrow\bar{\nu}_\not{e}$& $\sim$2-3$\sigma$ \\
%4. GALLEX/SAGE \cite{Kaether:2010ag,Abdurashitov:2009tn} & ${\nu}_e\rightarrow{\nu}_\not{e}$ & 2-3$\sigma$ \\
%\bottomrule
%\label{Karagiorgi:Table1}
%\end{tabular}
%\end{table}

\section*{Short-baseline Anomalies}

\subsection*{LSND}
The Liquid Scintillator Neutrino Detector (LSND) ran in the 1990's in a $\pi^+\rightarrow\mu^+$ decay-at-rest beam at Los Alamos National Lab \cite{Athanassopoulos:1996jb,Athanassopoulos:1997pv}. The experiment employed a liquid scintillator detector, sensitive to $\bar{\nu}_e$ from potential oscillations of $\bar{\nu}_{\mu}$ from $\mu^+$ decay-at-rest. With a mean neutrino energy of $\sim$30 MeV and a baseline of 50~m, LSND was sensitive to oscillations at $\Delta m^2\sim1$~eV$^2$. During its running, LSND observed a 3.8$\sigma$ excess of $\bar{\nu}_e$, identified by a double-coincidence of an electron from inverse beta decay in time with the beam plus a delayed neutron capture, which could be interpreted as $\bar{\nu}_\mu\rightarrow\bar{\nu}_e$ oscillations, with an oscillation probability of $<0.5$\%. When interpreted as two-neutrino oscillations, this points to a $\Delta m^2$ of 1.2~eV$^2$ and $\sin^22\theta_{\mu e}$ of 0.003. 

\subsection*{MiniBooNE}
The Mini Booster Neutrino Experiment (MiniBooNE) at Fermilab has been running in Fermilab's Booster Neutrino Beamline for more than 15 years. It has run with both $\nu_\mu$ and $\bar{\nu}_\mu$ beams produced from $\pi^+$ and $\pi^-$ decay in flight, respectively. The neutrinos were identified as electron- or muon-like in the detector through their distinct cherenkov light topologies. With a mean neutrino energy of a $\sim$600~MeV and a baseline of 540~m, MiniBooNE was sensitive to oscillations at a $\Delta m^2$ similar to LSND, but it was subject to entirely different systematics, both due to the tenfold increase in neutrino energy, and due to the different detection technique used for $\nu_e$ identification. During its first neutrino beam running, MiniBooNE observed an excess of $\nu_e$ events at relatively low energy \cite{AguilarArevalo:2008rc}; the same excess was seen, albeit with lower significance, in antineutrino beam running \cite{Aguilar-Arevalo:2013pmq}. The most recent MiniBooNE results \cite{Aguilar-Arevalo:2018gpe}, inclusive of all data collected to date, find the neutrino and antineutrino mode results consistent with each other, and a combined analysis yields an excess signal significance of 4.8$\sigma$. When interpreted as (anti)$\nu_\mu\rightarrow$~(anti)$\nu_e$ two-neutrino oscillations, this excess points to a best-fit $\Delta m^2$ of 0.041~eV$^2$, and $\sin^2 2\theta_{\mu e}$ of 0.96. There is, however, a large allowed phase-space overlap with LSND.

\subsection*{Reactor Anomaly}
In 2011, a re-evaluation of detailed physics involved in nuclear beta-decay of fission fragments in reactors led to the realization that, previously, reactor-based experiments had under-predicted the expected reactor $\bar{\nu}_e$ fluxes by a few~\%. This included experiments performed at very short baselines from reactors ($<100$~m), which were thus found to observe effective $\bar{\nu}_e$ deficits consistent with $\bar{\nu}_e$ disappearance at $\Delta m^2>1$~eV$^2$ and $\sin ^2 2\theta_{ee}\sim 5-10$\% (assuming two-neutrino oscillations). This ``Reactor Anomaly'' (RA) is discussed in detail in \cite{Hayes,Sonzogni,Onillon}, and its significance remains to be determined more precisely, since more accurate flux treatments are warranted. In particular, if this anomaly were due to oscillations, one might expect that deficits would be isotope-independent; recently, Daya Bay has carried out isotopic evolution measurements which have shown that current model predictions for $^{235}$U are off, unlike $^{239}$U measurements \cite{An:2017osx}, suggesting isotope-dependent effects being at least partly responsible for this deficit. Regardless, oscillation analyses considering fits to ``free fluxes'' and ``fixed fluxes with oscillations'' find that there is no clear data fit preference at the moment \cite{Dentler:2017tkw}. This warrants further flux corrections before a definitive statement can be made as to whether the RA is due to merely insufficient flux modeling, or potentially short-baseline oscillations, or, possibly, both. Multiple new results from very short baseline reactor experiments have been pouring in over the last two years (see detailed overviews in \cite{Han,Almazan,Serebrov,Mumm,Shitov}), but, as of yet, no clear picture of $\bar{\nu}_e$ disappearance has emerged. 

\subsection*{GALLEX/SAGE}
Prior to the RA, hints of $\nu_e$ disappearance had been provided by the GALLEX and SAGE solar neutrino experiments employing mega-curie $^{51}$Cr and $^{37}$Ar $\nu_e$ sources for detector calibration. The overall rates from multiple calibration measurements were consistent with each other and showed an overall deficit that is with $\nu_e$ disappearance \cite{Acero:2007su,Giunti:2010zu} at $\sim$2-3$\sigma$. Interpreted as two-neutrino oscillations, the deficit points to $\Delta m^2>1$~eV$^2$ and $\sin ^2 2\theta_{ee}\sim 10$\%.

\section*{Interpretations}
\begin{figure}[tbhp]
\centering
\includegraphics[width=.45\linewidth]{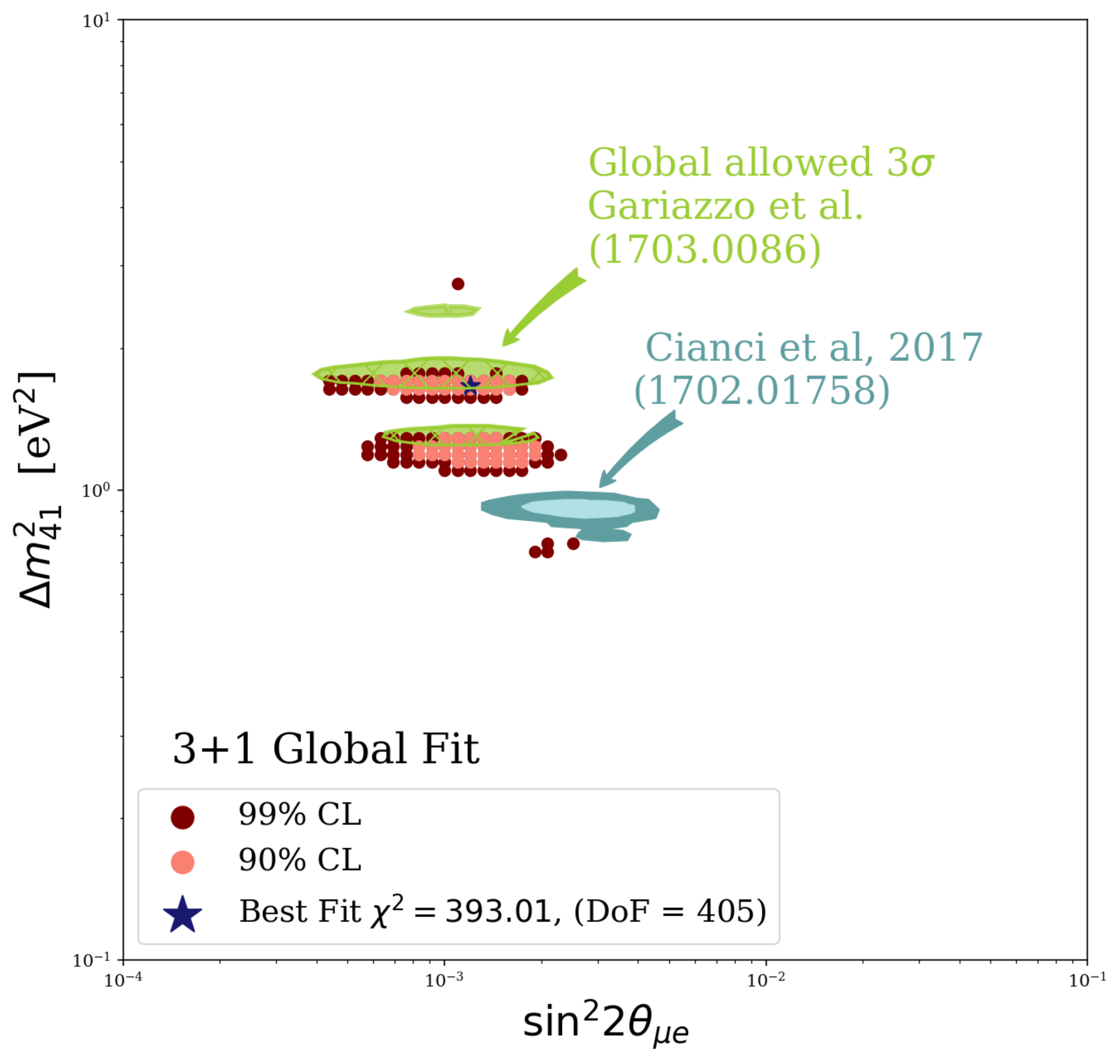}
\caption{Globally allowed regions at 90\% (light red) and 99\% (dark red) from 3+1 fits to short-baseline data sets. The best fit parameters correspond to $\Delta m^2_{41}=1.7$~eV$^2$ and $\sin^2 2\theta_{\mu e}=4|U_{e4}|^2|U_{\mu 4}|^2=0.0012$, where $U_{e4}=0.12$ and $U_{\mu4}=0.15$. Past global fit results are also overlaid for comparison. The figure is from \cite{Cianci}.}
\label{Karagiorgi:Figure1}
\end{figure}

A more specific framework within which all of the above signatures can fit is one referred to as 3 active + 1 sterile neutrino framework, or, 3+1. In this framework, a fourth neutrino mass eigenstate is assumed to exist, separated from the other three by a splitting of order 0.1-10~eV$^2$. Within this framework, a 3+1 fit to a global set of experimental data including anomalies and null observations from LSND, MiniBooNE, KARMEN, NOMAD, GALLEX/SAGE, Bugey, MINOS, CCFR84, CDHS, NEOS, DANSS, and atmospheric constraints (following the analysis methodology in \cite{Cianci:2017okw}) is found to be consistent with light sterile neutrino oscillations, yielding a $\chi^2$ probability of 66\% \cite{Cianci}, and the best fit parameters and allowed regions provided in Fig.~\ref{Karagiorgi:Figure1}. However, this agreement is merely apparent; within this framework, the appearance and disappearance amplitudes are correlated. %, assuming that the overall 4$\times$4 neutrino framework adheres to unitarity, and that the $\nu_e$ and $\nu_\mu$ mixing content of the fourth mass eigenstate is small. 
As such, the (anti)$\nu_e$ appearance and (anti)$\nu_e$ disappearance measurements, if interpreted as 3+1 oscillations, predict (anti)$\nu_\mu$ disappearance amplitudes of order 10\% \cite{Dentler:2018sju}. However, to this date, no such observation exists, and (anti)$\nu_\mu$ disappearance experiments with sensitivity to high-$\Delta m^2$ oscillations have placed stringent bounds for 3+1 models. This issue is referred to as ``tension'' among global data sets under the 3+1 framework. This tension motivates consideration of extended frameworks with additional sterile neutrinos, or, 3+$N$ \cite{Conrad:2012qt,Cianci:2017okw,Kopp:2013vaa}. In addition to providing more degrees of freedom, 3+$N$ also offers the possibility of different effective oscillation probabilities for neutrino vs.~antineutrino appearance, through new CP violation phases. Although they generally yield better (qualitatively) fits to global data sets, when examined more closely, 3+$N$ frameworks also reveal tension between appearance and disappearance data sets \cite{Dentler:2018sju}. 

In an attempt to alleviate this tension, other proposed interpretations involve models with non-standard neutrino interactions, extra dimensions, alternate dispersion relations, or neutrino decay (see, e.g.~\cite{Chu:2018gxk,Doring:2018cob,Bertuzzo:2018itn,Alvarez-Ruso:2017hdm}), often in combination with light sterile neutrinos. For example, \cite{Diaz} introduces a finite lifetime to the fourth mass eigenstate in a 3+1 model, resulting in decoherence in neutrino propagation and thereby no resonant matter effects, evading IceCube's stringent limits on $\nu_\mu$ disappearance. A significant contributor to the tension in 3+$N$ fits is the MiniBooNE low energy excess (LEE). The LEE is particularly special among short-baseline anomalies as it necessitates particularly large mixing amplitudes (because the signal is at low energy, where no significant (anti)$\nu_\mu$ flux exists). Because of that, people have also considered extended models to explain the MiniBooNE excess independently of the other anomalies. A more recent model which has received particular traction is one in which a sterile neutrino can be produced and subsequently decay in the MiniBooNE detector \cite{Ballett:2018ynz} producing an $e^+e^-$ pair, which, if significantly boosted, is indistinguishable from an $e^-$ produced in (anti)$\nu_e$ interactions in MiniBooNE. MicroBooNE has been running since 2015 and aims to resolve whether the LEE is truly due to $\nu_e$'s (such as from $\nu_\mu\rightarrow\nu_e$ appearance) or due to neutrino-induced single photon background, or due to some other yet-unknown process. MicroBooNE LEE analyses are currently ongoing \cite{Guenette,neutrino2018singlegamma,Yates:2017lxa,neutrino2018nue}.

\section*{Status and Outlook}

From the RA perspective, the new results from NEOS \cite{Han}, STEREO \cite{Almazan}, Neutrino-4 \cite{Serebrov}, PROSPECT \cite{Mumm}, and DANSS \cite{Shitov} at the moment disfavor the RA best fit at 2-5$\sigma$, depending on the data set \cite{Maricic}. Among those, Neutrino-4 \cite{Serebrov:2018vdw,Serebrov} provides the strongest hints for $\bar{\nu}_e$ disappearance with rather large mixing parameters of $\Delta m^2=7.22$~eV$^2$ and $\sin^2 2\theta_{ee}=0.35$. Taken at face value, this large mixing is inconsistent with Daya Bay, RENO, and Double Chooz measurements \cite{Dentler:2018sju}; furthermore, PROSPECT's data set provides a poor fit for the Neutrino-4 best fit point \cite{Mumm}. NEOS/DANSS provide very weak hints for a $\Delta m^2=1.73/1.4$~eV$^2$ and $\sin^2 2\theta_{ee}=0.14/0.05$ \cite{Han,Shitov}. 

A number of future experiments, including SoLiD \cite{Verstraeten}, CHANDLER \cite{Park}, ANGRA \cite{Chimenti}, ISMRAN \cite{Mulmule}, VIDARR \cite{Coleman}, and other Japanese reactor near field detectors \cite{Nakajima} are expected to come online and provide additional measurements to disambiguate what seems to be a convoluted neutrino flavor picture. Particular emphasis is being placed on improving and constraining reactor flux predictions \cite{Maricic}. New short-baseline accelerator-based experiments are also expected to come online, including SBN \cite{Antonello:2015lea}, and potentially IsoDAR \cite{Alonso:2017fci}. %SBN will begin three-detector concurrent operations by mid 2020. 
There are also sterile neutrino searches being conducted with long-baseline accelerator-based and atmospheric neutrino experiments, including MINOS/MINOS+, NOvA, T2K, IceCube/DeepCore, and Super-K \cite{Dentler:2018sju}. In the future, sterile neutrino searches should also be possible with neutrino-nucleus coherent elastic scattering at COHERENT and CE$\nu$NS \cite{Cebrera}.

Aside from experimental tests with increased precision, the neutrino phenomenology community is also resorting to the exploration of new theoretical models as a source of one or more of the anomalies, as described in the previous section. What will also be necessary is a rethinking of statistical approaches followed in global fits, in particular to more accurately quantify fit quality and tension among data sets. It may be necessary to resort to a weighted statistical treatment of inputs, and/or the adoption of frequentist methods for global fits. 

%\section*{Guide to using this template}
%If you have a question while using this template on Overleaf, please contact us at aap2018@llnl.gov. 

%\subsection*{Author Affiliations}
%Include the institution for each author. Use lower case letters to match authors with institutions, as shown in the example. 

%\subsection*{Submitting Manuscripts}
%This project is shared with the AAP2018 organizing committee. When your proceeding is finished please contact aap2018@llnl.gov . No other submission steps are needed. The committee will contact you if there are any missing elements to your proceedings.

%\subsection*{Manuscript Length}
%The length of the proceeding are not to exceed two pages in length including any figures or tables you wish to include.

%\subsection*{References}
%References should be cited in numerical order as they appear in text; this will be done automatically via bibtex, e.g. \cite{belkin2002using} and \cite{berard1994embedding,coifman2005geometric}. All references should be included in the main manuscript file.  Please use inspire \url{http://inspirehep.net/} for your references; they will be tabulated at the end of the proceedings and do not count toward your two page limit. 

%\subsection*{Labeling}

%\subsection*{Single column equations}

%Authors may use 1- or 2-column equations in their %article, according to their preference. For example, 
%\begin{eqnarray}
%(x+y)^3&=&(x+y)(x+y)^2\nonumber\\
%       &=&(x+y)(x^2+2xy+y^2) \nonumber \label{eqn:example} \\
%       &=&x^3+3x^2y+3xy^3+x^3. 
%\end{eqnarray}
%To allow an equation to span both columns, use the \verb|\begin{figure*}...\end{figure*}| environment mentioned above for figures.

%Note that the use of the \verb|widetext| environment for equations is not recommended, and should not be used. 
%% Uncomment following for example of equation spanning %two columns :
%% \begin{figure*}[bt!]
%% \begin{align*}
%% (x+y)^3&=(x+y)(x+y)^2\\
%%       &=(x+y)(x^2+2xy+y^2) \numberthis %%\label{eqn:example} \\
%%       &=x^3+3x^2y+3xy^3+x^3. 
%% \end{align*}
%% \end{figure*}

% \matmethods{Please describe your materials and methods here. This can be more than one paragraph, and may contain subsections and equations as required. Authors should include a statement in the methods section describing how readers will be able to access the data in the paper. 

% \subsection*{Subsection for Method}
% Example text for subsection.
% }

% \showmatmethods{} % Display the Materials and Methods section

\section*{Summary}
The amassing of anomalous (anti)$\nu_e$ excesses and deficits at $L/E\sim1$~m/MeV from (anti)$\nu_\mu$ and (anti)$\nu_e$ sources, respectively, motivates extensions to the three-neutrino model, in the form of light sterile neutrino oscillations. However, such interpretations are in conflict with null (anti)$\nu_\mu$ disappearance searches at short baselines. This has driven the community to resort to improving the metrics used to quantify fit quality, considering alternative phenomenological interpretations, and deploying new experimental tests with unprecedented sensitivity to such potential oscillations. All of those efforts in concert will be necessary to resolve these anomalies.%, and to remove ambiguities that their implications introduce on future searches for CP violation and neutrino mass hierarchy determination in the neutrino sector,  cosmological observations, and beyond.

\acknow{The author acknowledges the U.S.~National Science Foundation for their support.}

\showacknow{} % Display the acknowledgments section

% Bibliography
% \bibliography{aap2018-participant00x}

% \end{document}

%% file: AAP-Nakajima/aap2018-participant00x.tex
% \documentclass[9pt,twocolumn,twoside,lineno]{aap2018}
%\documentclass{article}
% \usepackage[utf8]{inputenc}

% \title{AAP-PersonalizedTemplate}
% \author{bergevin1 }
% \date{September 2018}

% \begin{document}

% \maketitle

% \section{Introduction}

% \end{document}

% \documentclass[9pt,twocolumn,twoside,lineno]{pnas-new}
% Use the lineno option to display guide line numbers if required.

% \templatetype{aap2018proceedings} % Choose template 
% {pnasresearcharticle} = Template for a two-column research article
% {pnasmathematics} %= Template for a one-column mathematics article
% {pnasinvited} %= Template for a PNAS invited submission

\title{Reactor neutrino monitor experiments in Japan}

% Use letters for affiliations, numbers to show equal authorship (if applicable) and to indicate the corresponding author
\author[a,1]{K.~Nakajima}
\author[a]{T.~Akama}
\author[b]{H.~Furuta}
\author[g]{Y.~Hino}
\author[a]{A.~Hirota}
\author[a]{Y.~Ikeyama}
\author[c]{S.~Iwata}
\author[d]{T.~Kawasaki}
\author[d]{T.~Konno}
\author[e]{H.~Miyata}
\author[f]{H.~Ono}
\author[d]{A.~Shibata}
\author[a]{K.~Shimizu}
\author[g]{F.~Suekane}
\author[a]{Y.~Tamagawa}
\author[d]{T.~Torizawa}
\author[g]{R.~Ujiie}
\author[f]{M.~Watanabe}

\affil[a]{Graduate School of Engineering, University of Fukui, Fukui 910-8507, Japan}
\affil[b]{High Energy Accelerator Research Organization (KEK), Tsukuba 305-0801, Japan}
\affil[c]{Tokyo Metropolitan College of Industrial Technology, Tokyo 140-0011, Japan}
\affil[d]{School of Science, Kitasato University, Sagamihara 252-0373, Japan}
\affil[e]{Faculty of Science, Niigata University, Niigata 950-2181, Japan}
\affil[f]{School of Life Dentistry at Niigata, The Nippon Dental University, Niigata 951-8580, Japan}
\affil[g]{Research Center for Neutrino Science, Tohoku University, Sendai 980-8578, Japan}

% Please give the surname of the lead author for the running footer
\leadauthor{Nakajima}

% Please include corresponding author, author contribution and author declaration information
% \authorcontributions{Please provide details of author contributions here.}
% \authordeclaration{Please declare any conflict of interest here.}
% \equalauthors{\textsuperscript{1}A.O.(Author One) and A.T. (Author Two) contributed equally to this work (remove if not applicable).}
\correspondingauthor{\textsuperscript{1}E-mail: nkyohei@u-fukui.ac.jp}

% Keywords are not mandatory, but authors are strongly encouraged to provide them. If provided, please include two to five keywords, separated by the pipe symbol, e.g:
% \keywords{Keyword 1 $|$ Keyword 2 $|$ Keyword 3 $|$ ...} 

\begin{abstract}
Nuclear reactors are gradually restarting in Japan after the Fukushima Daiichi nuclear power station accident. There are seven operating reactors in Oct.~2018. In this situation, we made a consortium of reactor monitor experiments among Japanese researchers and share the research progress regularly. The PANDA detector, which consists of 100 plastic scintillator arrays with a total mass of 1~ton, was developed and is ready to perform a measurement. The background measurement was performed at the campus of Kitasato University in 2017, and the background rate on the ground level was measured to be about 7,000 events/day. As a next step, we plan to perform a measurement in Ohi nuclear power plant next spring. There are 4 nuclear reactors (3.4~GW$\mathrm{_{th}} \times $ 4 units) and 2 of them are in operation. We have two criteria of measurement location, whose distance to the reactor core are about 45~m or about 100~m. We plan to take a data for a 1 month in reactor-on and in reactor-off, respectively. If we perform a measurement at the distance of 45~m from the reactor core, the number of expected neutrino signal is about 20 events/day, and then the significance of the signals will be around 4 sigma level. Except for the PANDA experiment, we are also developing the detector with Gd-loaded liquid scintillator which has an capability of pulse shape discrimination. The status and plan of reactor neutrino monitor experiments in Japan is described.
\end{abstract}

% \dates{This manuscript was compiled on \today}
\doi{\url{https://neutrinos.llnl.gov/workshops/aap2018}}

% \begin{document}

\maketitle
\thispagestyle{firststyle}
\ifthenelse{\boolean{shortarticle}}{\ifthenelse{\boolean{singlecolumn}}{\abscontentformatted}{\abscontent}}{}

% If your first paragraph (i.e. with the \dropcap) contains a list environment (quote, quotation, theorem, definition, enumerate, itemize...), the line after the list may have some extra indentation. If this is the case, add \parshape=0 to the end of the list environment.
%\dropcap{T}his template is provided to help you summarize your work in the correct format.  Instructions for use are provided below. 

\section*{Introduction}
In nuclear reactors, a large amount of $\bar{\nu}_e$ are generated by beta decay of fission products. The reactor neutrino monitor is an experiment with a neutrino detector installed near the reactor and we attempt to monitor the reactor operation remotely by measuring neutrino signals via inverse beta decay reaction. The SONGS experiment have demonstrated the effectiveness of reactor neutrino monitor \cite{ref:SONGS} and some reactor neutrino monitor experiments are in progress. Since the neutrino flux differs depending on the fission isotopes such as $^{235}$U and $^{239}$Pu, there is a possibility that fuel composition can be monitored by measuring the neutrino energy. Besides these objectives, reactor neutrino monitors will lead to an understanding of so-called ``5 MeV bump'' \cite{ref:RENO1, ref:RENO2, ref:DayaBay, ref:DoubleChooz} and ``reactor anomaly'' \cite{ref:anomaly} which are two problems in the field of reactor neutrino experiments. The reactor neutrino monitor experiment is a meaningful research from these aspects.

\section*{Nuclear reactors and neutrino monitor experiments}
There were 54 nuclear reactors in operation before the Fukushima Daichi nuclear disaster in 2011, supplying ~30\% of the country’s electric power. In 2013, new stricter safety regulations were established to withstand earthquakes and tsunami. All nuclear reactors had stopped after then, making it difficult to restart reactors and to perform a measurement in the nuclear power plant. In Oct.2018, 9 reactors in five nuclear power plants met the new standards and 7 reactors are in operation, supplying about 3\% of the country’s electric power. It was decided that 19 reactors will be decommissioned, however, it is planned that nuclear energy will account for about 20\% of energy output by 2030, with 30 nuclear reactors in operation.

The reactor neutrino monitors have been developed at each institute in Japan. There are two types of detectors; plastic scintillator and liquid scintillator. All Japanese researchers related to the reactor neutrino monitor organized a consortium to achieve the measurement in the nuclear power plant where the safety regulation is strict, and the informations and technologies are shared among us. We firstly plan to measure neutrinos by plastic scintillator from the aspect of safety. We will also continue the liquid scintillator development which has an capability of pulse shape discrimination between $\gamma$-rays and neutrons.

\section*{PANDA experiment}

\subsection*{Detector}
\begin{figure}[tbhp]
\centering
\includegraphics[width=.8\linewidth]{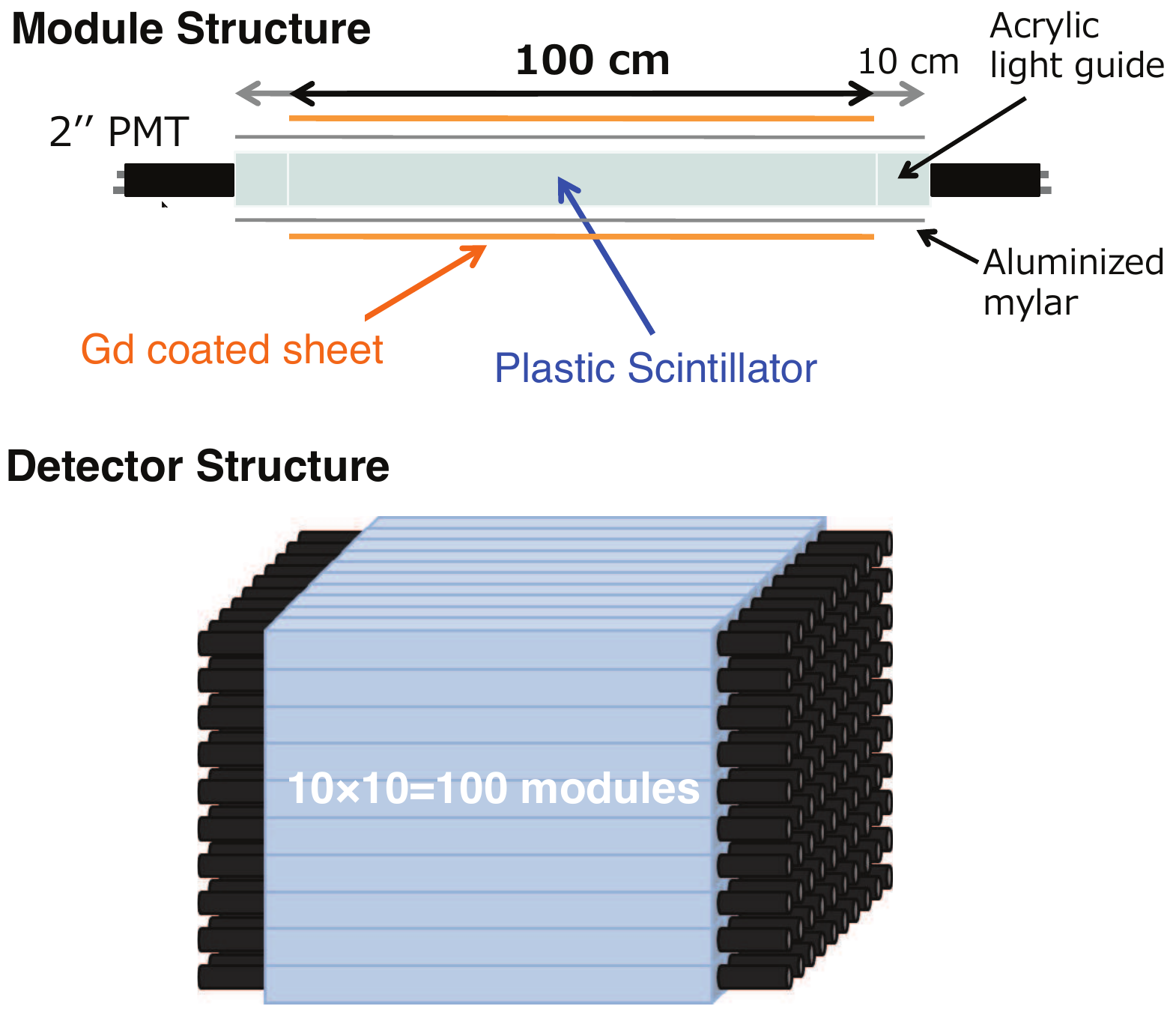}
\caption{A schematic view of the PANDA detector. The size of plastic scintillator in a module is 100 cm in length and 10 cm in width and height. The detector consists of 100 modules and a total mass of plastic scintillator is about 1 ton.}
\label{fig:detectorP}
\end{figure}

The PANDA (Plastic Anti-Neutrino Detector Array) detector is made of plastic scintillator arrays. The structure of a module and a whole detector is shown in Fig.~\ref{fig:detectorP}. A Gd coated sheet is set on the surface of plastic scintillator, and an average neutron thermalization time is about 60 $\mu$sec. The PANDA detector has been developed since 2006, and a full set of 100 modules with a total mass of 1 ton (PANDA100) were constructed in 2016.

\subsection*{Measurement at nuclear power plant in 2012}
A measurement was performed with 36 modules (PANDA36) at Ohi nuclear power plant in 2012, before the establishment of new safety regulation. Reactor power was 3.4~GW$_{\mathrm{th}}$ and distance to reactor core was 35.9~m. Measurement period for reactor-on and reactor-off was 30 days and 34 days, respectively. As a result, event rate difference between reactor-on and reactor-off was measured to be 21.8$\pm$11.4 events/day. A significance of neutrino signals was 2$\sigma$ in the PANDA36 modules~\cite{ref:oguri}. It is important to know the background level in the PANDA100 to estimate a significance of neutrino signals.

\subsection*{Background measurement in 2017}
Background measurement was performed in the campus of Kitasato University to understand the background level of the PANDA100 on the ground. The measurement date was from Aug.31 to Sep.8, 2017. A detector was set in a 20 feet container which was set outside of the building. An effect of water shield with 24~cm thickness was checked by the measurement. An event rate below 3 MeV (ambient $\gamma$-rays region) decreased by about 30\%, however, neutron events (correlated events) increased by about 20\% due to muon spallation in the water. Since the neutron background is more significant for the measurement, we will not use water shield.

Since the neutrino signals is observed by inverse beta decay, we applied delayed coincidence technique to observe signals. As a result, the background rate was measured to be about 7,000 events/day. It was also confirmed that the largest background was multi-neutrons induced by cosmic muon, from the shape of prompt and delayed energy spectra.

\section*{Plan of next measurement}
We are asking to electric power company (KEPCO; Kansai Electric Power Company) about the measurement in Ohi nuclear power plant, which is the same location of the PANDA 36 measurement in 2012. Ohi nuclear power plant had supplied the second largest electrical output among nuclear power plants in Japan. There are 4 reactors with thermal power of 3.4~GW$_{th}$ for each reactor. Reactor unit~1 and unit~2 stopped and they were decided to be decommissioned. Reactor unit~3 and unit~4 have been in operation since spring 2018. We plan to perform a measurement outside of the reactor building of unit~3 or unit~4 next spring.

%Five reactors are currently in operation under this situation and two of them are Takahama reactors in Fukui which is close to the author's institute. In addition, seven PWR reactors were permitted to restart. We plan to perform the environmental radiation measurement at Takahama or Ohi reactor in Fukui.

\begin{figure}[tbhp]
\centering
\includegraphics[width=.6\linewidth]{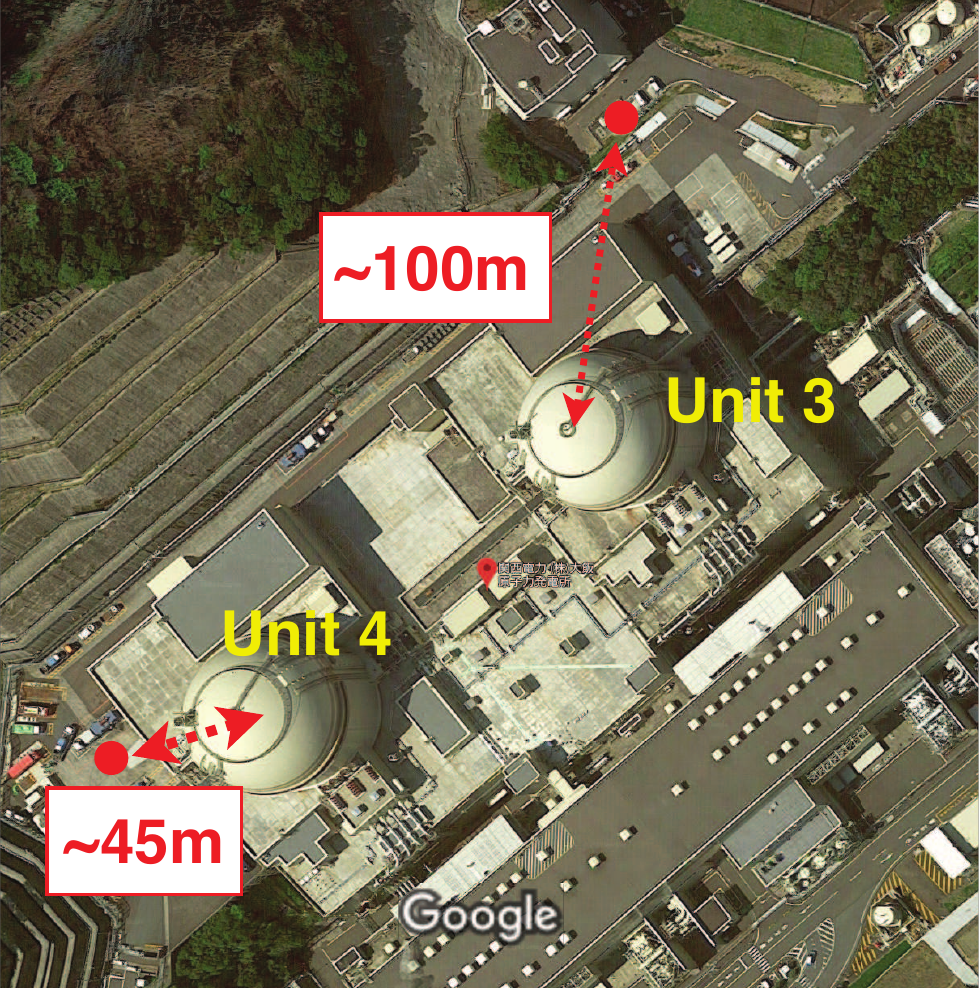}
\caption{Criteria of measurement location in Ohi nuclear power plant. There are two criteria; one is about 45~m from unit 4 and the other is about 100~m from unit 3, respectively.}
\label{fig:location}
\end{figure}

There are two criteria of measurement location where the lashing apparatus is available, as shown in Fig.~\ref{fig:location}. In the new safety regulation, we should fix a car or a container using lashing apparatus, which is against for tornado. Distance from the core is about 45~m or about 100~m. We plan to measure for 1 month in reactor-on and in reactor-off, respectively. Measurement condition and expected rate of signals and backgrounds are summarized in Table.~\ref{tab:condition}, with the measurement result of the PANDA36. Significance will be about 4$\sigma$ at 45~m and about 1$\sigma$ at 100~m.

\begin{table}%[tbhp]
\centering
\caption{Results of the PANDA36 and measurement conditions of the PANDA100 at Ohi nuclear power plant}
\begin{tabular}{lrrr}
Detector & PANDA36 & PANDA100 & PANDA100 \\
\midrule
Target Mass [kg] & 360 & 1,000 & 1,000\\
Distance from core [m] & 36 & 45 & 100\\
Efficiency [\%] & 3.2 & 9.2 & 9.2\\
Expected signals [/day] & 19 & 98 & 20\\
Expected backgrounds [/day] & $\sim$3,000 & $\sim$7,000 & $\sim$7,000\\
\bottomrule
\label{tab:condition}
\end{tabular}
% \addtabletext{nomenclature for the TSs refers to the numbered species in the table.}
\end{table}

\section*{Background measurement with liquid scintillator at Ohi}
PSD capabilityWe performed an ambient background measurement at Ohi nuclear power plant with a set of detectors; NaI (2 inches) for $\gamma$-rays, liquid scintillator (BC501, 3L) which has a PSD capability for fast neutrons, and plastic scintillators for cosmic muons. This may be the first measurement using liquid scintillator at the nuclear power plant in the new safety standards. The measurement location was 100~m from unit~3, and measurement was performed for 2~days. Afs a next step, we plan to install a few tens litter of Gd-loaded liquid scintillator.

\section*{Conclusion}
After the accident of nuclear power plant in Japan, reactors are gradually restarting, but there is a hurdle to perform a measurement in the nuclear power plant due to the strict regulation. A consortium of the research for reactor neutron monitor was formed in this situation. In the PANDA experiment, which consists of plastic scintillator modules with a total mass of 1 ton, a background measurement was performed capability, and the background event rate was measured to be about 7,000 events per day. We plan to install the PANDA detector at Ohi nuclear power plant next spring. If we measure a data for 1 month in reator-on and in reactor-off, a significance of signals is expected to be about 4$\sigma$ at 45 m from the core.

% \matmethods{Please describe your materials and methods here. This can be more than one paragraph, and may contain subsections and equations as required. Authors should include a statement in the methods section describing how readers will be able to access the data in the paper. 

% \subsection*{Subsection for Method}
% Example text for subsection.
% }

% \showmatmethods{} % Display the Materials and Methods section

% \acknow{Please include your acknowledgments and any support information here.}

% \showacknow{} % Display the acknowledgments section

% % Bibliography
% \bibliography{aap2018-participant00x}

% \end{document}

%% file: AAP-Chimenti/aap2018-participant00x.tex
% \documentclass[9pt,twocolumn,twoside,lineno]{aap2018}
% \templatetype{aap2018proceedings}
% \usepackage{siunitx}

\title{Status of the Neutrinos Angra Experiment}

\author[a]{Anjos,J. C.}
\author[a]{Cernicchiaro,G.}
\author[b,*]{Chimenti,P.}
\author[c]{Costa,I. A.}
\author[d]{Gonzales,L. F. G.}
\author[e]{Guedes,G. P.}
\author[d]{Kemp,E.}
\author[a]{Lima Júnior,H. P.}
\author[c]{Lopes,G. S.}
\author[c]{Nóbrega,R. A.}
\author[f]{Pepe,I. M.}
\author[f]{Ribeiro,D. B. dos S.}
\author[c]{Souza,D. M.}
\affil[a]{Brazilian Center for Physical Research, Rio de Janeiro-RJ, Brazil}
\affil[b]{State University of Londrina, Londrina-PR, Brazil}
\affil[c]{Federal University of Juiz de Fora, Juiz de Fora-MG, Brazil}
\affil[d]{University of Campinas - UNICAMP, Campinas-SP, Brazil}
\affil[e]{State University of Feira de Santana, Feira de Santana-BA, Brazil}
\affil[f]{Federal University of Bahia, Salvador-BA, Brazil}

\leadauthor{P. Chimenti} 
\correspondingauthor{\textsuperscript{2}E-mail: pietro.chimenti\@gmail.com}


\begin{abstract}
The Angra Neutrino Experiment aims at monitoring the Angra-II power plant by a non-invasive measurement of its emitted neutrinos. 
A commercial container, refurbished for housing the detector, is placed at surface in proximity of the reactor dome wall, at about 30m from the reactor core. 
The detector itself has an active volume of about \SI{1.42}{m^3} filled with a water solution of \SI{0.2}{\percent} Gd. 
The detector has been installed at the Angra site in the second half of 2017 and is now being deployed. 
A first neutrino measurement is expected during the first semester of 2019.
\end{abstract}

\doi{\url{https://neutrinos.llnl.gov/workshops/aap2018}}

% \begin{document}

\maketitle
\thispagestyle{firststyle}
\ifthenelse{\boolean{shortarticle}}{\ifthenelse{\boolean{singlecolumn}}{\abscontentformatted}{\abscontent}}{}

\dropcap{T}his is the edited text of a talk presented at Livermore on October 10, 2018 as part of the Applied Antineutrino Physics 2018 Workshop. 

\section*{Introduction: motivation and design}

The Angra site was proposed in 2006 as a suitable location for a theta1-3 measurement. 
Later on the international community has opted for concentrating the efforts along three experiments: Daya Bay, Double-Chooz and Reno.
Still the Brazilian community found the interest of developing a neutrino detector for Nuclear non-proliferation purposes: the Neutrinos Angra Collaboration \cite{Anjos:2005pg,Alvarenga:2016lgi} was formed.

The plant operators have been extremely cooperative since the very beginning, imposing only restriction on the detector design on security bases: in particular no flammable liquid was allowed in proximity of the nuclear reactor. 
This excluded the use of organic liquid scintillators (water based liquid scintillators had been developed several years later). 
A water Cerenkov detector was therefore considered as a viable option for our case. 
This kind of detector in fact profits from the following features:
\begin{itemize}
\item it is of very simple construction
\item it is insensitive to cosmogenic fast neutron as a neutrino background
\item it is insensitive to a large fraction of environmental gammas (only the high energy tail of this radiation may produce a signal interacting with the detector).
\end{itemize}

We designed the detector following with the following guidelines: a target, with inner fiducial volume of about \SI{1.42}{m^3} of a \SI{0.2}{\percent} water solution of Gd, instrumented with Hamamatsu Photomultiplier Tubes (PMTs) R5912 with special potting for water Cerenkov applications.
These PMTs have been chosen on the basis of previous positive experience in other experiments. 
In order to increase the light collection efficiency, inner faces of the target tank are covered with GORE\textsuperscript{TM} foils. 
The inner volume is surrounded by tanks of pure water, instrumented with PMTs and with inner surfaces covered with tyvek\textsuperscript{\textregistered}, acting as both shield against background radiation and active veto.
GORE foils for the inner tank have been chosen due to higher reflectivity. 

\section*{Electronics and data acquisition}

The 40 PMTs are powered by a CAEN SY4527 supply. 
Electric signals from photo-electrons (p.e.) are conditioned by front-end electronics boards, custom made by the collaboration \cite{Lopes:2018,Paschoal:2018}.
These boards have been designed to have for a single p.e. (at a typical $10^7$ gain):
\begin{itemize}
    \item 27 ns rise time
    \item 84 ns fall time
    \item 75 ns FWHM
    \item 25.6 mV pulse height
    \item good linearity up to 50 p.e.
    \item a discriminated output.
\end{itemize}

Conditioned signals are digitized by other custom boards (called NDAQ).
These boards, designed to read-out 8 channels each, will operate with the following configuration:
\begin{itemize}
    \item 125 MHz sampling rate (8 ns sampling time)
    \item 100 samples per trigger
    \item -1.25 V to 1.25 V dynamic range
    \item 8-bits vertical resolution (about 9.76 V/bit).
\end{itemize}
NDAQ boards are also equipped with 81 ps resolution Time-to-Digital converters allowing precise timing of the incoming pulses.

Discriminated output from the front-end boards are sent to a commertial ALTERA FPGA board implementing a majority trigger logic.
Once a trigger signal is formed, data from NDAQ modules are sent to a single board computer (ROP) running on the VME crate.
These data are then transferred via Gigabit Ethernet to a local PC controlling the data acquisition.

Completed runs are pre-processed locally and sent to computing clusters at CBPF and Unicamp.
All the data acquisition process can be controlled and partially configured remotely.

\section*{Simulation and expected results}

Considering the fiducial volume of about 1.42 m$^3$ and the reactor-detector mean distance of about 30 m we expect a neutrino interaction rate of about 5070 events/day.
Main backgrounds include cosmic particles (muons, electromagnetic and hadronic components) and enviromental gammas.
A preliminary geant4 simulation \cite{Chimenti:2014} has been developed in order to understand the expected neutrino signal as well as the background, and therefore the signal over background ratio.
Critical inputs of the simulation are related to the optical properties such as water absorption length of Cherenkov light (wavelength dependent), reflectivity of GORE surfaces and quantum efficiency of PMTs.
Some of these quantities depends on the actual working condition of the detector but educated guesses have been made.
A more realistic simulation will be tuned against the real working condition of the detector once the commissioning phase is over.
Simulated events are selected on the basis of the number of p.e. in the prompt and delayed pulses as well as their time difference in order to optimize the neutrino detection.
The expected neutrino detection efficiency is about 40\%.
Still a considerable rate of background events survives the selection making necessary background subtraction techniques by reactor on-off comparisons. 
On the basis of these assumptions we expect to be able to observe a highly significant reactor on/off difference with 24h of data.

\section*{Commissioning and plans}

\begin{figure}[tbhp]
\centering
\includegraphics[width=1.0\linewidth]{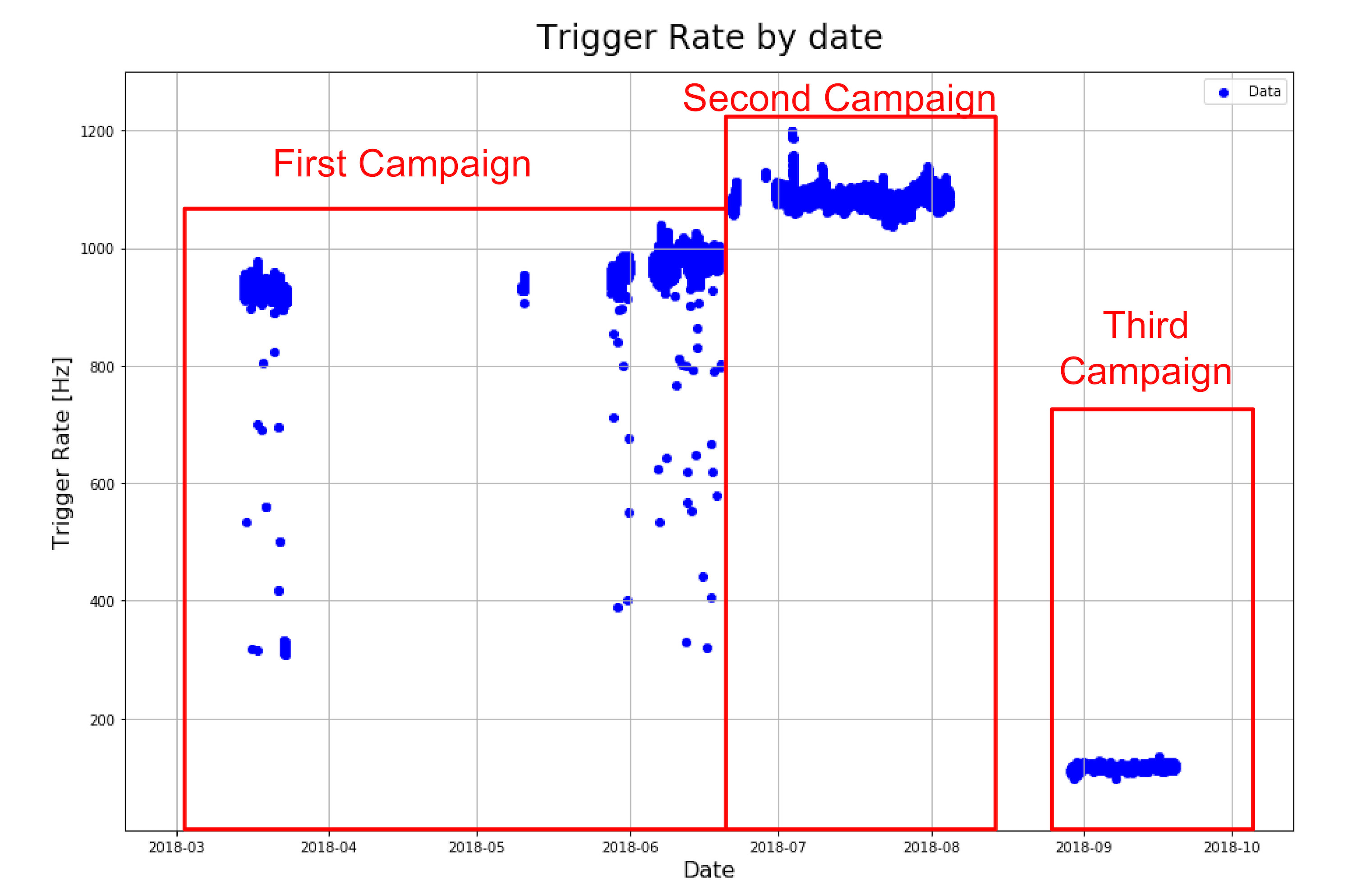}
\caption{Trigger rate during the commissioning campaigns.}
\label{fig:commissioning}
\end{figure}

The Neutrinos Angra detector has been installed in the experimental site in September 2017.
The electronics has been installed in January 2018.
We are devoting the entire 2018 to improve the data acquisition system and trigger configuration.
Three commissioning campaigns (see figure \ref{fig:commissioning}) have been done until now in order to fix issues related to the trigger and veto condition, data acquisition stability and data quality assessment. 

A reactor off period is scheduled for February 2019, we plan therefore to start a data taking campaign early in 2019.

%References should be cited in numerical order as they appear in text; this will be done automatically via bibtex, e.g. \cite{belkin2002using} and \cite{berard1994embedding,coifman2005geometric}. All references should be included in the main manuscript file.  Please use inspire \url{http://inspirehep.net/} for your references; they will be tabulated at the end of the proceedings and do not count toward your two page limit. 

\acknow{The ANGRA collaboration acknowledge the financial support
from CNPq, FAPESP and FINEP. The ANGRA collaboration also thanks Eletronuclear. P.Chimenti would like to thank CNPq for the financial support to participate to the AAP2018 workshop.}

\showacknow{} % Display the acknowledgments section

% Bibliography
% \bibliography{aap2018-participant00x}

% \end{document}

%% file: AAP-Coleman/aap2018-participant00x.tex
%\documentclass[9pt,twocolumn,twoside,lineno]{aap2018}
%\documentclass{article}
% \usepackage[utf8]{inputenc}

% \title{The VIDARR Detector}
% \author{bergevin1 }
% \date{September 2018}

% \begin{document}

% \maketitle

% \section{Introduction}

% \end{document}

% \documentclass[9pt,twocolumn,twoside,lineno]{pnas-new}
% Use the lineno option to display guide line numbers if required.

%\usepackage{siunitx}
%\usepackage{wrapfig}
%\usepackage{epsfig,epstopdf}
%\usepackage{graphicx, color}
%\usepackage{subfig}
% \usepackage{caption}

%\usepackage{subcaption}
% \usepackage{wrapfig}
%\usepackage{caption}
%\usepackage[font={small,it}]{caption}

%\captionsetup[figure]{font=small,it}
%\templatetype{aap2018proceedings} % Choose template 
% {pnasresearcharticle} = Template for a two-column research article
% {pnasmathematics} %= Template for a one-column mathematics article
% {pnasinvited} %= Template for a PNAS invited submission

\title{The VIDARR Detector}

% Use letters for affiliations, numbers to show equal authorship (if applicable) and to indicate the corresponding author
\author[a]{J.Coleman}
\author[a]{R. Collins}
\author[a]{G. Holt} 
\author[a]{C. Metelko} 
\author[a]{M. Murdoch} 
\author[a]{Y. Schnellbach} 
\author[b]{R. Mills}

\affil[a]{University of Liverpool, Merseyside, L69 7ZE}
\affil[b]{National Nuclear Laboratory, Sellafield,  Cumbria, CA20 1PG}
%\affil[c]{Affiliation Three}

% Please give the surname of the lead author for the running footer
\leadauthor{J. Coleman}

% Please include corresponding author, author contribution and author declaration information
% \authorcontributions{Please provide details of author contributions here.}
% \authordeclaration{Please declare any conflict of interest here.}
% \equalauthors{\textsuperscript{1}A.O.(Author One) and A.T. (Author Two) contributed equally to this work (remove if not applicable).}
\correspondingauthor{\textsuperscript{2} E-mail: coleman@liverpool.ac.uk}

% Keywords are not mandatory, but authors are strongly encouraged to provide them. If provided, please include two to five keywords, separated by the pipe symbol, e.g:
% \keywords{Keyword 1 $|$ Keyword 2 $|$ Keyword 3 $|$ ...} 

\begin{abstract}

%Please provide an abstract of no more than 250 words in a single paragraph. Abstracts should explain to the general reader the major contributions of the article. 
The project presented here aims to provide a reliable, autonomous device for the monitoring of reactor anti-neutrino emissions in a safeguards context.  A “drop-in” deployment with minimal intrusion was demonstrated with the device   at Wylfa Power Station, Anglesey, UK. The detector and associated readout and services are designed to be housed in a 20 ft ISO freight container. This technology is highly suitable for use on reactor sites thanks to the use of non-toxic materials with a high flash point and a robust mechanical design. After a successful field trial the detector is undergoing a series of improvements.

\end{abstract}

% \dates{This manuscript was compiled on \today}
\doi{\url{https://neutrinos.llnl.gov/workshops/aap2018}}

%\begin{document}

\maketitle
\thispagestyle{firststyle}
\ifthenelse{\boolean{shortarticle}}{\ifthenelse{\boolean{singlecolumn}}{\abscontentformatted}{\abscontent}}{}

% If your first paragraph (i.e. with the \dropcap) contains a list environment (quote, quotation, theorem, definition, enumerate, itemize...), the line after the list may have some extra indentation. If this is the case, add \parshape=0 to the end of the list environment.

\section*{VIDARR}

\dropcap{T}he detector  utilises the same technology from the ND280 electromagnetic calorimeter (ECal) of the T2K particle physics experiment~\cite{Allan:2013ofa}. Upgrades to the design include a new multi-pixel photon counters (MPPC) variant, and designed for purpose electronics.

\begin{figure}[tbhp]
\centering
\includegraphics[width=.8\linewidth]{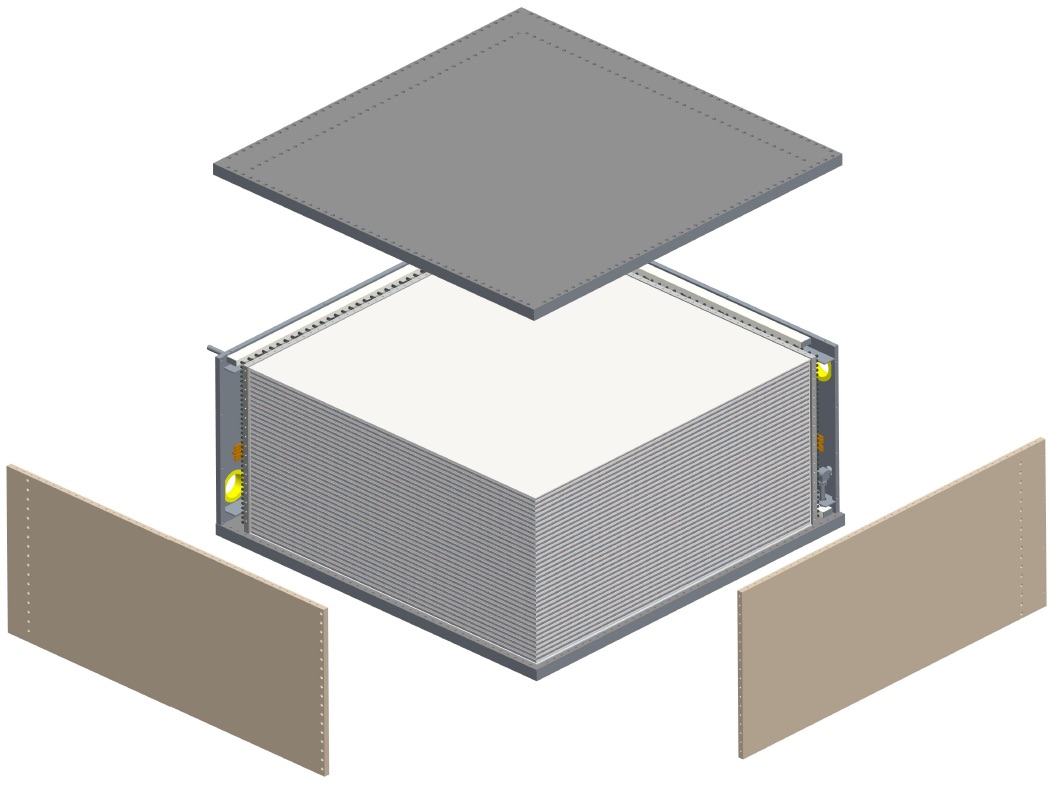}
\caption{A cutaway of the VIDARR detector, illustrating the basic mechanical design.}
\label{fig:frog1}
\end{figure}

The design goals of the detector were based upon the requirements in the 2008 IAEA report~\cite{IAEA}. These include, inert construction,  avoiding liquids,  and low cost. These requirements are satisfied through the use of extruded plastic scintillator and the use of low-voltage MPPCs. Above ground operation is enabled through the segmentation of the scintillator  and portability  arises from being housed inside a freight container only requiring a single power socket. The design has proven to be robust with the ECal surviving the 2011 earthquake off the Pacific coast of Tōhoku in Japan 

After the deployment at Wylfa, the detector was returned to the University of Liverpool for a series of upgrades, working in conjunction with John Caunt Scientific Ltd.  The upgrade relies upon the same requirements from the design goals of the initial iteration~\cite{Carroll:2018kad}, additional operational experience and relevant advances in the associated technology.

\begin{figure}[tbhp]
\centering
\includegraphics[width=.85\linewidth]{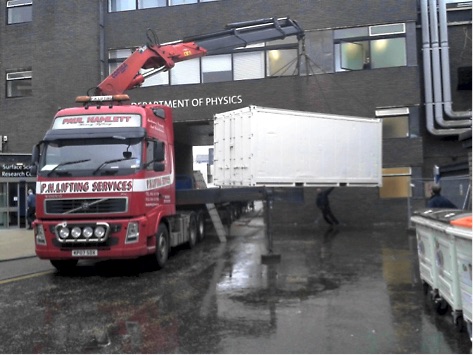}
\caption{Loading of the shipping container onto the flat-bed truck, via a HIAB, at the University of Liverpool.}
\label{fig:frog2}
\end{figure}

\section*{Upgrades}
The detector consists of planes of extruded plastic
scintillator bars with a Ti0$_2$ reflective coating. The bars are orientated to form 
a square and are stacked in a hodoscope topology. A wavelength shifting fibre acts as a light guide through the centre of each bar, and guides the scintillation
photons to the MPPCs located at one end of each fibre.

Additional scintillator bars have been added as part of the upgrade, and the detector
is increased in size to a total of 70 planes
giving an active volume of \SI{1.6}{m^3} and active mass of
approx. 1.5 tonnes. Taking advantage of developments in MPPC technology
since they were first deployed for T2K, results in an MCCP gain
of $2~\times 10^6$, almost a order of magnitude improvement
over their predecessors. The new sensors also give a reduction
in the dark noise rate and cross-talk probability
both by an order of magnitude. A significant improvement
has also been made in photon-detection efficiency,
increasing from 20\% to above 40\%.

\begin{figure}[tbhp]
\centering
\includegraphics[width=.5\linewidth, angle =270 ]{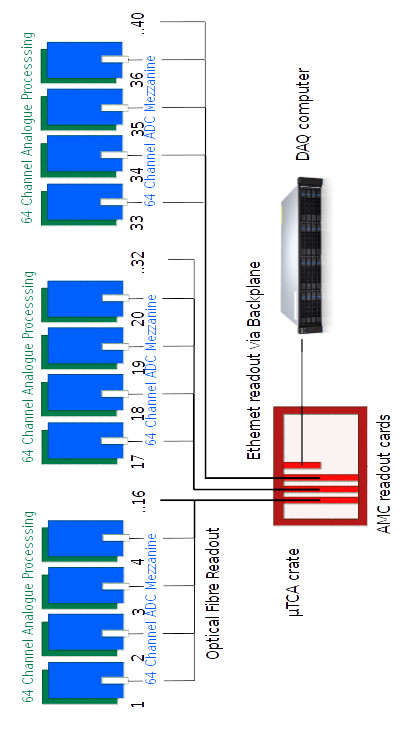}
\caption{An overview of the readout schema for the VIDARR Detector.}
\label{fig:readout}
\vspace{-5mm}
\end{figure}

A complete replacement of the readout electronics
has been developed, replacing those adapted from
Trip-T based readout systems  for the near detector
(ND280) of T2K. Replacing the previous system designed for pulsed
beams, allows for 100\% live readout, improved triggering,
and data taking capability. An overview of the upgraded readout system is shown in figure \ref{fig:readout}. The Front end electronics consist of Analogue
Processing, the signals from which are Digitized, sent, 
and time synchronised to the FC7s utilising
synchronous multi-gigabit links over the optical fibres. The FC7s are synchronised using the $\mu$TCA backplane. The trigger primitives are collated, and a trigger issued. Data from the FC7 is first buffered into DDR3 memory before being transmitted to the DAQ computer
via the $\mu$TCA backplane. A fuller description of the readout electronics can be found in~\cite{Metelko}.

\section*{Performance}

\begin{figure}[!bhp]
 \includegraphics[width=0.23\textwidth]{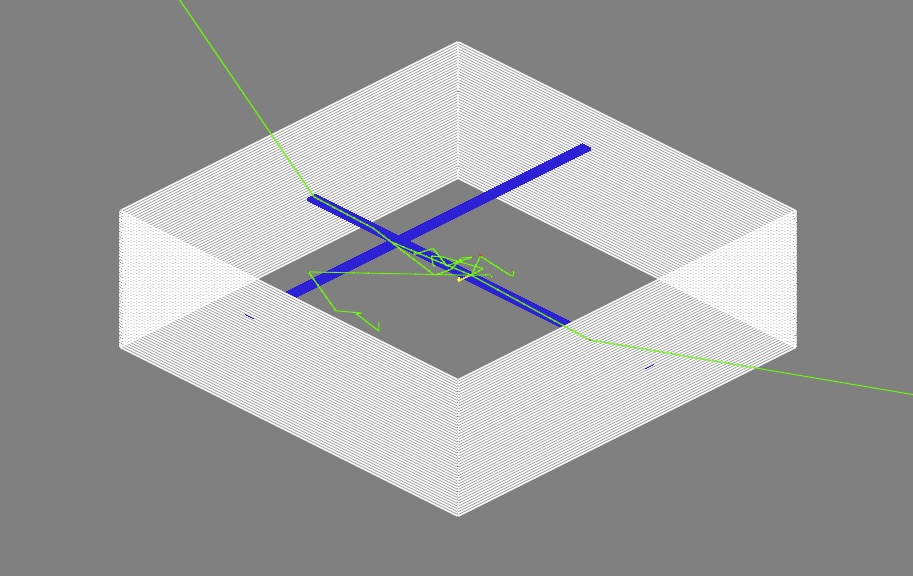}
\includegraphics[width=0.23\textwidth]{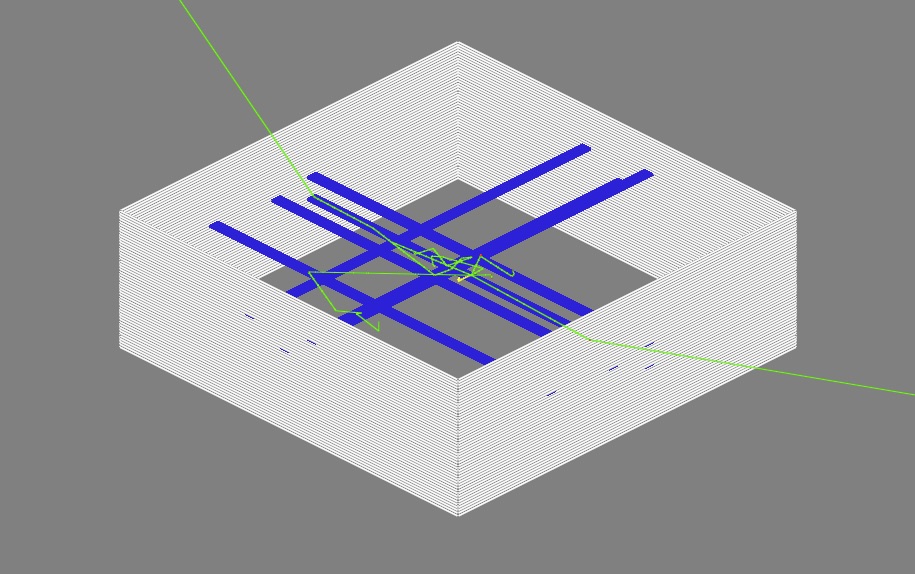}
\caption{An example neutron interaction: (left) high threshold 700 keV, with low number of channels; (right) reduced threshold 100 keV, the number of Channels is high, and hits are spatially separated.}
\label{fig:thresholds}
\end{figure}

The increased photo-detection efficiency and a lower
dark noise rate from the newer MPPCs, combined with improvements in data-rates allows the trigger threshold to be set to \SI{100}{keV}. Resulting in an enhancement of 
the neutron trigger. Figure \ref{fig:thresholds} shows a simulation of
the number of channels triggered with thresholds of
\SI{700}{keV} and \SI{100}{keV} respectively, greatly enhancing
the neutron particle identification over the initial version of the detector. 

The overall upgraded performance of the detector contributes to a reduced background and increased detection rate of anti-neutrinos, due to a 50\% increase in cross-section from the additional instrumented mass, a  further 12\% from improved geometry for the containment of the Gd Interactions. Further gains are expected from the enhanced  performance of the MPPCs and the associated electronic readout and trigger.

\section*{Reactor Simulations}

To support the development of this detector technology for reactor monitoring and to understand its capabilities, the National Nuclear Laboratory modelled the Wylfa graphite moderated and natural uranium fuelled reactor with existing codes used to support Magnox reactor operations and waste management~\cite{refId0}. The 3D multi-physics code PANTHER~\cite{Hutt} was used to determine the individual powers of each of the 49248 fuel elements during the year and a half period of monitoring from reactor records. The WIMS/TRAIL/FISPIN code route~\cite{WIMS} was then used to determine the radionuclide inventory of each nuclide on a daily basis in each element. These nuclide inventories were then used with the BTSPEC~\cite{BTSPEC} code to determine the anti-neutrino spectra and source strength using JEFF-3.1.1~\cite{JEFF} data. Finally the anti-neutrino source from the reactor for each day during the year and a half of monitored reactor operation was calculated. Work is currently ongoing converting these into a form suitable for inclusion in GEANT-4 simulations of the detector.

\begin{figure}[!thp]
\centering
\includegraphics[width=.8\linewidth]{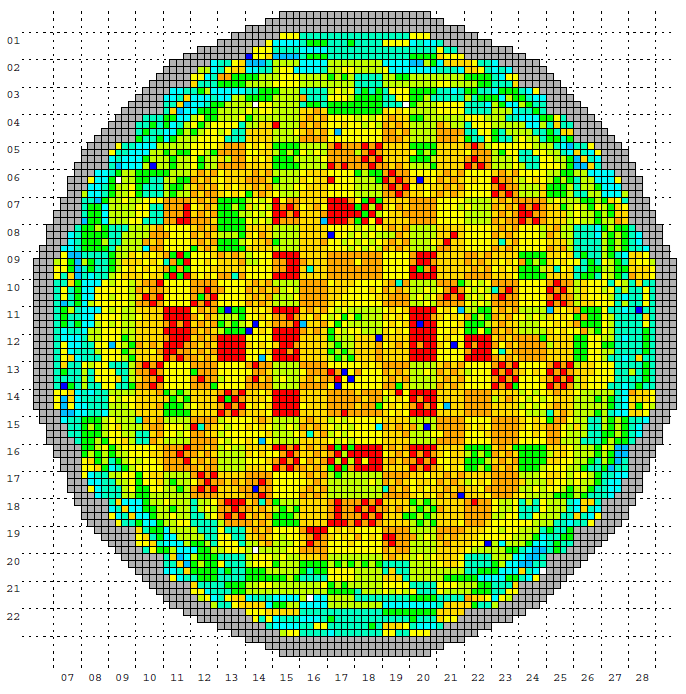}
\caption{An illustrative example from the Reactor simulation  showing the channel powers of the reactor as calculated by PANTHER at one point during the irradiation.}
\label{fig:frog3}
\end{figure}

\section*{Status}
The VIDARR detector is currently being developed and characterized. The upgrades will be completed, and the detector will be ready for deployment in 2020. 

This series of upgrades moves the project towards an industrial demonstrator fully compliant with IAEA recommendations.

\acknow{\\This work was supported by Innovate-UK, the STFC Innovations Partnership Scheme (IPS), the Royal Society of Edinburgh, and the Royal Society. The authors also wish to thank the UK NNL for funding the reactor simulation work and co-funding an EPSRC NGN studentship. We are grateful for the support and contributions of the Particle Physics group and the mechanical workshop from the Physics Department at the University of Liverpool. We would like to thank the members T2K-UK for their advice and the loan of relevant components. These proceedings are an update of reports presented at the AAP in previous years ~\cite{Metelko, Schnellbach}, and the proceedings for ANIMMA~\cite{refId0}.}

\showacknow{} % Display the acknowledgments section

% Bibliography
%\bibliography{aap2018-participant00x}

%ESARDA from MATT\\

%\end{document}

%% file: AAP-Askins/aap2018-participant00x.tex
\graphicspath{{./figures}}

% \templatetype{aap2018proceedings} % Choose 

\title{Water Cherenkov Monitor for Antineutrinos (WATCHMAN)}

% Use letters for affiliations, numbers to show equal authorship (if applicable) and to indicate the corresponding author
\author[ ]{Morgan Askins\textsuperscript{\normalfont a,b,1}, for the AIT-WATCHMAN collaboration}
%\author[b,1,2]{Author Two} 
%\author[a]{Author Three}

\affil[a]{University of California, Berkeley}
\affil[b]{Lawrence Berkeley National Laboratory}
%\affil[b]{Affiliation Two}
%\affil[c]{Affiliation Three}

% Please give the surname of the lead author for the running footer
\leadauthor{Askins} 

% Please include corresponding author, author contribution and author declaration information
% \authorcontributions{Please provide details of author contributions here.}
% \authordeclaration{Please declare any conflict of interest here.}
% \equalauthors{\textsuperscript{1}A.O.(Author One) and A.T. (Author Two) contributed equally to this work (remove if not applicable).}
% \correspondingauthor{\textsuperscript{2}To whom correspondence should be addressed. E-mail: author.two\@email.com}
\correspondingauthor{\textsuperscript{1}Corresponding author: M. Askins, E-mail: maskins@berkeley.edu}

% Keywords are not mandatory, but authors are strongly encouraged to provide them. If provided, please include two to five keywords, separated by the pipe symbol, e.g:
% \keywords{Keyword 1 $|$ Keyword 2 $|$ Keyword 3 $|$ ...} 

\begin{abstract}
    WATCHMAN is a kiloton-scale gadolinium-doped water Cherenkov detector, currently
    under development, to be placed 25 km from the Hartlepool Nuclear Power Station at
    the Boulby mine at a depth of 1070 m. The experiment is part of a nuclear nonproliferation effort with
    the goal of demonstrating the use of anti-neutrino detection for continuous nuclear reactor
    monitoring. WATCHMAN is a single component of the Advanced Instrumentation Testbed
    initiative, which will provide benefit to the scientific community through the
    research and development of advanced detector technologies such as water-based
    liquid scintillator and fast photosensors at a large scale. The detector is
    scheduled to be running by 2022 with an initial 2-year measurement program
    followed by deployment of next-generation detector technologies.
\end{abstract}

% \dates{This manuscript was compiled on \today}
\doi{\url{https://neutrinos.llnl.gov/workshops/aap2018}}

% \begin{document}

\maketitle
\thispagestyle{firststyle}
\ifthenelse{\boolean{shortarticle}}{\ifthenelse{\boolean{singlecolumn}}{\abscontentformatted}{\abscontent}}{}
% If your first paragraph (i.e. with the \dropcap) contains a list environment 
% (quote, quotation, theorem, definition, enumerate, itemize...), the line 
% after the list may have some extra indentation. If this is the case, 
% add \parshape=0 to the end of the list environment.

\dropcap{N}uclear power reactors have both the capability of bringing an abundance of
energy to the world as well as producing the fissile material required to create
highly destructive weapons. As a means to take advantage of the inherent
benefits of nuclear power, it is the International Atomic Energy Agency's (IAEA)
intent to inhibit the use of nuclear power for military use through peace
agreements and monitoring of world reactors while promoting the use of nuclear
energy. In the situation where direct access to a nuclear power facility may
not be available, the monitoring of such reactors through alternative means
is required. Anti-neutrinos from $\beta$-decay of nuclear fission products provide a unique
means to profile the state of a nuclear power reactor at long stand-off distances,
all the while being impossible to shield against.

\section*{Introduction}
A primary requirement for anti-neutrino detection at the mid- and far-field range
is that the technology used is not prohibitively expensive when scaled up to match
the lower rates at increased stand-off distances.
The AIT-WATCHMAN collaboration intends to demonstrate the detection of reactor
anti-neutrinos in a Gadolinium-doped water Cherenkov detector as a means to show
that this particular technology is ideal in this scenario when compared to standard
liquid scintillator anti-neutrino detectors. Anti-neutrinos are detected by their
interaction on hydrogen nuclei through inverse beta decay:
\begin{equation}
    \bar{\nu}_e + \mathrm{p} \rightarrow \mathrm{n} + \mathrm{e}^+.
\end{equation}
By doping the water with 0.1\% Gd by mass, the capture time of the emitted neutron
is reduced from approximately 200 $\mu$s to 30 $\mu$s (depending on final Gd concentration)
due to the significant thermal neutron capture cross-section of Gd .
This leads to fewer accidental coincidences from backgrounds, which results in
a greater sensitivity to anti-neutrinos. The de-excitation of the resulting
excited state of ${}^{158}$Gd results in a higher energy deposit than on water (8 MeV cascade
as opposed to a 2.2 MeV $\gamma$), resulting in a higher detection
efficiency. Detection of neutron capture in Gd-doped water was demonstrated
in a water Cherenkov test stand at Lawrence Livermore National Laboratory \cite{llnl} 
as well as in an acrylic container submerged in the Super-Kamiokande detector \cite{superk}.
The technology was further explored by the WATCHMAN collaboration to look
for cosmogenic radionuclides in the Kimbalton mine \cite{Dazeley:2015uyd}.
The optical clarity and long term stability of Gd-doped water has been
demonstrated using EGADS, which has proven the ability to separate Gadolinium
from water with no loss of material while maintaining optical properties
consistent with pure water \cite{egads}.

\begin{figure}[t]
    \centering
    \includegraphics[width=0.8\linewidth]{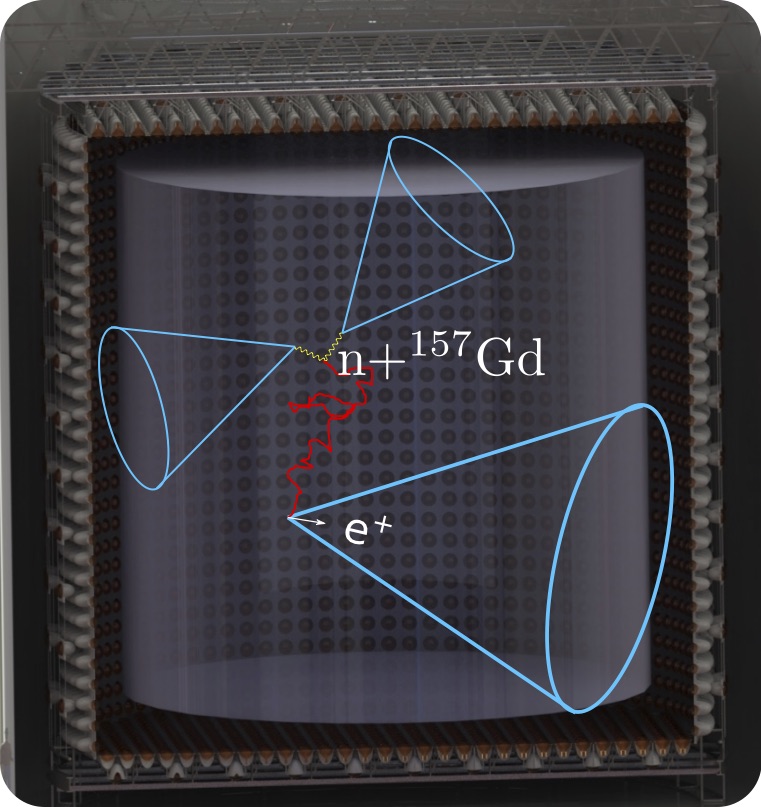}
    \caption{Preliminary design of the WATCHMAN detector, showing a $\sim$ 1 kton
    fiducial volume surrounded by an array of photomultiplier tubes.}
    \label{askins:Figure1}
\end{figure}

\section*{WATCHMAN Detector}
The WATCHMAN baseline design is a cylindrical tank approximately 20 meters in diameter
and 20 meters tall, which will house around 4,000 high quantum efficiency 10-inch
photomultiplier tubes (PMT). A preliminary design of this detector is shown in figure
\ref{askins:Figure1}.To reduce the rate of radioactivity produced by natural $^{214}$Bi
and $^{208}$Tl in the PMT glass, low background PMTs are being considered, which
may reduce the PMT background by a factor of 10 compared to normal PMTs.
The detector will be located near the Boulby Underground Laboratory at the Boulby
Mine in the United Kingdom, which provides a low background environment with
$\sim$ 2800 meters water equivalent overburden. The site is located 25 km from
the Hartlepool Nuclear Power Station, which houses two 1500 MWt advanced gas-cooled
reactors. The target for WATCHMAN is to demonstrate a $~$1 kton fiducial volume
with 0.1\% Gadolinium by mass.

\section*{Backgrounds}
The backgrounds in WATCHMAN can be categorized into two groups: coincident
backgrounds, which produce both a prompt and delayed interaction-- 
mimicking the anti-neutrino signal--and random coincidence from high rate 
single component sources. The primary sources of true coincidence signals
will come from anti-neutrinos from other reactors around the world, neutron
spallation products from muons going through the rock wall, and muon-induced
radionuclides. Radioactivity in the detector components and the target medium will
produce single events, which could randomly mimic the coicidence signal due
to their very high rates. The photomultiplier tubes (despite being low
activity) produce the single most abundant source of radioactivity and
are the limiting component when choosing a fiducial volume. Figure \ref{askins:Figure2}
shows the expected backgrounds per week after event selection, along with the signal
from one reactor core; the second core is treated as an additional
background.

\section*{Reactor Monitoring Analysis Modes}
\begin{figure}[t]
    \centering
    \includegraphics[width=1.0\linewidth]{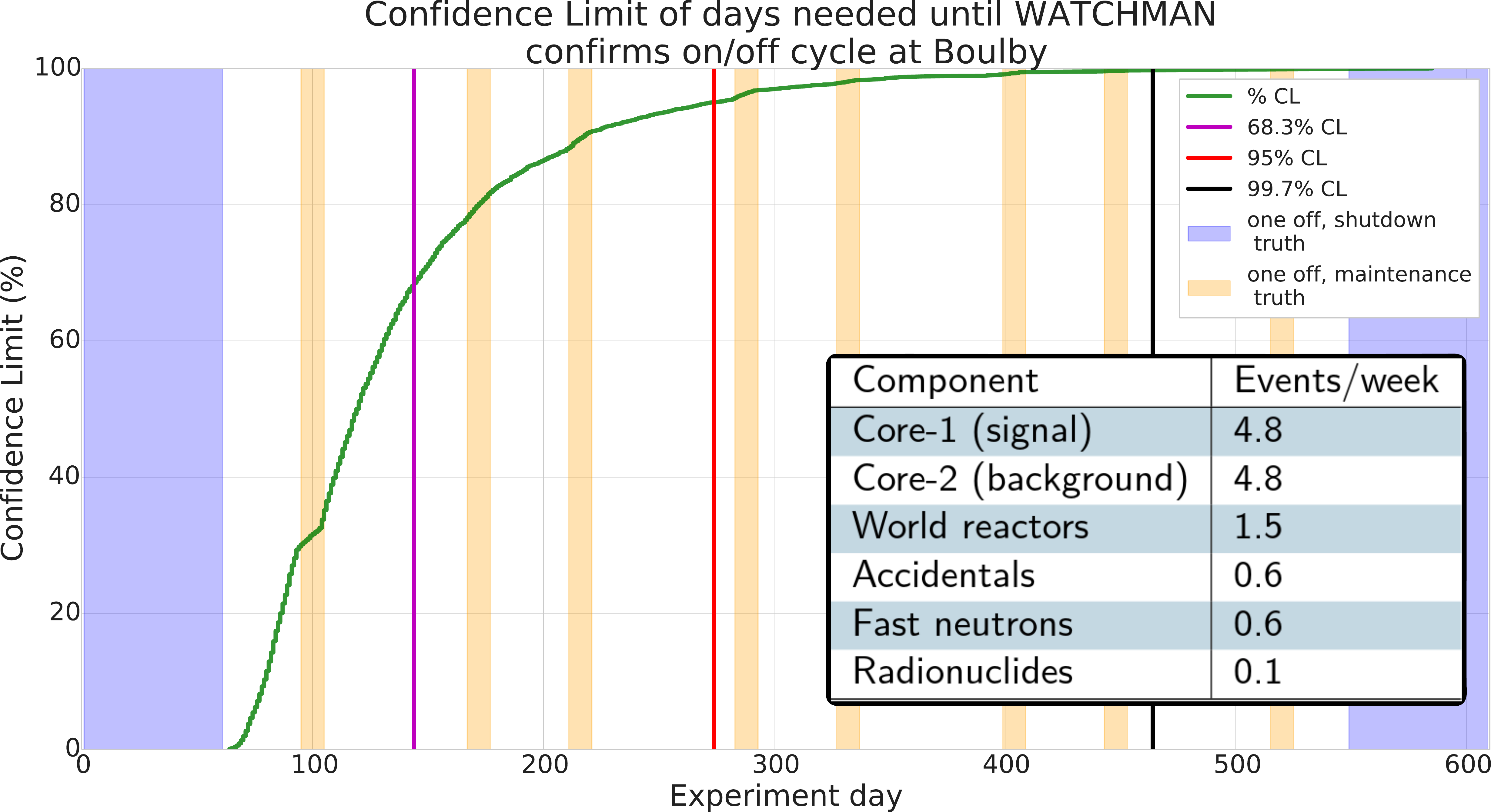}    \caption{Monte Carlo studies performed by T. Pershing indicating the required dwell
    time to distinguish the on/off event rate of two known power reactors
    at a 25 km stand-off. The studies indicate a difference can
    be seen to better than 3$\sigma$ after 300 days at 95\% C.L.}
    \label{askins:Figure2}
\end{figure}

The two-reactor configuration of the Hartlepool reactor allows for
multiple analysis modes, which demonstrate different real-life scenarios.
The first is to perform an analysis with knowledge of both reactor cores,
representing a scenario where one is simply confirming the power cycle of
a reactor facility. Another scenario is to identify an unknown reactor
in the presence of known reactors, where WATCHMAN would try to distinguish
the one-core hypothesis from the two-core hypothesis. Finally, an analysis
can be done in which neither core is known and WATCHMAN must detect the
presence of both reactors together without knowledge of their refueling
cycles. An analysis study was performed using Monte Carlo simulations
produced by the software package RAT-PAC \cite{ratpac} assuming knowledge
of the Hartlepool operation schedule. The time to confirm
observation of the reactor on/off cycle is shown in figure \ref{askins:Figure2}.
Results indicate that a 3$\sigma$ observation will be seen just before 300
days at 95\% confidence.

\section*{Advanced Instrumentation Testbed (AIT)}
The future of the WATCHMAN experiment beyond the initial project goal
is to provide a platform to test future detector technologies, which
could be used in future large scale deployments such as the Theia
experiment \cite{theia}. The exact configuration for WATCHMAN after
the initial reactor monitoring has not been finalized, but the most
appealing options involve filling the detector with water-based liquid
scintillator. Doing so would allow for the largest-scale test of
water-based liquid scintillator and is important in proving the technology's
scalability. Since the primary motivation for using water-based liquid
scintillator is to achieve a detector medium that can be scaled to megaton
size, has high light yield (thus good energy resolution), and retains
the direction information from Cherenkov radiation, the deployment of
fast photosensors would be desirable. Both of these technologies are under development
and will be used in the ANNIE experiment \cite{annie} prior to WATCHMAN.

\section*{Conclusion}
The WATCHMAN experiment is currently under development with preliminary
designs currently being studied. A conceptual design will be completed
during the first half of 2019 with the excavation to begin shortly
afterwards (summer 2019). Installation of the detector will start in
spring 2021 followed by commissioning and calibration. Data taking
will begin some time in 2022 and will continue until at least 2024.
Research and development towards water-based liquid scintillator and fast 
photosensors in WATCHMAN will occur concurrently.

% \showmatmethods{} % Display the Materials and Methods section

% \acknow{Please include your acknowledgments and any support information here.}
\acknow{LLNL-PROC-785244. This work was performed under the auspices of the U.S. Department of Energy by Lawrence Livermore National Laboratory under Contract DE-AC52-07NA27344.}

\showacknow{} % Display the acknowledgments section

% Bibliography
% \bibliography{biblio}

% \end{document}

%% file: AAP-MartiMagro/aap2018-participant00x.tex
% \documentclass[9pt,twocolumn,twoside,lineno]{aap2018}

% \templatetype{aap2018proceedings} % Choose template 
% {pnasresearcharticle} = Template for a two-column research article
% {pnasmathematics} %= Template for a one-column mathematics article
% {pnasinvited} %= Template for a PNAS invited submission

\title{SuperK-Gd}

% Use letters for affiliations, numbers to show equal authorship (if applicable) and to indicate the corresponding author
\author{Llu\'is Mart\'i-Magro}
%\author[b,1,2]{Author Two} 
%\author[a]{Author Three}

\affil{Kamioka Observatory, Institute for Cosmic Ray Research, University of Tokyo.}
%\affil[b]{Affiliation Two}
%\affil[c]{Affiliation Three}

% Please give the surname of the lead author for the running footer
\leadauthor{Mart\'i-Magro}

% Please include corresponding author, author contribution and author declaration information
% \authorcontributions{Please provide details of author contributions here.}
% \authordeclaration{Please declare any conflict of interest here.}
% \equalauthors{\textsuperscript{1}A.O.(Author One) and A.T. (Author Two) contributed equally to this work (remove if not applicable).}
\correspondingauthor{\textsuperscript{2}martillu@suketto.icrr.u-tokyo.ac.jp}

% Keywords are not mandatory, but authors are strongly encouraged to provide them. If provided, please include two to five keywords, separated by the pipe symbol, e.g:

\begin{abstract}
Super-Kamiokande (SK) started collecting data in 1996 and since then it has produced outstanding results in atmospheric and solar neutrino oscillations, as well as in proton decay searches. SK also has the best limits in searches for the diffuse supernova neutrino background but these studies are limited by irreducible backgrounds. In 2004 GADZOOKS! was proposed: add gadolinium (Gd) to the SK ultra-pure water. Gd has the largest thermal neutron capture cross section and produces a gamma cascade of about 8 MeV. To prove the feasibility of GADZOOKS! an R$\&$D project was funded in 2009, EGADS. The project achieved good results and in June 2015 the SuperK-Gd project was approved. In June 2018, the refurbishment of the SK tank begun and will be completed in early 2019.

Here we give a brief account of the benefits of adding Gd, the tests conducted at EGADS and the refurbishment work done so far.
\end{abstract}

% \dates{This manuscript was compiled on \today}
\doi{\url{https://neutrinos.llnl.gov/workshops/aap2018}}

% \begin{document}

\maketitle
\thispagestyle{firststyle}
\ifthenelse{\boolean{shortarticle}}{\ifthenelse{\boolean{singlecolumn}}{\abscontentformatted}{\abscontent}}{}

% If your first paragraph (i.e. with the \dropcap) contains a list environment (quote, quotation, theorem, definition, enumerate, itemize...), the line after the list may have some extra indentation. If this is the case, add \parshape=0 to the end of the list environment.
\dropcap{S}uper-Kamiokande (SK) has been running for over 20 years with outstanding results from atmospheric and solar neutrino oscillation, to proton decay or diffuse supernova neutrino background (DSNB) searches. Adding the capability of neutron tagging would allow to reduce the backgrounds in many studies. For this reason, SK started in June 2018 its refurbishment. 

\section*{Motivation}

The original motivation for GADZOOKS!~\cite{Beacom:2003nk} was the search of the DSNB, the neutrinos from all the past core collapse supernovae (SNe) in the history of the universe. All six types of neutrinos are produced by SNe but they are most likely detected by inverse beta decay (IBD): $\bar{\nu}_{e} p^+ \rightarrow e^+ n $. Observing DSNB is limited by two irreducible backgrounds~\cite{Bays:2011si}. At low energies by the decay of invisible muons (below Cherenkov threshold) and at high energies by atmospheric neutrinos. Because these two backgrounds do not produce neutrons, they could be greatly reduced if we had neutron tagging capabilities. 

When a neutron is produced in SK, a capture on hydrogen after about 200 $\mu$s usually follows and produces a single gamma of 2.2 MeV. This single gamma cannot be detected efficiently. The efficiency to detect an IBD (prompt and delayed neutron capture combined) is about 13$\%$~\cite{Zhang:2013tua}. At EGADS this efficiency is above 80$\%$ when loaded with 0.2 $\%$ of Gd sulfate.

There are many other situations where neutron tagging can be very useful. To pick up an example, proton decay could benefit too since in most of decay channels there is no accompanying neutron.

\section*{EGADS}

EGADS (Evaluating Gadolinium's Action on Detector Systems) was funded to ensure that loading Gd into the otherwise ultra-pure water of SK would not pose a danger for the detector and the current analyses. It basically consists in a 200-ton tank with 240 photo-multipliers, dedicated water filtration systems to purify Gd loaded water, a data acquisition system and other ancillary equipment. It had to demonstrate the following goals were achieved:

\begin{itemize}

\item The filtration system can achieve and maintain a good water quality while keeping the Gd concentration in water constant.
\item Current analyses will not be negatively affected.
\item Gd sulfate has no adverse effects on detector components.
\item Gd can be added/removed in an efficient and economical way.
\item The now visible neutron background, and specially that from impurities in the Gd sulfate itself, does not represent a problem.

\end{itemize}

In addition, a method to stop the water leak in the SK tank had to be developed in order to avoid releasing Gd into the environment while safe to use with Gd.

\begin{figure*}[tbhp]
\centering
\includegraphics[width=1.\linewidth]{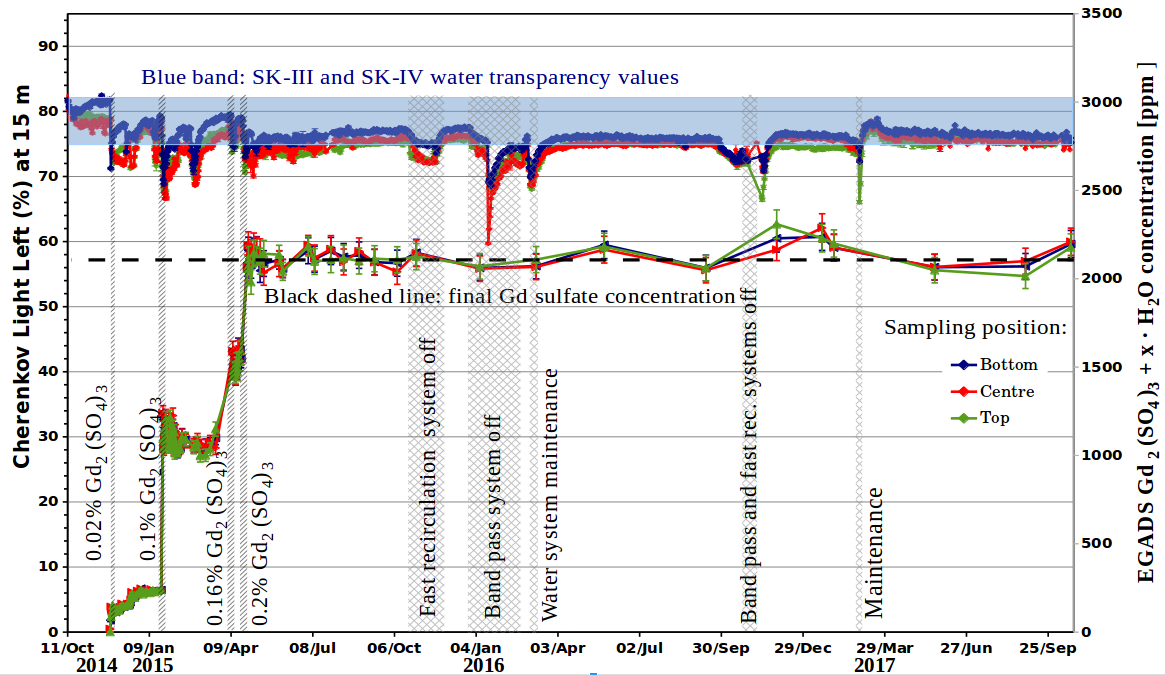}
\caption{Water transparency and Gd sulfate concentration measurements for three different sampling positions: top, centre and bottom of the EGADS tank.}
\label{Marti-Magro:transparency}
\end{figure*}

Figure~\ref{Marti-Magro:transparency} shows the water transparency and the Gd sulfate concentrations. The transparency is shown as the amount of detected Cherenkov light after travelling 15 meters, the typical distance a signal photon travels in SK. This is measured in three positions: top, centre and bottom of the EGADS tank. The blue band represents the typical SK-III and SK-IV values. The Gd sulfate concentration is also monitored in the same positions. The horizontal dashed line shows the final target concentration value.

After each Gd sulfate loading (vertical dashed bands) there is a momentarily drop in the transparency values at all positions. They recover and if the running conditions are kept normal then stay within the blue band.

The Gd sulfate concentration increases after each loading and is quickly homogeneous throughout the EGADS detector. Moreover, after the final loading the concentration is constant, i.e. there is no Gd lost after about 2.5 years.

Figure~\ref{Marti-Magro:transparency} shows that Gd sulfate is transparent to Cherenkov light, the filtration system can remove impurities but keep Gd and that Gd sulfate dissolves homogeneously throughout the detector. At the end of that run, EGADS was emptied. Gd was removed from solution using a resin. The Gd concentration left was measured to be below our detection threshold ($<$ 0.5 ppb).

Once the tank was empty, we examined the tank walls and structure, as well as the photo-multipliers and their supporting structures. Every material looked perfect with no sign of deterioration.

\section*{Neutron background}

The existence of radio impurities in the Gd sulfate could be the source of backgrounds for DSNB and other current analyses. The typical impurities found in our Gd sulfate batches are show in table~\ref{Marti-Magro:Radio-limits}. We studied the radio-purity levels required for each study and then we collaborated with companies to produce Gd sulfate with those goals in mind. These limits are also shown in the table, where the most stringent limits come from DSNB and solar analyses. If no number is given (-) it means that the corresponding requirement is less restrictive. Until now one company has been able to produce Gd sulfate according to this goals while others are still trying to improve their results.

\begin{table}%[tbhp]
\centering
\caption{Table example.}
\begin{tabular}{ccccc}
Chain & Sub-chain & Typical & DSNB & Solar \\
      &           & (mBq/Kg)& (mBq/Kg)& (mBq/Kg) \\
\midrule
 $^{238}$U  & $^{238}$U              &  50        &  $<$ 5    &   -       \\
            & $^{226}$Ra             &  5         &   -       & $<$ 0.5   \\
 $^{232}$Th & $^{232}$Th             &  10        &   -       & $<$ 0.05  \\
            & $^{228}$Th             &  100       &   -       & $<$ 0.05  \\
 $^{235}$U  & $^{235}$U              &  32        &   -       & $<$ 3     \\
            & $^{227}$Ac/$^{227}$Th  &  300       &   -       & $<$ 3     \\

\bottomrule
\label{Marti-Magro:Radio-limits}
\end{tabular}
% \addtabletext{nomenclature for the TSs refers to the numbered species in the table.}
\end{table}

\section*{Detector refurbishment}

\subsection*{Leak fix} To stop the leak we developed a two layer strategy: use two different materials with different mechanical properties. The first one is BIO-SEAL 197 by Thin Film Technology Inc. This material can easily fill cracks and is mechanically very strong, albeit rigid. The second material is MineGuard-C by Hodogawa company. This material was chosen to overcoat BIO-SEAL 197 because is more flexible, allows larger displacements and has lower Rn emanation levels.

Multiple tests including fatigue tests have been conducted with these materials to ensure the mechanical properties meet our requirements. During refurbishment work, samples of these materials were taken in order to measure contamination from radio-impurities. 

\subsection*{Photo-multiplier replacement} About 140 photo-multipliers (PMTs) were replaced in the inner detector (ID) while about 200 PMTs were replaced in the outer detector (OD). Newly developed PMTs for Hyper-Kamiokande were used for the replacement in the ID. For the OD most of the replacements were done in the top region (about 100 PMTs).

\subsection*{Tyvek replacement} Tyvek sheets separate optically the ID and the OD. Their surfaces are black in the ID and white in the OD, to absorb and reflect light. These sheets have been largely replaced by new ones. 

\subsection*{New water filtration system} A new hall was excavated near SK and new water filtration systems suitable for Gd loaded water installed. 

SK is currently being filled with pure water. Some work in the tank top is left and it is expected to be finished in January 2019. Several time schedules for Gd loading are being considered in coordination with the T2K collaboration.

%\subsection*{Labeling}
%
%Figures and Tables should be labelled and referenced in the standard way using the \verb|\label{firstAuthorLastName:FigureX}| and \verb|\ref{firstAuthorLastName:FigureX}| commands.

% Figure \ref{fig:frog} shows an example of how to insert a column-wide figure. To insert a figure wider than one column, please use the \verb|\begin{figure*}...\end{figure*}| environment. Figures wider than one column should be sized to 11.4 cm or 17.8 cm wide. Use \verb|\begin{SCfigure*}...\end{SCfigure*}| for a wide figure with side captions.

% \matmethods{Please describe your materials and methods here. This can be more than one paragraph, and may contain subsections and equations as required. Authors should include a statement in the methods section describing how readers will be able to access the data in the paper. 

% \subsection*{Subsection for Method}
% Example text for subsection.
% }

% \showmatmethods{} % Display the Materials and Methods section

%\acknow{Please include your acknowledgments and any support information here.}
%\showacknow{} % Display the acknowledgments section

% Bibliography
% \bibliography{aap2018-participant00x}

% \end{document}

%% file: AAP-Hill/aap2018-participant00x.tex
% \documentclass[9pt,twocolumn,twoside,lineno]{aap2018}

% \templatetype{aap2018proceedings} % Choose template 
% \usepackage{enumitem}

\title{The Versatile Test Reactor Overview}

% Use letters for affiliations, numbers to show equal authorship (if applicable) and to indicate the corresponding author
\author[a,b]{Tony S. Hill}

\affil[a]{Idaho State University}
\affil[b]{Idaho National Laboratory}

% Please give the surname of the lead author for the running footer
\leadauthor{Hill}

% Please include corresponding author, author contribution and author declaration information
% \authorcontributions{Please provide details of author contributions here.}
% \authordeclaration{Please declare any conflict of interest here.}
% \equalauthors{\textsuperscript{1}A.O.(Author One) and A.T. (Author Two) contributed equally to this work (remove if not applicable).}
\correspondingauthor{Tony Hill E-mail: Tony.Hill@inl.gov}

% Keywords are not mandatory, but authors are strongly encouraged to provide them. If provided, please include two to five keywords, separated by the pipe symbol, e.g:
% \keywords{VTR$|$ Versatile Test Reactor $|$ Sterile neutrinos} 

\begin{abstract}
The DOE Office of Nuclear Energy has initiated preliminary R\&D to develop the Versatile Test Reactor (VTR) program.  The mission of the VTR program is to provide leading edge capability for accelerated testing and qualification of advanced fuels and materials, enabling the U.S. to regain and sustain technology leadership in the area of advanced reactor systems.  The goal of the VTR program is to enable a fast spectrum test reactor that can begin operations by 2026.  The VTR is an essential tool for supporting a new generation of high-value experiments, tests and validation.  The VTR Experimental R\&D effort is focused on maximizing experimental outcomes by developing a broad array of experimental capabilities, as well as providing the most accurate and precise experimental boundary conditions and validation opportunities as possible, which may require additional instrumentation outside the nominal instrumentation and control suite required for safe reactor operations.  Given the aggressive schedule of the VTR program, one focus of the Experimental R\&D effort is to expeditiously identify potentially valuable scientific and engineering opportunities that are only possible or highly optimized when included in the baseline design of the VTR system.

\end{abstract}

% \dates{This manuscript was compiled on \today}
\doi{\url{https://neutrinos.llnl.gov/workshops/aap2018}}

% \begin{document}

\maketitle
\thispagestyle{firststyle}
\ifthenelse{\boolean{shortarticle}}{\ifthenelse{\boolean{singlecolumn}}{\abscontentformatted}{\abscontent}}{}

% If your first paragraph (i.e. with the \dropcap) contains a list environment (quote, quotation, theorem, definition, enumerate, itemize...), the line after the list may have some extra indentation. If this is the case, add \parshape=0 to the end of the list environment.
\dropcap{T}The VTR is an essential tool for supporting a new generation of high-value experiments, tests and validation opportunities.  The goal of the VTR Experimental Program is to maximize experimental outcomes by providing a broad array of experimental capabilities, as well as providing the most accurate and precise experimental boundary conditions and validation opportunities as possible, which may require additional instrumentation outside the nominal instrumentation and control suite required for safe reactor operations.  Given the aggressive schedule of the VTR program, one focus of the Experimental R\&D effort is to identify potentially valuable scientific instrumentation and determine the impact on the VTR design.  As an example, VTR University R\&D funding was awarded to Georgia Institute of Technology to develop and design a VTR specific ex-core power measurement system based on inverse beta decay (IBD) detector technologies, originally developed for the international safeguards mission.  The system will be designed for operation in an engineered environment, well outside the reactor vessel, to avoid sensor degradation or drift, which is common to instrumentation in the harsh environment inside the vessel.  Such a system may provide an accurate, precise, reliable, and independent reactor power history over the entire VTR operational timeline.  The integration of IBD power monitor data in the VTR data analytics layer may provide independent and foundational support in minimizing experimental systematic uncertainties associated with the challenge of accurately reconstructing power of a pool-type sodium fast reactor.  If a decision is made to include such a system, the required infrastructure will need to be included in the plant design.  

The VTR will produce over $10^{19}$ electron antineutrinos per second at full power and could possibly support advanced development of reactor antineutrino detection systems by providing appropriate testing galleries close enough to the core to quickly impress signals on new detector designs, including those based on coherent neutrino scattering, or for direct inter-comparisons between various detection systems to understand systematic performance differences.  Interest in such infrastructure will need to be communicated to the VTR program in an appropriate time frame in order to be considered for inclusion in the plant layout.  

The baseline fuel being studied for the VTR core is $80LEU_{5\%}20PU_{RG}10Zr$, where 99.5\% of the fissions occur in five isotopes. The isotopic fission ratios are significantly different from LWR systems and are relatively stable during an anticipated 100-day irradiation cycle (see Table \ref{Hill:FissionFractions}).  Initial plans for power operations do not include load following, which results in a linear decrease in power of $\sim2\%$ over a run cycle.  Nominal operations are anticipated to include $\sim20$ days for reconfiguration between irradiations.  The unique fission ratios, the relative stability of the system, along with adequate power-off periods, is a perfect laboratory for studying the response of antineutrino detectors and provides a unique window on the Pu line shapes in support of advanced safeguard systems and potentially contributing to the resolution of the existing reactor antineutrino IBD line-shape discrepency between data and simulations.

\begin{table}%[htbp]
\centering
\caption{Isotopic fission fractions expected in the VTR baseline fuel at the beginning and end of a 100 day power cycle once the core loading has reached equilibrium.}
\begin{tabular}{lrrr}
Isotope & Begin & End & Relative Change \\
\midrule
U235 & 13.2\% & 12.8\% & -3.7\% \\
U238 & 12.6\% & 12.7\% & 1.5\% \\
Pu239 & 61.8\% & 61.8\% & 0.02\% \\
Pu240 & 8.2\% & 8.4\% & 2.2\% \\
Pu241 & 3.7\% & 3.8\% & 2.3\% \\
\bottomrule
\label{Hill:FissionFractions}
\end{tabular}
% \addtabletext{nomenclature for the TSs refers to the numbered species in the table.}
\end{table}

The tension between observed and predicted reactor antineutrino rates is commonly referred to as the reactor antineutrino anomaly and remains an active field of research.  The anomaly may be related to deficiencies in the nuclear data that support the rate predictions or may be related to deficiencies in our fundamental understanding of the neutrino sector in the standard model.  The VTR may provide opportunities to make substantial progress on both fronts.  Reactor antineutrinos are produced in the successive decays of excited fission products as they make their way towards stability.  The fission fragment source term is isotope and neutron spectra dependent.  The nuclear data available contains yield estimates from many of the actinides but are typically limited to one or two incident neutron energies.  First order interpolation and extrapolations are typically utilized for numeric predictions.  To support antineutrino measurements at the VTR, integral fission product yield measurements can be carried out using a rabbit system and a mass spectrometer, such as ICP-MS, to provide valuable constraints for all the fissionable isotopes in the VTR spectrum. The beta decay spectrum is also an important ingredient, as it is correlated to the antineutrino spectrum through the available transition energy.  Most of the pertinent beta spectra data available today were measured at ILL in a very thermalized spectrum.  New beta spectra measurements carried out in the VTR, combined with VTR fission product yield measurements, can certainly increase the confidence in predictions and provide direct comparisons in spectra and rates between the betas and antineutrinos.  A back of the envelope calculation suggests that a beta spectrometer at the VTR would provide data rates about 10\% of that seen at ILL without considering any specific spectrometer enhancements. The business case for this added capability needs to be fully delineated for further evaluation.

\begin{figure}[tbhp]
\centering
\includegraphics[width=.9\linewidth]{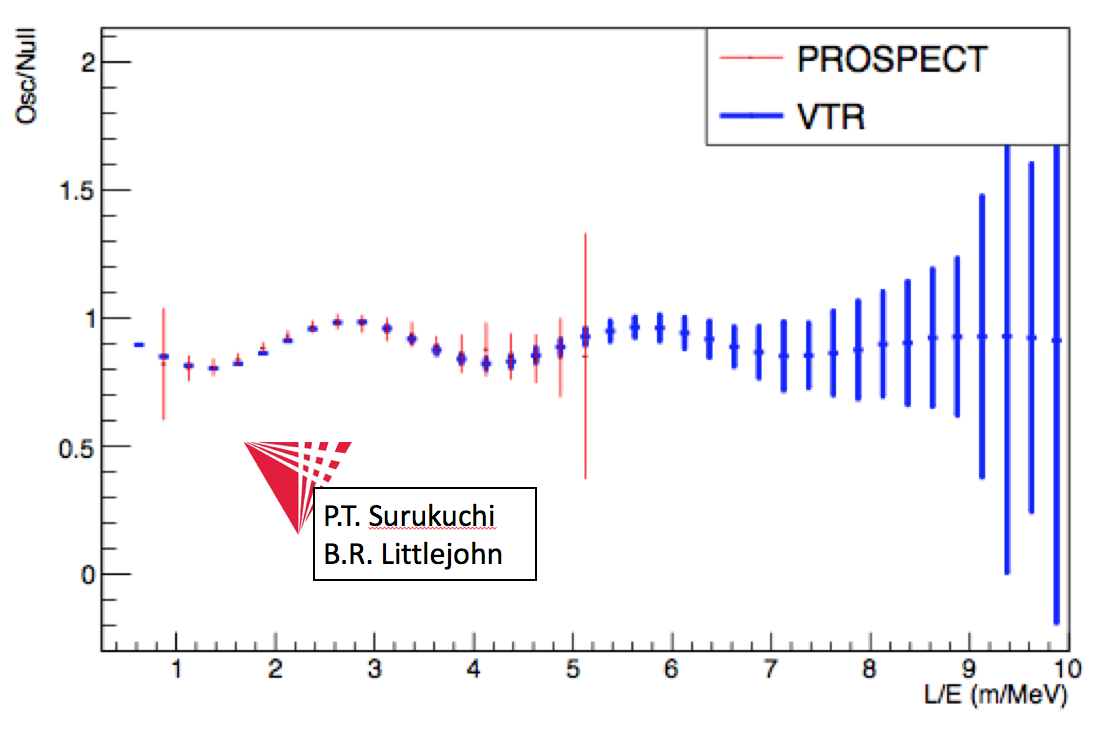}
\caption{Estimated VTR L/E oscillation sensitivity compared to current PROSPECT result shows nearly twice the L/E coverage and represents a good situation for addressing a wide range of coverage of 3+1, or 3+2, or osc+decay, or other non-standard oscillatory patterns.}
\label{Hill:Oscillations}
\end{figure}

\begin{figure}[tbhp]
\centering
\includegraphics[width=1.1\linewidth]{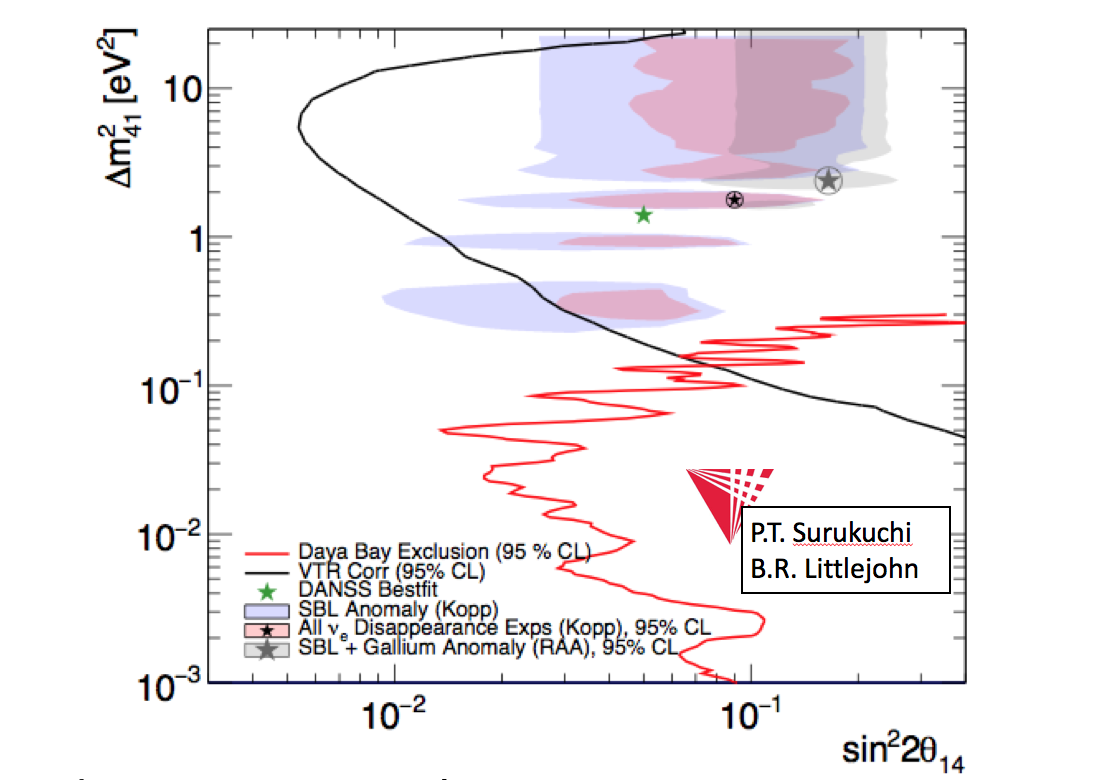}
\caption{The estimated 3+1 sensitivity at VTR covers the entire Kopp region at the 95\% confidence level.}
\label{Hill:Kopp}
\end{figure}

The reactor antineutrino anomaly may also be due to physics beyond the standard model.  Neutrino research over the years has certainly determined that at least two of the three known “light” neutrinos have mass and their relative mass differences and mixing angles are known to some precision.  The existence of yet undiscovered “heavy sterile” neutrinos is plausible with extensions to the standard model framework and if they exist, could be responsible for the reactor antineutrino anomaly.  The reactor rate anomaly suggests a specific range for the mixing angle; other constraints, including those from cosmology, suggest a mass difference of order 1 eV.  Due to detection resolution, the signature for these exotic neutrinos can only be accessed at very short distances from the source.  Many short baseline reactor experiments have collected, are collecting, or will be collecting IBD data with limited statistics and baseline coverage, looking for subtle differences in the neutrino energy distributions as a function of distance. All of these experiments have baseline coverage restrictions due to the limited access to existing reactor cores.  The collective results to date are intriguing, neither statistically confirming or completely excluding their existence, leading to the development of more refined and definitive experiments.  The VTR may provide the perfect opportunity to resolve this important fundamental physics issue.  If one considers a single IBD detector system below the VTR core that covers a baseline from 4 to 17 meters, the discovery space is significantly enhanced over current or planned experiments, such as PROSPECT \cite{Ashenfelter:2018iov}.  PROSPECT collaborators at the Illinois Institute of Technology provided an estimate of the impact for a large PROSPECT-type system below the VTR core (see Figures \ref{Hill:Oscillations} and \ref{Hill:Kopp}).  The inputs and assumptions for the calculations are:
\begin{itemize}
\item Detector
\begin{itemize}[itemsep=-1ex,topsep=-2ex]
\item	2m x 2m x 13m system (52 tonnes)
\item	Baseline from 4 to 17 meters
\item	95 x 18, 15cm x 15cm x 2m segments
\item	Assumed ‘best’ (center) PROSPECT segments energy response for all segments
\item	Assumed signal:background = 1:3
\end{itemize}
\item	Reactor
\begin{itemize}[itemsep=-1ex,topsep=-2ex]
\item	0.4m wide and 0.5m high core (HFIR)
\item	Assumed 10 years running at 100 MW, or 3.3 years at 300MW
\end{itemize}
\item	Assumed Uncertainties
\begin{itemize}[itemsep=-1ex,topsep=-2ex]
\item	Signal and cosmic background statistics
\item	PROSPECT IBD energy response uncertainty (nonlinearity, scaling, resolution), treated as uncorrelated between each segment
\item	2\% segment-uncorrelated IBD rate uncertainty
\end{itemize}
\end{itemize}

As a new advanced reactor system that is still on the drawing board, the VTR may provide unique opportunities to support a variety of interests, by design, from fundamental physics discoveries to advanced safeguards detector development.  However, the aggressive timeline to construction requires interested parties to develop and share their ideas with the VTR management as soon as possible for consideration.  Many opportunities exist but time is short.

\acknow{The author would like to thank PROSPECT collaborators B.R.~Littlejohn and P.T.~Surukuchi at Illinois Institute of Technology for the estimated VTR sensitivity plots.}

\showacknow{} % Display the acknowledgments section

% Bibliography
% \bibliography{aap2018-participant00x}

% \end{document}

%% file: AAP-Carr/aap2018-participant00x.tex
% \documentclass[9pt,twocolumn,twoside,lineno]{aap2018}
%\documentclass{article}
% \usepackage[utf8]{inputenc}

% \title{AAP-PersonalizedTemplate}
% \author{bergevin1 }
% \date{September 2018}

% \begin{document}

% \maketitle

% \section{Introduction}

% \end{document}

% \documentclass[9pt,twocolumn,twoside,lineno]{pnas-new}
% Use the lineno option to display guide line numbers if required.

% \templatetype{aap2018proceedings} % Choose template 
% {pnasresearcharticle} = Template for a two-column research article
% {pnasmathematics} %= Template for a one-column mathematics article
% {pnasinvited} %= Template for a PNAS invited submission

\title{Nuclear explosion monitoring: Can neutrinos add value to the global system?}

% Use letters for affiliations, numbers to show equal authorship (if applicable) and to indicate the corresponding author
\author[a,1]{Rachel Carr}

\affil[a]{Department of Nuclear Science and Engineering, Massachusetts Institute of Technology}

% Please give the surname of the lead author for the running footer
\leadauthor{Carr}

% Please include corresponding author, author contribution and author declaration information
% \authorcontributions{Please provide details of author contributions here.}
% \authordeclaration{Please declare any conflict of interest here.}
% \equalauthors{\textsuperscript{1}A.O.(Author One) and A.T. (Author Two) contributed equally to this work (remove if not applicable).}
\correspondingauthor{\textsuperscript{1}recarr@mit.edu}

% Keywords are not mandatory, but authors are strongly encouraged to provide them. If provided, please include two to five keywords, separated by the pipe symbol, e.g:
% \keywords{Keyword 1 $|$ Keyword 2 $|$ Keyword 3 $|$ ...} 

\begin{abstract}
Hundreds of seismic, hydroacoustic, infrasound, and radionuclide sensors monitor the earth for evidence of nuclear explosions. Since the Comprehensive Nuclear Test Ban Treaty (CTBT) opened for signatures two decades ago, some analysts have asked whether neutrino detection could provide a useful complement to this network. In principle, detecting neutrino emissions could uncover a clandestine nuclear weapon test missed by other technologies, or clarify the nuclear nature and fission yield of a suspected test event. In practice, these capabilities would require very large neutrino detectors, potentially beyond the scale of any built or currently planned. The current cost scale for a single, regional-coverage neutrino detector is roughly \$1B. While neutrino detectors may find practical use in monitoring nuclear reactors, it remains difficult to see where they could add value, for a reasonable cost, to the strong existing network for detecting nuclear explosions.
\end{abstract}

% \dates{This manuscript was compiled on \today}
\doi{\url{https://neutrinos.llnl.gov/workshops/aap2018}}

% \begin{document}

\maketitle
\thispagestyle{firststyle}
\ifthenelse{\boolean{shortarticle}}{\ifthenelse{\boolean{singlecolumn}}{\abscontentformatted}{\abscontent}}{}

% If your first paragraph (i.e. with the \dropcap) contains a list environment (quote, quotation, theorem, definition, enumerate, itemize...), the line after the list may have some extra indentation. If this is the case, add \parshape=0 to the end of the list environment.

\dropcap{E}xplosive testing has been important to the development of nuclear weapons, and nations and international agencies retain an interest in knowing about nuclear explosions occurring anywhere in the world. While most nuclear weapons states stopped explosive testing by the late 1990s, the Comprehensive Nuclear Test Ban Treaty (CTBT) is not yet in force. Nuclear explosions have occurred as recently as 2017, with the sixth nuclear test in North Korea. 

Since the CTBT was negotiated in the 1990s, nations have cooperated to develop a global system for detecting and characterizing nuclear explosions. Seismic, hydroacoustic, infrasound, and radionuclide sensors form a network watching for signals from nuclear explosions anywhere in the world, down to a detection threshold of 0.5 kton or lower \cite{NAP12849}. In principle, the neutrino emissions from nuclear fission explosions could complement this existing system by:
\begin{itemize}
\item Detecting explosions that would otherwise go undiscovered, namely, explosions below the current $\sim 0.5$ kton detection threshold; 
\item Confirming that a suspected explosion event included fission yield, helping to exclude alternative explanations such as a chemical explosion or earthquake;
\item Estimating the fission yield of the explosion and, in combination with seismic yield estimates, indicating if the explosion contained a substantial fusion yield, or if the seismic signal had been intentionally masked by underground cavity engineering.
\end{itemize}

An important question is: Could any of these hypothetical neutrino applications add value to the existing explosion monitoring system, at a reasonable cost? To respond, it is useful to consider the basic physics of the neutrino signal, prospects for detection, and the relative benefits of neutrino detection compared to existing approaches.
\setcounter{secnumdepth}{0}
\section*{Neutrino signal from a nuclear explosion}

A nuclear explosion generates approximately $10^{24}$ neutrinos per kiloton of fission yield \cite{Bernstein2001}. The dominant mechanism for neutrino production is the same as that in nuclear reactors (note that the main fusion reactions typical of nuclear weapons do not produce neutrinos). After a $^{235}$U, $^{239}$Pu, or other nucleus fissions, the resulting nuclear fragments undergo several beta decays, each releasing a neutrino. For a rough comparison with with neutrino rates in reactor-based experiments, it is helpful to consider the energy units: 1 kton $\approx$ 1 GWh.

In contrast to a nearly steady-state nuclear reactor, a nuclear explosion releases neutrinos in a concentrated burst. Our recent simulation indicates that about 60\% of the neutrino emission in a detectable energe range (above inverse beta decay (IBD) threshold) occurs in the first 10 s \cite{Carr:2017uuq}. Compared to the equilibrium flux from reactors, neutrinos emitted in the first 10 s post-explosion have a higher mean energy, due to the lifetime-endpoint anticorrelation of beta decays. See Ref. \cite{Carr:2017uuq} for details about the signal simulation. Observing IBD in water-based detectors is currently the most conceivable approach to detecting neutrinos from nuclear explosions. See \cite{Askins:2015bmb} for perspective on using large, Gd-doped water Cherenkov detectors for neutrinos from fission.

\section*{Hypothetical use cases}

\subsection*{A. Detect explosions otherwise missed}

Theoretically, a large water-based detector could identify neutrinos from a nuclear explosion that was not detected by the existing monitoring network. A 2001 study focused on this possibility \cite{Bernstein2001}. The authors estimated that detecting a significant signal (taken to be 10 neutrinos) from a 1-kton explosion at 100-km standoff would require a detector 60 times the size of the largest existing water detector, Super-Kamiokande. The study concluded that ``while antineutrino detectors are in theory very attractive'' for detecting otherwise missed explosions, ``engineering difficulties and ultimately physics limitations severely proscribe actual applications.'' This conclusion stands in 2018, and is indeed strengthened by the fact that non-neutrino techniques now put the global explosion detection threshold below 1 kton.

\subsection*{B. Confirm nuclear nature of explosion}

Seismic observations can now identify the time and place of a suspected explosion to within a few seconds and tens of kilometers, depending on the explosion size and location \cite{epicenter}. Using the suspected detonation time as an analysis trigger reduces the number of events needed to attribute a neutrino signal to a fission explosion, compared to the previous case of detection \textit{per se}. We explored this possibility for Gd-doped water Cherenkov detectors, accounting for estimated background levels \cite{Carr:2017uuq}. A significant neutrino signal could play a role similar to the detection of airborne radionuclides, which currently help to confirm the nuclear nature of suspected explosions.

\begin{figure}[t]
\centering
\includegraphics[width=\linewidth]{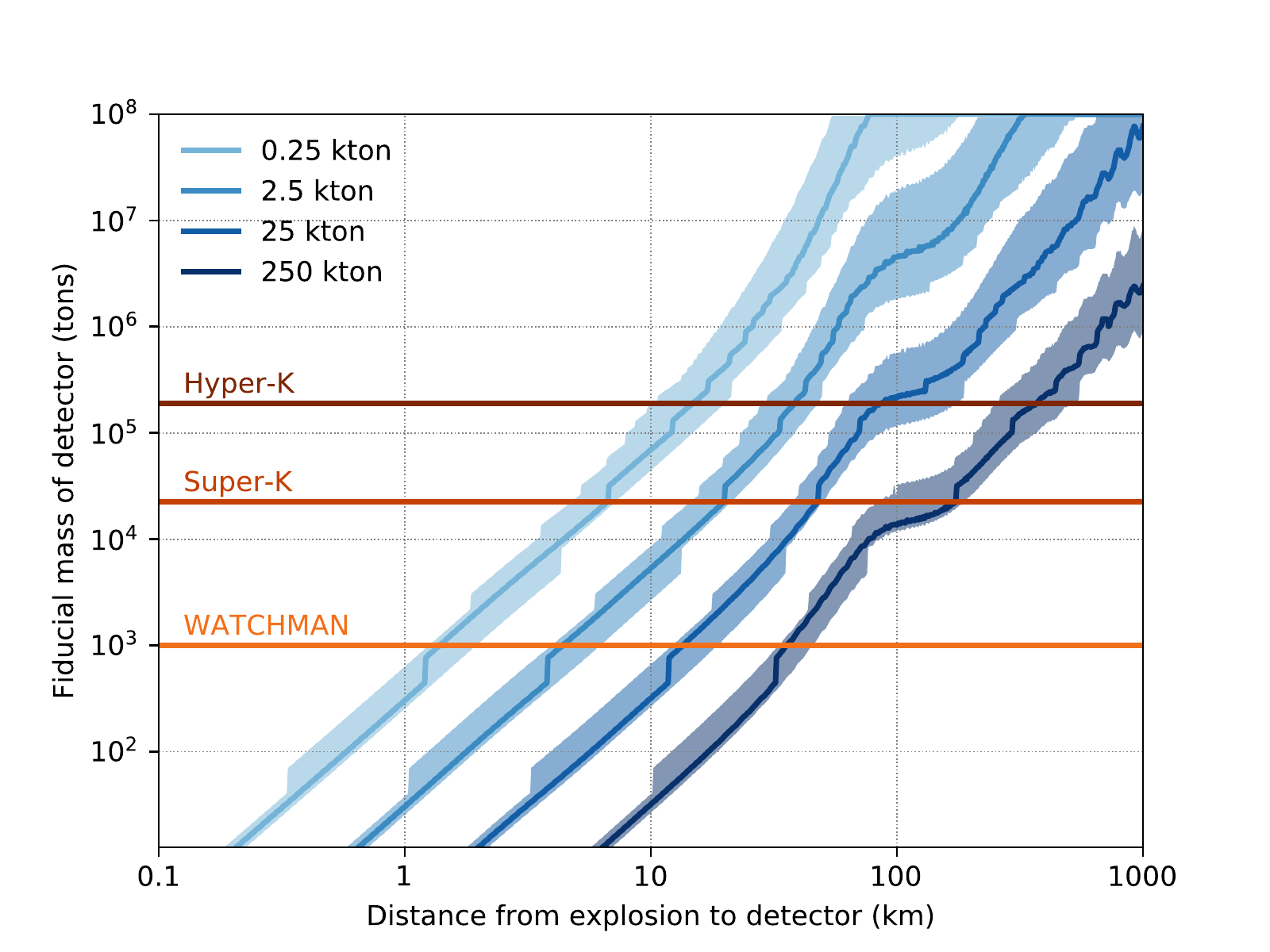}
\caption{Mass of a Gd-doped water Cherenkov detector capable of providing 90\% probability of confirming fission yield for a suspect event at 99\% CL, for various true yields (blue curves). The step discontinuities come from the small number of discrete events required when backgrounds are low. The smooth waves come from flavor oscillations. See Ref. \cite{Carr:2017uuq} for details.}
\label{fig:sens}
\end{figure}

It is possible to confirm a nuclear explosion with less neutrino detector mass than it takes to detect an otherwise missed explosion. However, the requisite detector size is still on the megaton scale or larger for cases of greatest practical value. Figure \ref{fig:sens} shows the mass of a Gd-doped water Cherenkov detector needed to achieve a 90\% probability of confirming fission yield at 99\% CL for various explosion sizes, as a function of standoff. A detector of the proposed Hyper-Kamiokande size could confirm the nuclear nature of a 25-kton explosion at a standoff of 100 km. The cost scale of Hyper-Kamiokande approaches \$1B \cite{hyperk}. A detector about 10 times the proposed Hyper-Kamiokande size would be needed to confirm the nuclear nature of a 250-kton explosion (near some estimates for the total yield of the largest North Korean nuclear test \cite{norsar}) at 900 km (the distance from Kamioka, Japan to the North Korean test site).

\subsection*{C. Estimate fission yield; indicate presence of fusion yield or seismic decoupling}

Neutrinos offer a theoretically attractive way to estimate fission yield, because the total neutrino emission is nearly proportional to fission yield. By contrast, seismic signal magnitude depends heavily on the depth of an underground explosion and other geological factors \cite{NAP12849}. However, placing a seismic-competitive constraint on fission yield requires a stronger signal than confirming the presence of fission. Therefore, this application would require detectors larger than those depicted in Fig. \ref{fig:sens}.

In principle, discrepancy between neutrino-based and seismic-based yield estimates could reveal the presence of some fusion yield (if the seismic-based, total-energy yield is known to be larger than the neutrino-based, fission-only yield) or intentional masking of the seismic signal by underground cavity engineering (if the seismic signal is significantly smaller than the neutrino signal). The stringent requirements on both the neutrino and seismic signals strongly limit these applications.

\section*{Conclusions}

The effectiveness of the existing explosion detection network, particularly its ability to detect explosions worldwide to a threshold of 0.5 kton or lower, leaves little room for neutrino detectors to add value. Any added capability from neutrino detectors would come at large cost, comparable to the largest projects in basic neutrino science. The key constraint comes from the basic physics of the situation: although many neutrinos are produced in the source, they rarely interact, making detection from a long distance feasible only in a very large and expensive detector. While large science detectors may have some incidental sensitivity to nuclear explosions in their regions, it is difficulty to foresee building neutrino detectors specifically for nuclear explosion monitoring.

\acknow{Adam Bernstein and Ferenc Dalnoki-Veress, coauthors of \cite{Carr:2017uuq}, made crucial contributions to the signal modeling and sensitivity studies. Michael Foxe and Theodore Bowyer provided valuable discussion on the existing nuclear explosion monitoring network.}

\showacknow{} % Display the acknowledgments section

% Bibliography
% \bibliography{aap2018-participant00x}

% \end{document}

%% file: AAP-Johnston/aap2018-participant00x.tex
% \documentclass[9pt,twocolumn,twoside,lineno]{aap2018}
%\documentclass{article}
% \usepackage[utf8]{inputenc}

% \title{AAP-PersonalizedTemplate}
% \author{bergevin1 }
% \date{September 2018}

% \begin{document}

% \maketitle

% \section{Introduction}

% \end{document}

% \documentclass[9pt,twocolumn,twoside,lineno]{pnas-new}
% Use the lineno option to display guide line numbers if required.

% \templatetype{aap2018proceedings} % Choose template 
% {pnasresearcharticle} = Template for a two-column research article
% {pnasmathematics} %= Template for a one-column mathematics article
% {pnasinvited} %= Template for a PNAS invited submission

\title{Ricochet and Prospects for Probing New Physics with Coherent Elastic Neutrino Nucleus Scattering}

% Use letters for affiliations, numbers to show equal authorship (if applicable) and to indicate the corresponding author

\author[a,1]{Joseph Johnston for the Ricochet Collaboration}
\author[b]{Bradley J. Kavanagh}

\affil[a]{Laboratory for Nuclear Science, Massachusetts Institute of Technology, Cambridge, MA, USA}
\affil[b]{GRAPPA, University of Amsterdam, Science Park 904, 1098 XH Amsterdam, The Netherlands}

% Please give the surname of the lead author for the running footer
\leadauthor{Johnston}

% Please include corresponding author, author contribution and author declaration information
% \authorcontributions{Please provide details of author contributions here.}
% \authordeclaration{Please declare any conflict of interest here.}
% \equalauthors{\textsuperscript{1}A.O.(Author One) and A.T. (Author Two) contributed equally to this work (remove if not applicable).}
\correspondingauthor{\textsuperscript{1}To whom correspondence should be addressed. E-mail: author.jpj13\@mit.edu}

% Keywords are not mandatory, but authors are strongly encouraged to provide them. If provided, please include two to five keywords, separated by the pipe symbol, e.g:

\begin{abstract}
Coherent Elastic Neutrino-Nucleus Scattering (CEvNS) is a Standard Model process that was recently detected by the COHERENT collaboration. Ricochet is an experiment in development that aims to detect CEvNS at a nuclear reactor with Germanium and Zinc bolometers. We present the design that will allow Ricochet to detect CEvNS, and explore the enhanced background rejection capability of superconducting Zinc due to pulse shape discrimination. We demonstrate the potential of CEvNS detection at a reactor to probe new physics. The cases of a neutrino magnetic moment, generic non-standard couplings of neutrinos to quarks, and non-standard couplings with a massive mediator are considered. We find that degeneracy in the non-standard couplings can be broken by combining multiple detector materials, and that massive mediator models are more strongly constrained by reactor experiments when lower energies are probed.
\end{abstract}

% \dates{This manuscript was compiled on \today}
\doi{\url{https://neutrinos.llnl.gov/workshops/aap2018}}

% \begin{document}

\maketitle
\thispagestyle{firststyle}
\ifthenelse{\boolean{shortarticle}}{\ifthenelse{\boolean{singlecolumn}}{\abscontentformatted}{\abscontent}}{}

% If your first paragraph (i.e. with the \dropcap) contains a list environment (quote, quotation, theorem, definition, enumerate, itemize...), the line after the list may have some extra indentation. If this is the case, add \parshape=0 to the end of the list environment.
\dropcap{C}oherent Elasic Neutrino-Nucleus Scattering (CEvNS) is a Standard Model process first predicted in 1974 \cite{PhysRevD.9.1389}:

\begin{align}
\label{eq:CEvNS}
\frac{\mathrm{d}\sigma_{\nu-N}}{\mathrm{d}E_R} = \frac{G_F^2}{4\pi} Q_W^2 m_N \left(1 - \frac{m_N E_R}{2 E_\nu^2}\right) F^2(E_R)\,,
\end{align}

where $M_N$ is the mass of the nucleus, N(Z) is the number of neutrons (protons), $F(q^2)$ is the nuclear form factor, and $Q_W = N-Z(1-4\sin^2(\theta_W))$. The  cross section is large due to $\approx N^2$ scaling, but detection of the low recoil energies is difficult. CEvNS was recently discovered by the COHERENT collaboration at the Spallation Neutron Source  (SNS) \cite{Akimoveaao0990}.

Measurement of CEvNS has numerous possible applications, such as low energy measurement of $\sin^2(\theta_W)$ \cite{Lindner2017}, understanding of the neutrino floor in direct dark matter detection experiments \cite{PhysRevD.89.023524}, and searches for sterile neutrinos \cite{PhysRevD.85.013009}. CEvNS also allows detection of neutrinos below the 1.8 MeV IBD threshold, enabling reactor monitoring applications such as detecting a breeder blanket at a fast reactor \cite{Huber2016}.

\section*{Ricochet}

The Ricochet experiment \cite{0954-3899-44-10-105101} will use cryogenic Ge and Zn bolometers in order to detect CEvNS at a nuclear reactor. The masses will be 500 g each of Zn and Ge in phase 1, and 5 kg in phase 2. In order to achieve the desired low thresholds, the mass will be split into an array of $\approx10$ g detectors, several of which have already been fabricated. Initial pulses have been taken with one Zn crystal, and are currently being analyzed.

\begin{figure}[tbhp]
\centering
\includegraphics[width=.6\linewidth]{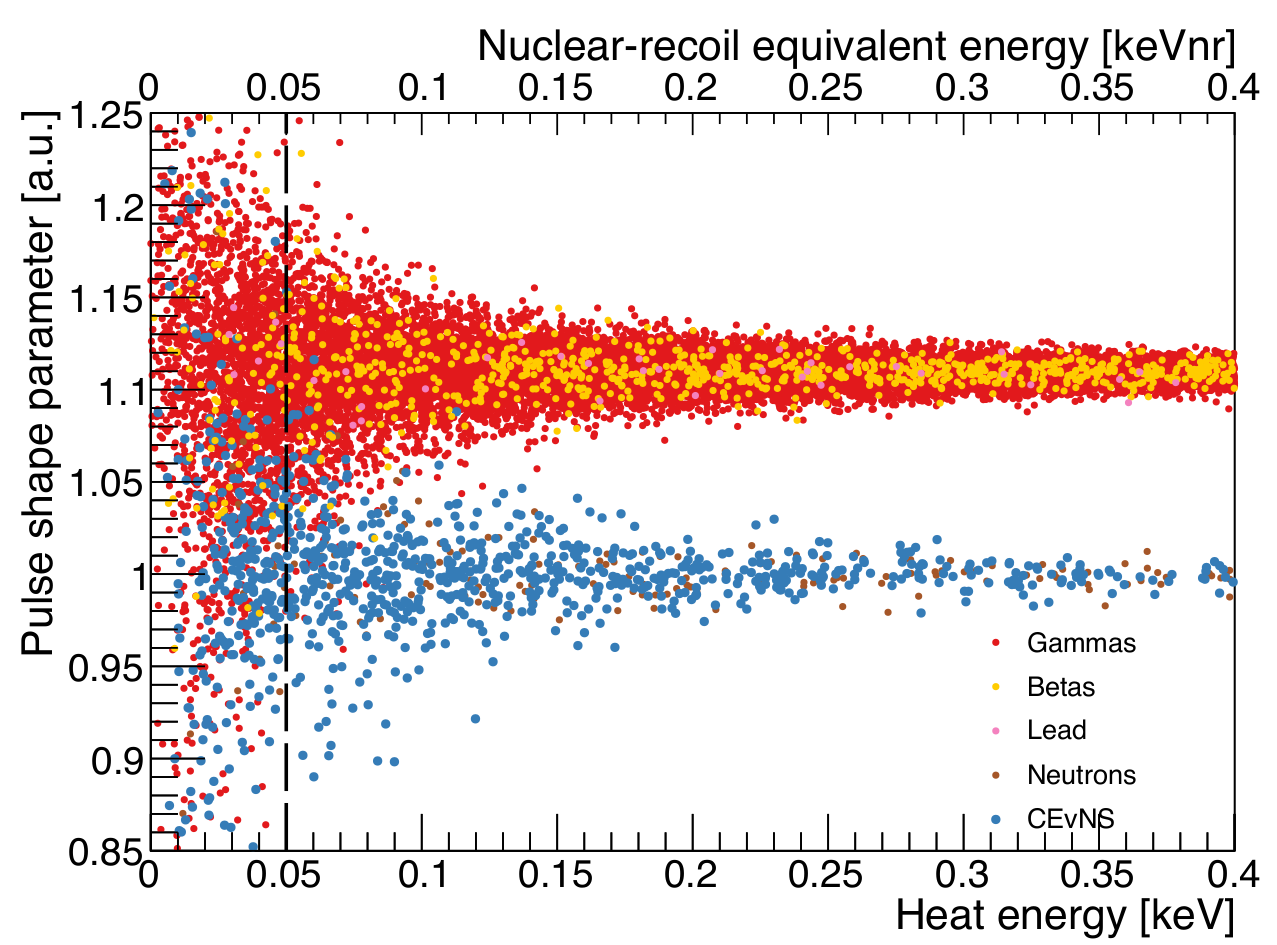}
\caption{Simulation of five years of Ricochet data assuming a thermalization rate five times slower for electromagnetic backgrounds.}
\label{fig:psd}
\end{figure}

Superconducting Zn bolometers are expected to provide strong background discrimination power. An event will deposit energy in a superconducting bolometer via quasiparticle excitations and phonons. Electromagnetic events are expected to create more quasiparticles than phonons. Calculations indicate that quasiparticle lifetimes are long compared to phonon collection times, meaning that pulses for electromagnetic events will be longer. Fig. \ref{fig:psd} shows a simulation of recoil energy vs pulse shape discrimination parameter for five years of Ricochet data, demonstrating strong discrimination power.

Scaling Ricochet to 10 kg with gram-sized detectors will require reading out hundreds of signals. A multiplexing SQUID array is in development in order to make this possible. Each detector is read out using a transition edge sensor (TES) with a DC current through it. The flux from the TES is read with a rf-SQUID, whose inductance is dependent on the magnetic flux passing through it, which is then coupled to a resonator. This means that the TES resistance (and by extension event energy) is tied to the frequency of the signal read out. Each TES, rf-SQUID, and resonator combination is tuned to a different frequency, then all are connected in parallel.

Ricochet will likely operate at the Chooz reactor complex in France, consisting of two reactor cores with a total power of 8.5 GW. There are two possible locations at the complex, respectively located 80 m and 400 m from the reactor, with 10 m.w.e. and 120 m.w.e. overburden. The dominant backgrounds will be cosmogenic neutrons and internal radioactivity. The expected background rate at the 400 m baseline site is 1.5 events/kg/day in the 0.1 to 1 keV region of interest. The expected signal rate is 0.5 events/kg/day \cite{0954-3899-44-10-105101}.

\section*{Prospects for New Physics Searches}

CEvNS detection at a reactor can be used to probe new physics beyond the Standard Model (BSM) \cite{Billard2018}. This potential for BSM searches was considered for a combination of Ge, Zn, Si \cite{conus}, CaWO$_4$, and Al$_2$O$_3$. CaWO$_4$, and Al$_2$O$_3$ are considered because $\nu$-cleus is an experiment in development that will aim to achieve a very low threshold and excellent background rejection by using very small CaWO$_4$ and Al$_2$O$_3$ detectors \cite{Strauss:2017cuu}. 5 kg each of Zn, Ge, and Si with a 10 eV threshold, 68 g CaWO$_4$ with a 7 eV threshold, and 44 g Al$_2$O$_3$ with a 4 eV threshold are assumed, with all detectors located at the Chooz reactor complex.

A Compton background of 100 events/kg/day/keV in Ge is included, with a factor of $10^{-3}$ discrimination power in all detectors. The neutron background from the 400 m site is also used \cite{0954-3899-44-10-105101}. It is assumed to be 10 times larger for the 80 m site due to decreased overburden, and $\nu$-cleus is assumed to have a factor of 0.1 discrimination power.

In minimal Standard Model extensions, a Dirac neutrino can obtain a neutrino magnetic moment (NMM) up to $3.2\times 10^{-19}[m_\nu/\text{ 1 eV}] \mu_B$ \cite{PhysRevLett.45.963}, and new physics contributions can increase the NMM to $10^{-12}\mu_B$ \cite{Shrock:1974nd}, \cite{Lee:1977tib}, \cite{Shrock:1982sc}, \cite{Pal:1991pm}, \cite{Balantekin:2006sw}, \cite{PhysRevLett.95.151802}, \cite{Bell:2006wi}, \cite{Lindner:2017uvt}. Then an additional term is added to Standard Model CEvNS:

    \[
    \frac{\mathrm{d}\sigma_{\nu-N}^\mathrm{mag.}}{\mathrm{d}E_R} = \frac{\pi \alpha^2 \mu_\nu^2 Z^2}{m_e^2} \left(\frac{1}{E_R} - \frac{1}{E_\nu} + \frac{E_R}{4E_\nu^2}\right) F^2(E_R)
  \]
  
\begin{figure}[tbhp]
\centering
\includegraphics[width=.6\linewidth]{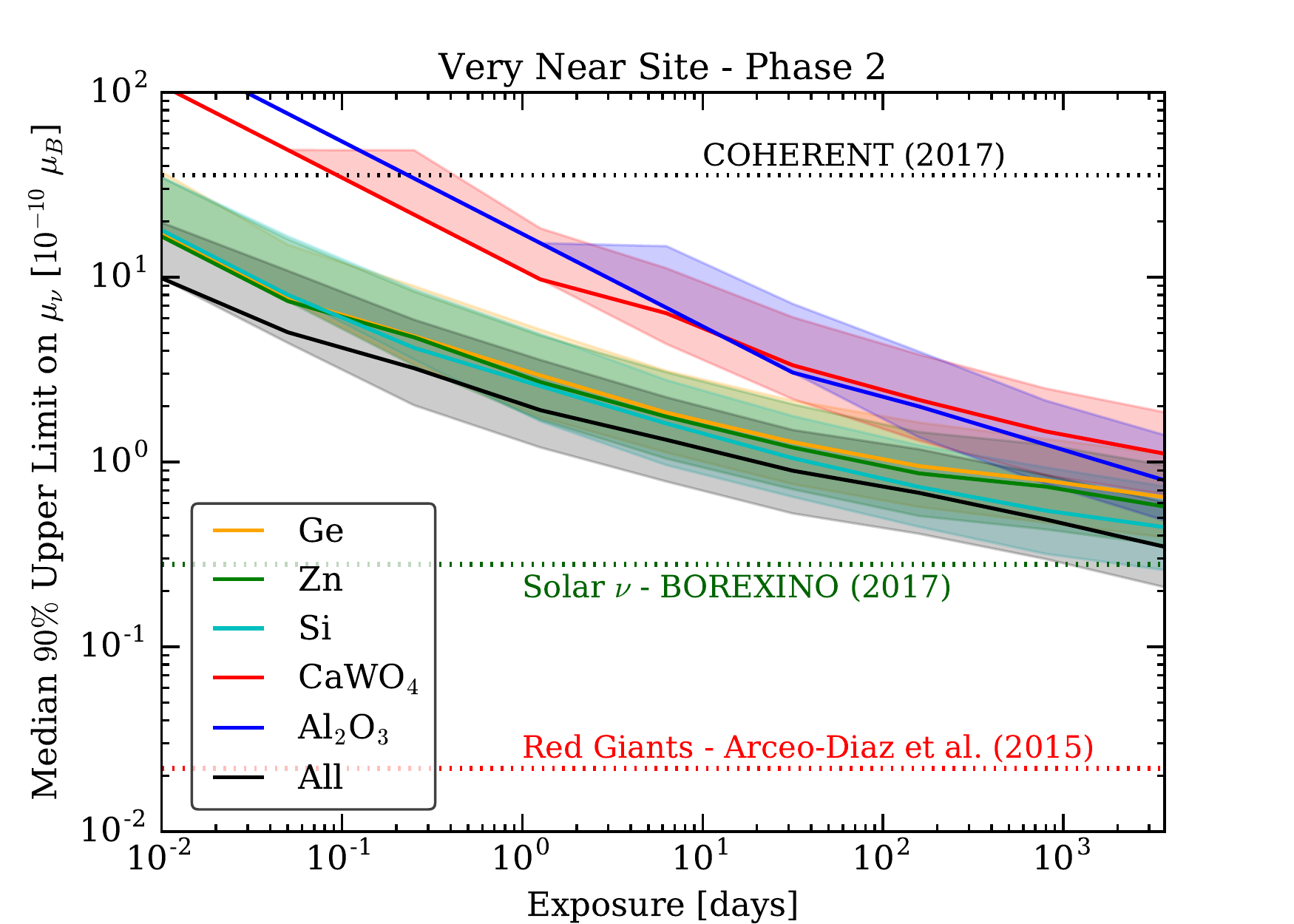}
\caption{Projected 90\% CL upper bounds on neutrino magnetic moment vs exposure, plus the current COHERENT bound, the leading terrestrial bound from BOREXINO \cite{Borexino:2017fbd}, and the leading bound from energy loss in red giants \cite{Arceo-Diaz:2015pva}.}
\label{fig:munu_bounds}
\end{figure}

  Fig. \ref{fig:munu_bounds} shows the bounds that can be placed at the 80 m site as a function of exposure.

A generic vector modification to the neutrino-nucleus interaction with two neutrino legs and two quark legs will modify the weak nuclear charge \cite{Lindner2017}:

\[
\begin{split}
\label{eq:Q_NSI}
Q_W^2 \rightarrow Q_\mathrm{NSI}^2 &= 4 [ N\left(-\frac{1}{2} + \epsilon_{ee}^{uV} + 2 \epsilon_{ee}^{dV}\right) + \\
&\qquad Z \left(\frac{1}{2} - 2 \sin^2\theta_W + 2 \epsilon_{ee}^{uV} + \epsilon_{ee}^{dV}\right)]^2\\
&\qquad+ 4  \left[ N(\epsilon_{e \tau}^{uV} + 2 \epsilon_{e \tau}^{dV}) + Z(2 \epsilon_{e \tau}^{uV} + \epsilon_{e \tau}^{dV})\right]^2\\
\end{split}
\]

\begin{figure}[tbhp]
\centering
\includegraphics[width=.6\linewidth]{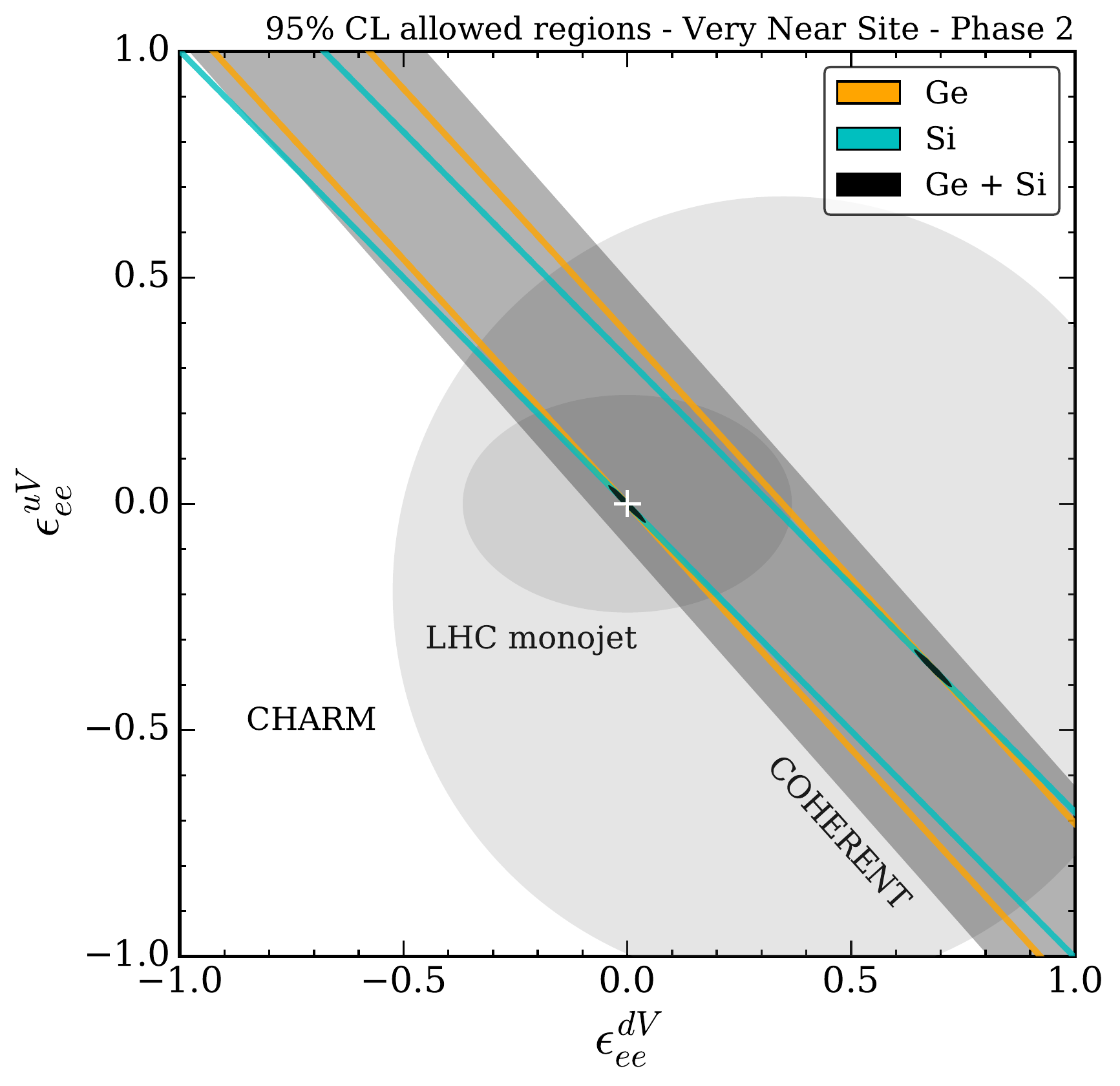}
\caption{Projected 95\% CL allowed regions on non-standard neutrino interactions for near site, in flavor conserving $\nu_e\rightarrow\nu_e$ case.}
\label{fig:nsi_generic}
\end{figure}

The degeneracy between up and down quark couplings makes it difficult to extract the neutrino mass ordering with oscillation experiments \cite{PhysRevD.94.055005}.  This degeneracy can be broken by detecting CEvNS in multiple detector materials with different $N/Z$ ratios, as shown in Fig. \ref{fig:nsi_generic}.

If the above non-standard interaction is mediated by a massive scalar mediator, then an additional term is added to the Standard Model CEvNS cross section ($Q_{\phi} \approx (15.1 \,Z + 14\, N)g_q$) \cite{Bertuzzo2017}:

\[\frac{\mathrm{d}\sigma_{\phi}}{\mathrm{d}E_R} = \frac{(g_\nu)^2 Q_{\phi}^2}{4\pi} \frac{E_R m_N^2}{E_\nu^2 (q^2 + m_\phi^2)^2} F^2(E_R)\]

The largest modification to the cross section occurs at low recoil energies. Detection at a nuclear reactor probes lower energy neutrinos (3 MeV) than the SNS (30 MeV). This enables reactor experiments to place stronger bounds on the coupling strength, especially at mediator masses below around 100 MeV.  Similar behavior occurs for a vector mediator \cite{Billard2018}.

\begin{figure}[tbhp]
\centering
\includegraphics[width=.7\linewidth]{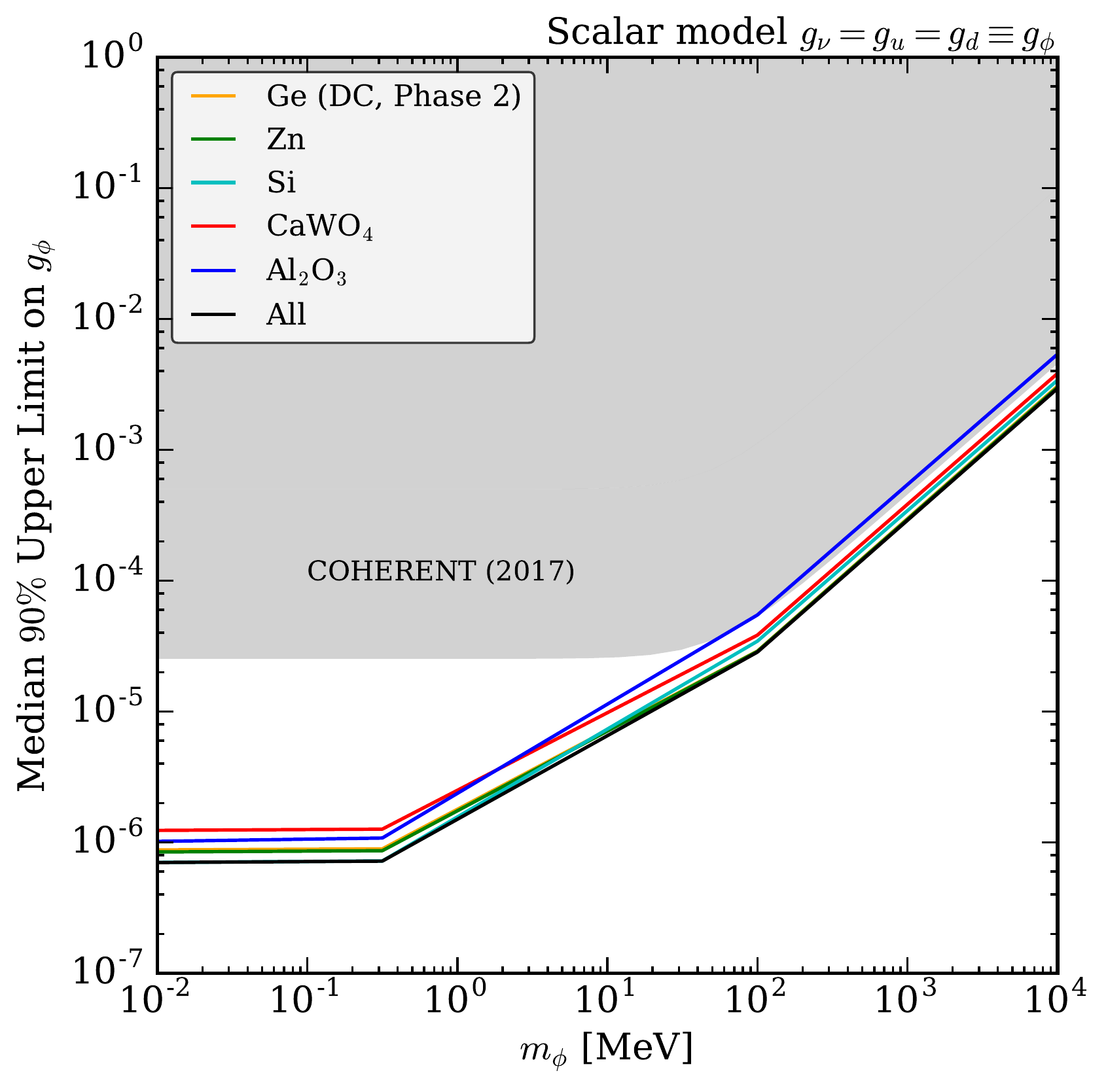}
\caption{Projected 90\% upper limits on the couplings of a new scalar mediator.}
\label{fig:nsi_scalar}
\end{figure}

\section*{Conclusion}

CEvNS detection at a reactor can be used to probe new physics, including a neutrino magnetic moment and non-standard couplings to quarks. Degeneracy in non-standard couplings can be broken by combining different target materials. Detection at a reactor places the strongest bounds on massive mediator models, especially at mediator masses below 100 MeV.

% \matmethods{Please describe your materials and methods here. This can be more than one paragraph, and may contain subsections and equations as required. Authors should include a statement in the methods section describing how readers will be able to access the data in the paper. 

% \subsection*{Subsection for Method}
% Example text for subsection.
% }

% \showmatmethods{} % Display the Materials and Methods section

\acknow{We wish to thank the Heising-Simons Foundationand, the MIT MISTI-France Program, and the National Science Foundation (PHY-1806251) for their support of this work. BJK acknowledges funding from the European Research Council (Erc) under the EUSeventh Framework Programme (FP7/2007-2013)/ErcStarting Grant (agreement n. 278234— ‘NewDark’ project) and from the NWO through the VIDI research program ”Probingthe Genesis of Dark Matter” (680-47-532).}

\showacknow{} % Display the acknowledgments section

% Bibliography
% \bibliography{aap2018-participant00x}

% \end{document}

%% file: AAP-Mabe/aap2018-participant00x.tex
% \documentclass[9pt,twocolumn,twoside,lineno]{aap2018}
%\documentclass{article}
% \usepackage[utf8]{inputenc}

% \title{AAP-PersonalizedTemplate}
% \author{bergevin1 }
% \date{September 2018}

% \begin{document}

% \maketitle

% \section{Introduction}

% \end{document}

% \documentclass[9pt,twocolumn,twoside,lineno]{pnas-new}
% Use the lineno option to display guide line numbers if required.

% \templatetype{aap2018proceedings} % Choose template 
% {pnasresearcharticle} = Template for a two-column research article
% {pnasmathematics} %= Template for a one-column mathematics article
% {pnasinvited} %= Template for a PNAS invited submission

\title{Plastic Scintillator Development at LLNL}

% Use letters for affiliations, numbers to show equal authorship (if applicable) and to indicate the corresponding author
\author{Andrew N. Mabe{\textsuperscript{1}}}
\author{M. Leslie Carman} 
\author{Andrew M. Glenn}
\author{Steven A. Dazeley}
\author{Natalia P. Zaitseva}
\author{Stephen A. Payne}

\affil{Lawrence Livermore National Laboratory}

% Please give the surname of the lead author for the running footer
\leadauthor{Mabe} 

% Please include corresponding author, author contribution and author declaration information
% \authorcontributions{Please provide details of author contributions here.}
% \authordeclaration{Please declare any conflict of interest here.}
% \equalauthors{\textsuperscript{1}A.O.(Author One) and A.T. (Author Two) contributed equally to this work (remove if not applicable).}
\correspondingauthor{\textsuperscript{1}E-mail: mabe2@llnl.gov}

% Keywords are not mandatory, but authors are strongly encouraged to provide them. If provided, please include two to five keywords, separated by the pipe symbol, e.g:

\begin{abstract}
The development of plastics with pulse shape discrimination (PSD) provided new capabilities that required much in-depth research to refine. Herein we describe results from extensive optimization studies which have led to the development of PSD plastics with improved scintillation performance and physical properties. Due to the large concentration of primary dye required to manifest optimal PSD properties in plastic scintillators, the physical stability can be limited and subject to mechanical deformation, especially in larger volume samples. Practical solutions have been developed to address these issues, resulting in physically stable scintillators with robust mechanical properties. Performance deterioration with increasing size is also addressed. At large sizes, physical and performance characteristics are much more sensitive to preparation conditions and compositional alterations as compared with small scintillators, and efforts to improve these properties are described. Finally, efforts to incorporate both aromatic and nonaromatic lithium compounds into PSD plastics are summarized. 
\end{abstract}

% \dates{This manuscript was compiled on \today}
\doi{\url{https://neutrinos.llnl.gov/workshops/aap2018}}

% \begin{document}

\maketitle
\thispagestyle{firststyle}
\ifthenelse{\boolean{shortarticle}}{\ifthenelse{\boolean{singlecolumn}}{\abscontentformatted}{\abscontent}}{}

%\section*{Introduction}
\dropcap{D}ifficulties associated with liquid scintillators stimulated the development of solid-state materials that can serve as replacements for liquids. Solid state scintillators such as plastics are a relatively inexpensive alternative to liquids and can be fabricated in a variety of shapes and sizes. Traditional plastic scintillators were fabricated without pulse shape discrimination properties until a plastic that showed modest PSD., plastic 77, developed by Brooks et. al., was reported~\cite{BROOKS1960}. This plastic was found to be unstable and no further developments in PSD plastics occurred for several decades. A report in 2012 described that the delayed component can be enhanced by increasing the amount of primary dye (e.g., PPO) in the plastic, resulting in PSD~\cite{ZAITSEVA201288}. When this plastic became available commercially, produced by Eljen Technology under the name EJ-299, many researchers reported characterizations of the material and stated that it exhibited poorer scintillation properties than traditional PSD liquids such as EJ-309~\cite{POZZI201319,CESTER2014202,HARTMAN2015137,LIAO2015150,LAWRENCE201416,1748-0221-9-06-P06014, WOOLF201580}. This work presents results of multi-year studies intended to improve the physical and mechanical properties of PSD plastics, as well as efforts to improve the scintillation performance of PSD plastics to the level of PSD liquids~\cite{ZAITSEVA201897}. These improvements are directly applicable to PSD plastics loaded with thermal neutron capture nuclides such as lithium-6~\cite{ZAITSEVA2013747,MABE201680}. The results from incorporating lithium compounds into the newly developed PSD plastics are described.

\begin{figure}[tbhp]
\centering
\includegraphics[width=.75\linewidth]{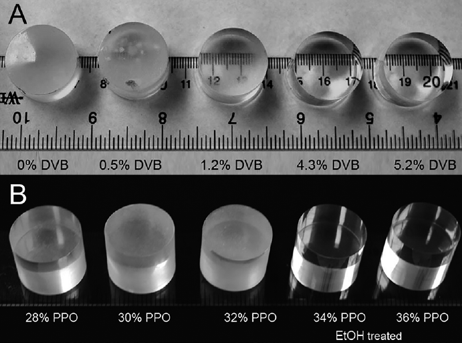}
\caption{Images of plastics showing improved physical stability. A) Plastics aged over 4 years with increasing concentration of crosslinker. Higher concentrations of crosslinker improve physical stability and resistance to crashing. B). Leaching is observed in plastics containing 28 – 32\% PPO. Treating the surface with ethanol eliminates leaching, thereby permitting loads of PPO up to 36\% without any visible degradation.}
\label{fig:mabe1}
\end{figure}

\begin{figure}[tbhp]
\centering
\includegraphics[width=.75\linewidth]{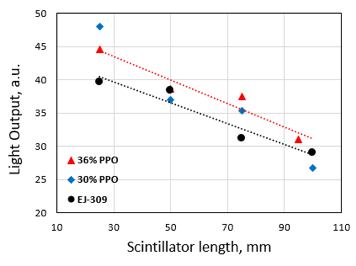}
\includegraphics[width=.75\linewidth]{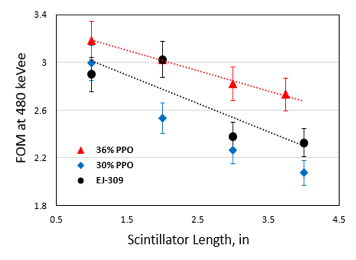}
\caption{Scintillation performance of PSD plastics containing 30\% PPO, 36\% PPO, and EJ-309 liquid scintillator as functions of length. Top: light output; bottom: PSD figure of merit in the region 450 – 510 keVee. Figure of merit is defined as in~\cite{ZAITSEVA201288}.}
\label{fig:mabe2}
\end{figure}

\section*{Results and Discussion}

Due to the high concentration of PPO required to produce PSD, several degradation mechanisms that do not occur in standard plastic scintillators can negatively impact both scintillation performance and mechanical stability. Plastics can undergo mechanical deformation over time, the dyes leach and crystallize on the surface, and the interior of the plastic can become cloudy. Many of the problems with physical stability have been addressed by adding divinylbenzene (DVB) to provide crosslinking within the plastic. Crosslinking eliminates mechanical deformation and improves resistance to dye leaching and precipitation. Further, leaching has been practically eliminated by treating the surface of freshly fabricated PSD plastics with ethanol. This dissolves any residual material on the surface and prevents the dyes from crystallizing on the surface. The mechanism by which ethanol treatment eliminates leaching has not yet been fully elucidated. These modifications have permitted the production of stable plastics containing up to 40\% PPO, compared to the previous maximum of 30\%. Images of the physically stabilized plastics are shown in Fig.~\ref{fig:mabe1}. Additionally, it was found that the new secondary dye 7-diethylamino-4-methylcoumarin (MDAC) produces PSD plastics with significantly improved PSD and light output compared to traditional secondary dyes. These improvements have resulted in the production of a new commercial PSD plastic, EJ-276, which shows performance on the level of the commercial liquid scintillator EJ-309. Figure~\ref{fig:mabe2} shows the scintillation performance of LLNL PSD plastics containing 30\% PPO, 36\% PPO, and EJ-309 and demonstrates that PSD plastics can have performance exceeding that of commercial liquid scintillators. Attenuation lengths of these plastics is approximately 18 cm.

\begin{figure*}[tbhp]
\centering
\includegraphics[width=0.65\linewidth]{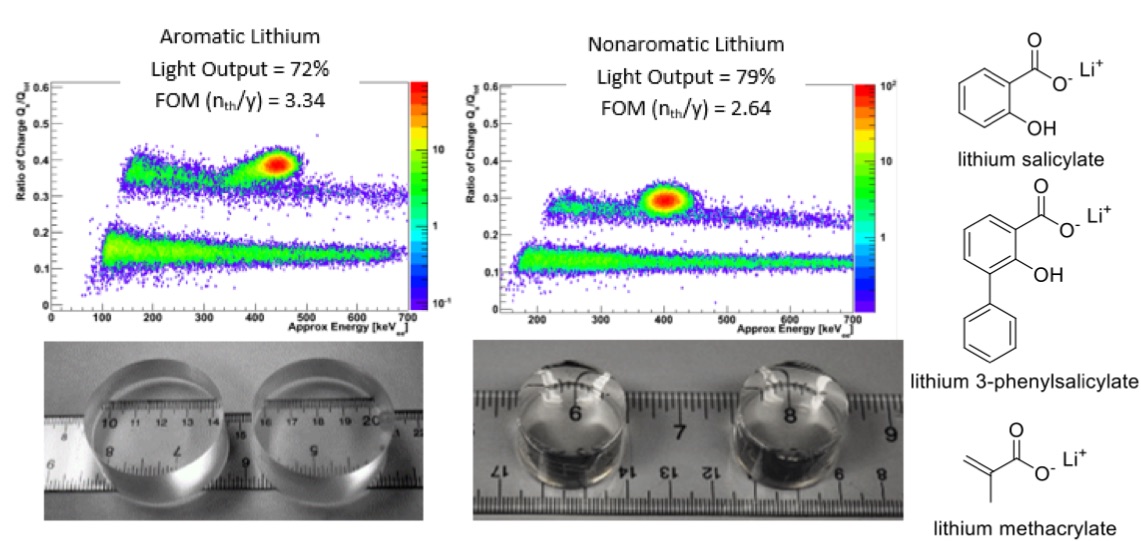}
\caption{PSD plots of plastics containing aromatic lithium (left) and nonaromatic lithium (right). PSD plots are responses to $^{252}$Cf moderated with 3\,inch HDPE and are show with optimized time gates. Molecular structures of aromatic compounds lithium salicylate and lithium 3-phenylsalicylate and nonaromatic lithium methacrylate are shown.}
\label{fig:mabe3}
\end{figure*}

Two strategies are implemented to incorporate lithium into PSD plastics to impart sensitivity to thermal energy neutrons. The first strategy involves dispersing aromatic lithium compounds into a polystyrene (PS)-polymethylmethacrylate (PMMA) copolymer. The aromatic compounds which can be successfully incorporated into PSD plastics are based on lithium salicylate. The addition of PMMA into the PSD plastic reduces the light output and PSD primarily by diluting the number density of active aromatic, or scintillation-producing, centers. This strategy permits incorporation of lithium up to about 0.4\%. A problem with this strategy is that the lithium compounds have high absorption in the region of PPO emission and interfere with the scintillation process, reducing light output. In order to circumvent this issue, we incorporated a nonabsorbing, nonaromatic lithium compounds in the PSD plastic. This was accomplished by dispersing the lithium compounds in a polystyrene-polymethacrylic acid copolymer. While the nonaromatic lithium salts do not absorb scintillation light from either PPO or the secondary dye, the presence of methacrylic acid quenches the light output more than what is expected from simple dilution. The nature of this quenching mechanism is still under investigation. Results from representative plastics bearing each type of lithium compound are shown in Figure~\ref{fig:mabe3}. The light output of the plastic containing aromatic lithium is 72\% of the unloaded plastic, whereas that with the nonaromatic lithium compound is 79\% of the unloaded plastic. Though the light output of the plastic containing nonaromatic lithium is higher, the figure of merit around the thermal neutron spot is higher for the plastic containing the aromatic lithium compound.

\section*{Conclusions}

We report the results of extensive optimization of PSD plastics and multi-year aging studies. PSD plastics with higher loads of PPO (36\% as compared to previous maximum of 30\%) by adding crosslinker and treating the surface with ethanol. PSD plastics produced with the new secondary dye MDAC have significantly higher PSD and light output as compared to those produced with traditional secondary dyes. These improvements have resulted in plastics that have performance comparable to common commercial liquid scintillators such as EJ-309 and have resulted in the new commercial PSD plastic EJ-276. The improvements in the standard PSD plastics provide improvements in lithium-loaded PSD plastics. Strategies have been implemented to disperse both aromatic and nonaromatic lithium compounds in PSD plastics. While plastics containing aromatic lithium compounds have better PSD, plastics with nonaromatic lithium compounds have higher light output and are easier to produce.

\acknow{This work was conducted under the auspices of the U. S. Department of Energy by Lawrence Livermore National Laboratory under Contract DE-AC52-07NA27344. This work was funded by the Department of Energy Office of Nonproliferation Research and Development (NA-22) and the LLNL Laboratory Directed Research and Development program.}

\showacknow{} % Display the acknowledgments section

% Bibliography
% \bibliography{aap2018-participant00x}

% \end{document}

%% file: AAP-Yeh/aap2018-participant00x.tex
% \documentclass[9pt,twocolumn,twoside,lineno]{aap2018}
%\documentclass{article}
% \usepackage[utf8]{inputenc}

% \title{AAP-PersonalizedTemplate}
% \author{bergevin1 }
% \date{September 2018}

% \begin{document}

% \maketitle

% \section{Introduction}

% \end{document}

% \documentclass[9pt,twocolumn,twoside,lineno]{pnas-new}
% Use the lineno option to display guide line numbers if required.

% \templatetype{aap2018proceedings} % Choose template 
% {pnasresearcharticle} = Template for a two-column research article
% {pnasmathematics} %= Template for a one-column mathematics article
% {pnasinvited} %= Template for a PNAS invited submission

\title{BNL Material Development}

% Use letters for affiliations, numbers to show equal authorship (if applicable) and to indicate the corresponding author
\author{Minfang Yeh}

\affil{Brookhaven National Laboratory, Upton, NY11790, USA}

% Please give the surname of the lead author for the running footer
\leadauthor{Yeh}

% Please include corresponding author, author contribution and author declaration information
% \authorcontributions{Please provide details of author contributions here.}
% \authordeclaration{Please declare any conflict of interest here.}
% \equalauthors{\textsuperscript{1}A.O.(Author One) and A.T. (Author Two) contributed equally to this work (remove if not applicable).}
\correspondingauthor{\textsuperscript{2}To whom correspondence should be addressed. E-mail: yeh@bnl.gov}

% Keywords are not mandatory, but authors are strongly encouraged to provide them. If provided, please include two to five keywords, separated by the pipe symbol, e.g:
% \keywords{Keyword 1 $|$ Keyword 2 $|$ Keyword 3 $|$ ...} 

\begin{abstract}
Summary of liquid scintillator detector Research and Development at the Brookhaven National Laboratory, Neutrino and Nuclear Chemistry Group.
\end{abstract}

% \dates{This manuscript was compiled on \today}
\doi{\url{https://neutrinos.llnl.gov/workshops/aap2018}}

% \begin{document}

\maketitle
\thispagestyle{firststyle}
\ifthenelse{\boolean{shortarticle}}{\ifthenelse{\boolean{singlecolumn}}{\abscontentformatted}{\abscontent}}{}

% If your first paragraph (i.e. with the \dropcap) contains a list environment (quote, quotation, theorem, definition, enumerate, itemize...), the line after the list may have some extra indentation. If this is the case, add \parshape=0 to the end of the list environment.
%\dropcap{} 

\section*{Scintillator Physics}
%\subsection*{Author Affiliations}

Significant progress has been made in use of liquid scintillators (LS), especially metal-loaded LS (M-doped LS), for different neutrino experiments over the past decade.  The different types of scintillator detectors in use for different nuclear and particle physics experiments have common requirements in long-term chemical stability, high light output, high light transmission through long path-lengths, superior pulse-shape-discrimination capability and very low levels of colored impurities and of natural radioactive contaminants, such as Uranium, Thorium, and Radium. The quantity of selected scintillator should be scalable to many tons with appropriate compatibility.  The nominal properties in terms of light-yield and attenuation length for scintillator and Cherenkov detectors are compared in Figure 1. Being able to satisfy many of these requirements depends on R\&D in scintillator mechanism and nuclear chemistry.  

\begin{figure}[tbhp]
\centering
\includegraphics[width=0.95\linewidth]{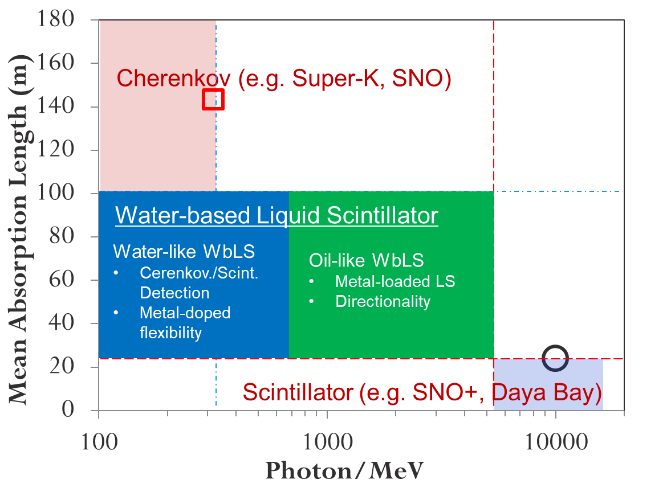}
\caption{Comparison of photon-yield and optical attenuation length for water Cherenkov and scintillator detector.}
%\label{fig:frog}
\end{figure}
\subsection*{Brookhaven National Laboratory}

The major thrust of research conducted at BNL has become the development of next-generation scintillator detectors for future frontier physics experiments including double beta decay, dark matter, reactor antineutrino and neutrino beam physics with other implications in nonproliferation science and medical physics. The BNL scintillator R\&D program has led to identificatrion of new environment-friendly liquid scintillator (e.g. linear alkyl benzene), improvement of chemical formulation for preparing M-doped LS, and development of purification and synthesis with large-scale facility.  An example of LAB purification using pilot-scale instrumentation (40 liters per hour) is presented in Figure 2.

\begin{figure}[tbhp]
\centering
\includegraphics[width=0.95\linewidth]{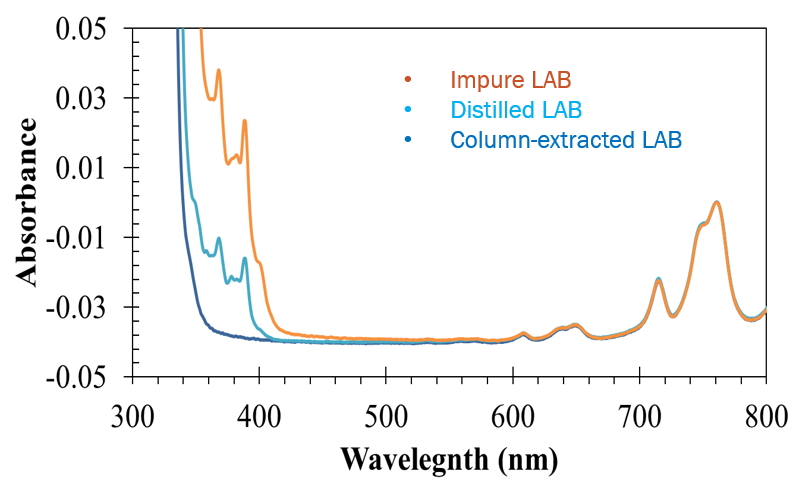}
\caption{An illustration of scintillator purification}
%\label{fig:frog}
\end{figure}

\section*{Metal-doped Liquid Scintillator}

Much of research in metal-doped LS has involved either solvent extraction to extract the organometallic complex into the LS or preparation of the solid organometallic complex in aqueous media followed by its dissolution in the organic LS.  These methods have been applicable to many of the different full-scale neutrino detectors since they all will contain many tons of M-LS.  All scintillator development conducted at BNL generates prototype liquids produced in quantity of 1-1000 liters.  Once tested to be feasible, liquid is delivered to off-site locations for detector deployment.  We are also developing new water-based metal-loaded methods that mix aqueous metallic ion (particularly hydrophilic elements) with organic solvent directly.  

\subsection*{PROSPECT}

(Precision Oscillation and Spectrum Experiment) aims to answer the new neutrino puzzle of sterile neutrinos and to explorer new physics beyond the standard model with high precision measurements of reactor neutrino spectrum/flux.  PROSPECT uses 6Li-doped water-based liquid scintillator (LiLS) to probe potential oscillation in close distance to the reactor core (10m).  A high ability of pulse shape discrimination of scintillator to reject the fast neutron and ambient gamma backgrounds associated with the detector on surface is essential to the success of the project.  Approximately 5000 liters of LiLS prepared in di-isopropylnaphthalene (DIN)-based scintillator (EJ309) consistent with the required performance were prepared for PROSPECT in six months.  The quality of synthesized Li-LS (795±15 PE/MeV and >1.36 correlated S/B ratio) surpassed the accepted quality criteria and were used for PROSPECT detector at ORNL.  Yet the LiLS developed for PROSPECT satisfied the detector needs in absorbance, light yield, and PSD, a further improvement to reduce its scattering effect has been continuing even after detector deployment.  A 2nd generation of LiLS with improved optical transmission is shown in Figure 3.  

\begin{figure}[tbhp]
\centering
\includegraphics[width=0.95\linewidth]{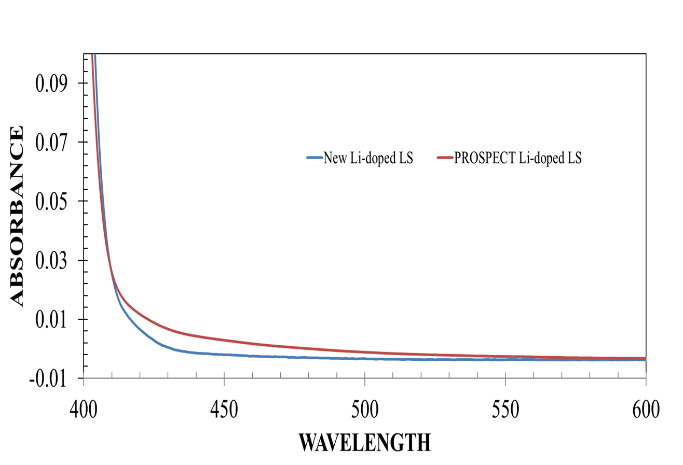}
\caption{An optical-improved Li-doped LS}
%\label{fig:frog}
\end{figure}

\section*{Directional Scintillator Detector}
Addition of directionality to isotropic scintillator emission is the key feature for future development of scintillator detectors.  Several chemical approaches, such as slowing fluorescence decay time or reducing fluorescence component in formulated scintillation liquid, to separate or enhance Cherenkov imaging from scintillation emission have been proposed by BNL.  A new R\&D of water-based liquid scintillator (WbLS) began in 2010 aiming to develop a next-generation large Cherenkov scintillation detector medium that is capable of low-energy detection with Cherenkov directionality.  To search for rare event interaction (e.g. geo-neutrino or proton decay) often requires a large scintillator detector at the scale of many kilotons. However, this large scintillator application could be limited by its chemical cost and increasing safety concerns associated with large liquid handing.  The novel WbLS development is a cost-effective approach with the capability to probe physics below Cherenkov threshold and has preliminary success in laboratory-sized environment.  

\subsection*{WATCHMAN}

(WATer CHerenkov Monitor of Anti-Neutrinos) is designed to monitor the reactor fuel (ON/OFF) using Gadolinium-doped water to permit investigation of advanced antineutrino-based, stand-off methods.  WATCHMAN plans to make use of a test-bed, the Advanced Instrumentation Testbed (AIT), to explore the feasibility of WbLS as an antineutrino detection medium, to improve sensitivity to the existence and operation of nuclear reactors.  The project is currently supported by NNSA.  BNL group is developing a low-doping scintillation water with scintillation and Cherenkov detection features to be deployed at the AIT-detector.  A R\&D plan is proposed to investigate the feasibility, in terms of formulation, characterization, production and deployment, of several kilotons of WbLS in the planned WATCHMAN detector at the Boulby Underground Laboratory, U.K.  The developing technologies associated with this R\&D proposal could also benefit other nuclear and particle physics experiment, such as THEIA, along with further implications in safeguard and medical physics.

\section*{Large-scale Scintillator Production Facility}

\subsection*{LZ}

(LUX-ZEPLIN) searches for the existence of a WINP through its scatter depositing 5-50 keV of energy in the central volume of LXe.  Such signals could be mimicked from ambient backgrounds, such as gamma rays with energies in the few MeV range and neutrons from (alpha,n) reactions or from cosmic ray origin.  BNL developed and proposed a 20-ton, high radiopurity (U/Th below 1 part per trillion) Gadolinium-doped liquid scintillator to be used as an outer detector to reject these background events.  LZ passed CD1/3a in March 2015 and CD2/3b in Apr 2016.  We are responsible of  purification, production, and transportation of scintillator at BNL and filling at SURF.  The construction of scintillator production facility was completed in summer 2018 (Figure 4).

\begin{figure}[tbhp]
\centering
\includegraphics[width=.78\linewidth]{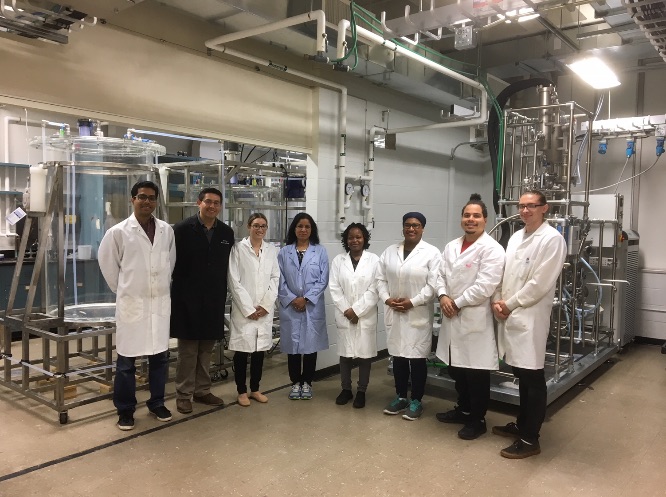}
\caption{BNL liquid scintillator development and production facility}
%\label{fig:frog}
\end{figure}

\section*{Summary}

In all, the research at BNL focuses on developing and applying chemical methodologies for frontier physics experiments, nonproliferation, and medical physics.  Several developed technologies have been adapted by industries.  We have developed varied M-LS, involving organometallic complexation and water-based extraction to load different kinds of metals in various scintillating solvents, such as pseudocumene (PC), Linear Alkyl Benzene (LAB), and di-isopropylnaphthalene (DIN).  These formulation developments are being translated into processes that can be applied at the multi-ton chemical scale detectors.  The principal of water-based liquid scintillator has been demonstrated at liter-sized scale in a laboratory environment.  A further development to test the WbLS performance in a kiloton-scale detector is continuing at BNL in collaboration with other national laboratories and universities.

\section*{Acknowledgement}
This work, conducted at Brookhaven National Laboratory, was supported by the U.S. Department of Energy (DOE) under Contract No. DE-SC0012704. 

% \end{document}

%% file: AAP-OrebiGann/aap2018-participant00x.tex
% \documentclass[9pt,twocolumn,twoside,lineno]{aap2018}
%\documentclass{article}
% \usepackage[utf8]{inputenc}

% \title{AAP-PersonalizedTemplate}
% \author{bergevin1 }
% \date{September 2018}

% \begin{document}

% \maketitle

% \section{Introduction}

% \end{document}

% \documentclass[9pt,twocolumn,twoside,lineno]{pnas-new}
% Use the lineno option to display guide line numbers if required.

% \templatetype{aap2018proceedings} % Choose template 
% {pnasresearcharticle} = Template for a two-column research article
% {pnasmathematics} %= Template for a one-column mathematics article
% {pnasinvited} %= Template for a PNAS invited submission

\title{Large-Scale Water-Based Liquid Scintillator Detector R\&D}

% Use letters for affiliations, numbers to show equal authorship (if applicable) and to indicate the corresponding author
\author[a,b]{Gabriel D. Orebi Gann}
\author{the Theia Collaboration}

\affil[a]{University of California, Berkeley, Department of Physics}
\affil[b]{Lawrence Berkeley National Laboratory, Nuclear Science Division}

% Please give the surname of the lead author for the running footer
\leadauthor{Orebi Gann}

% Please include corresponding author, author contribution and author declaration information
% \authorcontributions{Please provide details of author contributions here.}
% \authordeclaration{Please declare any conflict of interest here.}
% \equalauthors{\textsuperscript{1}A.O.(Author One) and A.T. (Author Two) contributed equally to this work (remove if not applicable).}
\correspondingauthor{\textsuperscript{1}To whom correspondence should be addressed. E-mail: gorebigann\@lbl.gov}

% Keywords are not mandatory, but authors are strongly encouraged to provide them. If provided, please include two to five keywords, separated by the pipe symbol, e.g:
% \keywords{Keyword 1 $|$ Keyword 2 $|$ Keyword 3 $|$ ...} 

\begin{abstract}
A detector capable of discriminating Cherenkov and scintillation signals would be capable of an unprecedented level of particle and event identification and, hence, background rejection for a broad spectrum of both applied and fundamental physics topics.  Use of the newly developed water-based liquid scintillator (WbLS) target medium facilitates this possibility in a number of ways, including the ability to scale the relative magnitudes of each signal, and to delay the scintillation signal relative to the prompt Cherenkov light.
This paper describes the ongoing technical work to realise such a detector.
\end{abstract}

% \dates{This manuscript was compiled on \today}
\doi{\url{https://neutrinos.llnl.gov/workshops/aap2018}}

% \begin{document}

\maketitle
\thispagestyle{firststyle}
\ifthenelse{\boolean{shortarticle}}{\ifthenelse{\boolean{singlecolumn}}{\abscontentformatted}{\abscontent}}{}

% If your first paragraph (i.e. with the \dropcap) contains a list environment (quote, quotation, theorem, definition, enumerate, itemize...), the line after the list may have some extra indentation. If this is the case, add \parshape=0 to the end of the list environment.

\dropcap{T}he  \textsc{Theia}  detector~\cite{asdc, theia} leverages a tried and tested methodology in combination with novel, cutting-edge technology.  
\textsc{Theia} would combine the use of a 30--100-kton WbLS target, doping with a number of potential isotopes, high efficiency and ultra-fast timing photosensors, and a deep underground location.  %A potential site is the Long Baseline Neutrino Facility (LBNF) far site, where \textsc{Theia} could operate in conjunction with the liquid argon tracking detector proposed by DUNE~\cite{dune}.  
The basic elements of this detector are being developed now in experiments such as WATCHMAN~\cite{Askins:2015bmb}, ANNIE~\cite{annie} and SNO+~\cite{snopl}.  

\section*{Detector Concept}

A large-scale WbLS detector such as \textsc{Theia} can achieve an impressively broad program of physics topics, with enhanced sensitivity beyond that of previous detectors.  Much of the program hinges on the capability to separate prompt Cherenkov light from delayed scintillation.  This separation provides many key benefits, including:
\begin{itemize}
\item The ring-imaging capability of a pure water Cherenkov detector (WCD).  This enables a long-baseline program in a scintillation-based detector, with the additional benefit of low-threshold detection of hadronic events.

\item  Direction reconstruction using prompt Cherenkov photons.   This allows statistical identification of events such as solar neutrinos, which offer a rich physics program in their own right, as well as forming a background to many rare-event searches, including NLDBD and nucleon decay.

\item Low thresholds and good energy and vertex resolution using the abundant scintillation light.

\item Detection of sub-Cherenkov threshold scintillation light.  This provides excellent particle identification, including enhanced neutron tagging, detection of sub-Cherenkov threshold particles such as kaons in nucleon decay searches, and separation of atmospheric neutrino-induced neutral current backgrounds for inverse beta decay searches.

\end{itemize}

One of the most powerful aspects of \textsc{Theia} is the flexibility:  in the target medium itself, and even in the detector configuration.  The WbLS target can be tuned to meet the most critical physics goals at the time by modifying features of the target cocktail, including: the fraction of water vs scintillator; the choice of wavelength shifters and secondary fluors; and the choice of loaded isotope.  There is also the potential to construct a bag to contain isotope, and perhaps a higher scintillator-fraction target, in the centre of the detector, building on work by KamLAND-Zen~\cite{klz} and Borexino~\cite{bor}.  

\section*{Physics Program}

\textsc{Theia} targets a broad physics program~\cite{theia}, including a next-generation neutrinoless double beta decay search capable of reaching into the normal hierarchy region of phase space, sensitivity to solar neutrinos~\cite{richiegdog}, supernova neutrinos, nucleon decay searches, and measurement of the neutrino mass hierarchy and CP violating phase.   

Sensitivity to antineutrinos is enhanced due to the impressive background rejection capabilities.  Antineutrino detection via inverse beta decay (IBD) provides a coincidence signal: a prompt positron, followed by a 2.2-MeV $\gamma$ from neutron capture on hydrogen.  The scintillator component of the WbLS target allows for high neutron detection efficiency, estimated to surpass that even of Gd-loaded water detectors, thus reducing the single-event background that limits such experiments.  A dominant background for pure LS detectors is from neutral current interactions of cosmic muons on carbon, which cause a nuclear recoil followed by potential capture of liberated neutrons.  In a pure LS detector this can mimic the coincidence of the IBD signal.  However, access to the Cherenkov signal would allow excellent discrimination between the nuclear recoil and the positron of the IBD signal.

This capability opens up the potential for a high-statistics measurement of geo-neutrinos in a complementary geographical location.  Existing data are from the KamLAND~\cite{KLgeo} and Borexino~\cite{Bgeo} experiments, located in Asia and Europe, respectively.  A measurement in North America would provide additional information on the relative contributions of the crust and mantle.  A high sensitivity search for DSNB would also be possible, as well as sensitivity to the antineutrino signal from potential supernovae.

\section*{Ongoing Detector Development}

The R\&D program for \textsc{Theia} strongly leverages existing efforts and synergy with other programs, such as WATCHMAN~\cite{wm}, ANNIE~\cite{annie}, SNO+~\cite{snopl} and others.  Ongoing work includes WbLS development at BNL~\cite{wbls}, purification and compatibility studies at UC Davis, characterization and optimization with the CHESS detector at UC Berkeley and LBNL~\cite{chess, chess2}, fast photon sensor development at ANL, U Chicago, Iowa State and others ~\cite{mcp,lappd,lappd2,lappd3}, development of reconstruction algorithms~\cite{elagin, elagin2}, and potential nanoparticle loading in NuDot at MIT~\cite{nudot}. This paper focuses on the WbLS purification and characterization ongoing at UC Davis and UC Berkeley.

\subsection*{Nanofiltration}
Deionization is critical to maintain optical transparency of the target medium.  To deionize WbLS, the water and LS components must first be separated, and recombined after the filtration process.  US Davis is developing a membrane filtration process to separate the water and heavy ions from the LS, allowing the water to be filtered.  This process must be scalable to large volumes, such that the entire \textsc{Theia} detector could be turned over in a reasonable time frame, and have no impact on the LS optical properties or light yield.  Successful separation has been achieved, and the final product has been demonstrated to have both light yield and absorption consistent with the initial material.  The flow rate is sufficiently high for a molecular weight cut off greater than 1000, which has been observed to achieve the necessary separation.

\subsection*{Cherenkov-Scintillation Separation}
The CHESS experiment in Berkeley is a ring-imaging detector designed to demonstrate the capability of identifying the prompt Cherenkov signal from a scintillating liquid, using both photon detection time and event topology.  The detector is described in detail in~\cite{chess}.  The primary source is vertical-going cosmic muons, selected to be within six degrees of vertical.  The design of the detector allows Cherenkov photons to be identified by the known cone-like topology of the light produced from charged particle interactions.  Initial studies using CHESS have demonstrated successful identification of Cherenkov photons from both pure linear alkyl benzene (LAB), and LAB loaded with 2g/L of PPO.  This fluor complicates Cherenkov photon detection both by increasing the scintillation output by roughly an order of magnitude, and by shortening the scintillation time constant, thus increasing overlap in hit-time with the prompt Cherenkov peak.  Cherenkov detection efficiency of 83$\pm$3 (96$\pm$2) \% was achieved using hit time (charge) in pure LAB, and 70$\pm$3 (63$\pm$8) \% in LAB/PPO, with scintillation contamination of 11$\pm$1 (6$\pm$3) \% and 36$\pm$5 (38$\pm$4) \%, respectively~\cite{chess2}.

\begin{figure}[tbhp]
\centering
\includegraphics[width=.8\linewidth]{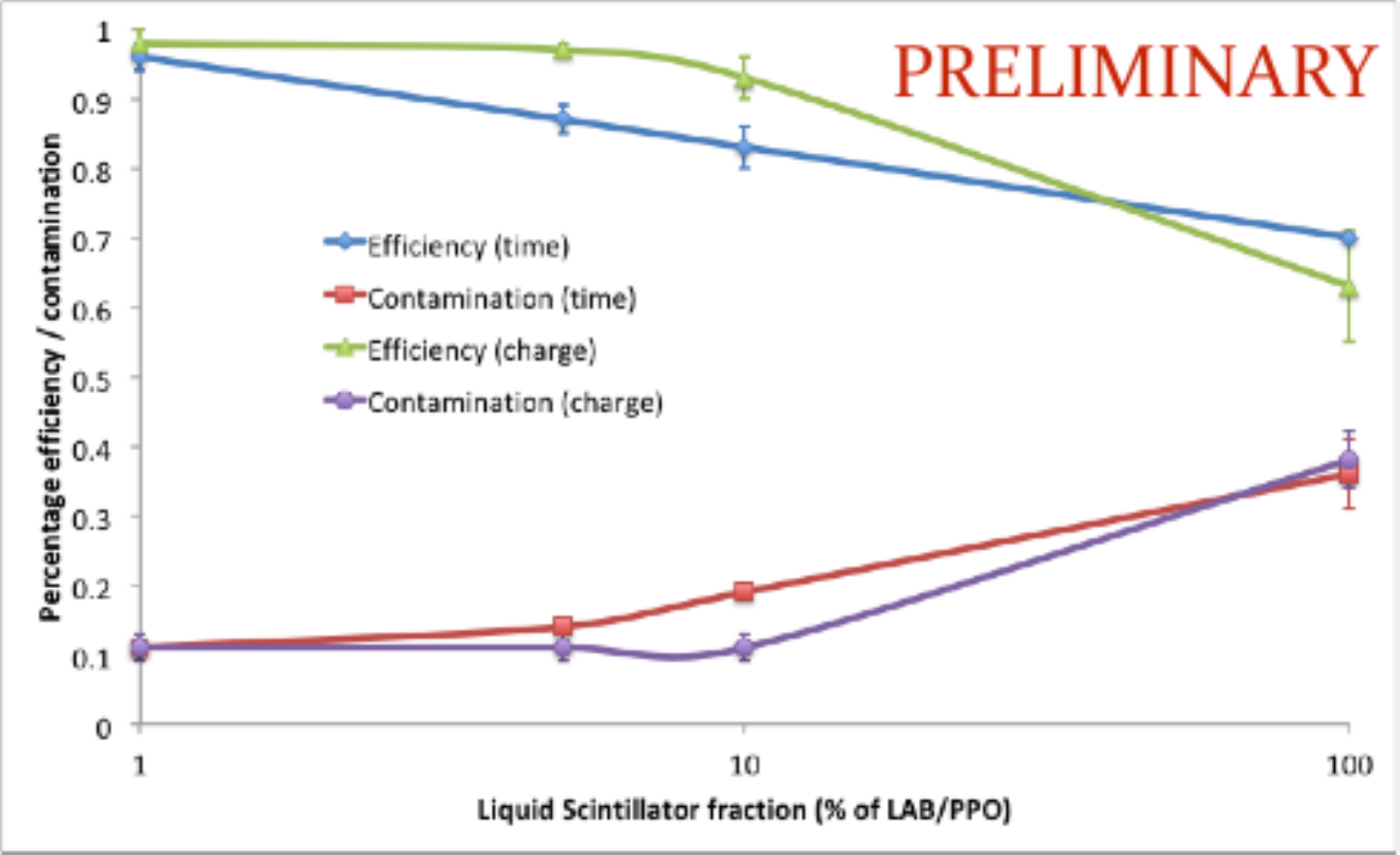}
\caption{Cherenkov detection efficiency and scintillation contamination for WbLS samples using both a hit-time based and prompt charge based method.}
\label{fig:WbLS}
\end{figure}

Samples of WbLS, provided by Dr. Yeh's group at BNL, were deployed in the CHESS detector.   
The Cherenkov detection efficiency and scintillation contamination were measured using through-going cosmic muons, as described in ~\cite{chess2}, as a function of the LS fraction in the target.  Fig.~\ref{fig:WbLS} shows the preliminary results.  As expected, the separation is more efficient at lower LS fractions.  However, even at 10\% the Cherenkov photon detection efficiency is better than 90\%, with approximately 10\% contamination of scintillation light.

By simplifying the setup and using a $^{90}$Sr $\beta$ source in place of cosmic muons, the time profile of scintillation light can be measured.  The low-energy $\beta$ source is important to move the setup into the single photoelectron regime for per-PMT detection, in order to be sensitive to the late tail of the scintillation profile.  Using a full, high precision Monte Carlo model of the detector, including full optical properties, photon propagation, calibrated DAQ simulation and PMT response, the intrinsic scintillation time profile can be extracted.  This method was tested by deploying LAB/PPO and the resulting time profile seen to agree incredibly well with that predicted by the Monte Carlo model, assuming a three-exponential decay taken from ~\cite{queens} and a single rise time as measured in~\cite{chess2}.  Deployment of a range of WbLS cocktails (with 1--10\% LS) showed uniformly faster scintillation time profiles, and also increased light yield compared to a direct scaling from LAB/PPO.  This is consistent with an increased concentration of PPO in the WbLS mixture.  

These results demonstrate the sensitivity of the CHESS detector to the microphysical behaviour and formulation of the WbLS cocktail.  These results, and further R\&D moving forwards, will be a critical ingredient in the optimization of the WbLS target medium for experimental use by projects such as ANNIE, WATCHMAN and \textsc{Theia}, since the time profile directly affects  the efficiency of Cherenkov signal extraction.

\section*{Summary}

 Use of the novel, potentially inexpensive WbLS target  allows construction of a precision detector on a massive scale.   Successful identification of Cherenkov light in a scintillating detector would result in unprecedented background-rejection capability and signal detection efficiency via directionality and sub-Cherenkov threshold particle identification.  This low-threshold, directional detector could achieve a fantastically broad physics program, combining conventional neutrino physics with rare-event searches in a single, large-scale detector.
The flexibility of the WbLS target, of the options for isotope loading, and even of the detector configuration is a crucial aspect of \textsc{Theia}'s design.  As the field evolves, \textsc{Theia} has the unique ability to adapt to new directions in the scientific program, making it a powerful instrument of discovery that could transform the next-generation of experiments.  
Further studies hinge on additional R\&D, including the Cherenkov detection efficiency enhancement provided by deployment of fast photon sensors, and demonstration of quenching and particle ID capabilities in WbLS.

\acknow{This material is based upon work supported by the Director, Office of Science, of the U.S. Department of Energy under Contract No. DE-AC02-05CH11231 and by the U.S. Department of Energy, Office of Science, Office of Nuclear Physics, under Award Number DE-SC0010407.
Research conducted at Lawrence Berkeley National Laboratory is supported by the Laboratory Directed Research and Development Program of Lawrence Berkeley National Laboratory under U.S. Department of Energy Contract No. DE-AC02-05CH11231, and the University of California, Berkeley.  
}

\showacknow{} % Display the acknowledgments section

% \end{document}

%% file: AAP-Mendenhall/aap2018-participant00x.tex
% \documentclass[9pt,twocolumn,twoside,lineno]{aap2018}
%\documentclass{article}
% \usepackage[utf8]{inputenc}

% \title{AAP-PersonalizedTemplate}
% \author{bergevin1 }
% \date{September 2018}

% \begin{document}

% \maketitle

% \section{Introduction}

% \end{document}

% \documentclass[9pt,twocolumn,twoside,lineno]{pnas-new}
% Use the lineno option to display guide line numbers if required.

% \templatetype{aap2018proceedings} % Choose template 
% {pnasresearcharticle} = Template for a two-column research article
% {pnasmathematics} %= Template for a one-column mathematics article
% {pnasinvited} %= Template for a PNAS invited submission

\newcommand{\nuebar}{\ensuremath{\overline{\nu}_e}}
\newcommand{\Psp}{{\textsc{Prospect}}}

\title{Near-surface backgrounds for ton-scale IBD detectors}

% Use letters for affiliations, numbers to show equal authorship (if applicable) and to indicate the corresponding author
\author[a]{Michael P. Mendenhall for the \Psp\ Collaboration}

\affil[a]{Nuclear and Chemical Sciences Division, Lawrence Livermore National Laboratory}

% Please give the surname of the lead author for the running footer
\leadauthor{Mendenhall}

% Please include corresponding author, author contribution and author declaration information
% \authorcontributions{Please provide details of author contributions here.}
% \authordeclaration{Please declare any conflict of interest here.}
% \equalauthors{\textsuperscript{1}A.O.(Author One) and A.T. (Author Two) contributed equally to this work (remove if not applicable).}
\correspondingauthor{\textsuperscript{2}To whom correspondence should be addressed. E-mail: mpmendenhall\@llnl.gov}

% Keywords are not mandatory, but authors are strongly encouraged to provide them. If provided, please include two to five keywords, separated by the pipe symbol, e.g:

\begin{abstract}
Compact Inverse Beta Decay (IBD) detectors need to run near their \nuebar\ source,
    often at a minimal-overburden site.
Without many mWE of overburden, cosmic fast neutron backgrounds are the
    dominant source of IBD-mimic neutron-correlated detector interactions.
With an appropriate combination of detector capabilities, demonstrated by the
    \Psp\ experiment, high-precision \nuebar\ measurements can be performed
    at a surface-level site.
\end{abstract}

% \dates{This manuscript was compiled on \today}
\doi{\url{https://neutrinos.llnl.gov/workshops/aap2018}}

% \begin{document}

\maketitle
\thispagestyle{firststyle}
\ifthenelse{\boolean{shortarticle}}{\ifthenelse{\boolean{singlecolumn}}{\abscontentformatted}{\abscontent}}{}

%This project is shared with the AAP2018 organizing committee. When your proceeding is finished please contact aap2018@llnl.gov . No other submission steps are needed. The committee will contact you if there are any missing elements to your proceedings.

%The length of the proceeding are not to exceed two pages in length including any figures or tables you wish to include.

%References should be cited in numerical order as they appear in text; this will be done automatically via bibtex, e.g. \cite{belkin2002using} and \cite{berard1994embedding,coifman2005geometric}. All references should be included in the main manuscript file.  Please use inspire \url{http://inspirehep.net/} for your references; they will be tabulated at the end of the proceedings and do not count toward your two page limit. 

% If your first paragraph (i.e. with the \dropcap) contains a list environment (quote, quotation, theorem, definition, enumerate, itemize...), the line after the list may have some extra indentation. If this is the case, add \parshape=0 to the end of the list environment.
\dropcap{D}etectors to enable near-field reactor monitoring applications must go wherever the reactors are, indicating compact designs (few-meter sizes at few-ton scales) at minimally-shielded sites.
The \Psp\ detector provides a working example of such a system, performing precision \nuebar\ measurements at a surface-level site.
\Psp\ data confirms intuition from simulations on the source of, and mitigation for, IBD-mimic backgrounds.

\subsection*{Near-surface backgrounds come from cosmic neutrons}

The tens-of-$\mu$s timescale for neutron-capture-correlated events is highly distinctive: much longer than ns-scale electromagnetic interactions, while much shorter than typical ms-scale accidental coincidences.
IBD-mimic backgrounds are unlikely without neutrons present.
The cosmic-ray-induced fast neutron background is the dominant source for neutrons in near-surface environments, until many meters water-equivalent overburden attenuates the direct fast neutron flux below the level of secondary local neutron production by muon spallation.

IBD-mimic events primarily come from two mechanisms:
\begin{enumerate}
    \item Correlated capture of two thermal neutrons, when the first is misidentified as an IBD prompt positron, or
    \item Recoil followed by capture of a fast neutron, when the recoil interaction is misidentified as an IBD positron.
\end{enumerate}
The first category of backgrounds is typically associated with multi-neutron-producing spallation showers near the detector active volume, while the second category comes from tens-of-MeV fast neutrons entering the active volume.

\subsection*{A combination of detector capabilities reject backgrounds}

\begin{figure}
\centering
\includegraphics[width=\linewidth]{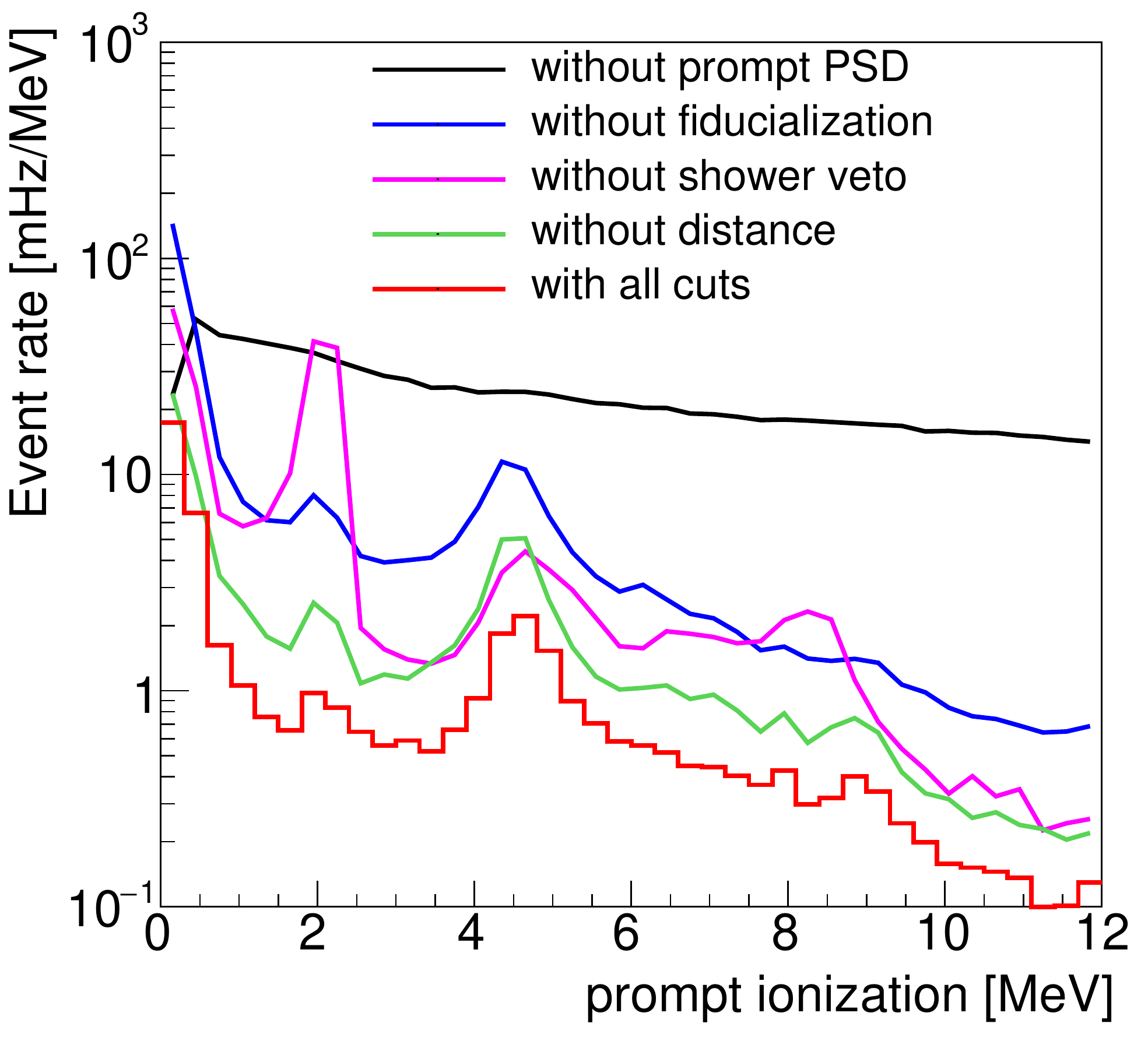}
\caption{\Psp\ IBD-like reactor-off correlated backgrounds, with various detector capabilities removed from analysis.}
\label{Mendenhall:Figure1}
\end{figure}

Multiple detector capabilities work in tandem to reject backgrounds.
Figure~\ref{Mendenhall:Figure1} shows the reactor-off correlated IBD-like background rate in \Psp, and the increased background rates when various information is removed from the analysis.

Ability to identify fast-neutron-induced recoil components in the prompt event is most critical, providing up to 2 orders of magnitude in background suppression.
Pulse shape discrimination (PSD) identifies recoils, in combination with detector segmentation to isolate proton recoils from being hidden by inelastic recoil gammas.

Fiducialization provides an order-of-magnitude suppression in backgrounds.
Events occurring deeper in the detector interior are less likely to ``lose information'' from the active volume that assists in event identification.

A hadronic shower veto around high-energy tracks, fast neutron recoils, or multiple thermal captures strongly suppresses backgrounds from the thermal-neutron-pair channel.
Good efficiency in thermal neutron capture identification (provided by the $^6$Li capture tag), along with recoil PSD, contribute to this cut.

Finally, the distance between prompt and delayed interactions provides a $\sim 2\times$ background suppression factor, enabled by detector position sensitivity.

\begin{figure}
\centering
\includegraphics[width=\linewidth]{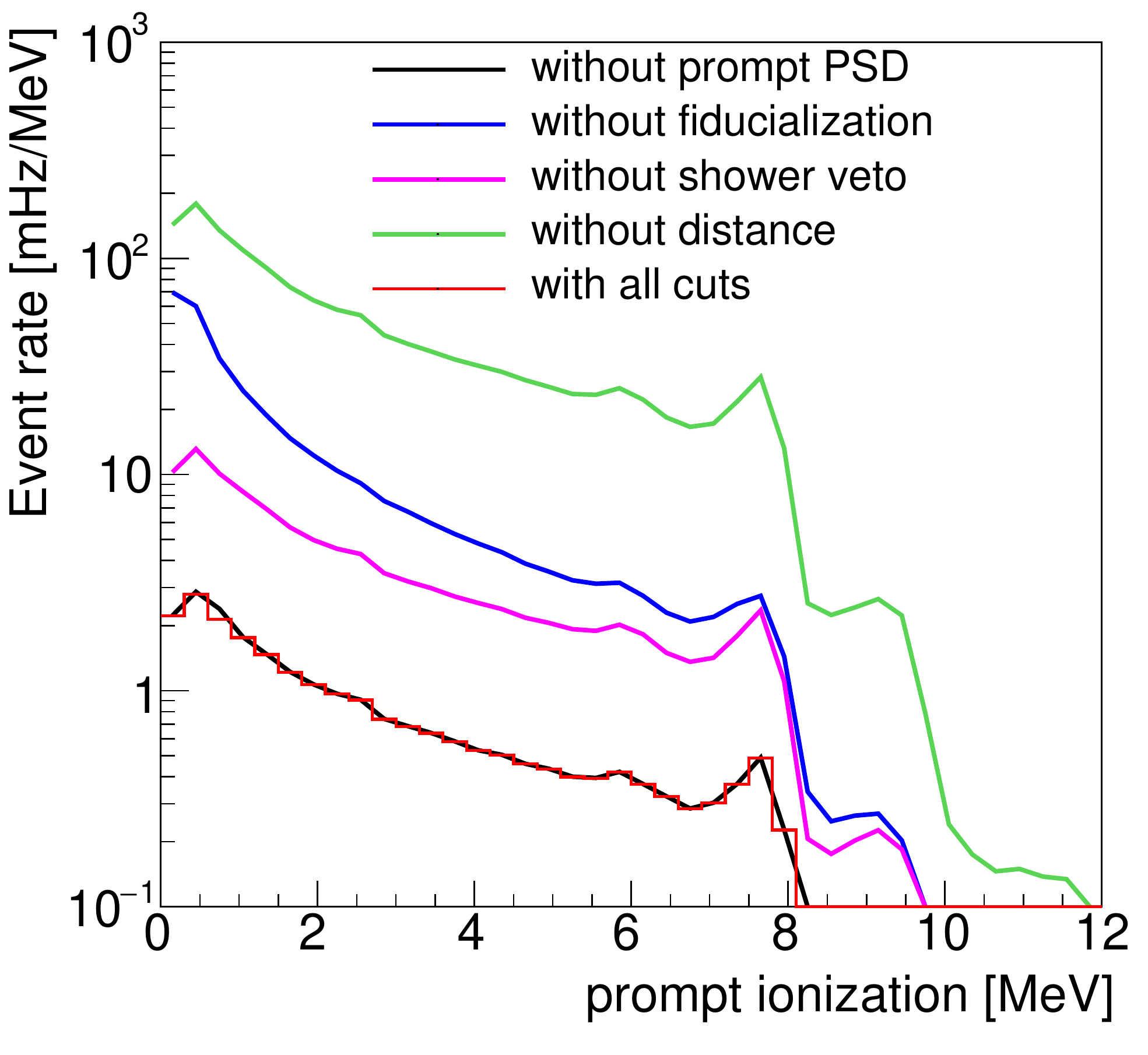}
\caption{\Psp\ reactor-on accidental backgrounds, with various detector capabilities removed from analysis.}
\label{Mendenhall:Figure2}
\end{figure}

Figure \ref{Mendenhall:Figure2} demonstrates detector capabilities for accidental backgrounds rejection.
When the reactor is on, \Psp\ receives a high rate of energetic gammas from neutron capture on iron in building structures (reactor-off accidentals are far lower).

Distance between prompt and delayed events provides the largest suppression of accidentals, followed by fiducialization.
The hadronic shower veto significantly reduces the load of neutron capture singles with which gammas can accidentally pair.

\subsection*{Simulation is a good guide to detector performance}

\texttt{Geant4}-based simulations of the cosmic fast neutron flux incident on \Psp\ reproduce the observed correlated background distributions.
Simulations, run through a detector response model and analyzed without reliance on MC truth data, can provide reliable guidance for detector design and performance projections.

%\subsection*{Conclusions}

% \matmethods{Please describe your materials and methods here. This can be more than one paragraph, and may contain subsections and equations as required. Authors should include a statement in the methods section describing how readers will be able to access the data in the paper. 

% \subsection*{Subsection for Method}
% Example text for subsection.
% }

% \showmatmethods{} % Display the Materials and Methods section

\acknow{LLNL-PROC-788379. This work was performed under the auspices of the U.S. Department of Energy by Lawrence Livermore National Laboratory under Contract DE-AC52-07NA27344.}

\showacknow{} % Display the acknowledgments section

% Bibliography
% \bibliography{aap2018-participant00x}

% \end{document}

%% file: AAP-Mulmule/aap2018-participant00x.tex
\graphicspath{{./AAP-Mulmule/}}

% \documentclass[9pt,twocolumn,twoside,lineno]{aap2018}
% \usepackage{float}

% \templatetype{aap2018proceedings} % Choose template 
% {pnasresearcharticle} = Template for a two-column research article
% {pnasmathematics} %= Template for a one-column mathematics article
% {pnasinvited} %= Template for a PNAS invited submission

\title{Exploring anti-neutrino event selection and background reduction techniques for ISMRAN}

% Use letters for affiliations, numbers to show equal authorship (if applicable) and to indicate the corresponding author
\author[a,b,*]{D.~Mulmule}
\author[a]{P.K.~Netrakanti} 
\author[a]{S.P.~Behera}
\author[a]{D.K.~Mishra}
\author[a]{V.~Jha} 
\author[a,b]{L.M.~Pant}
\author[a,b]{B.K.~Nayak}

\affil[a]{Nuclear Physics Division, Bhabha Atomic Research Centre, Trombay, Mumbai-400085}
\affil[b]{Homi Bhabha National Institute, Anushakti Nagar, Mumbai - 400094}

% Please give the surname of the lead author for the running footer
\leadauthor{Mulmule}

% Please include corresponding author, author contribution and author declaration information
% \authorcontributions{Please provide details of author contributions here.}
% \authordeclaration{Please declare any conflict of interest here.}
% \equalauthors{\textsuperscript{1}A.O.(Author One) and A.T. (Author Two) contributed equally to this work (remove if not applicable).}
\correspondingauthor{\textsuperscript{*}D. Mulmule. E-mail: dhruvm@barc.gov.in}

% Keywords are not mandatory, but authors are strongly encouraged to provide them. If provided, please include two to five keywords, separated by the pipe symbol, e.g:
% \keywords{Keyword 1 $|$ Keyword 2 $|$ Keyword 3 $|$ ...} 

\begin{abstract}
The Indian Scintillator Matrix for Reactor Anti-Neutrino detection (ISMRAN), is a $\sim$ 1 ton by weight (1 $\mathrm{m^{3}}$ volume) anti-neutrino ($\bar{\nu}_{e}$) detection setup being developed for purpose of monitoring nuclear reactors. It is an above-ground detector which uses an array (10$\times$10) of 100 plastic scintillator (PS) bars as the core detector and will be housed inside the Dhruva research reactor hall at Bhabha Atomic Research Centre, $\sim$13 m from core of the 100 $\mathrm{MW_{th}}$ output reactor. The segmented geometry of PS bars can prove useful to filter correlated $\bar{\nu}_{e}$ events based on their multiplicity and sum energy signature. In this talk, we present the simulation studies performed to understand the characteristics of these parameters for pure $\bar{\nu}_{e}$ events. Also, background studies using the 10 cm Lead and 10 cm borated polyethylene shield are presented for the 16$\%$ volume prototype detector - mini-ISMRAN setup.
\end{abstract}
%\vspace{-1.0cm}
% \dates{This manuscript was compiled on \today}
\doi{\url{https://neutrinos.llnl.gov/workshops/aap2018}}

% \begin{document}

\maketitle
\thispagestyle{firststyle}
\ifthenelse{\boolean{shortarticle}}{\ifthenelse{\boolean{singlecolumn}}{\abscontentformatted}{\abscontent}}{}

% If your first paragraph (i.e. with the \dropcap) contains a list environment (quote, quotation, theorem, definition, enumerate, itemize...), the line after the list may have some extra indentation. If this is the case, add \parshape=0 to the end of the list environment.
%\dropcap{T}his template is provided to help you summarize your work in the correct format.  Instructions for use are provided below. 
%\vspace{-1.2cm}
\section*{Introduction: Anti-neutrino detection using ISMRAN}
\vspace{-0.1cm}
The basic element of ISMRAN is an EJ-200 plastic scintillator (PS) bar of dimensions 10 cm $\times$ 10 cm $\times$ 100 cm, wrapped with $\mathrm{Gd_{2}O_{3}}$ (areal density: $\mathrm{4.8~mg/cm^{2}}$) coated  aluminized mylar foil on the periphery. Two 3" PMTs are used at either end of the PS bar to readout scintillation output of the detector. 100 such bars in a 10$\times$10 array will constitute the final setup along with a DAQ system using high sampling rate digitizers to read the PMT outputs. An electron anti-neutrino ($\bar{\nu}_{e}$) will interact inside ISMRAN volume via inverse beta decay (IBD) process to produce a positron and a neutron. The prompt signal from ionization loss and annihilation $\gamma$-rays of positron and delayed signal from neutron capture $\gamma$-rays are expected to span multiple bars and have characteristic energy signatures. The presence of a variety of cosmogenic and reactor hall specific backgrounds necessitate understanding of these $\bar{\nu}_{e}$ prompt and delayed signal characteristics. 
\begin{figure}[H]
\centering
\includegraphics[width=.75\linewidth]{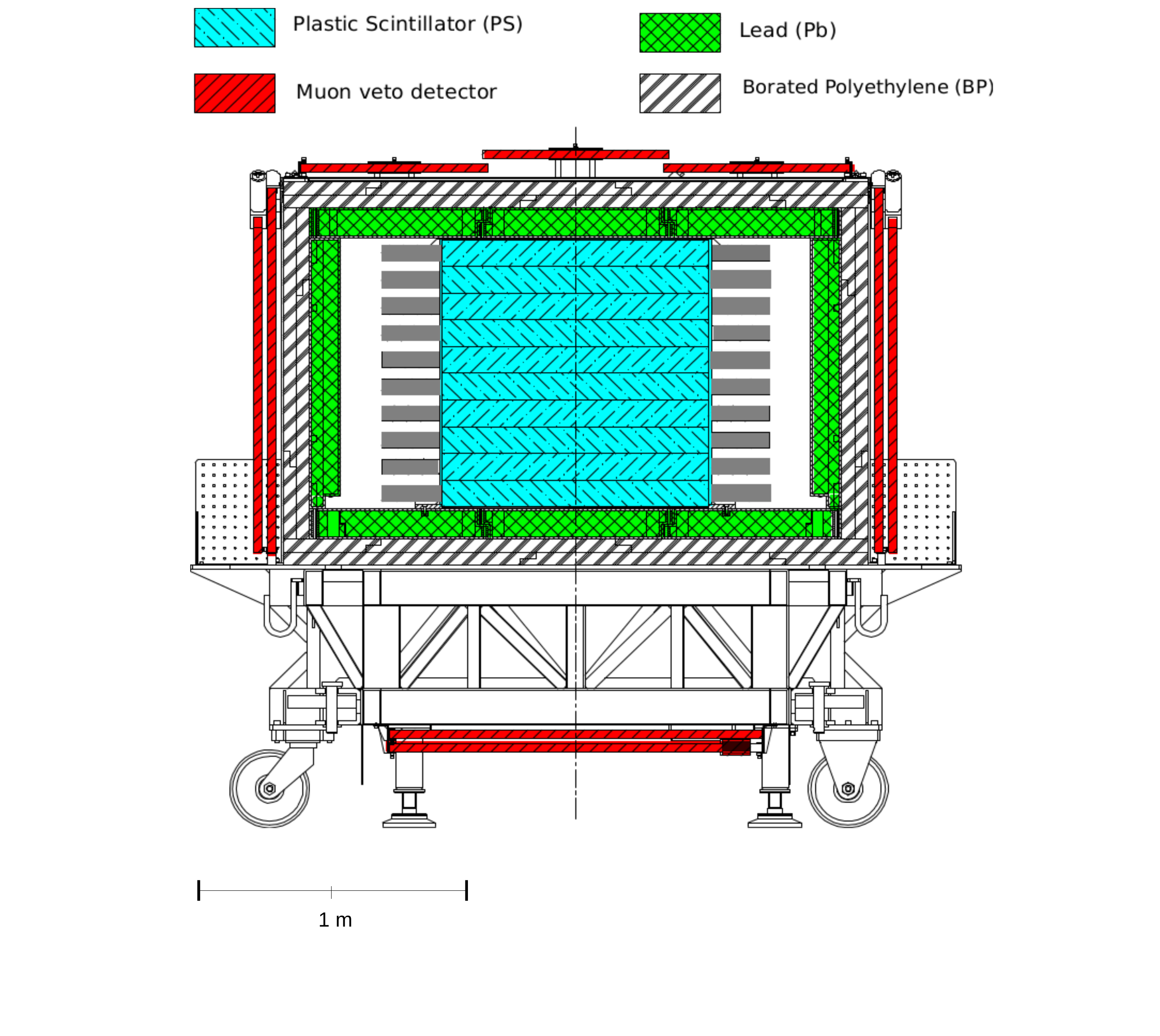}
\vspace{-0.5cm}
\caption{Proposed ISMRAN detector setup comprising of shielding trolley and 100 PS bars. The major components of the setup are listed in their respective colors at top}
\label{Trolley}
\end{figure}
ISMRAN uses a passive shielding of 10 cm thick borated polyethylene (BP) and 10 cm thick Lead (Pb) to reduce the external $\gamma$-ray and neutron background. Additionally, due to the segmented geometry of such a detector, event multiplicity - $\mathrm{N_{bars}}$ i.e. number of PS bars with energy deposit and the sum total of these deposited energies, can form a basis for filtering out signal events~\cite{OGURI201433}. The final mobile trolley setup will also include an active shielding of PS covering all six faces to veto out cosmic muon activity as shown in Fig.~\ref{Trolley}. Detailed characterization of PS bars with digital DAQ and studies to optimize the shielding are performed using the unshielded prototype detector - mini-ISMRAN of 16 PS bars in the laboratory environment. This setup is currently taking data in the reactor environment under both ON and OFF conditions inside the shielding~\cite{MULMULE2018104}.
\vspace{-0.1cm}
%\vspace{-0.5cm}
\section*{Background measurements in reactor environment}
\begin{figure}[H]
\centering
\includegraphics[width=.75\linewidth]{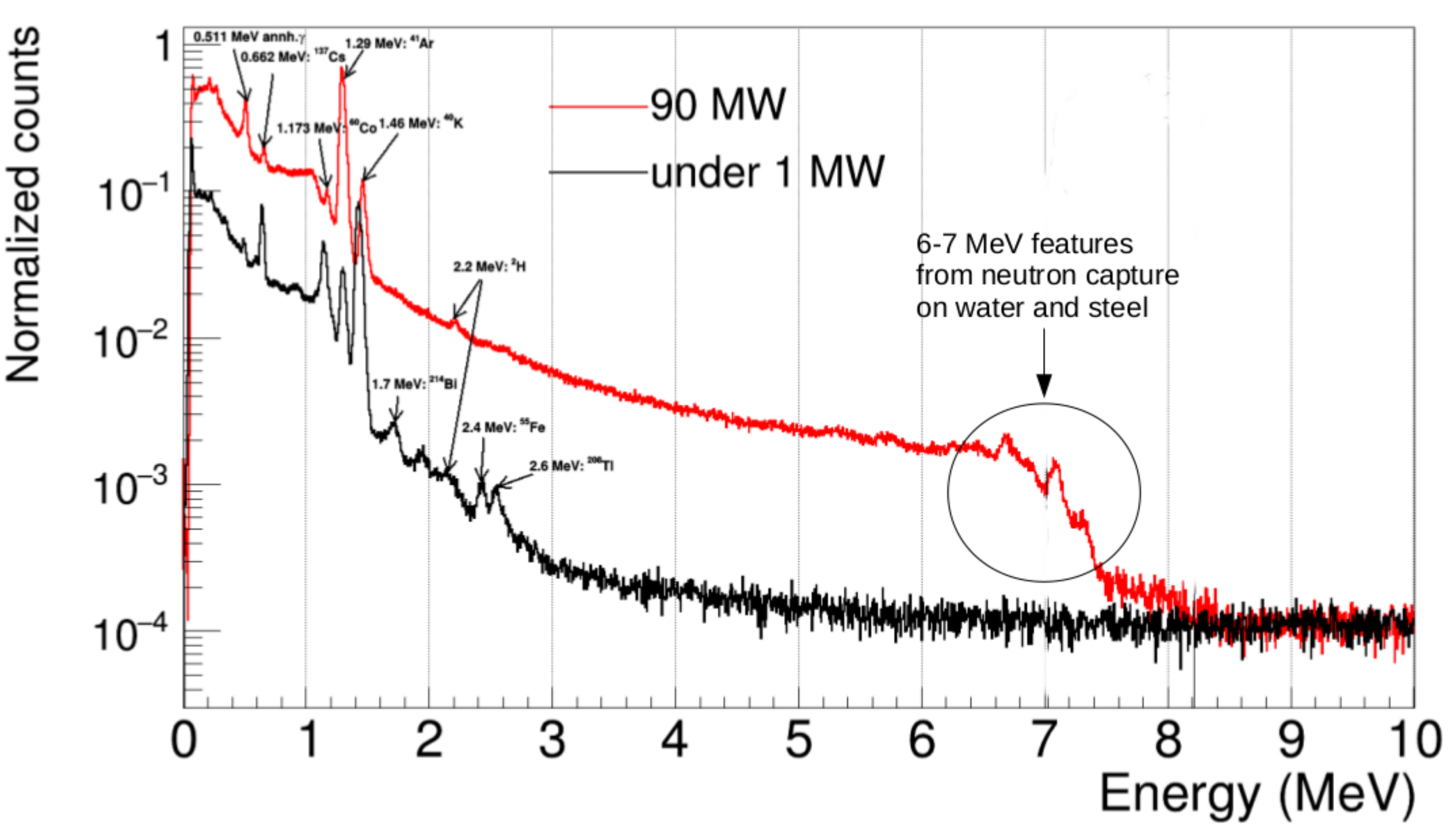}
\vspace{-0.3cm}
\caption{Reactor $\gamma$-ray background in ON and OFF conditions in a $\mathrm{CeBr_{3}}$ scintillator}
\label{CeBR}
\end{figure}

\vspace{-0.1cm}
A number of $\gamma$-ray and neutron backgrounds are expected at the ISMRAN location in Dhruva reactor hall. High energy $\gamma$-ray activity, predominantly from neutron capture on surrounding materials during reactor ON time, and residual and natural background activities during OFF time are observed from a measurement performed using a 2 inch $\mathrm{CeBr_{3}}$ scintillator, as seen in Fig~\ref{CeBR}.
%\begin{table}[bh]
%\centering
%\caption{Reactor ON background rates in a single PS bar in mini-ISMRAN}
%\begin{tabular}{ll}
%Detector configurations & Count rates (Hz) \\
%\midrule
%No shielding (Single PS bar) & $\sim$ 24000 \\
%10 cm thick lead shield & $\sim$ 2000 \\
%10 cm thick lead shield + 10 cm thick B.P. & $\sim$ 500\\
%10 cm thick lead shield + 10 cm thick B.P. \\ ($N_{bars}$ = 2, time window < 40 ns) & $\sim$ 10\\
%\bottomrule
%\label{Rates}
%\end{tabular}
%\end{table}
\vspace{-0.1cm}
Also, cosmogenic muon and neutron induced activity is present irrespective of the location and reactor status and is more pronounced for above-ground setups. The overall background rates in reactor ON measured for the 10 cm $\times$ 10 cm PS bars in the mini-ISMRAN setup reduce from $\sim$24000 Hz under no shielding to $\sim$500 Hz with 10 cm BP and 10 cm Pb shielding. If an added criteria of coincidence between two bars, as expected for a prompt-like event, is imposed this rate falls down to $\sim$10 Hz.
\vspace{-0.32cm}
%\subsection*{Tables}
%Please use the following format for tables:
% \addtabletext{nomenclature for the TSs refers to the numbered species in the table.}
%\provsubsodeecto tion*{Author Affiliations}
%Include the institution for each author. Use lower case letters to match authors with institutions, as shown in the example. 
\section*{IBD event simulations in ISMRAN}
\vspace{-0.1cm}
Monte carlo simulations using the GEANT4 toolkit are performed for pure IBD events in ISMRAN. The parameterization used for $\bar{\nu}_{e}$ input energy spectrum is as per reference~\cite{Mueller:2011nm}, while calculations of cross section are as per reference~\cite{Mention:2011rk} and the fission fractions for different isotopes are taken from reference~\cite{PhysRevD.78.111103}. A threshold $\mathrm{E_{th}}$ > 0.2 MeV is applied on the energy deposited in each bar in simulation, as used in the calibrated measurement data to achieve spectral uniformity among different bars.
\begin{figure}[H]
\centering
\includegraphics[width=.7\linewidth]{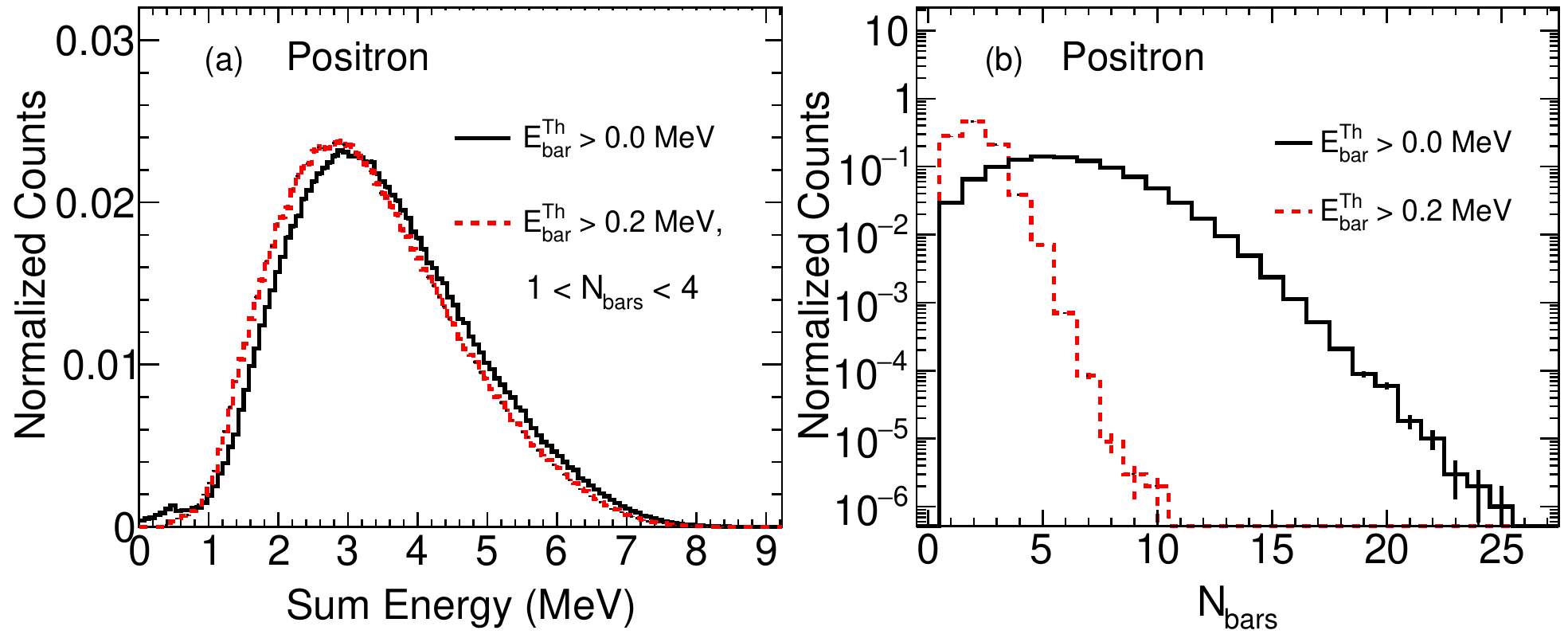}
\vspace{-0.2cm}
\caption{Sum energy (a) and multiplicity (b) distributions in ISMRAN for prompt positron events for different threshold :$\mathrm{E_{th}}$ and $\mathrm{N_{bars}}$ criteria}
\label{SimuPos}
\end{figure}
Figure~\ref{SimuPos}(a) shows the sum energy and Fig~\ref{SimuPos}(b) the $\mathrm{N_{bars}}$ values for prompt positron event. The sum energy spectrum closely follows the input $\bar{\nu}_{e}$ energy spectrum as the positron is expected to carry almost all of the $\bar{\nu}_{e}$ energy. Comparison of two cases :$\mathrm{E_{th}}$ > 0.0 MeV and $\mathrm{E_{th}}$ > 0.2 MeV with 1 < $\mathrm{N_{bars}}$ < 4 shows sum energy reconstructed to slightly lower values, but the distortion to the spectrum due to the threshold and $\mathrm{N_{bars}}$ selection is not significant. Also, the requirement of $\mathrm{E_{th}}$ > 0.2 MeV is seen to reduce the $\mathrm{N_{bars}}$ range significantly. These criteria are expected to reduce both the noise and uncorrelated background component significantly. 
\begin{figure}[H]
\centering
\includegraphics[width=.7\linewidth]{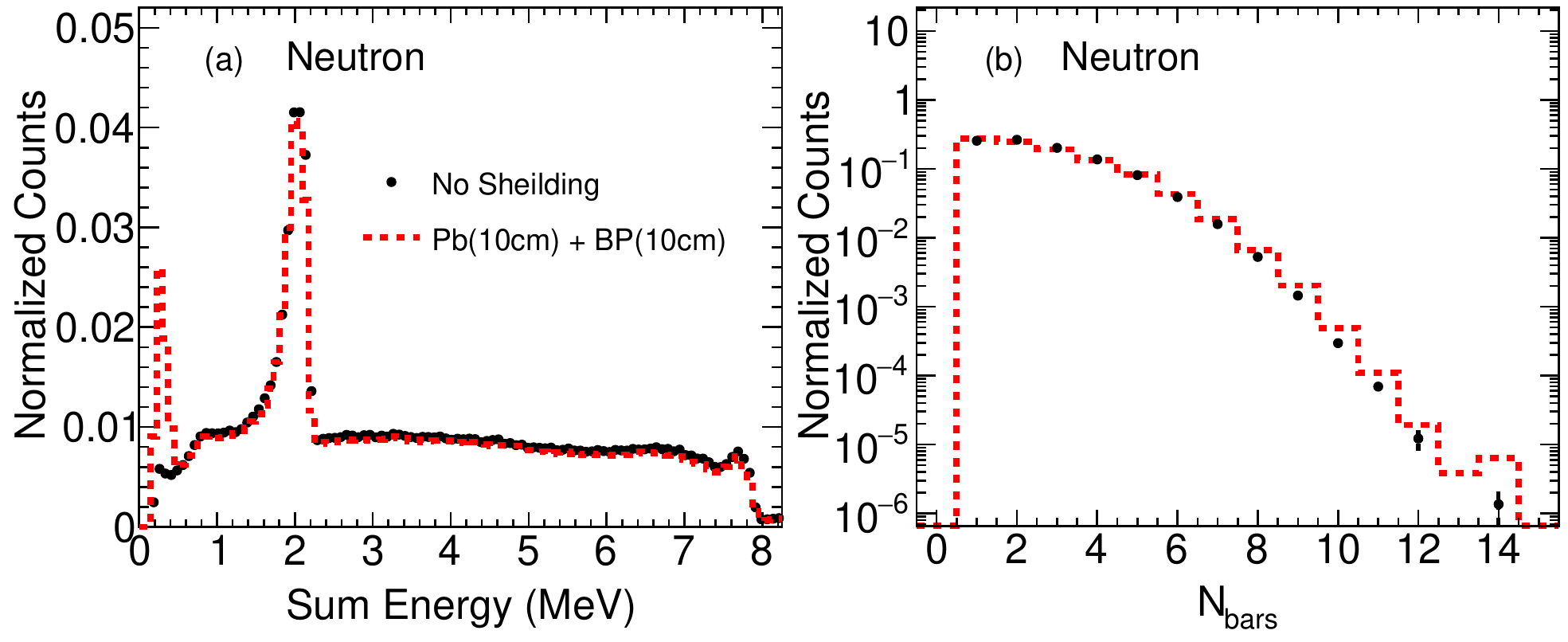}
\vspace{-0.2cm}
\caption{Sum energy (a) and multiplicity (b) distributions in ISMRAN for delayed neutron capture events under no shielding and full shielding scenarios}
\label{NeuPos}
\end{figure}
\begin{figure*}[b]
\centering
\includegraphics[width=.7\linewidth]{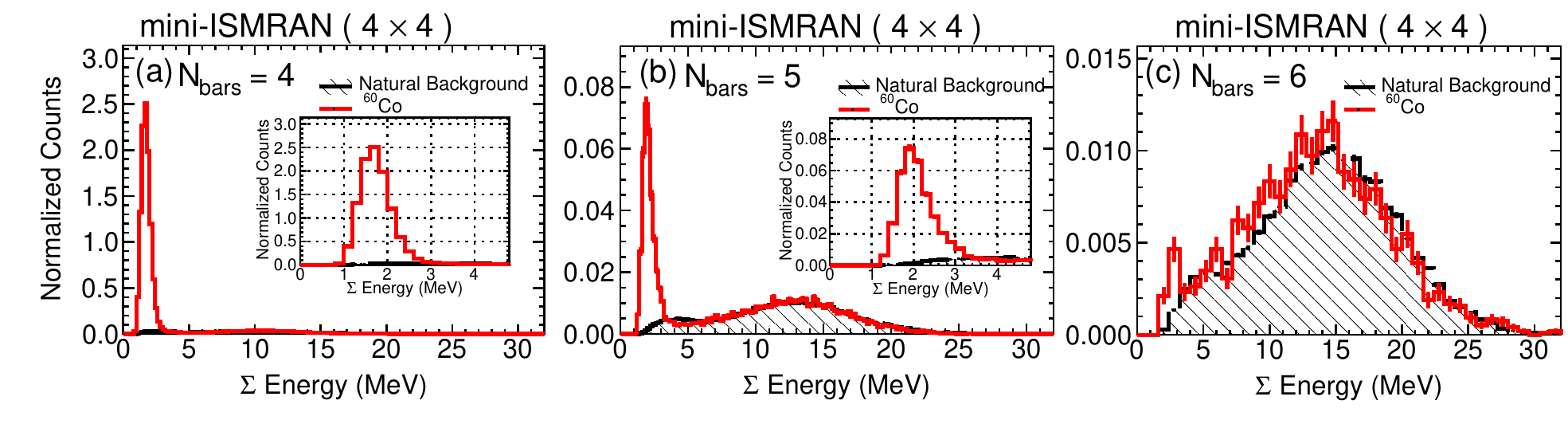}
\vspace{-0.2cm}
\caption{The sum energy distribution within 40 ns time window, for (a) $\mathrm{N_{bars}}$ = 4, (b) $\mathrm{N_{bars}}$ = 5 and (c) $\mathrm{N_{bars}}$ = 6, for the ${}^{60}$Co (solid histogram) and natural background (filled histogram) events. The insets in panel (a) and (b) shows the zoomed x-axis of the sum energy distribution for the ${}^{60}$Co source and natural background.}
\label{Co60}
\end{figure*}
Similar study for the delayed neutron capture event, shows a continuum-like reconstructed energy spectrum, due to the finite acceptance of the matrix to Gd-capture $\gamma$-rays, as seen in Figure~\ref{NeuPos}(a). Here, both the unshielded and fully shielded setups are compared to bring forth the effect of shielding in the form of a very slight reduction in counts at higher energies and a peak at $\sim$0.3 MeV, likely to be due to captures near the periphery of the PS volume. The $\mathrm{N_{bars}}$ plot in Figure~\ref{NeuPos}(b) shows substantial number of events with higher multiplicities unlike the prompt events. The mean of exponential time difference distribution between IBD prompt and delayed pairs has been observed in simulation to be $\sim$68 $\mu$s allowing additional rejection of uncorrelated background component. With these set of selection criteria, a $\bar{\nu}_{e}$ detection efficiency of $\sim$16$\%$ is expected for ISMRAN leading to an event rate of $\sim$60 per day.
\vspace{-0.2cm}
\section*{Sum Energy and Multiplicity for ${}^{60}$Co $\gamma$-rays}
\vspace{-0.1cm}
Correlated event selection based on expected sum energy and optimal choice of multiplicity is tested in the laboratory environment using ${}^{60}$Co source placed at the center of the prototype mini-ISMRAN detector. The comparison of the sum energy distributions in the PS bars, between time normalized source and `no-source' or natural background data is performed. The summation is performed for successive time windows of 40 ns, which allows complete inclusion of the correlated $\gamma$-ray event in time. Sum energy spectra for  different multiplicity events are obtained, with three representative cases shown in the fig~\ref{Co60}. Lower multiplicities, $N_{bars}$ = 4 and below, suffer from incomplete containment of the 1.17 MeV and 1.33 MeV correlated ${}^{60}$Co $\gamma$-rays leading to reconstruction of energies to lower values than expected but with higher S:B ratio. Whereas, higher multiplicities, $N_{bars}$ = 6 and above, allowed energy reconstruction closer to expected value but with drastic fall in S:B. The case of $N_{bars}$ = 5 is found to be optimal, with energy reconstruction closer to the expected $\sim$2.5 MeV energy and a significant signal component above the background. The mean sum energy of `no-source' data is observed to be higher than signal in all cases but scales with the background-like high energy component in source data.
\vspace{-0.1cm}
\section*{Summary and Outlook}
\vspace{-0.1cm}
The sum energy and multiplicity distributions obtained from simulated IBD events provide a useful first level selection for filtering $\bar{\nu}_{e}$-like correlated events. Measurement results from data taken with ${}^{60}$Co source inside mini-ISMRAN in laboratory, support this approach. Simulations incorporating background with the pure IBD events are in progress to further refine these selections and to arrive at more realistic $\bar{\nu}_{e}$ detection efficiency and event rates. The data from shielded mini-ISMRAN setup under both reactor ON and OFF conditions is being analyzed to understand background and to identify $\bar{\nu}_{e}$-like events. Further background reduction possible with additional layer of materials like high-density polythene is also being studied.
\vspace{-0.1cm}
% Uncomment following for example of equation spanning two columns :
% \begin{figure*}[bt!]
% \begin{align*}
% (x+y)^3&=(x+y)(x+y)^2\\
%       &=(x+y)(x^2+2xy+y^2) \numberthis \label{eqn:example} \\
%       &=x^3+3x^2y+3xy^3+x^3. 
% \end{align*}
% \end{figure*}

% \matmethods{Please describe your materials and methods here. This can be more than one paragraph, and may contain subsections and equations as required. Authors should include a statement in the methods section describing how readers will be able to access the data in the paper. 

% \subsection*{Subsection for Method}
% Example text for subsection.
% }

% \showmatmethods{} % Display the Materials and Methods section

\acknow{We are thankful to the Center for Design and Manufacture, BARC, for taking up the design and fabrication of detector trolley and Research reactor services division, BARC, for the logistical support.}

\showacknow{} % Display the acknowledgments section

%% file: AAP-Danielson/aap2018-participant00x.tex
\title{Directional Detection of Antineutrinos}

% Use letters for affiliations, numbers to show equal authorship (if applicable) and to indicate the corresponding author
\author[ ]{Daine L. Danielson\textsuperscript{{\normalfont a,b,1}} for the AIT-WATCHMAN Collaboration}

\affil[a]{Los Alamos National Laboratory, }
\affil[b]{University of California, Davis}
% \affil[b]{Lawrence Livermore National Laboratory}

% Please give the surname of the lead author for the running footer
\leadauthor{Danielson}

% Please include corresponding author, author contribution and author declaration information
% \authorcontributions{Please provide details of author contributions here.}
% \authordeclaration{Please declare any conflict of interest here.}
% \equalauthors{\textsuperscript{1}A.O.(Author One) and A.T. (Author Two) contributed equally to this work (remove if not applicable).}
\correspondingauthor{\textsuperscript{1}To whom correspondence should be addressed. E-mail: dldanielson@ucdavis.edu}

% Keywords are not mandatory, but authors are strongly encouraged to provide them. If provided, please include two to five keywords, separated by the pipe symbol, e.g:

\begin{abstract}
\noindent Antineutrino interactions convey information about the direction of antineutrino sources. Various potential applications of this directionality are summarized. We present one such application in the context of nuclear nonproliferation, for detecting the presence of an unknown reactor in the vicinity of another, known reactor. Potential refinements to this and comparable applications are discussed. Finally, some possible future directions towards event-by-event antineutrino directionality reconstruction are reviewed.
\end{abstract}

% \dates{This manuscript was compiled on \today}
\doi{\url{https://neutrinos.llnl.gov/workshops/aap2018}}

% \begin{document}
\maketitle
\thispagestyle{firststyle}
\ifthenelse{\boolean{shortarticle}}{\ifthenelse{\boolean{singlecolumn}}{\abscontentformatted}{\abscontent}}{}

\dropcap{B}oth antineutrino-electron elastic scattering and inverse beta decay (IBD) interactions carry information about the direction of the incident antineutrino. In elastic scattering, the scattered electron's Cherenkov light boosts into a forward cone along the axis of the incident antineutrino. In inverse beta decay, momentum conservation throws the outgoing neutron in approximately the same direction as the incident antineutrino. At reactor antineutrino energies the weak cross section biases the outgoing positron into the opposite direction, giving a supplemental, higher-order statistical directional effect \cite{PhysRevD.60.053003}. Several applications arise from the reconstruction of these directional cues.

\subsection*{Antineutrino Directionality Applications}
Since (anti)neutrinos constitute the first signal to reach Earth following a supernova, electron elastic scattering directionality offers early pointing information for observers aiming to capture the event.

On Earth, the study of geoneutrino inverse beta decay directionality may reveal insights into the radioactive composition of the planet's crust and mantle.

Towards nuclear nonproliferation, inverse beta decay directionality has already been demonstrated in reconstructing the direction of the Chooz reactor using the Double Chooz far detector \cite{Gomez}. Electron elastic scattering directionality may also support reactor monitoring efforts \cite{HELLFELD2017130}. 

A recent study led by this author has analyzed the applications of a Double Chooz-like technique in a nonproliferation monitoring context. The results of this investigation are summarized below.

\subsection*{IBD Directionality in Gd-doped Liquid Scintillator}
Directional reconstruction in a gadolinium-doped liquid scintillator detector exploits the double coincidence event topology of inverse beta decay detection to reconstruct a vector pointing toward the antineutrino source. The outgoing positron generates a short scintillation track ($\sim$0.5 mm for reactor antineutrinos) and Cherenkov light, before annihilating into two antiparallel 511 keV gamma rays. The initially forward-going neutron thermalizes in the medium and then captures on a Gd nucleus, which subsequently relaxes via a cascade of 8 MeV gamma rays following $\sim$30 $\upmu$s after the positron's prompt signal.

A directional vector is obtained from the statistical average of vectors connecting each prompt vertex to its associated delayed vertex. No one such vector points reliably to the source, because the neutron loses much of its forward-kinematic directionality to collisions during thermalization, and the positron's backward weak-cross-section bias is not a leading-order effect. Naturally, however, given sufficient event statistics the average of all such vectors converges towards alignment with the incident antineutrino flux direction.

The next section presents an application of this technique in a nuclear nonproliferation context.

\section*{Gd-doped Liquid Scintillator Monitor for Antineutrinos}\label{GLSMAN}
We have investigated the sensitivity of a particular hypothetical gadolinium-doped liquid scintillator detector with 80\% IBD detection efficiency, using directional information and event-rate information to detect the presence of an unknown reactor. Backgrounds are neglected, to set a baseline for the performance of the underlying detector technology and analysis method.

The detector under consideration employs a 1 kT fiducial cylinder based on the detector geometry of WATCHMAN (Water Cherenkov Monitor for Antineutrinos), an upcoming antineutrino monitoring experiment whose design and deployment will be detailed in a forthcoming white paper (also see Morgan Askins's contribution in these Proceedings \cite{askins}). Unlike WATCHMAN, however, this detector employs a fiducial mass of mineral oil doped with 0.1\% gadolinium, and scintillator. This hypothetical detector is dubbed ``GLSMAN''.

\subsection*{Mid-Field Monitoring Scenario}
\begin{figure}[tbhp]
\centering
\includegraphics[width=0.72\linewidth]{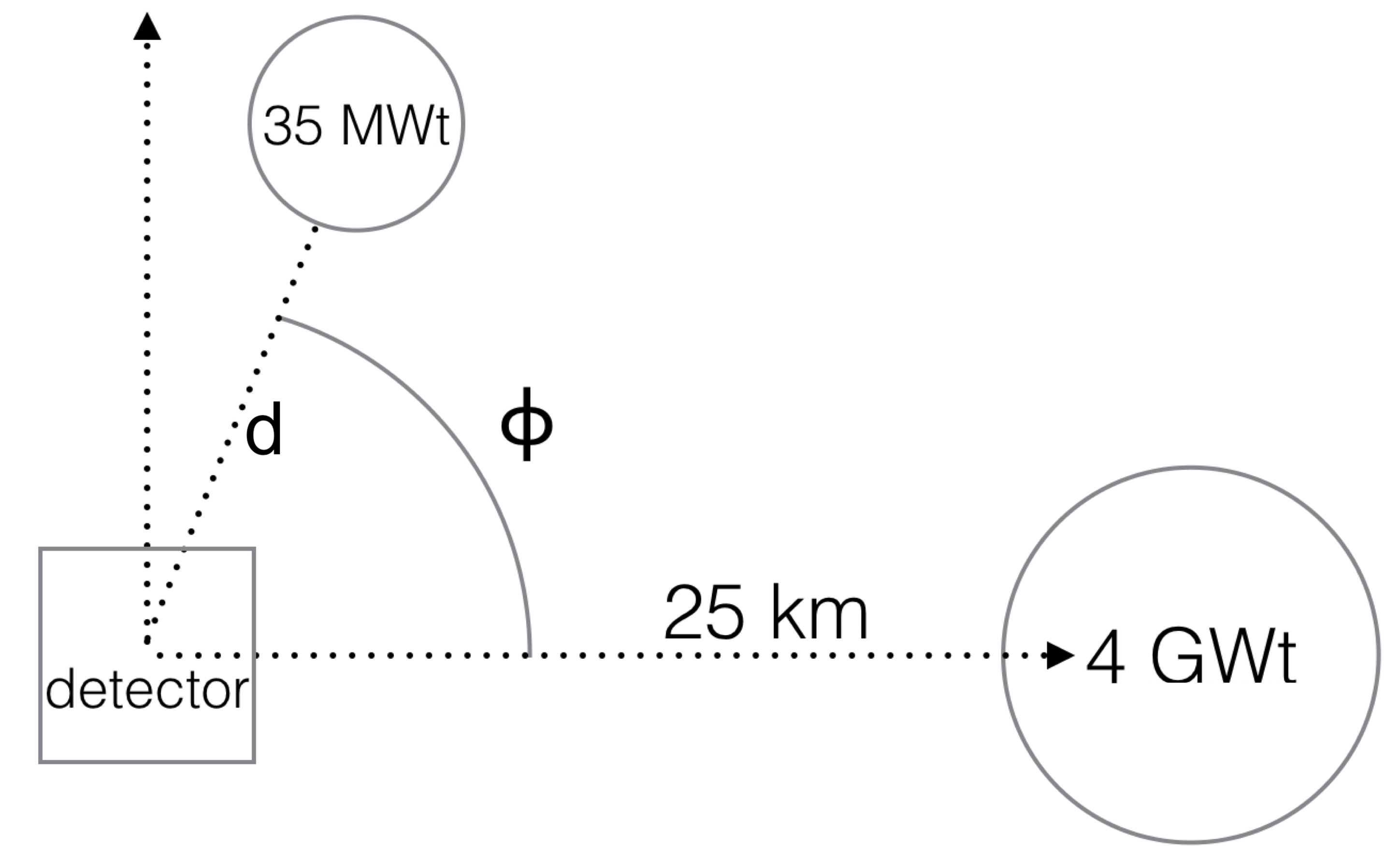}
\caption{A mid-field reactor monitoring scenario: a known 4 GWt reactor sits 25 km in the positive $x$-direction, with a second, unknown 35 MWt reactor at a standoff $d$, with an azimuthal separation $\phi$. The reactor fluxes follow reference \cite{RevModPhys.74.297}.}
\label{danielson:Figure1}
\end{figure}

Figure \ref{danielson:Figure1} summarizes the monitoring scenario under study.

Next follows a brief summary of the results obtained for the sensitivity of GLSMAN to detect the presence of the unknown reactor. A complete derivation and findings will appear in a forthcoming AIT-WATCHMAN publication led by this author, ``Detecting a Second, Unknown Reactor with a 1 kT Cylinder of GdLS for Mid-Field Nonproliferation Monitoring.''

\subsection*{Results}
\begin{figure}[tbhp]
\centering
\includegraphics[width=0.96\linewidth]{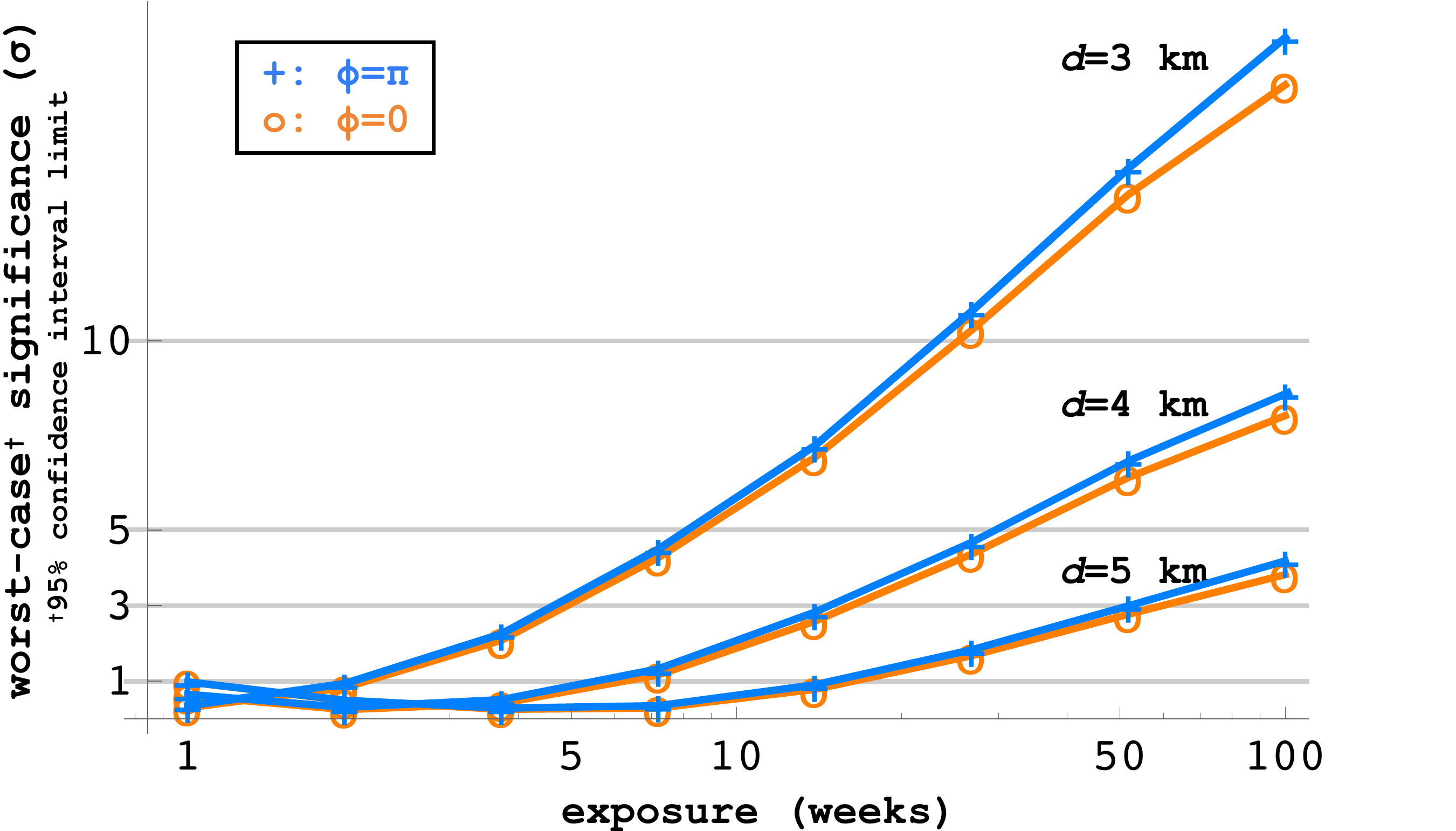}
\caption{Rejection of the single-reactor (null) hypothesis over time in the presence of an unknown 35 MWt reactor at $d \in \{3,4,5\}$ km for $\phi \in \{0,\pi\}$, given a known 4 GWt reactor at 25 km. The lower limit of the 95\% confidence interval is plotted, as integrated downward from $\sigma=+\infty$, thus giving a worst-case statistical bound.}
\label{danielson:Figure2}
\end{figure}

For the detector under consideration without backgrounds, the time to achieve $3\sigma$ detection of the unknown reactor depends more strongly on its standoff than on directional discrimination, but directionality can provide a speedup. Figure \ref{danielson:Figure2} shows that $3\sigma$ detection is likely (95\% CL) within 5 weeks at an unknown-reactor standoff of 3 km, 15 weeks ($\phi = \pi$) to 16 weeks ($\phi = 0$) at 4 km, and 52 weeks ($\phi=\pi$) to 60 weeks ($\phi=0$) at 5 km. Improvements in direction reconstruction could accelerate detection for $\phi > 0$. The next section reviews various directionality improvements to IBD detectors.

\section*{Improvements}
\subsection*{Near-Term Improvements}
Double Chooz has demonstrated the sensitivity of gadolinium-doped liquid scintillator detectors to IBD directionality. There are, however, alternative detector compositions that should offer an improvement in directional sensitivity.

Lithium-6 can substitute for gadolinium as a neutron capturing dopant to improve the accuracy of reconstruction of the neutron capture vertex. $^6$Li emits a $\sim$2.73 MeV triton and a $\sim$2.05 MeV alpha particle, whose subsequent light yields originate significantly closer to the neutron capture site in the detector medium than gadolinium's. These more massive capture products do, however, reduce the total visible energy. 

Water-based liquid scintillator, when coupled with LAPPDs instead of photomultiplier tubes, presents another avenue for improved direction reconstruction. Water offers lower light attenuation than mineral oil, while improved timing resolution enables independent reconstruction of positron Cherenkov cones \cite{wbls}.

Segmented detectors provide significantly better directional reconstruction than their monolithic counterparts by enabling precise localization of each IBD vertex to within a particular detector segment. Timing information can further localize the interaction site. Two segmented geometries have been successfully demonstrated: bundle designs (e.g. PROSPECT, PANDA, and Palo Verde), and lattice designs (e.g. NuLat, LENS, and CHANDLER).

\subsection*{Potential Future Directions}
Next-generation design proposals may significantly improve antineutrino direction detection in future experiments. Each of the following proposals suggests the possibility of event-by-event directional reconstruction---a vast improvement over statistical pointing.

SANTA (Segmented Antineutrino Tomography Apparatus) consists of two cube-segmented planes, separated by a substantial gap space, with a thin target plane in the center of the gap \cite{PhysRevLett.114.071802}. Such a detector would enable directional reconstruction on an event-by-event basis, but at the costs of dramatically reduced event rate due to its low fiducial volume, and of narrowed angular acceptance limited by the off-axis dimensions of the planes, and their separation.

In hopes of overcoming these limitations of SANTA, John Learned at the University of Hawaii has suggested a `hybrid' detector concept aiming to combine the design principles of SANTA with a bundled-segmented geometry. This tantalizing concept requires further study to determine its viability.

Another design proposal aiming to provide event-by-event directionality, and sub-cm spatial resolution, is the Hydrogenous TPC \cite{1748-0221-9-07-P07002}. This design employs an organic liquid target under an electric field to drift electrons up to a noble gas layer at the surface, terminating at an anode grid. It appears to be a very challenging detector technology to realize, and requires further study.

Each of these forward-looking proposals present unique technical challenges and design limitations, but taken together, hint at the possibility of event-by-event directional reconstruction in antineutrino experiments and applications.

\section*{Conclusions}
Antineutrino direction reconstruction present applications in a variety of fields. We have analyzed the performance of a Double Chooz-like technique applied to a problem in nuclear nonproliferation. Lithium-6 doping, water-based liquid scintillator with LAPPD sensors, and segmented geometries offer improved direction sensitivity, suggesting further improvements to this and comparable applications. Looking to the future, novel design concepts suggest the possibility of event-by-event direction reconstruction, meriting further study, and hinting at significant advances yet to come in direction-sensitive applications.

\matmethods{
We have developed an analytical model describing the positron and neutron vertex distribution and vertex reconstruction error, the derivation of which will appear in a forthcoming publication led by this author. Geant4 \cite{AGOSTINELLI2003250} validates this model for the case of a single reactor and perfect vertex resolution.

A hypothesis test was applied to various two-reactor scenarios in our analytical model, given a null hypothesis modeling the one known reactor. Systematic uncertainties were considered as given in Table \ref{danielson:Table1}.
\begin{table}[H]%[H] add [H] placement to break table across pages
		\caption{Systematic Uncertainties in Expected IBD Event Count}
			\begin{tabular}{r l}
			 	     source & systematic uncertainty\\
				\midrule
				thermal power & $2\%$ \\
				IBD cross section & $0.5\%$ \cite{ibd} \\
		        reactor neutrino anomaly & $2\%$  \\
		        oscillation parameters & See reference \cite{CAPOZZI2016218} \\
		        		       \bottomrule
			\end{tabular} \label{danielson:Table1}
	\end{table}
}
% \showmatmethods{} % Display the Materials and Methods section
\acknow{LLNL-PROC-789112. This work was performed under the auspices of the U.S. Department of Energy by Lawrence Livermore National Laboratory under Contract DE-AC52-07NA27344.}
\showacknow{} % Display the acknowledgments section

% Bibliography
% \bibliography{aap2018-participant00x}

% \end{document}

%% file: AAP-JohnLearnedTalk/talk.tex
\title{Truthiness and Neutrinos;
A Discussion of scientific truth in relation to neutrinos and their applications}
\author{John Gregory Learned}
\affil{University of Hawaii}
\leadauthor{Learned} 

\correspondingauthor{John Learned E-mail: jgl@phys.hawaii.edu}
\maketitle
\noindent
{\it Transcript of the lecture presented at the conference dinner.}
\newline
\newline

Truthiness was quixotically invented by Stephen Colbert more than a dozen years ago to refer to notions that seemed true, independent of "facts".  

We have witnessed in our country now a blurring of the lines between what most consider to be facts, and wishes, or even lies.  This blurring of the lines between demonstrable facts and irresponsible statements, including the claims of 'fake news' is however not unique to the present USA, neither in time nor space.

Worldwide reactionary movements are chipping at rolling-back the basis of our modern technology and indeed at our social interactions. Perhaps much of this is driven by the accelerating pace of technology and consequent rapid changes in lifestyles (e.g. the dissolution of the elemental family).

A fair amount of the current anti-science sentiment in fact traces back to the grand days of social upheaval in the late 1960's, with mind altering chemicals flowing freely, and the youth questioning all.

Particularly in academic circles, people in such disciplines as social anthropology began to suspect that their field rested on shaky ground, maybe all nonsense.  The foggy French philosophes piled on with questions about the meaning of history as written, literature criticism, and of other intellectual endeavors.  This led to the post-modernist era, and in the extreme, calling into question all of science as simply a social construct.

To most physicists the idea that the "laws" of physics might represent mere social constructs of our language and ways-of-thinking, seems patently absurd, a statement I would not expect to be challenged here tonight.

But I think we probably all can agree that the emphasis and order of matters studied has everything to do with our culture. This is most exemplified by the application of science to war.  As well as those ideas studied, those topics forbidden or not considered has all to do with culture and religion. (Example: whether the earth orbits the sun or vice versa).

All that said, and upon which it seems safe to presume that here amongst practicing (faithful?) physicists not much objection will be heard.
Yet, given the national turmoil it may be worthwhile to think a bit about our philosophical foundations.  For example, just what is "truth", and how do we know it when we see it.  And how does science differ from religion, both being in some ways "matters of faith".

\section*{Truth}
Nobody here wants to hear a recitation about the history of philosophy, but let us dabble a bit.  Please bear with me for a little while, as I recall to you a few things about those philosophers you read in school and have long forgotten (as have I).

Wikipedia (do I have to say, I love Wikepedia for finding references?) says: "Some philosophers view the concept of truth as basic, and unable to be explained in any terms that are more easily understood than the concept of truth itself. Commonly, truth is viewed as the correspondence of language or thought to an independent reality, in what is sometimes called the correspondence theory of truth. "

Much of philosophical discussion hinges on the supposed binary nature of truth, true or false. In our modern quantum age we now understand that there is at least another possibility, simply that a particular truth is not incontrovertibly either true or false... as with Schrodinger’s cat prior to opening the box.  So much for many philosophical arguments….

How do we know things? Epistemology... not to scare you.  Many would say that all we know, we take from our senses.  Yet there is a little problem in that some things we "know" are internal, such as some mathematical ideas internally generated, and independent of the environs (or can they really be so?).

Some question our sense of being “real” and ability to make choices (ontology).  For example, do we live in a simulation?  I find this line of thought unproductive and dispiriting.  We are here now and might as well play the game and enjoy what we can, leaving a trail of being good to others.

For most of us the trail of truth is followed by the scientific method.  This contrasts with the trail of religion, wherein a truth is accepted ab initio, unchallenged and not susceptible to revision.

\section*{The Scientific Method}
There is much nonsense written about the so-called "Scientific Method" (recall Karl Popper). Most of us heard in school that there are six or seven steps: 
\begin{description}[font=$\bullet$\scshape\bfseries]
    \item Make an observation
    \item Formulate a Question
    \item Form a hypothesis
    \item Test hypothesis  (something missing: build the apparatus, upon which many of us spend most of our time, jgl).
    \item Record data (and analyze same) 
    \item Draw Conclusions
    \item Replicate
    \item Communicate results (often neglected in such lists) (peer review)
\end{description}
% - Make an observation
% - Formulate a Question 
% - Form a hypothesis
% - Test hypothesis  (something missing: build the apparatus, upon which many of us spend most of our time, jgl).
% - Record data (and analyze same) 
% - Draw Conclusions
% - Replicate
% - Communicate results (often neglected in such lists) (peer review)

As we practitioners all know, matters in science are seldom so linear...

Take for a further example of foolish academic punditry, the famous Thomas Kuhn, noted author of "The structure of scientific revolutions"…

“Thomas Kuhn said that the scientist generally has a theory in mind before designing and undertaking experiments so as to make empirical observations, and that the "route from theory to measurement can almost never be traveled backward". This implies that the way in which theory is tested is dictated by the nature of the theory itself, which led Kuhn (1961, p. 166) to argue that ‘once it has been adopted by a profession ... no theory is recognized to be testable by any quantitative tests that it has not already passed’.” (Wikipedia: Scientific Method)  This constitutes, I hope we can all agree, nonsense.  Think of Quantum Mechanics...

On the other hand Paul Feyerabend similarly examined the history of science, and was led to deny that science is genuinely a methodological process. In his book Against Method he argues that scientific progress is not the result of applying any particular method. In essence, he says that “for any specific method or norm of science, one can find a historic episode where violating it has contributed to the progress of science.”

Thus, if believers in scientific method wish to express a single universally valid rule, Feyerabend jokingly suggests, it should be 'anything goes'. (By which we mean anything which passes the test or replicability and such.)

However, criticisms such as this led to the strong programme, a radical approach to the sociology of science (“radical relativism”).

Postmodernists assert that scientific knowledge is simply another discourse (note that this term has special meaning in this context) and not representative of any form of fundamental truth; realists in the scientific community maintain that scientific knowledge does reveal real and fundamental truths about reality.

We all recall the use of parody to make fun of some of the fancy Post Moderns... take for example the wonderful article of physics theorist Alan Sokol of NYU in 1996. "Transgressing the Boundaries: Towards a Transformative Hermeneutics of Quantum Gravity", was published in the Social Text spring/summer 1996 "Science Wars" issue. It proposed that quantum gravity is a social and linguistic construct.\footnote{https://en.wikipedia.org/wiki/Sokal\_affair\#cite\_note-3}

{\it He said "The results of my little experiment demonstrate, at the very least, that some fashionable sectors of the American academic Left have been getting intellectually lazy. The editors of Social Text liked my article because they liked its conclusion: that "the content and methodology of postmodern science provide powerful intellectual support for the progressive political project" [sec. 6]. They apparently felt no need to analyze the quality of the evidence, the cogency of the arguments, or even the relevance of the arguments to the purported conclusion" \footnote{https://en.wikipedia.org/wiki/Sokal\_affair\#cite\_note-9}.}

In sum, for them the results were truthy enough!

BTW, Jacques Derrida (Delouze, Foucault...) and their deconstructionist friends have lost much luster... gone but certainly not forgotten from academe, particularly in many English Literature Departments.  Florida law professor Stanley Fish who was much embarrassed by the Sokol affair, continues to write foolish blather in the NYT on topics in science, religion and philosophy, and is taken seriously. (Yet I have read some of his good ideas about changing people's minds.)

And BTW, this sort of controversy is not gone and forgotten.... you may have noted in the last week that three scholars gulled 7 academic journals into publishing hoaxed papers on "grievance studies" (WSJ, 10/5/18).  (Whatever Grievance Studies are?) As the Harvard psychologist Steven Pinker tweeted, “Is there any idea so outlandish that it won’t be published in a Critical/PoMo/Identity/‘Theory’ journal?”  (NYT 10/01/18)

Lest we physicists become too smug, the NYT also noted that "With stories like this in the news, it’s hardly a surprise that according to a recent Pew poll, political party affiliation predicts whether one believes universities are having a positive or a negative effect on the country."

To me it is very frightening that much of our country and much of the world no longer (if indeed ever it did) holds up science and indeed education as lighting the way to a better future.  
Dark ages here we come, maybe?

Some interesting topics I wanted to broach, but skip for lack of time:
\begin{description}[font=$-$\scshape\bfseries]
\item	Relation with mathematics, and just what is Math and how does it relate to truth?
\item	Statistics and physics… what is statistical truth?
\end{description}

\section*{Junk Science and Pathological Science}
There is much confusion in the public sector about what really constitutes real and trustworthy Science.  Junk Science, often what we would call real science, but as designated by truthy folks (such as Global Warming). And we must confront pseudoscience, such as was often produced by tobacco and sugar companies.  We all need to be out there explaining science and debunking nonsense at every opportunity.  But we cannot stand on claims of authority.  We need to educate, charm and even occasionally directly confront.

An issue other than simple denial, wishful thinking, or outright cheating relates to Pathological Science (discussed by Irving Langmuir, Chemistry Nobel)...  (This is different from simple scientific misconduct, fraud, and more dangerous.)

{\it The maximum effect that is observed is produced by a causative agent of 
\begin{description}[font=$\bullet$\scshape\bfseries]
    \item  Barely detectable intensity, and the magnitude of the effect is substantially 
independent of the intensity of the cause
 \item  The effect is of a magnitude that remains close to the limit of detectability 
or, many measurements are necessary because of very low statistical 
significance of the results (ahem)
 \item  There are claims of great accuracy
 \item  Fantastic theories contrary to experience are suggested
 \item  Criticisms are met by ad hoc excuses thought up on the spur of the moment
(And often secrecy about sharing raw data and permitting external investigations... jgl)
 \item  The ratio of supporters to critics rises up to somewhere near 50\% and 
then falls gradually to oblivion
\end{description}
}
(Think of N-Rays by Blondhot, and Cold Fusion by Pons and Fleischman, but also the foundations of gravity wave research by Joe Weber and his wrong claim of detecting G waves). (Some claims of the discovery of Dark Matter may fit this description, BTW.)

Comment: The gravity-wave case is an example of how, not fake, but simply wrong science (by Joe Weber about his incorrect observation of signals) can lead to interesting and even revolutionary results (the now successful campaign to detect waves from gravitational inspirals). Some of our current topics, such as the hints at sterile neutrinos and their interpretation by enthusiasts may well be in that category, despite honesty and good will by all involved....  We presume the scientific process will prevail however.

[ Interesting related subject: Should one read the old works?  In most disciplines it is a waste of time! (Except for the historical stories, and noting the evolution of ideas.) In physics we can learn all we need to know about GR by reading modern books and papers and never reading old Albert!]

BTW, most of modern science really got started with the formation of the Royal Society in London in late 1660, under royal charter from Charles II.  It was organized by Freemasons, and the key to success was the Masonic tradition of avoiding religion and politics in their deliberations (really?), and under the motto "Nullis in verba" (take nobody’s word).  Note also  that a large fraction of early Royal Society considerations revolved around "the longitude problem"... which brought together everything from celestial mechanics to studies of the vacuum. So social/economic concerns indeed provided motivation. The point of relevance for us is that until then (Renaissance era), religion and ancient Aristotle had dominated much of scientific investigations (e.g. Galileo).

An aside: JGL's theory of the scientific method (not sure if this is unique but I put it out there):
Science on the large scale is a random walk.  Think of Markov processes.  Science proceeds through those who study at least the most recent work, with few delving into the past (this is directed mostly at other disciplines, such as oxymoronic "social science"). Workers on the forefront then flounder about trying all manner of ideas, until one shows promise of working better than others; and then effort is put into the that direction resulting in progress. (This is what I call building the Yellow Brick Road of scientific progress... no map, it just happens).  

An example is SUSY... a great idea of a beautiful symmetry, but one which we now know seems everywhere violated and which has produced nothing of verifiable particle physics import.  Another related example, the efforts of our many friends continuing the search for the WIMP Miracle, Dark Matter particles in the hundred GeV range.  These are also examples of looking under the streetlight, since better, or even heavily competing, theories were/are not on the market, possibly allowing illumination  elsewhere.

Nuclear theory provides another an example... a patchwork of phenomenological attempts to systematize nuclear structure.  But of course we have QCD and particularly lattice QCD, which seems to work: one can carry out the calculations and find a few basic parameters of particle interactions, nuclei and quark groupings.  We can in principle calculate the phenomenological laws which have been deduced from nuclear data, but these are generally too complex to accomplish.

Moreover, nuclear theory is perhaps just the next step on the complexity ladder... Consider atomic physics.  We think we know practically all about the quantum mechanics of atoms, but elements above the first few are simply too complex for much calculation without significant approximations.

More interesting is that we nowadays know about emergent complexity.  It is simply true that we would perhaps never get from atomic level QM to DNA and the astounding phenomena exhibited by cells.  New structure and unprecedented behavior emerge as one goes up the complexity ladder. Every step of going to a more complex system we find (mostly) unpredicted emergent and often amazing phenomenon.  Hence Science must nearly independently investigate the way the universe works at many levels.

In sum, I submit to you that science largely proceeds by fumbling about, trying all possible alternatives until something fits the emerging facts.  Brilliant leaps are made surely, such as old Albert's greatest idea of the equivalence principle (for which he did not immediately understand the consequences either).

I found some support for this from Bas van Fraassen who says ``I claim that the success of current scientific theories is no miracle. It is not even surprising to the scientific (Darwinist) mind. For any scientific theory is born into a life of fierce competition, a jungle red in tooth and claw. Only the successful theories survive—the ones which in fact latched on to actual regularities in nature.'' (The Scientific Image, 1980).  (It would seem that the demise of SUSY represents such a casualty of such!)

This view I think makes great sense, that there is a sort of Darwinian evolution in scientific ideas as well as species! How we sample the ``landscape'' remains an interesting discussion. Maybe we observe an evolution of memes, perhaps some good and some bad.

\section*{Neutrinos, Truth and Applications}
I am supposed to say something about neutrinos tonight, so better get at it... What set me off on this tour was the notion that in our modern society where truthy folk question many more or less well established scientific truths.  I have heard from some contrarians that "you scientists have beliefs too, not just us religious folks".  Let me dispense with that: our beliefs are always tentative and conditioned on continued consistency with nature.  Most strongly, the importance of replicability to science most clearly distinguishes what we believe from religion, wherein you take it or leave it, and most predictions, if any, fail.  I hope this does not insult or upset anyone, but such is my position and though not often articulated, that of most scientists.

I have encountered often enough people who say, "neutrinos, bah humbug, just some nonsense to cover up inadequacies of your physics", or the like.  Now we all here know very well we can make neutrinos at an accelerator at a given precise time, and predictably see some interactions appear, at a time consistent with the speed of light, several hundred miles away. (And, those interactions pointing in the right direction).
In fact, as most of you recall, an experiment (OPERA) in Italy found events in 2011 which apparently travelled faster than light from CERN in France, some thousand miles away.  This set off a huge reconsideration of theory, and remarkably no credible theory did emerge to explain how neutrinos could be superluminal without implying huge contradictions to many areas (such as lack of causality!). Everyone relaxed when the experimenters did indeed find a faulty fiber-optic connection in their apparatus and the scientific process did demonstrably work well.

(A slightly worrisome issue is the tendency to keep looking over data if a contrary or divergent answer is initially found, and then to stop when consistency with other results is discovered. CERN collider experiments often show more consistency than expected by random fluctuations…).

So, as we all know, neutrinos started as Pauli's hypothesis to save conservation of energy (and spin) and were a considerable embarrassment to the theorists who deemed them to be undetectable. (This modesty seems to have dissipated in recent times, I note... witness modern string theory).  
Clyde Cowan and Fred Reines and company in the 1950's observed neutrinos, little neutral siblings to electrons, emanating from reactors. Then suspicions about neutral siblings to the muon and later to the tau were similarly borne out.  Note that these all pass any reasonable test of being "real".  We can and do count them… we predict and they appear (albeit infrequently)!   (And even so some of their properties remain mysterious, as much discussed at this and other meetings).

But, are neutrinos good for anything?  Indeed they are as we have heard and will hear more at this meeting tomorrow. First, neutrinos explain much about the functioning of stars and supernovae, and even the remnant light element mix from the Big Bang! Of more immediate import as we have heard much here, since nuclear reactors are inescapably un-shieldable sources of stupendous numbers of (electron anti-) neutrinos (10$^{21}$ /sec from a typical power reactor), we can peer inside the reactor's furnace to determine what and how much is burning.  This constitutes x-raying without the need for an x-ray source tube.  Same story for both the earth and the sun!  We know the sun is still burning, despite the fact that photons take eons to randomly diffuse out.  We can also (and no other way) learn about the fuel for driving the Earth's convection... the generative cause of continental motion, earthquakes and volcanoes. People have also talked of using neutrinos for the fastest possible signaling for stock market trading. (BTW, I have actually been approached to work on this, and have rejected the offer).

Neutrinos are as real as your wine and like many abstruse findings in fundamental physics, they are now, 50 years after discovery, being put to harness.  We here are lucky to get to participate in these adventures, which without fundamental guidance from theory are truly exploring new realms of the universe!

I conclude with a joke, which is said to have practically drawn tears of laughter from Einstein and Oppenheimer. One neutrino asks the other neutrino, weaving about in space: “Can’t you move straight? You must be drunk again!” The other neutrino protests vehemently: ``What do you expect? Can’t you see that I am getting soaked in a gravitational field?'' (From {brainpickings.org/2014/04/16/random-walk-in-science-humor}). [{\it The point here is simply that we in science think quite differently from most, and even that is a function of space and time. This is a joke that tickled old Albert and Oppy, but unsurprisingly did not resonate with our dinner group of physicists, and probably would not make sense to others! I fear my point was too obscure… that our scientist’s arcane understanding of the world does not speak to our society generally, and illustrating that we must work at communicating to the larger population.}]

\section*{Conclusion} 

A more serious ending calls to your attention that we scientists, physicists and neutrino aficionados in particular, need to pay attention to the fact of being under attack in present USA… attack from the truthiness right, the foolish PoMo left, and even (not discussed), the ignorant indecisive middle.  Civilization and intellectual progress have ever been a thin container of burgeoning chaos. We have the privilege to study the enigmatic neutrino and help not only with the grand game of puzzling out the universe, but also make contributions to important matters such as arms control, and even disparate scientific studies in understanding the sun and earth.

{\it Post Presentation Note: This represents the draft of the talk but not exactly as presented.  I had only quick glances at the notes, and inserted some levity on the fly, and skipped some as written above, not captured here. The text is perhaps quite somber, and has a most serious intent, but the presentation involved a lot of laughter as well.}

%% file: main.bbl
\begin{thebibliography}{193}
\providecommand{\natexlab}[1]{#1}
\providecommand{\url}[1]{\texttt{#1}}
\expandafter\ifx\csname urlstyle\endcsname\relax
  \providecommand{\doi}[1]{doi: #1}\else
  \providecommand{\doi}{doi: \begingroup \urlstyle{rm}\Url}\fi

\bibitem[Mueller et~al.(2011)]{Mueller:2011nm}
Th.~A. Mueller et~al.
\newblock {Improved Predictions of Reactor Antineutrino Spectra}.
\newblock \emph{Phys. Rev.}, C83:\penalty0 054615, 2011.
\newblock \doi{10.1103/PhysRevC.83.054615}.

\bibitem[Huber(2011)]{Huber:2011wv}
Patrick Huber.
\newblock {On the determination of anti-neutrino spectra from nuclear
  reactors}.
\newblock \emph{Phys. Rev.}, C84:\penalty0 024617, 2011.
\newblock \doi{10.1103/PhysRevC.85.029901, 10.1103/PhysRevC.84.024617}.
\newblock [Erratum: Phys. Rev.C85,029901(2012)].

\bibitem[Mention et~al.(2011)Mention, Fechner, Lasserre, Mueller, Lhuillier,
  Cribier, and Letourneau]{Mention:2011rk}
G.~Mention, M.~Fechner, Th. Lasserre, Th.~A. Mueller, D.~Lhuillier, M.~Cribier,
  and A.~Letourneau.
\newblock {The Reactor Antineutrino Anomaly}.
\newblock \emph{Phys. Rev.}, D83:\penalty0 073006, 2011.
\newblock \doi{10.1103/PhysRevD.83.073006}.

\bibitem[Kopp et~al.(2013)Kopp, Machado, Maltoni, and Schwetz]{Kopp:2013vaa}
Joachim Kopp, Pedro A.~N. Machado, Michele Maltoni, and Thomas Schwetz.
\newblock {Sterile Neutrino Oscillations: The Global Picture}.
\newblock \emph{JHEP}, 1305:\penalty0 050, 2013.
\newblock \doi{10.1007/JHEP05(2013)050}.

\bibitem[An et~al.(2016)]{An:2015nua}
Feng~Peng An et~al.
\newblock {Measurement of the Reactor Antineutrino Flux and Spectrum at Daya
  Bay}.
\newblock \emph{Phys. Rev. Lett.}, 116\penalty0 (6):\penalty0 061801, 2016.
\newblock \doi{10.1103/PhysRevLett.116.061801, 10.1103/PhysRevLett.118.099902}.
\newblock [Erratum: Phys. Rev. Lett.118,no.9,099902(2017)].

\bibitem[Abe et~al.(2014)]{Abe:2014bwa}
Y.~Abe et~al.
\newblock {Improved measurements of the neutrino mixing angle $\theta_{13}$
  with the Double Chooz detector}.
\newblock \emph{JHEP}, 1410:\penalty0 086, 2014.
\newblock \doi{10.1007/JHEP1502(2015)074, 10.1007/JHEP10(2014)086}.
\newblock [Erratum: JHEP1502,074(2015)].

\bibitem[Seo et~al.(2018)]{Seo:2016uom}
S.~H. Seo et~al.
\newblock {Spectral Measurement of the Electron Antineutrino Oscillation
  Amplitude and Frequency using 500 Live Days of RENO Data}.
\newblock \emph{Phys. Rev.}, D98\penalty0 (1):\penalty0 012002, 2018.
\newblock \doi{10.1103/PhysRevD.98.012002}.

\bibitem[Ashenfelter et~al.(2016)]{Ashenfelter:2015uxt}
J.~Ashenfelter et~al.
\newblock {The PROSPECT Physics Program}.
\newblock \emph{J. Phys.}, G43\penalty0 (11):\penalty0 113001, 2016.
\newblock \doi{10.1088/0954-3899/43/11/113001}.

\bibitem[HFI()]{HFIR_model}
Modeling and depletion simulations for a high flux isotope reactor cycle with a
  representative experiment loading
  https://info.ornl.gov/sites/publications/files/pub60920.pdf.
\newblock Technical report.

\bibitem[Ashenfelter et~al.(2018{\natexlab{a}})]{Ashenfelter:2018zdm}
J.~Ashenfelter et~al.
\newblock {The PROSPECT Reactor Antineutrino Experiment}.
\newblock 2018{\natexlab{a}}.

\bibitem[Bernstein et~al.(2010)Bernstein, Baldwin, Boyer, Goodman, Learned,
  Lund, Reyna, and Svoboda]{Bernstein:2009ab}
A.~Bernstein, G.~Baldwin, B.~Boyer, M.~Goodman, J.~Learned, J.~Lund, D.~Reyna,
  and R.~Svoboda.
\newblock {Nuclear Security Applications of Antineutrino Detectors: Current
  Capabilities and Future Prospects}.
\newblock \emph{Sci. Global Secur.}, 18:\penalty0 127--192, 2010.
\newblock \doi{10.1080/08929882.2010.529785}.

\bibitem[Bernstein et~al.(2002)Bernstein, Wang, Gratta, and
  West]{Bernstein:2001cz}
Adam Bernstein, Yi-fang Wang, Giorgio Gratta, and Todd West.
\newblock {Nuclear reactor safeguards and monitoring with anti-neutrino
  detectors}.
\newblock \emph{J. Appl. Phys.}, 91:\penalty0 4672, 2002.
\newblock \doi{10.1063/1.1452775}.

\bibitem[Carr et~al.(2018{\natexlab{a}})]{Carr:2018tak}
Rachel Carr et~al.
\newblock {Neutrino-based tools for nuclear verification and diplomacy in North
  Korea}.
\newblock 2018{\natexlab{a}}.

\bibitem[Ashenfelter et~al.(2018{\natexlab{b}})]{Ashenfelter:2018iov}
J.~Ashenfelter et~al.
\newblock {First search for short-baseline neutrino oscillations at HFIR with
  PROSPECT}.
\newblock 2018{\natexlab{b}}.

\bibitem[Abreu et~al.({\natexlab{a}})Abreu, Amhis, Arnold, Ban, Beaumont,
  Bongrand, Boursette, Buhour, Castle, Clark, Coupé, Cucoanes, Cussans, Roeck,
  D'Hondt, Durand, Fallot, Fresneau, Ghys, Giot, Guillon, Guilloux, Ihantola,
  Janssen, Kalcheva, Kalousis, Koonen, Labare, Lehaut, Mermans, Michiels,
  Moortgat, Newbold, Park, Petridis, Piñera, Pommery, Popescu, Pronost,
  Rademacker, Reynolds, Ryckbosch, Ryder, Saunders, Shitov, Schune, Scovell,
  Simard, Vacheret, Dyck, Mulders, van Remortel, Vercaemer, Waldron, Weber, and
  Yermia]{JINST_12_P04024}
Y.~Abreu, Y.~Amhis, L.~Arnold, G.~Ban, W.~Beaumont, M.~Bongrand, D.~Boursette,
  J.M. Buhour, B.C. Castle, K.~Clark, B.~Coupé, A.S. Cucoanes, D.~Cussans,
  A.~De Roeck, J.~D'Hondt, D.~Durand, M.~Fallot, S.~Fresneau, L.~Ghys, L.~Giot,
  B.~Guillon, G.~Guilloux, S.~Ihantola, X.~Janssen, S.~Kalcheva, L.N. Kalousis,
  E.~Koonen, M.~Labare, G.~Lehaut, J.~Mermans, I.~Michiels, C.~Moortgat,
  D.~Newbold, J.~Park, K.~Petridis, I.~Piñera, G.~Pommery, L.~Popescu,
  G.~Pronost, J.~Rademacker, A.~Reynolds, D.~Ryckbosch, N.~Ryder, D.~Saunders,
  Yu.A. Shitov, M.-H. Schune, P.R. Scovell, L.~Simard, A.~Vacheret, S.~Van
  Dyck, P.~Van Mulders, N.~van Remortel, S.~Vercaemer, A.~Waldron, A.~Weber,
  and F.~Yermia.
\newblock A novel segmented-scintillator antineutrino detector.
\newblock \emph{Journal of Instrumentation}, 12\penalty0 (04):\penalty0 P04024,
  {\natexlab{a}}.

\bibitem[Huber(2017)]{Huber2016xis}
Patrick Huber.
\newblock Neos data and the origin of the 5 mev bump in the reactor
  antineutrino spectrum.
\newblock \emph{Phys. Rev. Lett.}, 118\penalty0 (4):\penalty0 042502, 2017.
\newblock \doi{10.1103/PhysRevLett.118.042502}.

\bibitem[Dentler et~al.(2017{\natexlab{a}})Dentler, Hern{\'a}ndez-Cabezudo,
  Kopp, Maltoni, and Schwetz]{Dentler2017}
Mona Dentler, {\'A}lvaro Hern{\'a}ndez-Cabezudo, Joachim Kopp, Michele Maltoni,
  and Thomas Schwetz.
\newblock Sterile neutrinos or flux uncertainties? --- status of the reactor
  anti-neutrino anomaly.
\newblock \emph{Journal of High Energy Physics}, 2017\penalty0 (11):\penalty0
  99, Nov 2017{\natexlab{a}}.
\newblock ISSN 1029-8479.
\newblock \doi{10.1007/JHEP11(2017)099}.
\newblock URL \url{https://doi.org/10.1007/JHEP11(2017)099}.

\bibitem[Hendricks et~al.(2008)]{MCNPX}
John~S. Hendricks et~al.
\newblock \emph{{MCNPX EXTENSIONS - Version 2.5.0}}.
\newblock Los Alamos National Laboratory, LA-UR-05-2675, 2008.

\bibitem[Fallot et~al.(2012)Fallot, Giot, Kalcheva, and Koonen]{pred}
M.~Fallot, L.~Giot, S.~Kalcheva, and E.~Koonen.
\newblock Br2 reactor coupled mcnpx \& mure simulation for the solid experiment
  (antineutrino detection).
\newblock \emph{PRL 109}, 101504:\penalty0 6, 2012.

\bibitem[Agostinelli et~al.(2003{\natexlab{a}})]{geant4}
S.~Agostinelli et~al.
\newblock Geant4: A simulation toolkit.
\newblock \emph{Nucl. Instrum. Meth.}, A506:\penalty0 250--303,
  2003{\natexlab{a}}.

\bibitem[Abreu et~al.({\natexlab{b}})Abreu, Amhis, Arnold, Ban, Beaumont,
  Bongrand, Boursette, Castle, Clark, Coupé, Cussans, Roeck, D'Hondt, Durand,
  Fallot, Ghys, Giot, Guillon, Ihantola, Janssen, Kalcheva, Kalousis, Koonen,
  Labare, Lehaut, Manzanillas, Mermans, Michiels, Moortgat, Newbold, Park,
  Pestel, Petridis, Piñera, Pommery, Popescu, Pronost, Rademacker, Ryckbosch,
  Ryder, Saunders, Schune, Simard, Vacheret, Dyck, Mulders, van Remortel,
  Vercaemer, Verstraeten, Weber, and Yermia]{JINST_13_P05005}
Y.~Abreu, Y.~Amhis, L.~Arnold, G.~Ban, W.~Beaumont, M.~Bongrand, D.~Boursette,
  B.C. Castle, K.~Clark, B.~Coupé, D.~Cussans, A.~De Roeck, J.~D'Hondt,
  D.~Durand, M.~Fallot, L.~Ghys, L.~Giot, B.~Guillon, S.~Ihantola, X.~Janssen,
  S.~Kalcheva, L.~N. Kalousis, E.~Koonen, M.~Labare, G.~Lehaut, L.~Manzanillas,
  J.~Mermans, I.~Michiels, C.~Moortgat, D.~Newbold, J.~Park, V.~Pestel,
  K.~Petridis, I.~Piñera, G.~Pommery, L.~Popescu, G.~Pronost, J.~Rademacker,
  D.~Ryckbosch, N.~Ryder, D.~Saunders, M.-H. Schune, L.~Simard, A.~Vacheret,
  S.~Van Dyck, P.~Van Mulders, N.~van Remortel, S.~Vercaemer, M.~Verstraeten,
  A.~Weber, and F.~Yermia.
\newblock Performance of a full scale prototype detector at the br2 reactor for
  the solid experiment.
\newblock \emph{Journal of Instrumentation}, 13\penalty0 (05):\penalty0 P05005,
  {\natexlab{b}}.

\bibitem[Abreu et~al.({\natexlab{c}})Abreu, Amhis, Beaumont, Bongrand,
  Boursette, Castle, Clark, Coupé, Cussans, Roeck, Durand, Fallot, Ghys, Giot,
  Graves, Guillon, Henaff, Hosseini, Ihantola, Jenzer, Kalcheva, Kalousis,
  Labare, Lehaut, Manley, Manzanillas, Mermans, Michiels, Moortgat, Newbold,
  Park, Pestel, Petridis, Piñera, Popescu, Ryckbosch, Ryder, Saunders, Schune,
  Settimo, Simard, Vacheret, Vandierendonck, Dyck, Mulders, Remortel,
  Vercaemer, Verstraeten, Viaud, Weber, and Yermia]{JINST_13_P09005}
Y.~Abreu, Y.~Amhis, W.~Beaumont, M.~Bongrand, D.~Boursette, B.~C. Castle,
  K.~Clark, B.~Coupé, D.~Cussans, A.~De Roeck, D.~Durand, M.~Fallot, L.~Ghys,
  L.~Giot, K.~Graves, B.~Guillon, D.~Henaff, B.~Hosseini, S.~Ihantola,
  S.~Jenzer, S.~Kalcheva, L.~N. Kalousis, M.~Labare, G.~Lehaut, S.~Manley,
  L.~Manzanillas, J.~Mermans, I.~Michiels, C.~Moortgat, D.~Newbold, J.~Park,
  V.~Pestel, K.~Petridis, I.~Piñera, L.~Popescu, D.~Ryckbosch, N.~Ryder,
  D.~Saunders, M.-H. Schune, M.~Settimo, L.~Simard, A.~Vacheret,
  G.~Vandierendonck, S.~Van Dyck, P.~Van Mulders, N.~Van Remortel,
  S.~Vercaemer, M.~Verstraeten, B.~Viaud, A.~Weber, and F.~Yermia.
\newblock Optimisation of the scintillation light collection and uniformity for
  the solid experiment.
\newblock \emph{Journal of Instrumentation}, 13\penalty0 (09):\penalty0 P09005,
  {\natexlab{c}}.

\bibitem[Abreu et~al.(2018{\natexlab{a}})Abreu, Amhis, Beaumont, Bongrand,
  Boursette, Castle, Clark, Coupé, Cussans, Roeck, Durand, Fallot, Ghys, Giot,
  Graves, Guillon, Henaff, Hosseini, Ihantola, Jenzer, Kalcheva, Kalousis,
  Labare, Lehaut, Manley, Manzanillas, Mermans, Michiels, Moortgat, Newbold,
  Park, Pestel, Petridis, Piñera, Popescu, Ryckbosch, Ryder, Saunders, Schune,
  Settimo, Simard, Vacheret, Vandierendonck, Dyck, Mulders, Remortel,
  Vercaemer, Verstraeten, Viaud, Weber, and Yermia]{RO}
Y.~Abreu, Y.~Amhis, W.~Beaumont, M.~Bongrand, D.~Boursette, B.~C. Castle,
  K.~Clark, B.~Coupé, D.~Cussans, A.~De Roeck, D.~Durand, M.~Fallot, L.~Ghys,
  L.~Giot, K.~Graves, B.~Guillon, D.~Henaff, B.~Hosseini, S.~Ihantola,
  S.~Jenzer, S.~Kalcheva, L.~N. Kalousis, M.~Labare, G.~Lehaut, S.~Manley,
  L.~Manzanillas, J.~Mermans, I.~Michiels, C.~Moortgat, D.~Newbold, J.~Park,
  V.~Pestel, K.~Petridis, I.~Piñera, L.~Popescu, D.~Ryckbosch, N.~Ryder,
  D.~Saunders, M.-H. Schune, M.~Settimo, L.~Simard, A.~Vacheret,
  G.~Vandierendonck, S.~Van Dyck, P.~Van Mulders, N.~Van Remortel,
  S.~Vercaemer, M.~Verstraeten, B.~Viaud, A.~Weber, and F.~Yermia.
\newblock Deployment of the readout system for the solid neutrino detector.
\newblock \emph{NSS/MIC}, submitted, 2018{\natexlab{a}}.

\bibitem[Abreu et~al.(2018{\natexlab{b}})Abreu, Amhis, Beaumont, Bongrand,
  Boursette, Castle, Clark, Coupé, Cussans, Roeck, Durand, Fallot, Ghys, Giot,
  Graves, Guillon, Henaff, Hosseini, Ihantola, Jenzer, Kalcheva, Kalousis,
  Labare, Lehaut, Manley, Manzanillas, Mermans, Michiels, Moortgat, Newbold,
  Park, Pestel, Petridis, Piñera, Popescu, Ryckbosch, Ryder, Saunders, Schune,
  Settimo, Simard, Vacheret, Vandierendonck, Dyck, Mulders, Remortel,
  Vercaemer, Verstraeten, Viaud, Weber, and Yermia]{QA}
Y.~Abreu, Y.~Amhis, W.~Beaumont, M.~Bongrand, D.~Boursette, B.~C. Castle,
  K.~Clark, B.~Coupé, D.~Cussans, A.~De Roeck, D.~Durand, M.~Fallot, L.~Ghys,
  L.~Giot, K.~Graves, B.~Guillon, D.~Henaff, B.~Hosseini, S.~Ihantola,
  S.~Jenzer, S.~Kalcheva, L.~N. Kalousis, M.~Labare, G.~Lehaut, S.~Manley,
  L.~Manzanillas, J.~Mermans, I.~Michiels, C.~Moortgat, D.~Newbold, J.~Park,
  V.~Pestel, K.~Petridis, I.~Piñera, L.~Popescu, D.~Ryckbosch, N.~Ryder,
  D.~Saunders, M.-H. Schune, M.~Settimo, L.~Simard, A.~Vacheret,
  G.~Vandierendonck, S.~Van Dyck, P.~Van Mulders, N.~Van Remortel,
  S.~Vercaemer, M.~Verstraeten, B.~Viaud, A.~Weber, and F.~Yermia.
\newblock Development of a quality assurance process for the phase 1 of the
  solid experiment,.
\newblock 2018{\natexlab{b}}.

\bibitem[Technology({\natexlab{a}})]{EJ-260}
Eljen Technology.
\newblock {EJ-260}.
\newblock
  \url{https://eljentechnology.com/products/plastic-scintillators/ej-260-ej-262},
  {\natexlab{a}}.
\newblock Accessed: 2018-11-12.

\bibitem[Grieb et~al.(2007)Grieb, Link, and Raghavan]{PhysRevD.75.093006}
C.~Grieb, J.~M. Link, and R.~S. Raghavan.
\newblock Probing active to sterile neutrino oscillations in the lens detector.
\newblock \emph{Phys. Rev. D}, 75:\penalty0 093006, May 2007.
\newblock \doi{10.1103/PhysRevD.75.093006}.
\newblock URL \url{https://link.aps.org/doi/10.1103/PhysRevD.75.093006}.

\bibitem[Technology({\natexlab{b}})]{EJ-426}
Eljen Technology.
\newblock {EJ-426}.
\newblock \url{https://eljentechnology.com/products/neutron-detectors/ej-426},
  {\natexlab{b}}.
\newblock Accessed: 2018-11-12.

\bibitem[Stewart and Erickson(2015)]{Christopher}
Christopher Stewart and Anna Erickson.
\newblock {Antineutrino analysis for continuous monitoring of nuclear reactors:
  Sensitivity study}.
\newblock \emph{J. Appl. Phys.}, 118:\penalty0 164902, 2015.
\newblock \doi{10.1063/1.4934638}.

\bibitem[Ko et~al.(2017)]{Ko:2016owz}
Y.J. Ko et~al.
\newblock {Sterile Neutrino Search at the NEOS Experiment}.
\newblock \emph{Phys. Rev. Lett.}, 118\penalty0 (12):\penalty0 121802, 2017.
\newblock \doi{10.1103/PhysRevLett.118.121802}.

\bibitem[Kim et~al.(2015)]{Kim:2015pba}
B.~R. Kim et~al.
\newblock {Pulse shape discrimination capability of metal-loaded organic liquid
  scintillators for a short-baseline reactor neutrino experiment}.
\newblock \emph{Phys. Scripta}, 90\penalty0 (5):\penalty0 055302, 2015.
\newblock \doi{10.1088/0031-8949/90/5/055302}.

\bibitem[Agostinelli et~al.(2003{\natexlab{b}})]{Agostinelli:2002hh}
S.~Agostinelli et~al.
\newblock {GEANT4: A Simulation toolkit}.
\newblock \emph{Nucl. Instrum. Meth.}, A506:\penalty0 250--303,
  2003{\natexlab{b}}.
\newblock \doi{10.1016/S0168-9002(03)01368-8}.

\bibitem[Abdurashitov et~al.(2006)]{Abdurashitov:2005tb}
J.~N. Abdurashitov et~al.
\newblock {Measurement of the response of a Ga solar neutrino experiment to
  neutrinos from an Ar-37 source}.
\newblock \emph{Phys. Rev.}, C73:\penalty0 045805, 2006.
\newblock \doi{10.1103/PhysRevC.73.045805}.

\bibitem[Giunti and Laveder(2011)]{Giunti:2010zu}
Carlo Giunti and Marco Laveder.
\newblock {Statistical Significance of the Gallium Anomaly}.
\newblock \emph{Phys. Rev.}, C83:\penalty0 065504, 2011.
\newblock \doi{10.1103/PhysRevC.83.065504}.

\bibitem[Athanassopoulos et~al.(1996)]{Athanassopoulos:1996jb}
C.~Athanassopoulos et~al.
\newblock {Evidence for anti-muon-neutrino ---> anti-electron-neutrino
  oscillations from the LSND experiment at LAMPF}.
\newblock \emph{Phys. Rev. Lett.}, 77:\penalty0 3082--3085, 1996.
\newblock \doi{10.1103/PhysRevLett.77.3082}.

\bibitem[Aguilar-Arevalo et~al.(2013)]{Aguilar-Arevalo:2013pmq}
A.~A. Aguilar-Arevalo et~al.
\newblock {Improved Search for $\bar \nu_\mu \rightarrow \bar \nu_e$
  Oscillations in the MiniBooNE Experiment}.
\newblock \emph{Phys. Rev. Lett.}, 110:\penalty0 161801, 2013.
\newblock \doi{10.1103/PhysRevLett.110.161801}.

\bibitem[Alekseev et~al.(2018)]{Alekseev:2018ijk}
I.~Alekseev et~al.
\newblock {Measurements of the Reactor Antineutrino with Solid State
  Scintillation Detector}.
\newblock \emph{Int. J. Mod. Phys. Conf. Ser.}, 46:\penalty0 1860044, 2018.
\newblock \doi{10.1142/S2010194518600443}.

\bibitem[Alekseev et~al.(2016)]{Alekseev:2016llm}
I.~Alekseev et~al.
\newblock {DANSS: Detector of the reactor AntiNeutrino based on Solid
  Scintillator}.
\newblock \emph{JINST}, 11\penalty0 (11):\penalty0 P11011, 2016.
\newblock \doi{10.1088/1748-0221/11/11/P11011}.

\bibitem[Gariazzo et~al.(2017)Gariazzo, Giunti, Laveder, and
  Li]{Gariazzo:2017fdh}
S.~Gariazzo, C.~Giunti, M.~Laveder, and Y.~F. Li.
\newblock {Updated Global 3+1 Analysis of Short-BaseLine Neutrino
  Oscillations}.
\newblock \emph{JHEP}, 06:\penalty0 135, 2017.
\newblock \doi{10.1007/JHEP06(2017)135}.

\bibitem[Serebrov et~al.(2015{\natexlab{a}})]{Serebrov:2015txp}
A.~P. Serebrov et~al.
\newblock {Creation of neutrino laboratory for carrying out experiment on
  search for a sterile neutrino at the SM-3 reactor}.
\newblock \emph{Tech. Phys.}, 60\penalty0 (12):\penalty0 1863--1871,
  2015{\natexlab{a}}.
\newblock \doi{10.1134/S106378421512018X}.
\newblock [Zh. Tekh. Fiz.60,no.12,128(2015)].

\bibitem[Serebrov et~al.(2015{\natexlab{b}})]{Serebrov:2015ros}
A.~P. Serebrov et~al.
\newblock {Neutrino-4 experiment on the search for a sterile neutrino at the
  SM-3 reactor}.
\newblock \emph{J. Exp. Theor. Phys.}, 121\penalty0 (4):\penalty0 578--586,
  2015{\natexlab{b}}.
\newblock \doi{10.1134/S1063776115100209}.
\newblock [Zh. Eksp. Teor. Fiz.148,no.4,665(2015)].

\bibitem[Brun et~al.(2015)]{Brun:2015aul}
E.~Brun et~al.
\newblock {TRIPOLI-4®, CEA, EDF and AREVA reference Monte Carlo code}.
\newblock \emph{Annals Nucl. Energy}, 82:\penalty0 151--160, 2015.
\newblock \doi{10.1016/j.anucene.2014.07.053}.

\bibitem[Schreckenbach et~al.(1981)Schreckenbach, Faust, von Feilitzsch, Hahn,
  Hawerkamp, and Vuilleumier]{Schreckenbach:1981wlm}
K.~Schreckenbach, H.~R. Faust, F.~von Feilitzsch, A.~A. Hahn, K.~Hawerkamp, and
  J.~L. Vuilleumier.
\newblock {Absolute measurement of the beta spectrum from 235 U fission as a
  basis for reactor antineutrino experiments}.
\newblock \emph{Phys. Lett.}, 99B:\penalty0 251--256, 1981.
\newblock \doi{10.1016/0370-2693(81)91120-5}.

\bibitem[Von~Feilitzsch et~al.(1982)Von~Feilitzsch, Hahn, and
  Schreckenbach]{VonFeilitzsch:1982jw}
F.~Von~Feilitzsch, A.~A. Hahn, and K.~Schreckenbach.
\newblock {EXPERIMENTAL BETA SPECTRA FROM PU-239 AND U-235 THERMAL NEUTRON
  FISSION PRODUCTS AND THEIR CORRELATED ANTI-NEUTRINOS SPECTRA}.
\newblock \emph{Phys. Lett.}, 118B:\penalty0 162--166, 1982.
\newblock \doi{10.1016/0370-2693(82)90622-0}.

\bibitem[Schreckenbach et~al.(1985)Schreckenbach, Colvin, Gelletly, and
  Von~Feilitzsch]{Schreckenbach:1985ep}
K.~Schreckenbach, G.~Colvin, W.~Gelletly, and F.~Von~Feilitzsch.
\newblock {DETERMINATION OF THE ANTI-NEUTRINO SPECTRUM FROM U-235 THERMAL
  NEUTRON FISSION PRODUCTS UP TO 9.5-MEV}.
\newblock \emph{Phys. Lett.}, 160B:\penalty0 325--330, 1985.
\newblock \doi{10.1016/0370-2693(85)91337-1}.

\bibitem[Hahn et~al.(1989)Hahn, Schreckenbach, Colvin, Krusche, Gelletly, and
  Von~Feilitzsch]{Hahn:1989zr}
A.~A. Hahn, K.~Schreckenbach, G.~Colvin, B.~Krusche, W.~Gelletly, and
  F.~Von~Feilitzsch.
\newblock {Anti-neutrino Spectra From $^{241}$Pu and $^{239}$Pu Thermal Neutron
  Fission Products}.
\newblock \emph{Phys. Lett.}, B218:\penalty0 365--368, 1989.
\newblock \doi{10.1016/0370-2693(89)91598-0}.

\bibitem[Haag et~al.(2014)Haag, Gütlein, Hofmann, Oberauer, Potzel,
  Schreckenbach, and Wagner]{Haag:2013raa}
N.~Haag, A.~Gütlein, M.~Hofmann, L.~Oberauer, W.~Potzel, K.~Schreckenbach, and
  F.~M. Wagner.
\newblock {Experimental Determination of the Antineutrino Spectrum of the
  Fission Products of $^{238}$U}.
\newblock \emph{Phys. Rev. Lett.}, 112\penalty0 (12):\penalty0 122501, 2014.
\newblock \doi{10.1103/PhysRevLett.112.122501}.

\bibitem[An et~al.(2017)]{An:2017osx}
F.~P. An et~al.
\newblock {Evolution of the Reactor Antineutrino Flux and Spectrum at Daya
  Bay}.
\newblock \emph{Phys. Rev. Lett.}, 118\penalty0 (25):\penalty0 251801, 2017.
\newblock \doi{10.1103/PhysRevLett.118.251801}.

\bibitem[Mampe et~al.(1978)Mampe, Schreckenbach, Maier, Braumandl, Larysz, and
  von Egidy]{BILL}
W.~Mampe, P.~Schreckenbach, K.~Jeuch, B.~P.~K. Maier, F.~Braumandl, J.~Larysz,
  and T.~von Egidy.
\newblock {Precision spectroscopy with reactor anti-neutrinos}.
\newblock \emph{Nucl. Instrum. Meth}, 154:\penalty0 127, 1978.

\bibitem[Kibédi et~al.(2008)Kibédi, Burrows, Trzhaskovskaya, Davidson, and
  Nestor]{BrIcc}
T.~Kibédi, T.W. Burrows, M.B. Trzhaskovskaya, P.M. Davidson, and C.W. Nestor.
\newblock Evaluation of theoretical conversion coefficients using bricc.
\newblock \emph{Nuclear Instruments and Methods in Physics Research Section A:
  Accelerators, Spectrometers, Detectors and Associated Equipment},
  589\penalty0 (2):\penalty0 202 -- 229, 2008.
\newblock ISSN 0168-9002.
\newblock \doi{https://doi.org/10.1016/j.nima.2008.02.051}.
\newblock URL
  \url{http://www.sciencedirect.com/science/article/pii/S0168900208002520}.

\bibitem[Schillebeeckx et~al.(2013)Schillebeeckx, Belgya, Borella, Kopecky,
  Mengoni, Quete, Szentmiklosi, Trešl, and Wynants]{Schill}
P.~Schillebeeckx, T.~Belgya, A.~Borella, S.~Kopecky, A.~Mengoni, C.R. Quete,
  L.~Szentmiklosi, I.~Trešl, and R.~Wynants.
\newblock Neutron capture studies of 206pb at a cold neutron beam.
\newblock \emph{Eur. Phys. J. A}, 49:\penalty0 143, 2013.

\bibitem[Tanabashi et~al.(2018)]{Tanabashi:2018oca}
M.~Tanabashi et~al.
\newblock {Review of Particle Physics}.
\newblock \emph{Phys. Rev.}, D98\penalty0 (3):\penalty0 030001, 2018.
\newblock \doi{10.1103/PhysRevD.98.030001}.

\bibitem[Abazajian et~al.(2012)]{Abazajian:2012ys}
K.~N. Abazajian et~al.
\newblock {Light Sterile Neutrinos: A White Paper}.
\newblock 2012.

\bibitem[Athanassopoulos et~al.(1998)]{Athanassopoulos:1997pv}
C.~Athanassopoulos et~al.
\newblock {Evidence for nu(mu) ---> nu(e) neutrino oscillations from LSND}.
\newblock \emph{Phys. Rev. Lett.}, 81:\penalty0 1774--1777, 1998.
\newblock \doi{10.1103/PhysRevLett.81.1774}.

\bibitem[Aguilar-Arevalo et~al.(2009)]{AguilarArevalo:2008rc}
A.~A. Aguilar-Arevalo et~al.
\newblock {Unexplained Excess of Electron-Like Events From a 1-GeV Neutrino
  Beam}.
\newblock \emph{Phys. Rev. Lett.}, 102:\penalty0 101802, 2009.
\newblock \doi{10.1103/PhysRevLett.102.101802}.

\bibitem[Aguilar-Arevalo et~al.(2018)]{Aguilar-Arevalo:2018gpe}
A.~A. Aguilar-Arevalo et~al.
\newblock {Significant Excess of ElectronLike Events in the MiniBooNE
  Short-Baseline Neutrino Experiment}.
\newblock \emph{Phys. Rev. Lett.}, 121\penalty0 (22):\penalty0 221801, 2018.
\newblock \doi{10.1103/PhysRevLett.121.221801}.

\bibitem[Hayes(2018)]{Hayes}
Anna Hayes.
\newblock Aap2018: Flux predictions - nuclear theory.
\newblock 2018.

\bibitem[Sonzogni(2018)]{Sonzogni}
Alejandro Sonzogni.
\newblock Aap2018: Flux predictions - nuclear data.
\newblock 2018.

\bibitem[Onillon(2018)]{Onillon}
Anthony Onillon.
\newblock Aap2018: Updated flux and spectral predictions relevant to the raa.
\newblock 2018.

\bibitem[Dentler et~al.(2017{\natexlab{b}})Dentler, Hernández-Cabezudo, Kopp,
  Maltoni, and Schwetz]{Dentler:2017tkw}
Mona Dentler, Álvaro Hernández-Cabezudo, Joachim Kopp, Michele Maltoni, and
  Thomas Schwetz.
\newblock {Sterile neutrinos or flux uncertainties? — Status of the reactor
  anti-neutrino anomaly}.
\newblock \emph{JHEP}, 11:\penalty0 099, 2017{\natexlab{b}}.
\newblock \doi{10.1007/JHEP11(2017)099}.

\bibitem[Han(2018)]{Han}
Bo-young Han.
\newblock Aap2018: Neos.
\newblock 2018.

\bibitem[Almazan(2018)]{Almazan}
Helena Almazan.
\newblock Aap2018: Stereo.
\newblock 2018.

\bibitem[Serebrov(2018)]{Serebrov}
Anatolii Serebrov.
\newblock Aap2018: Neutrino-4.
\newblock 2018.

\bibitem[Mumm(2018)]{Mumm}
Pieter Mumm.
\newblock Aap2018: Prospect.
\newblock 2018.

\bibitem[Shitov(2018)]{Shitov}
Yuri Shitov.
\newblock Aap2018: Danss.
\newblock 2018.

\bibitem[Acero et~al.(2008)Acero, Giunti, and Laveder]{Acero:2007su}
Mario~A. Acero, Carlo Giunti, and Marco Laveder.
\newblock {Limits on nu(e) and anti-nu(e) disappearance from Gallium and
  reactor experiments}.
\newblock \emph{Phys. Rev.}, D78:\penalty0 073009, 2008.
\newblock \doi{10.1103/PhysRevD.78.073009}.

\bibitem[Cianci et~al.(2017)Cianci, Furmanski, Karagiorgi, and
  Ross-Lonergan]{Cianci:2017okw}
Davio Cianci, Andy Furmanski, Georgia Karagiorgi, and Mark Ross-Lonergan.
\newblock {Prospects of Light Sterile Neutrino Oscillation and CP Violation
  Searches at the Fermilab Short Baseline Neutrino Facility}.
\newblock \emph{Phys. Rev.}, D96\penalty0 (5):\penalty0 055001, 2017.
\newblock \doi{10.1103/PhysRevD.96.055001}.

\bibitem[Cianci et~al.(in preparation)]{Cianci}
D.~Cianci et~al.
\newblock {Global Sterile Neutrino Oscillation Fits, SBN Reach, and DUNE
  Implications}.
\newblock in preparation.

\bibitem[Dentler et~al.(2018)Dentler, Hernández-Cabezudo, Kopp, Machado,
  Maltoni, Martinez-Soler, and Schwetz]{Dentler:2018sju}
Mona Dentler, Álvaro Hernández-Cabezudo, Joachim Kopp, Pedro A.~N. Machado,
  Michele Maltoni, Ivan Martinez-Soler, and Thomas Schwetz.
\newblock {Updated Global Analysis of Neutrino Oscillations in the Presence of
  eV-Scale Sterile Neutrinos}.
\newblock \emph{JHEP}, 08:\penalty0 010, 2018.
\newblock \doi{10.1007/JHEP08(2018)010}.

\bibitem[Conrad et~al.(2013)Conrad, Ignarra, Karagiorgi, Shaevitz, and
  Spitz]{Conrad:2012qt}
J.~M. Conrad, C.~M. Ignarra, G.~Karagiorgi, M.~H. Shaevitz, and J.~Spitz.
\newblock {Sterile Neutrino Fits to Short Baseline Neutrino Oscillation
  Measurements}.
\newblock \emph{Adv. High Energy Phys.}, 2013:\penalty0 163897, 2013.
\newblock \doi{10.1155/2013/163897}.

\bibitem[Chu et~al.(2018)Chu, Dasgupta, Dentler, Kopp, and
  Saviano]{Chu:2018gxk}
Xiaoyong Chu, Basudeb Dasgupta, Mona Dentler, Joachim Kopp, and Ninetta
  Saviano.
\newblock {Sterile Neutrinos with Secret Interactions -- Cosmological Discord?}
\newblock \emph{JCAP}, 1811\penalty0 (11):\penalty0 049, 2018.
\newblock \doi{10.1088/1475-7516/2018/11/049}.

\bibitem[Döring et~al.(2018)Döring, Päs, Sicking, and
  Weiler]{Doring:2018cob}
Dominik Döring, Heinrich Päs, Philipp Sicking, and Thomas~J. Weiler.
\newblock {Sterile Neutrinos with Altered Dispersion Relations as an
  Explanation for the MiniBooNE, LSND, Gallium and Reactor Anomalies}.
\newblock 2018.

\bibitem[Bertuzzo et~al.(2018)Bertuzzo, Jana, Machado, and
  Zukanovich~Funchal]{Bertuzzo:2018itn}
Enrico Bertuzzo, Sudip Jana, Pedro A.~N. Machado, and Renata
  Zukanovich~Funchal.
\newblock {Dark Neutrino Portal to Explain MiniBooNE excess}.
\newblock \emph{Phys. Rev. Lett.}, 121\penalty0 (24):\penalty0 241801, 2018.
\newblock \doi{10.1103/PhysRevLett.121.241801}.

\bibitem[Alvarez-Ruso and Saul-Sala(2017)]{Alvarez-Ruso:2017hdm}
Luis Alvarez-Ruso and Eduardo Saul-Sala.
\newblock {Radiative decay of heavy neutrinos at MiniBooNE and MicroBooNE}.
\newblock In \emph{{Proceedings, Prospects in Neutrino Physics (NuPhys2016):
  London, UK, December 12-14, 2016}}, 2017.

\bibitem[Diaz(2018)]{Diaz}
A.~Diaz.
\newblock Ichep2018: Updated miniboone neutrino oscillation results within the
  context of global fits to short-baseline neutrino data.
\newblock 2018.

\bibitem[Ballett et~al.(2018)Ballett, Pascoli, and
  Ross-Lonergan]{Ballett:2018ynz}
Peter Ballett, Silvia Pascoli, and Mark Ross-Lonergan.
\newblock {U(1)' mediated decays of heavy sterile neutrinos in MiniBooNE}.
\newblock 2018.

\bibitem[Guenette(2018)]{Guenette}
Roxanne Guenette.
\newblock Neutrino 2018: Microboone and the future sbn program.
\newblock 2018.

\bibitem[Acciarri et~al.(2018{\natexlab{a}})]{neutrino2018singlegamma}
R.~Acciarri et~al.
\newblock Microboone public note 1041: The microboone search for single photon
  events.
\newblock 2018{\natexlab{a}}.

\bibitem[Yates(2017)]{Yates:2017lxa}
Lauren~E. Yates.
\newblock {MicroBooNE Investigation of Low-Energy Excess Using Deep Learning
  Algorithms}.
\newblock In \emph{{Proceedings, Meeting of the APS Division of Particles and
  Fields (DPF 2017): Fermilab, Batavia, Illinois, USA, July 31 - August 4,
  2017}}, 2017.

\bibitem[Acciarri et~al.(2018{\natexlab{b}})]{neutrino2018nue}
R.~Acciarri et~al.
\newblock Microboone public note 1038: Electron-neutrino selection and
  reconstruction in the microboone lartpc using the pandora multi-algorithm
  pattern recognition.
\newblock 2018{\natexlab{b}}.

\bibitem[Maricic(2018)]{Maricic}
Jelena Maricic.
\newblock Aap2018: Meeting summary.
\newblock 2018.

\bibitem[Serebrov et~al.(2018)]{Serebrov:2018vdw}
A.P. Serebrov et~al.
\newblock {The first observation of effect of oscillation in Neutrino-4
  experiment on search for sterile neutrino}.
\newblock 2018.

\bibitem[Verstraeten(2018)]{Verstraeten}
Maja Verstraeten.
\newblock Aap2018: Solid.
\newblock 2018.

\bibitem[Park(2018)]{Park}
Jaewon Park.
\newblock Aap2018: Chandler.
\newblock 2018.

\bibitem[Chimenti(2018)]{Chimenti}
Pedro Chimenti.
\newblock Aap2018: Angra-nu.
\newblock 2018.

\bibitem[Mulmule(2018)]{Mulmule}
Dhruv Mulmule.
\newblock Aap2018: Exploring anti-neutrino event selection and background
  reduction techniques for ismran.
\newblock 2018.

\bibitem[Coleman(2018)]{Coleman}
Jonathan Coleman.
\newblock Aap2018: Vidaar.
\newblock 2018.

\bibitem[Nakajima(2018)]{Nakajima}
Kyohei Nakajima.
\newblock Aap2018: Status of reactor neutrino monitor experiments in japan.
\newblock 2018.

\bibitem[Antonello et~al.(2015)]{Antonello:2015lea}
M.~Antonello et~al.
\newblock {A Proposal for a Three Detector Short-Baseline Neutrino Oscillation
  Program in the Fermilab Booster Neutrino Beam}.
\newblock 2015.

\bibitem[Alonso and Nakamura(2017)]{Alonso:2017fci}
Jose~R. Alonso and K.~Nakamura.
\newblock {IsoDAR@KamLAND:A Conceptual Design Report for the Conventional
  Facilities}.
\newblock 2017.

\bibitem[Cebrera-Palmer(2018)]{Cebrera}
Belkis Cebrera-Palmer.
\newblock Aap2018: Coherent.
\newblock 2018.

\bibitem[et~al.(2009)]{ref:SONGS}
N.~S.~Bowden et~al.
\newblock Observation of the isotopic evolution of pressurized water reactor
  fuel using an antineutrino detector.
\newblock \emph{Journal of Applied Physics}, 105:\penalty0 064902, 2009.

\bibitem[et~al.(2015)]{ref:RENO1}
S-H~Seo et~al.
\newblock New results from reno and the 5 mev excess.
\newblock \emph{AIP Conference Proceedings}, 1666:\penalty0 080002, 2015.

\bibitem[et~al.(2016{\natexlab{a}})]{ref:RENO2}
J.H.~Choi et~al.
\newblock Observation of energy and baseline dependent reactor antineutrino
  disappearance in the reno experiment.
\newblock \emph{Physical Review Letters}, 116:\penalty0 211801,
  2016{\natexlab{a}}.

\bibitem[et~al.(2016{\natexlab{b}})]{ref:DayaBay}
F.~P.~An et~al.
\newblock Measurement of the reactor antineutrino flux and spectrum at daya
  bay.
\newblock \emph{Physical Review Letters}, 116:\penalty0 061801,
  2016{\natexlab{b}}.

\bibitem[ref()]{ref:DoubleChooz}
Measurement of $\theta_{13}$ in double chooz using neutron captures on hydrogen
  with novel background rejection techniques, author={Y. Abe et al.},
  journal={Journal of High Energy Physics}, volume={01}, pages={163},
  year={2016}, publisher={Springer}.

\bibitem[et~al.(2011)]{ref:anomaly}
G.~Mention et~al.
\newblock Reactor antineutrino anomaly.
\newblock \emph{Physical Review D}, 83:\penalty0 073006, 2011.

\bibitem[et~al.(2014)]{ref:oguri}
S.~Oguri et~al.
\newblock Reactor antineutrino monitoring with a plastic scintillator array as
  a new safeguards method.
\newblock \emph{Nuclear Instruments and Methods in Physics Research A},
  757:\penalty0 33--39, 2014.

\bibitem[Anjos et~al.(2006)Anjos, Barbosa, Zukanovich~Funchal, Kemp, Magnin,
  Nunokawa, Peres, Reyna, and Shellard]{Anjos:2005pg}
J.~C. Anjos, A.~F. Barbosa, R.~Zukanovich~Funchal, E.~Kemp, J.~Magnin,
  H.~Nunokawa, O.~L.~G. Peres, D.~Reyna, and R.~C. Shellard.
\newblock {Angra neutrino project: Status and plans}.
\newblock \emph{Nucl. Phys. Proc. Suppl.}, 155:\penalty0 231--232, 2006.
\newblock \doi{10.1016/j.nuclphysbps.2006.02.058}.

\bibitem[Alvarenga et~al.(2016)]{Alvarenga:2016lgi}
T.~A. Alvarenga et~al.
\newblock {Readout electronics validation and target detector assessment for
  the Neutrinos Angra experiment}.
\newblock \emph{Nucl. Instrum. Meth.}, A830:\penalty0 206--213, 2016.
\newblock \doi{10.1016/j.nima.2016.05.052}.

\bibitem[Lopes et~al.(2018)Lopes, Melo, Costa, Nóbrega, Anjos, Pepe,
  Lima~Junior, Cernicchiaro, Chimenti, Guedes, Gonzalez, Kemp, and
  Paschoal]{Lopes:2018}
G.~Lopes, D.~S. Melo, I.~A. Costa, R.~A. Nóbrega, J.~C.~C. Anjos, I.~M. Pepe,
  H.~P. Lima~Junior, G.~Cernicchiaro, P.~Chimenti, G.~P. Guedes, L.~F.~G.
  Gonzalez, E.~Kemp, and M.~Paschoal.
\newblock Signal simulation based on the characteristics of the neutrinos angra
  experiment's readout electronics.
\newblock In \emph{3rd International Symposium on Instrumentation Systems,
  Circuits and Transducers}, 2018.

\bibitem[Paschoal et~al.(2018)Paschoal, Lopes, Mela, Costa, Anjos, Lima~Junior,
  Cernicchiaro, Pepe, Ribeiro, Chimenti, Guedes, Gonzalez, Kemp, and
  Nóbrega]{Paschoal:2018}
M.~Paschoal, G.~Lopes, D.~S. Mela, I.~A. Costa, J.~C.~C. Anjos, H.~P.
  Lima~Junior, G.~Cernicchiaro, I.~M. Pepe, D.~B.~S. Ribeiro, P.~Chimenti,
  G.~P. Guedes, L.~F.~G. Gonzalez, E.~Kemp, and R.~A. Nóbrega.
\newblock v-angra readout electronics and target detector assessment using a
  cosmic rays based trigger.
\newblock In \emph{3rd International Symposium on Instrumentation Systems,
  Circuits and Transducers}, 2018.

\bibitem[Chimenti et~al.(2014)Chimenti, Valdiviesso, Souza, Anjos, Junqueira,
  and Abrahão]{Chimenti:2014}
P.~Chimenti, G.~Valdiviesso, M.~Souza, J.~C.~D. Anjos, T.~Junqueira, and
  T.~Abrahão.
\newblock Simulation of the angra neutrinos project i: primary generators, g4
  simulation and s/n for neutrino detection.
\newblock Neutrinos Angra Internal Note, 2014.

\bibitem[Allan et~al.(2013)]{Allan:2013ofa}
D.~Allan et~al.
\newblock {The Electromagnetic Calorimeter for the T2K Near Detector ND280}.
\newblock \emph{JINST}, 8:\penalty0 P10019, 2013.
\newblock \doi{10.1088/1748-0221/8/10/P10019}.

\bibitem[IAE(2008)]{IAEA}
Final report: Focused workshop on antineutrino detection for safeguards
  applications.
\newblock \emph{IAEA}, 2008.

\bibitem[Carroll et~al.(2018)]{Carroll:2018kad}
J.~Carroll et~al.
\newblock {Monitoring Reactor Anti-Neutrinos Using a Plastic Scintillator
  Detector in a Mobile Laboratory}.
\newblock 2018.

\bibitem[Metelko et~al.(2017)]{Metelko}
C.~Metelko et~al.
\newblock {Proceddings of the AAP}.
\newblock 2017.

\bibitem[{Mills, Robert W} et~al.(2018)]{refId0}
{Mills, Robert W} et~al.
\newblock Modelling of the anti-neutrino production and spectra from a magnox
  reactor.
\newblock \emph{EPJ Web Conf.}, 170:\penalty0 07008, 2018.
\newblock \doi{10.1051/epjconf/201817007008}.
\newblock URL \url{https://doi.org/10.1051/epjconf/201817007008}.

\bibitem[{Hutt, P. K.} et~al.(1991)]{Hutt}
{Hutt, P. K.} et~al.
\newblock The uk core performance code package.
\newblock \emph{Nuclear Energy 30.5}, pages 291--298, 1991.

\bibitem[WIM()]{WIMS}
{WIMS, TRAIL and FISPIN are reactor simulation codes distributed by the ANSWERS
  software service (https://www.answerssoftwareservice.com/). FISPIN10 being
  developed by the UK NNL.}

\bibitem[{Tobias, A.} et~al.(2002)]{BTSPEC}
{Tobias, A.} et~al.
\newblock {BTPLOT BTSPEC EXSPEC ORDTAB TABLST, Retrieval of ENDF/B Decay
  Spectra.}
\newblock 2002.

\bibitem[{Kellett, M.A.} et~al.(2009)]{JEFF}
{Kellett, M.A.} et~al.
\newblock {The JEFF-3.1/63.1. 1 radioactive decay data and fission yields
  sub-libraries}.
\newblock \emph{OECD/NEA JEFF report 20}, 2009.

\bibitem[Schnellbach et~al.(2018)]{Schnellbach}
Y.~Schnellbach et~al.
\newblock {Proceddings of the AAP }.
\newblock 2018.

\bibitem[Dazeley et~al.(2009)Dazeley, Bernstein, Bowden, and Svoboda]{llnl}
S.~Dazeley, A.~Bernstein, N.~S. Bowden, and R.~Svoboda.
\newblock {Observation of Neutrons with a Gadolinium Doped Water Cerenkov
  Detector}.
\newblock \emph{Nucl. Instrum. Meth.}, A607:\penalty0 616--619, 2009.
\newblock \doi{10.1016/j.nima.2009.03.256}.

\bibitem[Watanabe et~al.(2009)]{superk}
H.~Watanabe et~al.
\newblock {First Study of Neutron Tagging with a Water Cherenkov Detector}.
\newblock \emph{Astropart. Phys.}, 31:\penalty0 320--328, 2009.
\newblock \doi{10.1016/j.astropartphys.2009.03.002}.

\bibitem[Dazeley et~al.(2016)Dazeley, Askins, Bergevin, Bernstein, Bowden,
  Shokair, Jaffke, Rountree, and Sweany]{Dazeley:2015uyd}
S.~Dazeley, M.~Askins, M.~Bergevin, A.~Bernstein, N.~S. Bowden, T.~M. Shokair,
  P.~Jaffke, S.~D. Rountree, and M.~Sweany.
\newblock {A search for cosmogenic production of $\beta$-neutron emitting
  radionuclides in water}.
\newblock \emph{Nucl. Instrum. Meth.}, A821:\penalty0 151--159, 2016.
\newblock \doi{10.1016/j.nima.2016.03.014}.

\bibitem[Xu and Collaboration(2016)]{egads}
Chenyuan Xu and Super-Kamiokande Collaboration.
\newblock Current status of sk-gd project and egads.
\newblock \emph{Journal of Physics: Conference Series}, 718\penalty0
  (6):\penalty0 062070, 2016.
\newblock URL \url{http://stacks.iop.org/1742-6596/718/i=6/a=062070}.

\bibitem[rat()]{ratpac}
Rat (is an analysis tool) user’s guide.
\newblock \url{https://rat.readthedocs.io/en/latest/}.
\newblock Accessed: 2018-11-9.

\bibitem[Orebi~Gann(2015)]{theia}
Gabriel~D. Orebi~Gann.
\newblock {Physics Potential of an Advanced Scintillation Detector: Introducing
  THEIA}.
\newblock 2015.

\bibitem[Back et~al.(2017)]{annie}
A.~R. Back et~al.
\newblock {Accelerator Neutrino Neutron Interaction Experiment (ANNIE):
  Preliminary Results and Physics Phase Proposal}.
\newblock 2017.

\bibitem[Beacom and Vagins(2004)]{Beacom:2003nk}
John~F. Beacom and Mark~R. Vagins.
\newblock {GADZOOKS! Anti-neutrino spectroscopy with large water Cherenkov
  detectors}.
\newblock \emph{Phys. Rev. Lett.}, 93:\penalty0 171101, 2004.
\newblock \doi{10.1103/PhysRevLett.93.171101}.

\bibitem[Bays et~al.(2012)]{Bays:2011si}
K.~Bays et~al.
\newblock {Supernova Relic Neutrino Search at Super-Kamiokande}.
\newblock \emph{Phys. Rev.}, D85:\penalty0 052007, 2012.
\newblock \doi{10.1103/PhysRevD.85.052007}.

\bibitem[Zhang et~al.(2015)]{Zhang:2013tua}
H.~Zhang et~al.
\newblock {Supernova Relic Neutrino Search with Neutron Tagging at
  Super-Kamiokande-IV}.
\newblock \emph{Astropart. Phys.}, 60:\penalty0 41--46, 2015.
\newblock \doi{10.1016/j.astropartphys.2014.05.004}.

\bibitem[{National Research Council}(2012)]{NAP12849}
{National Research Council}.
\newblock \emph{{The Comprehensive Nuclear Test Ban Treaty: Technical Issues
  for the United States}}.
\newblock The National Academies Press, Washington, DC, 2012.
\newblock ISBN 978-0-309-14998-3.
\newblock \doi{10.17226/12849}.
\newblock URL
  \url{https://www.nap.edu/catalog/12849/the-comprehensive-nuclear-test-ban-treaty-technical-issues-for-the}.

\bibitem[Bernstein et~al.(2001)Bernstein, West, and Gupta]{Bernstein2001}
A.~Bernstein, T.~West, and V.~Gupta.
\newblock An assessment of antineutrino detection as a tool for monitoring
  nuclear explosions.
\newblock \emph{Science and Global Security}, 2001.

\bibitem[Carr et~al.(2018{\natexlab{b}})Carr, Dalnoki-Veress, and
  Bernstein]{Carr:2017uuq}
R.~Carr, F.~Dalnoki-Veress, and A.~Bernstein.
\newblock {Sensitivity of seismically-cued antineutrino detectors to nuclear
  explosions}.
\newblock \emph{Phys. Rev. Applied}, 10\penalty0 (2):\penalty0 024014,
  2018{\natexlab{b}}.
\newblock \doi{10.1103/PhysRevApplied.10.024014}.
\newblock [Phys. Rev. Applied.10,024014(2018)].

\bibitem[Askins et~al.(2015)]{Askins:2015bmb}
M.~Askins et~al.
\newblock {The Physics and Nuclear Nonproliferation Goals of WATCHMAN: A WAter
  CHerenkov Monitor for ANtineutrinos}.
\newblock 2015.

\bibitem[Bondar et~al.(2004)Bondar, Myers, Engdahl, and Bergman]{epicenter}
I.~Bondar, S.C. Myers, E.R. Engdahl, and E.A. Bergman.
\newblock Epicenter accuracy based on seismic network criteria.
\newblock \emph{Geophys. Jour. Int.}, 156:\penalty0 483 -- 496, 2004.

\bibitem[Normile(2018)]{hyperk}
Dennis Normile.
\newblock {Japan’s science ministry seeks large budget increase, prioritizing
  massive neutrino detector}.
\newblock \emph{Science}, 2018.

\bibitem[NORSAR(12 September 2017)]{norsar}
NORSAR.
\newblock {The nuclear explosion in North Korea on 3 September 2017: A revised
  magnitude assessment}, 12 September 2017.

\bibitem[Freedman(1974)]{PhysRevD.9.1389}
Daniel~Z. Freedman.
\newblock Coherent effects of a weak neutral current.
\newblock \emph{Phys. Rev. D}, 9:\penalty0 1389--1392, Mar 1974.
\newblock \doi{10.1103/PhysRevD.9.1389}.
\newblock URL \url{https://link.aps.org/doi/10.1103/PhysRevD.9.1389}.

\bibitem[Akimov et~al.(2017)Akimov, Albert, An, Awe, Barbeau, Becker, Belov,
  Brown, Bolozdynya, Cabrera-Palmer, Cervantes, Collar, Cooper, Cooper, Cuesta,
  Dean, Detwiler, Eberhardt, Efremenko, Elliott, Erkela, Fabris, Febbraro,
  Fields, Fox, Fu, Galindo-Uribarri, Green, Hai, Heath, Hedges, Hornback,
  Hossbach, Iverson, Kaufman, Ki, Klein, Khromov, Konovalov, Kremer, Kumpan,
  Leadbetter, Li, Lu, Mann, Markoff, Miller, Moreno, Mueller, Newby, Orrell,
  Overman, Parno, Penttila, Perumpilly, Ray, Raybern, Reyna, Rich, Rimal,
  Rudik, Scholberg, Scholz, Sinev, Snow, Sosnovtsev, Shakirov, Suchyta, Suh,
  Tayloe, Thornton, Tolstukhin, Vanderwerp, Varner, Virtue, Wan, Yoo, Yu,
  Zawada, Zettlemoyer, Zderic, and ]{Akimoveaao0990}
D.~Akimov, J.~B. Albert, P.~An, C.~Awe, P.~S. Barbeau, B.~Becker, V.~Belov,
  A.~Brown, A.~Bolozdynya, B.~Cabrera-Palmer, M.~Cervantes, J.~I. Collar, R.~J.
  Cooper, R.~L. Cooper, C.~Cuesta, D.~J. Dean, J.~A. Detwiler, A.~Eberhardt,
  Y.~Efremenko, S.~R. Elliott, E.~M. Erkela, L.~Fabris, M.~Febbraro, N.~E.
  Fields, W.~Fox, Z.~Fu, A.~Galindo-Uribarri, M.~P. Green, M.~Hai, M.~R. Heath,
  S.~Hedges, D.~Hornback, T.~W. Hossbach, E.~B. Iverson, L.~J. Kaufman, S.~Ki,
  S.~R. Klein, A.~Khromov, A.~Konovalov, M.~Kremer, A.~Kumpan, C.~Leadbetter,
  L.~Li, W.~Lu, K.~Mann, D.~M. Markoff, K.~Miller, H.~Moreno, P.~E. Mueller,
  J.~Newby, J.~L. Orrell, C.~T. Overman, D.~S. Parno, S.~Penttila,
  G.~Perumpilly, H.~Ray, J.~Raybern, D.~Reyna, G.~C. Rich, D.~Rimal, D.~Rudik,
  K.~Scholberg, B.~J. Scholz, G.~Sinev, W.~M. Snow, V.~Sosnovtsev, A.~Shakirov,
  S.~Suchyta, B.~Suh, R.~Tayloe, R.~T. Thornton, I.~Tolstukhin, J.~Vanderwerp,
  R.~L. Varner, C.~J. Virtue, Z.~Wan, J.~Yoo, C.-H. Yu, A.~Zawada,
  J.~Zettlemoyer, A.~M. Zderic, and .
\newblock Observation of coherent elastic neutrino-nucleus scattering.
\newblock \emph{Science}, 2017.
\newblock \doi{10.1126/science.aao0990}.

\bibitem[Lindner et~al.(2017{\natexlab{a}})Lindner, Rodejohann, and
  Xu]{Lindner2017}
Manfred Lindner, Werner Rodejohann, and Xun-Jie Xu.
\newblock Coherent neutrino-nucleus scattering and new neutrino interactions.
\newblock \emph{Journal of High Energy Physics}, 2017\penalty0 (3):\penalty0
  97, Mar 2017{\natexlab{a}}.
\newblock ISSN 1029-8479.
\newblock \doi{10.1007/JHEP03(2017)097}.
\newblock URL \url{https://doi.org/10.1007/JHEP03(2017)097}.

\bibitem[Billard et~al.(2014)Billard, Figueroa-Feliciano, and
  Strigari]{PhysRevD.89.023524}
J.~Billard, E.~Figueroa-Feliciano, and L.~Strigari.
\newblock Implication of neutrino backgrounds on the reach of next generation
  dark matter direct detection experiments.
\newblock \emph{Phys. Rev. D}, 89:\penalty0 023524, Jan 2014.
\newblock \doi{10.1103/PhysRevD.89.023524}.
\newblock URL \url{https://link.aps.org/doi/10.1103/PhysRevD.89.023524}.

\bibitem[Formaggio et~al.(2012)Formaggio, Figueroa-Feliciano, and
  Anderson]{PhysRevD.85.013009}
Joseph~A. Formaggio, E.~Figueroa-Feliciano, and A.~J. Anderson.
\newblock Sterile neutrinos, coherent scattering, and oscillometry measurements
  with low-temperature bolometers.
\newblock \emph{Phys. Rev. D}, 85:\penalty0 013009, Jan 2012.
\newblock \doi{10.1103/PhysRevD.85.013009}.
\newblock URL \url{https://link.aps.org/doi/10.1103/PhysRevD.85.013009}.

\bibitem[Bernadette K.~Cogswell(2016)]{Huber2016}
Patrick~Huber Bernadette K.~Cogswell.
\newblock Detection of breeding blankets using antineutrinos.
\newblock \emph{Science \& Global Security}, 24\penalty0 (2):\penalty0
  114--130, 2016.
\newblock URL
  \url{http://scienceandglobalsecurity.org/archive/2016/05/detection_of_breeding_blankets.html}.

\bibitem[Billard et~al.(2017)Billard, Carr, Dawson, Figueroa-Feliciano,
  Formaggio, Gascon, Heine, Jesus, Johnston, Lasserre, Leder, Palladino,
  Sibille, Vivier, and Winslow]{0954-3899-44-10-105101}
J~Billard, R~Carr, J~Dawson, E~Figueroa-Feliciano, J~A Formaggio, J~Gascon, S~T
  Heine, M~De Jesus, J~Johnston, T~Lasserre, A~Leder, K~J Palladino, V~Sibille,
  M~Vivier, and L~Winslow.
\newblock Coherent neutrino scattering with low temperature bolometers at chooz
  reactor complex.
\newblock \emph{Journal of Physics G: Nuclear and Particle Physics},
  44\penalty0 (10):\penalty0 105101, 2017.
\newblock URL \url{http://stacks.iop.org/0954-3899/44/i=10/a=105101}.

\bibitem[Billard et~al.(2018)Billard, Johnston, and Kavanagh]{Billard2018}
Julien Billard, Joseph Johnston, and Bradley~J. Kavanagh.
\newblock {Prospects for exploring New Physics in Coherent Elastic
  Neutrino-Nucleus Scattering}.
\newblock \emph{Journal of Cosmology and Astroparticle Physics},
  JCAP11(2018)016, November 2018.

\bibitem[Maneschg(2018)]{conus}
Werner Maneschg.
\newblock {The Status of CONUS}.
\newblock {Talk at XXVIII International Conference on Neutrino Physics and
  Astrophysics, 4-9 June 2018, Heidelberg, Germany}, June 2018.

\bibitem[Strauss et~al.(2017)]{Strauss:2017cuu}
R.~Strauss et~al.
\newblock {The $\nu$-cleus experiment: A gram-scale fiducial-volume cryogenic
  detector for the first detection of coherent neutrino-nucleus scattering}.
\newblock \emph{Eur. Phys. J.}, C77:\penalty0 506, 2017.
\newblock \doi{10.1140/epjc/s10052-017-5068-2}.

\bibitem[Fujikawa and Shrock(1980)]{PhysRevLett.45.963}
Kazuo Fujikawa and Robert~E. Shrock.
\newblock Magnetic moment of a massive neutrino and neutrino-spin rotation.
\newblock \emph{Phys. Rev. Lett.}, 45:\penalty0 963--966, Sep 1980.
\newblock \doi{10.1103/PhysRevLett.45.963}.
\newblock URL \url{https://link.aps.org/doi/10.1103/PhysRevLett.45.963}.

\bibitem[Shrock(1974)]{Shrock:1974nd}
R.~Shrock.
\newblock {Decay l0 ---> nu(lepton) gamma in gauge theories of weak and
  electromagnetic interactions}.
\newblock \emph{Phys. Rev.}, D9:\penalty0 743--748, 1974.
\newblock \doi{10.1103/PhysRevD.9.743}.

\bibitem[Lee and Shrock(1977)]{Lee:1977tib}
Benjamin~W. Lee and Robert~E. Shrock.
\newblock {Natural Suppression of Symmetry Violation in Gauge Theories: Muon -
  Lepton and Electron Lepton Number Nonconservation}.
\newblock \emph{Phys. Rev.}, D16:\penalty0 1444, 1977.
\newblock \doi{10.1103/PhysRevD.16.1444}.

\bibitem[Shrock(1982)]{Shrock:1982sc}
Robert~E. Shrock.
\newblock {Electromagnetic Properties and Decays of Dirac and Majorana
  Neutrinos in a General Class of Gauge Theories}.
\newblock \emph{Nucl. Phys.}, B206:\penalty0 359--379, 1982.
\newblock \doi{10.1016/0550-3213(82)90273-5}.

\bibitem[Pal(1992)]{Pal:1991pm}
Palash~B. Pal.
\newblock {Particle physics confronts the solar neutrino problem}.
\newblock \emph{Int. J. Mod. Phys.}, A7:\penalty0 5387--5460, 1992.
\newblock \doi{10.1142/S0217751X92002465}.

\bibitem[Balantekin(2006)]{Balantekin:2006sw}
A.~B. Balantekin.
\newblock {Neutrino magnetic moment}.
\newblock \emph{AIP Conf. Proc.}, 847:\penalty0 128--133, 2006.
\newblock \doi{10.1063/1.2234393}.
\newblock [,128(2006)].

\bibitem[Bell et~al.(2005)Bell, Cirigliano, Ramsey-Musolf, Vogel, and
  Wise]{PhysRevLett.95.151802}
Nicole~F. Bell, V.~Cirigliano, M.~J. Ramsey-Musolf, P.~Vogel, and Mark~B. Wise.
\newblock How magnetic is the dirac neutrino?
\newblock \emph{Phys. Rev. Lett.}, 95:\penalty0 151802, Oct 2005.
\newblock \doi{10.1103/PhysRevLett.95.151802}.
\newblock URL \url{https://link.aps.org/doi/10.1103/PhysRevLett.95.151802}.

\bibitem[Bell et~al.(2006)Bell, Gorchtein, Ramsey-Musolf, Vogel, and
  Wang]{Bell:2006wi}
Nicole~F. Bell, Mikhail Gorchtein, Michael~J. Ramsey-Musolf, Petr Vogel, and
  Peng Wang.
\newblock {Model independent bounds on magnetic moments of Majorana neutrinos}.
\newblock \emph{Phys. Lett.}, B642:\penalty0 377--383, 2006.
\newblock \doi{10.1016/j.physletb.2006.09.055}.

\bibitem[Lindner et~al.(2017{\natexlab{b}})Lindner, Radovčić, and
  Welter]{Lindner:2017uvt}
Manfred Lindner, Branimir Radovčić, and Johannes Welter.
\newblock {Revisiting Large Neutrino Magnetic Moments}.
\newblock 2017{\natexlab{b}}.

\bibitem[Agostini et~al.(2017)]{Borexino:2017fbd}
M.~Agostini et~al.
\newblock {Limiting neutrino magnetic moments with Borexino Phase-II solar
  neutrino data}.
\newblock \emph{Phys. Rev.}, D96\penalty0 (9):\penalty0 091103, 2017.
\newblock \doi{10.1103/PhysRevD.96.091103}.

\bibitem[Arceo-Díaz et~al.(2015)Arceo-Díaz, Schröder, Zuber, and
  Jack]{Arceo-Diaz:2015pva}
S.~Arceo-Díaz, K.~P. Schröder, K.~Zuber, and D.~Jack.
\newblock {Constraint on the magnetic dipole moment of neutrinos by the tip-RGB
  luminosity in $\omega$-Centauri}.
\newblock \emph{Astropart. Phys.}, 70:\penalty0 1--11, 2015.
\newblock \doi{10.1016/j.astropartphys.2015.03.006}.

\bibitem[Coloma and Schwetz(2016)]{PhysRevD.94.055005}
Pilar Coloma and Thomas Schwetz.
\newblock Generalized mass ordering degeneracy in neutrino oscillation
  experiments.
\newblock \emph{Phys. Rev. D}, 94:\penalty0 055005, Sep 2016.
\newblock \doi{10.1103/PhysRevD.94.055005}.
\newblock URL \url{https://link.aps.org/doi/10.1103/PhysRevD.94.055005}.

\bibitem[Bertuzzo et~al.(2017)Bertuzzo, Deppisch, Kulkarni, Perez~Gonzalez, and
  Funchal]{Bertuzzo2017}
Enrico Bertuzzo, Frank~F. Deppisch, Suchita Kulkarni, Yuber~F. Perez~Gonzalez,
  and Renata~Zukanovich Funchal.
\newblock Dark matter and exotic neutrino interactions in direct detection
  searches.
\newblock \emph{Journal of High Energy Physics}, 2017\penalty0 (4):\penalty0
  73, Apr 2017.
\newblock ISSN 1029-8479.
\newblock \doi{10.1007/JHEP04(2017)073}.
\newblock URL \url{https://doi.org/10.1007/JHEP04(2017)073}.

\bibitem[Brooks et~al.(1960)Brooks, Pringle, and Funt]{BROOKS1960}
F.~D. Brooks, R.~W. Pringle, and B.~L. Funt.
\newblock Pulse shape discrimination in a plastic scintillator.
\newblock \emph{IRE Trans. Nucl. Sci.}, NS-7:\penalty0 35, 1960.
\newblock \doi{https://doi.org/10.1016/j.nima.2018.01.093}.

\bibitem[Zaitseva et~al.(2012)Zaitseva, Rupert, Paweaczak, Glenn, Martinez,
  Carman, Faust, Cherepy, and Payne]{ZAITSEVA201288}
Natalia Zaitseva, Benjamin~L. Rupert, Iwona Paweaczak, Andrew Glenn, H.~Paul
  Martinez, Leslie Carman, Michelle Faust, Nerine Cherepy, and Stephen Payne.
\newblock Plastic scintillators with efficient neutron/gamma pulse shape
  discrimination.
\newblock \emph{Nuclear Instruments and Methods in Physics Research Section A:
  Accelerators, Spectrometers, Detectors and Associated Equipment},
  668:\penalty0 88 -- 93, 2012.
\newblock \doi{https://doi.org/10.1016/j.nima.2011.11.071}.

\bibitem[Pozzi et~al.(2013)Pozzi, Bourne, and Clarke]{POZZI201319}
S.A. Pozzi, M.M. Bourne, and S.D. Clarke.
\newblock Pulse shape discrimination in the plastic scintillator ej-299-33.
\newblock \emph{Nuclear Instruments and Methods in Physics Research Section A:
  Accelerators, Spectrometers, Detectors and Associated Equipment},
  723:\penalty0 19 -- 23, 2013.
\newblock \doi{https://doi.org/10.1016/j.nima.2013.04.085}.

\bibitem[Cester et~al.(2014)Cester, Nebbia, Stevanato, Pino, and
  Viesti]{CESTER2014202}
D.~Cester, G.~Nebbia, L.~Stevanato, F.~Pino, and G.~Viesti.
\newblock Experimental tests of the new plastic scintillator with pulse shape
  discrimination capabilities ej-299-33.
\newblock \emph{Nuclear Instruments and Methods in Physics Research Section A:
  Accelerators, Spectrometers, Detectors and Associated Equipment},
  735:\penalty0 202 -- 206, 2014.
\newblock \doi{https://doi.org/10.1016/j.nima.2013.09.031}.

\bibitem[Hartman et~al.(2015)Hartman, Barzilov, Peters, and
  Yates]{HARTMAN2015137}
J.~Hartman, A.~Barzilov, E.E. Peters, and S.W. Yates.
\newblock Measurements of response functions of ej-299-33a plastic scintillator
  for fast neutrons.
\newblock \emph{Nuclear Instruments and Methods in Physics Research Section A:
  Accelerators, Spectrometers, Detectors and Associated Equipment},
  804:\penalty0 137 -- 143, 2015.
\newblock \doi{https://doi.org/10.1016/j.nima.2015.09.068}.

\bibitem[Liao and Yang(2015)]{LIAO2015150}
Can Liao and Haori Yang.
\newblock Pulse shape discrimination using ej-299-33 plastic scintillator
  coupled with a silicon photomultiplier array.
\newblock \emph{Nuclear Instruments and Methods in Physics Research Section A:
  Accelerators, Spectrometers, Detectors and Associated Equipment},
  789:\penalty0 150 -- 157, 2015.
\newblock \doi{https://doi.org/10.1016/j.nima.2015.04.016}.

\bibitem[Lawrence et~al.(2014)Lawrence, Febbraro, Massey, Flaska, Becchetti,
  and Pozzi]{LAWRENCE201416}
Chris~C. Lawrence, Michael Febbraro, Thomas~N. Massey, Marek Flaska, F.D.
  Becchetti, and Sara~A. Pozzi.
\newblock Neutron response characterization for an ej299-33 plastic
  scintillation detector.
\newblock \emph{Nuclear Instruments and Methods in Physics Research Section A:
  Accelerators, Spectrometers, Detectors and Associated Equipment},
  759:\penalty0 16 -- 22, 2014.
\newblock \doi{https://doi.org/10.1016/j.nima.2014.04.062}.

\bibitem[Iwanowska-Hanke et~al.(2014)Iwanowska-Hanke, Moszynski, Swiderski,
  Sibczynski, Szczesniak, Krakowski, and Schotanus]{1748-0221-9-06-P06014}
J~Iwanowska-Hanke, M~Moszynski, L~Swiderski, P~Sibczynski, T~Szczesniak,
  T~Krakowski, and P~Schotanus.
\newblock Comparative study of large samples (2" x 2") plastic scintillators
  and ej309 liquid with pulse shape discrimination (psd) capabilities.
\newblock \emph{Journal of Instrumentation}, 9\penalty0 (06):\penalty0 P06014,
  2014.
\newblock URL \url{http://stacks.iop.org/1748-0221/9/i=06/a=P06014}.

\bibitem[Woolf et~al.(2015)Woolf, Hutcheson, Gwon, Phlips, and
  Wulf]{WOOLF201580}
Richard~S. Woolf, Anthony~L. Hutcheson, Chul Gwon, Bernard~F. Phlips, and
  Eric~A. Wulf.
\newblock Comparing the response of psd-capable plastic scintillator to
  standard liquid scintillator.
\newblock \emph{Nuclear Instruments and Methods in Physics Research Section A:
  Accelerators, Spectrometers, Detectors and Associated Equipment},
  784:\penalty0 80 -- 87, 2015.
\newblock ISSN 0168-9002.
\newblock \doi{https://doi.org/10.1016/j.nima.2014.10.067}.
\newblock URL
  \url{http://www.sciencedirect.com/science/article/pii/S0168900214012352}.
\newblock Symposium on Radiation Measurements and Applications 2014 (SORMA XV).

\bibitem[Zaitseva et~al.(2018)Zaitseva, Glenn, Mabe, Carman, Hurlbut, Inman,
  and Payne]{ZAITSEVA201897}
N.P. Zaitseva, A.M. Glenn, A.N. Mabe, M.L. Carman, C.R. Hurlbut, J.W. Inman,
  and S.A. Payne.
\newblock Recent developments in plastic scintillators with pulse shape
  discrimination.
\newblock \emph{Nuclear Instruments and Methods in Physics Research Section A:
  Accelerators, Spectrometers, Detectors and Associated Equipment},
  889:\penalty0 97 -- 104, 2018.
\newblock \doi{https://doi.org/10.1016/j.nima.2018.01.093}.

\bibitem[Zaitseva et~al.(2013)Zaitseva, Glenn, Martinez, Carman, Paweaczak,
  Faust, and Payne]{ZAITSEVA2013747}
Natalia Zaitseva, Andrew Glenn, H.~Paul Martinez, Leslie Carman, Iwona
  Paweaczak, Michelle Faust, and Stephen Payne.
\newblock Pulse shape discrimination with lithium-containing organic
  scintillators.
\newblock \emph{Nuclear Instruments and Methods in Physics Research Section A:
  Accelerators, Spectrometers, Detectors and Associated Equipment},
  729:\penalty0 747 -- 754, 2013.
\newblock \doi{https://doi.org/10.1016/j.nima.2013.08.048}.

\bibitem[Mabe et~al.(2016)Mabe, Glenn, Carman, Zaitseva, and Payne]{MABE201680}
Andrew~N. Mabe, Andrew~M. Glenn, M.~Leslie Carman, Natalia~P. Zaitseva, and
  Stephen~A. Payne.
\newblock Transparent plastic scintillators for neutron detection based on
  lithium salicylate.
\newblock \emph{Nuclear Instruments and Methods in Physics Research Section A:
  Accelerators, Spectrometers, Detectors and Associated Equipment},
  806:\penalty0 80 -- 86, 2016.
\newblock \doi{https://doi.org/10.1016/j.nima.2015.09.111}.

\bibitem[{et al.}(2014)]{asdc}
J.~R.~Alonso {et al.}
\newblock \emph{arXiv: 1409.5864 [hep-ex, nucl-ex]}, 2014.

\bibitem[{\it et al.}~(SNO+~Collaboration)(2016)]{snopl}
S.~Andringa {\it et al.}~(SNO+~Collaboration).
\newblock \emph{Adv. High Energy Phys. {\bf 2016}, 6194250}, 2016.

\bibitem[Collaboration(2013)]{klz}
KamLAND-Zen Collaboration.
\newblock \emph{Phys.\ Rev.\ Lett.\ {\bf 110}:062502}, 2013.

\bibitem[bor(G. Bellini {\emph et al.}, Phys. Rev. Lett. {\bf 107} 141302
  (2011); G. Bellini {\it et al.}, Phys. Rev. Lett. {\bf 108} 051302 (2012); G.
  Bellini {\it et al.}, Nature {\bf 512} 383-386 (2014))]{bor}
G. Bellini {\emph et al.}, Phys. Rev. Lett. {\bf 107} 141302 (2011); G. Bellini
  {\it et al.}, Phys. Rev. Lett. {\bf 108} 051302 (2012); G. Bellini {\it et
  al.}, Nature {\bf 512} 383-386 (2014).

\bibitem[ric(R. Bonventre and G. D. Orebi Gann, Eur. Phys. J. C 78:435
  (2018).)]{richiegdog}
R. Bonventre and G. D. Orebi Gann, Eur. Phys. J. C 78:435 (2018).

\bibitem[KLg(A. Gando \emph{et al.}, Phys. Rev. \textbf{D88} 033001
  (2013))]{KLgeo}
A. Gando \emph{et al.}, Phys. Rev. \textbf{D88} 033001 (2013).

\bibitem[Bge(G. Bellini \emph{et al.}, Phys. Lett. \textbf{B722} 295-300
  (2013))]{Bgeo}
G. Bellini \emph{et al.}, Phys. Lett. \textbf{B722} 295-300 (2013).

\bibitem[{et al.}(2015)]{wm}
M.~Askins {et al.}
\newblock \emph{arXiv: 1502.01132}, 2015.

\bibitem[Yeh(2015)]{wbls}
Minfang Yeh.
\newblock Water-based liquid scintillator detector.
\newblock Water Detector Workshop. Stony Brook University, 2015.

\bibitem[che(J. Caravaca et al., Phys. Rev. C \textbf{95}, 055801
  (2017))]{chess}
J. Caravaca et al., Phys. Rev. C \textbf{95}, 055801 (2017).

\bibitem[che(J. Caravaca et al., Eur. Phys. J. C \textbf{77}: 811
  (2017))]{chess2}
J. Caravaca et al., Eur. Phys. J. C \textbf{77}: 811 (2017).

\bibitem[mcp(B.~Adams, A.~Elagin, H.~Frisch, R.~Obaid, E.~Oberla, A.~Vostrikov,
  R.~Wagner, and M.~Wetstein, Nucl. Instum. Methods A732, 392 (2013).)]{mcp}
B.~Adams, A.~Elagin, H.~Frisch, R.~Obaid, E.~Oberla, A.~Vostrikov, R.~Wagner,
  and M.~Wetstein, Nucl. Instum. Methods A732, 392 (2013).

\bibitem[lap(B. Adams {\it et al.}, Rev. Sci. Instrum. 84 061301
  (2013).)]{lappd}
B. Adams {\it et al.}, Rev. Sci. Instrum. 84 061301 (2013).

\bibitem[lap(B.~W.~Adams {\it et al.} (LAPPD Collaboration), arXiv:1603.01843
  (2016), (Submitted to: JINST).)]{lappd2}
B.~W.~Adams {\it et al.} (LAPPD Collaboration), arXiv:1603.01843 (2016),
  (Submitted to: JINST).

\bibitem[lap(O. H. W. Siegmund, J. B. McPhate, J. V. Vallerga, A. S. Tremsin,
  H. E. Frisch, J. W. Elam, A. U. Mane, and R. G. Wagner, J. Instrum. 9, C04002
  (2014).)]{lappd3}
O. H. W. Siegmund, J. B. McPhate, J. V. Vallerga, A. S. Tremsin, H. E. Frisch,
  J. W. Elam, A. U. Mane, and R. G. Wagner, J. Instrum. 9, C04002 (2014).

\bibitem[ela(C. Aberle, A. Elagin, H. J. Frisch,M.Wetstein, and L.Winslow, J.
  Instrum. 9, P06012 (2014).)]{elagin}
C. Aberle, A. Elagin, H. J. Frisch,M.Wetstein, and L.Winslow, J. Instrum. 9,
  P06012 (2014).

\bibitem[ela(A. Elagin {\it et al.}, Nucl. Instrum. Methods {\bf A849}, 102
  (2017).)]{elagin2}
A. Elagin {\it et al.}, Nucl. Instrum. Methods {\bf A849}, 102 (2017).

\bibitem[nud(D. Gooding {\it et al.}, arXiv:1807.06634 (2018).)]{nudot}
D. Gooding {\it et al.}, arXiv:1807.06634 (2018).

\bibitem[que(H. M.O'Keeffe, E. O'Sullivan, and M. C. Chen, Nucl. Instum.
  Methods {A640}, 119 (2011).)]{queens}
H. M.O'Keeffe, E. O'Sullivan, and M. C. Chen, Nucl. Instum. Methods {A640}, 119
  (2011).

\bibitem[Oguri et~al.(2014)Oguri, Kuroda, Kato, Nakata, Inoue, Ito, and
  Minowa]{OGURI201433}
S.~Oguri, Y.~Kuroda, Y.~Kato, R.~Nakata, Y.~Inoue, C.~Ito, and M.~Minowa.
\newblock Reactor antineutrino monitoring with a plastic scintillator array as
  a new safeguards method.
\newblock \emph{Nuclear Instruments and Methods in Physics Research Section A:
  Accelerators, Spectrometers, Detectors and Associated Equipment},
  757:\penalty0 33 -- 39, 2014.
\newblock ISSN 0168-9002.
\newblock \doi{https://doi.org/10.1016/j.nima.2014.04.065}.
\newblock URL
  \url{http://www.sciencedirect.com/science/article/pii/S0168900214004781}.

\bibitem[Mulmule et~al.(2018)Mulmule, Behera, Netrakanti, Mishra, Kashyap, Jha,
  Pant, Nayak, and Saxena]{MULMULE2018104}
D.~Mulmule, S.P. Behera, P.K. Netrakanti, D.K. Mishra, V.K.S. Kashyap, V.~Jha,
  L.M. Pant, B.K. Nayak, and A.~Saxena.
\newblock A plastic scintillator array for reactor based anti-neutrino studies.
\newblock \emph{Nuclear Instruments and Methods in Physics Research Section A:
  Accelerators, Spectrometers, Detectors and Associated Equipment},
  911:\penalty0 104 -- 114, 2018.
\newblock ISSN 0168-9002.
\newblock \doi{https://doi.org/10.1016/j.nima.2018.10.026}.
\newblock URL
  \url{http://www.sciencedirect.com/science/article/pii/S0168900218313408}.

\bibitem[Zhan et~al.(2008)Zhan, Wang, Cao, and Wen]{PhysRevD.78.111103}
Liang Zhan, Yifang Wang, Jun Cao, and Liangjian Wen.
\newblock Determination of the neutrino mass hierarchy at an intermediate
  baseline.
\newblock \emph{Phys. Rev. D}, 78:\penalty0 111103, Dec 2008.
\newblock \doi{10.1103/PhysRevD.78.111103}.
\newblock URL \url{https://link.aps.org/doi/10.1103/PhysRevD.78.111103}.

\bibitem[Vogel and Beacom(1999)]{PhysRevD.60.053003}
P.~Vogel and J.~F. Beacom.
\newblock Angular distribution of neutron inverse beta decay,
  ${\overline{\ensuremath{\nu}}}_{e}+p\ensuremath{\rightarrow}{e}^{+}+n$.
\newblock \emph{Phys. Rev. D}, 60:\penalty0 053003, Jul 1999.
\newblock \doi{10.1103/PhysRevD.60.053003}.
\newblock URL \url{https://link.aps.org/doi/10.1103/PhysRevD.60.053003}.

\bibitem[Gomez(2015)]{Gomez}
H{\'e}ctor Gomez.
\newblock Neutrino directionality with double chooz: Latest results.
\newblock Neutrino Geoscience, 2015.

\bibitem[Hellfeld et~al.(2017)Hellfeld, Bernstein, Dazeley, and
  Marianno]{HELLFELD2017130}
D.~Hellfeld, A.~Bernstein, S.~Dazeley, and C.~Marianno.
\newblock Reconstructing the direction of reactor antineutrinos via electron
  scattering in gd-doped water cherenkov detectors.
\newblock \emph{Nuclear Instruments and Methods in Physics Research Section A:
  Accelerators, Spectrometers, Detectors and Associated Equipment},
  841:\penalty0 130 -- 138, 2017.
\newblock ISSN 0168-9002.
\newblock \doi{https://doi.org/10.1016/j.nima.2016.10.027}.
\newblock URL
  \url{http://www.sciencedirect.com/science/article/pii/S0168900216310555}.

\bibitem[Askins(2018)]{askins}
Morgan Askins.
\newblock Water cherenkov monitor for antineutrinos (watchman).
\newblock In \emph{Applied Antineutrino Physics 2018}, 2018.

\bibitem[Bemporad et~al.(2002)Bemporad, Gratta, and Vogel]{RevModPhys.74.297}
Carlo Bemporad, Giorgio Gratta, and Petr Vogel.
\newblock Reactor-based neutrino oscillation experiments.
\newblock \emph{Rev. Mod. Phys.}, 74:\penalty0 297--328, Mar 2002.
\newblock \doi{10.1103/RevModPhys.74.297}.
\newblock URL \url{https://link.aps.org/doi/10.1103/RevModPhys.74.297}.

\bibitem[Safdi and Suerfu(2015)]{PhysRevLett.114.071802}
Benjamin~R. Safdi and Burkhant Suerfu.
\newblock Directional antineutrino detection.
\newblock \emph{Phys. Rev. Lett.}, 114:\penalty0 071802, Feb 2015.
\newblock \doi{10.1103/PhysRevLett.114.071802}.
\newblock URL \url{https://link.aps.org/doi/10.1103/PhysRevLett.114.071802}.

\bibitem[Dawson and Kryn(2014)]{1748-0221-9-07-P07002}
J~V Dawson and D~Kryn.
\newblock Organic liquid tpcs for neutrino physics.
\newblock \emph{Journal of Instrumentation}, 9\penalty0 (07):\penalty0 P07002,
  2014.
\newblock URL \url{http://stacks.iop.org/1748-0221/9/i=07/a=P07002}.

\end{thebibliography}
